\documentclass[11pt]{article}

\usepackage[margin=1in]{geometry}
\usepackage{graphicx}
\newcommand{\train}{\hat{\mbox{train}}}
\usepackage{amsmath,amssymb,amsfonts,amsthm}
\usepackage[numbers,sort&compress]{natbib} 
\usepackage{hyperref}
\hypersetup{backref,colorlinks=true,citecolor=blue,linkcolor=blue,urlcolor=blue}

\usepackage{mathtools}
\mathtoolsset{showonlyrefs}
\usepackage{booktabs}

\usepackage{enumitem}

\usepackage[lined,ruled,linesnumbered]{algorithm2e}
\SetKw{KwSet}{Set}
\SetKwInOut{Input}{input}\SetKwInOut{Output}{output}

\newcommand{\SQRT}{$\sqrt{\mbox{lasso}}$}

\newcommand{\AICcv}{R-CV-2}
\newcommand{\AICrcv}{R-RCV-2}
\newcommand{\AICrmle}{R-RMLE-2}
\newcommand{\BICcv}{R-CV-logn}
\newcommand{\indicator}{\mathbf{1}}

\newcommand{\X}{\mathbb{X}}
\newcommand{\E}{\mathbb{E}}
\newcommand{\lassoL}{\hat{\beta}(\lambda)}

\newcommand{\lassoLcv}{\hat{\beta}^{(v)}(\lambda)}

\DeclareMathOperator{\df}{df}
\newcommand{\dfHat}{\hat{\df}}
\newcommand{\lamCV}{\hat\lambda_{CV}}

\renewcommand{\S}{\mathcal{S}}

\newcommand{\risk}{\textrm{R}}

\newcommand{\R}{\mathbb{R}}

\newcommand{\norm}[1]{\left\lVert #1 \right\rVert}

\DeclareMathOperator*{\argmin}{argmin}

\renewcommand{\hat}{\widehat}
\newcommand{\bstar}{\beta_{*}}
\newcommand{\sstar}{s_{*}}
\newcommand{\Sstar}{\S_{*}}
\newcommand{\EPconst}{M}

\usepackage{aliascnt}
\theoremstyle{plain}
\newtheorem{theorem}{Theorem}
\newaliascnt{lemma}{theorem}
\newtheorem{lemma}[lemma]{Lemma}
\aliascntresetthe{lemma}

\newaliascnt{cor}{theorem}

\aliascntresetthe{cor}

\newaliascnt{def}{theorem}

\aliascntresetthe{def}

\newtheorem{example}{Example}

\newcommand{\email}[1]{\href{mailto:#1}{#1}}

\begin{document}

\title{A study on tuning parameter selection for the high-dimensional lasso}
\author{Darren Homrighausen\\Department of Statistics\\Colorado State University\\
\email{darrenho@stat.colostate.edu} \and Daniel J. McDonald\\Department of Statistics\\Indiana
  University, Bloomington\\\email{dajmcdon@indiana.edu}}

\maketitle

\begin{abstract}
  High-dimensional predictive models, those with more measurements
  than observations, require regularization to be well defined,
  perform well empirically, and possess theoretical guarantees. The
  amount of regularization, often determined by tuning parameters, is
  integral to achieving good performance. One can choose the tuning
  parameter in a variety of ways, such as through resampling methods
  or generalized information criteria. However, the theory supporting
  many regularized procedures relies on an estimate for the variance
  parameter, which is complicated in high dimensions. We develop a
  suite of information criteria for choosing the tuning parameter in
  lasso regression by leveraging the literature on high-dimensional
  variance estimation. We derive intuition showing that existing
  information-theoretic approaches work poorly in this setting. We
  compare our risk estimators to existing methods with an extensive
  simulation and derive some theoretical justification. We find that
  our new estimators perform well across a wide range of simulation
  conditions and evaluation criteria.


\end{abstract}

\section{Introduction}

Suppose we are  given a data set, $Z_1,\ldots,Z_n$,
of paired observations including a covariate $X_i \in \mathbb{R}^p$
and its associated response $Y_i \in \mathbb{R}$ such that $Z_i^{\top}
= (X_i^{\top},Y_i) $.  Concatenating the covariates row-wise, we obtain the design matrix $\X = [X_1,\ldots,X_n]^{\top} \in
\mathbb{R}^{n \times p}$. We assume that the relationship between
the covariate and response is of the form
\begin{equation}
\label{eq:modelStatement}
  Y = \X \bstar + \epsilon,
\end{equation}
where $\epsilon \sim (0,\sigma^2I)$, meaning the entries of $\epsilon$
are mean zero with uncorrelated components each having variance
$\sigma^2$.  
%While many of these specifics can be obviated, this model
%connects more readily with the literature.

When $p > n$, estimation of the linear model requires some structural
assumptions on $\bstar$ for learning algorithms to possess
theoretical guarantees.  A common approach in this scenario is to
assume $\norm{\bstar}_q$ is small for some $q\geq 0$ and try to estimate $\bstar$ via penalized
least squares. We will focus mainly on the lasso
\begin{equation}
  \label{eq:optim}
  \hat{\beta}(\lambda) = \argmin_{\beta \in \R^p} \frac{1}{n}\norm{Y-\X\beta}_2^2 +\lambda\norm{\beta}_1,
\end{equation}
where $\lambda\geq 0$ is a tuning
parameter and $\norm{\cdot}_2$ and $\norm{\cdot}_1$ are the $\ell_2$-
(Euclidean) and $\ell_1$-norms respectively.  
Similar $M$-estimators with different penalties include, among others, ridge
regression, % \citep{hoerl1970ridge}, 
the group lasso
\citep{YuanLin2005}, and the smoothly clipped absolute deviation penalty
\citep[SCAD,][]{FanLi2001}. 
Though the focus of this paper is on lasso, 
we will occasionally
also reference ridge regression, 
\begin{align*}
  \hat{\beta}_{ridge}(\lambda) 
  &=\argmin_{\beta \in \R^p}
    \frac{1}{n}\norm{Y-\X\beta}_2^2 +\lambda\norm{\beta}^2_2\\ 
  &= (\X^\top\X+\lambda I_p)^{-1}\X^\top Y, 
\end{align*}
because it
has a closed form which can provide intuition.
% , though similar studies with
% different loss and penalty function combinations

For lasso, by convexity there is
always at least one solution to equation \eqref{eq:optim},
although if $\textrm{rank}(\X) < p$, there may be multiple minimizers \citep[see][for details]{Tibshirani2013}.  In this case,
we refer to `the' solution as the outcome of the particular minimization technique used
\citep[e.g.\ LARS,][]{EfronHastie2004}. For ridge regression, a unique solution always exists for
$\lambda>0$, although, for $\lambda$ small enough, numerical issues may
intercede. We will also consider
some modifications to equation \eqref{eq:optim} which attempt to eliminate the influence of tuning
parameters (see
\autoref{sec:modif-lasso-crit} for a more detailed description).

The theoretical optimality properties that exist in the 
literature for penalized regression
rely on appropriate tuning parameter selection. 
Under restrictions on the design matrix $\X$, the distribution of $\epsilon$,
and the sparsity pattern of $\bstar$, 
\cite{BuneaTsybakov2007} shows that, as long as the number of nonzero 
entries in  $\bstar$ does not increase too quickly,
the probability of making prediction errors with magnitude larger than $\sigma^2\log(p)/n$
goes to zero if $\lambda_n = a\sigma\sqrt{\log(p)/n}$ for some constant $a$.
Likewise, deviations in the distance between $\hat\beta(\lambda)$ and $\bstar$
of order larger than $\sigma\sqrt{\log(p)/n}$ have small probability.
While theoretical results of this type provide comfort that a data
analyst's procedure will eventually perform well given sufficient
data, they 
require the optimal $\lambda_n$ which depends on unknown quantities
such as $\sigma^2$, the noise distribution, and other constants.

In practice, many methods for empirically 
choosing $\lambda$ given a fixed dataset have been proposed. These methods can be lumped into
three broad categories: (1) generalized information criteria like AIC or BIC,
(2) resampling procedures such as cross-validation or the bootstrap, and
(3) reformulations of the lasso optimization problem (e.g. scaled
sparse regression or \SQRT).\footnote{There is some overlap between
  these categories. For example, generalized cross-validation can be thought of as
  either a resampling procedure or an information
  criterion.} 
In order to evaluate these approaches, we must be explicit as to
the properties we desire in our final estimator:  low prediction risk,
parameter estimation consistency, correct model selection, or simply
accurate estimates of the prediction risk.

The aim of this paper is to evaluate tuning parameter selection
procedures for high-dimensional lasso regression. To this end we (1) introduce
a suite of novel risk estimation methods that are simple to compute and perform well
empirically,
(2) contrast these new risk estimation methods with existing,
superficially similar GIC-based methods, 
and, lastly, (3) provide a comprehensive simulation
study over a broad range of data generating scenarios and estimation goals which
compares our procedure to existing methods. This investigation both justifies our
proposal and reveals deficiencies
for current high-dimensional approaches while also suggesting
interesting research directions, particularly the relationship between risk estimation and
high-dimensional variance estimation.

In \autoref{sec:tuning-param-select}, we discuss two broad categories of procedures for tuning parameter selection: 
cross-validation and generalized information
criteria.  We demonstrate that there is a significant difference between
using generalized information criteria in the
low-dimensional ($p<n$) versus the high-dimentional ($p > n$) regimes. 
\autoref{sec:sure} motivates and
introduces our proposed modification to Stein's unbiased risk
estimation using plug-in estimators for $\sigma^2$ and the
degrees-of-freedom for the lasso. It also discusses the different
versions of modern high-dimensional variance estimators which we
consider. In \autoref{sec:simulationConditions}, we present a
comprehensive simulation comparing our proposal with some existing
alternatives. We focus on the performance of the lasso, but we include
scaled-sparse regression, $\sqrt{\mbox{lasso}}$, and SCAD for comparison. 
We also demonstrate the methods on a genetics
dataset. Finally, \autoref{sec:theoretical-results} gives a
theoretical result, showing that under standard
assumptions our proposed
risk estimator converges to the true prediction risk at the parametric
rate. \autoref{sec:discussion} summarizes our recommendations
and suggest possible avenues for further research.

\textbf{Notation:} For any vector $\beta\in\R^p$, we denote $\S=\S(\beta) =
\{ j : \beta_j \neq 0\}$ and $\X_{\S(\beta)}$ to be the columns of the design matrix selected by
$\beta$. We write $\Sstar=\S(\bstar)$ and $\sstar=|\Sstar|$.  Also, for any square matrix $H$, define
the trace of $H$, $tr(H)$, to be the sum
of the diagonal entries.  Define the squared $\ell_2$-prediction risk of a coefficient vector 
$\beta$ to be
\begin{equation}
\label{eq:risk}
\risk_\beta = n^{-1}\E\norm{\X\beta - \X\bstar}_2^2,
\end{equation}
where the expectation is over the data $Z_1,\ldots,Z_n$.
Likewise, we define the training error to be
\begin{equation}
\train_\beta = n^{-1}\norm{\X\beta - Y}_2^2.
\end{equation} 
Throughout this paper, if a procedure $\beta$ is indexed by a tuning parameter $\lambda$,
we will write, for example, $\train_{\beta(\lambda)} \equiv
\train_\lambda$.

\section{Existing tuning parameter selection methods}
\label{sec:tuning-param-select}

In this section, we discuss existing procedures for tuning parameter
selection for lasso regression. In the context of regularized regression, risk estimation and tuning parameter selection
are often used interchangeably because any risk estimator can be used
to select tuning parameter(s). However, it is important for our
exposition to belabor the
distinction
for two reasons: (1) not all
tuning parameter selection procedures produce an estimate of the prediction
risk, and (2) we may wish to evaluate the quality of the selection
procedure by comparing model selection accuracy or parameter
consistency, metrics which don't require a risk estimate anyway. 
That is, we may ask if \SQRT{}, a tuning-free method which does not
estimate the prediction risk, produces better
estimates of $\bstar$ than the lasso with $\lambda$ selected by
cross-validation. As a preview of our results in
\autoref{sec:simulationResults}, the answer to this question is generally
no, but if we use GCV to select $\lambda$ instead, then this
conclusion is reversed. 
This section introduces existing tuning parameter selection
procedures, some of which estimate the prediction
risk---cross-validation, Stein's unbiased risk estimation (SURE), and
information criteria---while others do not.

\subsection{Cross-validation}
\label{sec:cross-validation}

Frequently \citep[for
example][]{HastieTibshirani2009,ZouHastie2007,HastieTibshirani2015},
the recommended technique for selecting $\lambda$ is through \emph{$K$-fold
cross-validation} (CV).  Letting $V_n = \{ v_1 , \ldots, v_{K} \}$ be
a partition of $\{1,\ldots,n\}$
\begin{equation*}
  \label{eq:cv-risk-main}
  CV(\lambda;V_n)
  = \frac{1}{K} \sum_{v \in V_n}
  \frac{1}{|v|} \sum_{r \in v} \left(Y_r - X_r^{\top}\lassoLcv
  \right)^2,
\end{equation*}
where $\lassoLcv$ is the lasso estimator in equation
\eqref{eq:optim} with the observations in the validation set
$v$ removed, and $|v|$ indicates the cardinality of the set $v$. We
define $\lamCV = \argmin_\lambda CV(\lambda;V_n)$.  Common choices for $K$ are $K=10$
or $K=n$. Cross-validation was shown to perform correct model
selection and lead to good prediction risk~\citep{HomrighausenMcDonald2016}.

Several adaptations of cross-validation for use with the lasso have been proposed.
One such method is Modified Cross-Validation
\citep[MCV,][]{YuFeng2014} which seeks to correct for a bias in CV induced
by the lasso penalty.  %We include this method in our investigation.
Generalized
cross-validation~\citep[GCV]{GolubHeath1979} is a much older
modification of cross-validation with some computational
benefits. It can also be viewed as an information criterion,
so we discuss it further in the next section.

% \attn{Implementation, move below}
% For this paper, we use {\tt glmnet} to fit CV and MCV and use {\tt lars} to find the sequence of models
% for CCV and to compute the lasso path for GCV.  Based on our simulation results, GCV, and CCV all dramatically underperform CV and MCV (See \autoref{fig:cvExample}
% for a plot of typical results, though we defer a precise definition of
% prediction risk and a discussion on the particulars of the simulation
% conditions to \autoref{sec:simulationConditions}). 
% Therefore, for simplicity we will not include either GCV or CCV in
% subsequent comparisons.

%The particular software used for finding a lasso solution is quite important. As 
%equation  \eqref{eq:optim} is convex (but no strictly so) when the penalty term is
%the $\ell_1$ norm, there exist many methods for solving the optimization problem (e.g.
%interior point methods).
%Additionally, there are more specialized methods that leverage the particular
%structure.  In particular, two widely used implementations are {\tt glmnet}, which uses coordinate
%descent, and
%{\tt lars}, which leverages the piece-wise linear nature of the lasso solution path.  The implementation 
%{\tt glmnet} is much faster and approximately finds a lasso solution along a grid of $\lambda$ values.  
%On the other hand, {\tt lars} 

\subsection{Generalized information criteria}
\label{sec:gic}

%The risk estimator $\hat R_\beta(\hat\sigma^2)$ in equation \eqref{eq:SURE}
% shares many
%similarities with generalized information criteria under Gaussian noise. 
%However, as we demonstrate below, a key component to
%defining GIC in high dimensions is knowledge of the scale parameter. 
%We discuss GIC in some detail here due to its frequent use in tuning parameter selection.
% Defining and developing GIC for high dimensional regression is an active
% field of research.  We focus in particular on GIC generated by assuming 
% a Gaussian likelihood.
% two commonly used GIC are formulated
%depending on whether the variance parameter is known. These have
%various names in the literature, and are an active area of
%research.
%Examples of GIC include 
%the Akaike information criterion (AIC) \citep{akaike1974new},
%generalized cross-validation (GCV),
%%\citep[GCV,][]{CravenWahba1978,Stone1977}, 
%and the Bayesian
%information criterion (BIC) \citep{schwarz1978estimating}, %fall under this area, %\citep[BIC,][]{Schwarz1978}, 
%as well as recent extensions
%\citep{wang2009shrinkage,chen2012extended,FanTang2013}.
%all fit into this category.
%Define $\RSS{\hat\beta} = \norm{Y - \X\hat\beta}_2^2$ to be the training error for
%an arbitrary estimator $\hat\beta$. If the estimator depends on a tuning parameter,
%we write $\RSS{\lambda} = \RSS{\lassoL}$ for simplicity. 

A common alternative to cross-validation
is to minimize a generalized information criterion (GIC). Define the \emph{degrees of freedom}
\citep{efron1986biased} 
of the prediction $\hat Y=\X\beta \in \R^n$ to be
\begin{equation*}
  \df = \frac{1}{\sigma^2} \sum_{i=1}^n \textrm{Cov}(\hat Y_i,Y_i),
  \label{eq:df}
\end{equation*}
where $\textrm{Cov}(\hat Y_i,Y_i) = \E\left[(\hat Y_i - \E\hat Y_i)(Y_i - \E Y_i)\right]$.

Referring to equation \eqref{eq:modelStatement}, if $\sigma^2$ is
unknown and $\epsilon$ is Gaussian, then a GIC takes the form
\begin{equation}
  \label{eq:GICsigUnknown}
  \textrm{info}(C_n,g) 
%  &= \log \mathcal{L}(\hat\beta,\
%  \hat\sigma^2) + C_n \; g(\df(\lambda))\\ 
  =\log\left(\train_\beta\right) +
  C_n \; g(\df),
\end{equation}
where $C_n$ depends only on $n$, and $g: [0,\infty)
\rightarrow \mathbb{R}$ is a fixed function. 
This GIC form is frequently suggested in  
the literature for choosing $\lambda$ in the lasso problem \citep[for example][]{BuhlmannGeer2011,
  WangLi2007,Tibshirani1996, FanLi2001,FlynnHurvich2013}, with $\df$
replaced by an estimator $\dfHat$.  We defer discussion of how to form $\dfHat$ for the lasso to 
\autoref{sec:sure}.  The choices
$C_n = 2/n$ or $C_n = \log(n)/n$ with $g(x) = x$ are commonly referred to as AIC and
BIC, respectively.  Additionally, generalized
cross-validation
is defined as
\begin{equation}
\textrm{GCV} = \frac{\train_\beta}{(1-\df/n)^2}.
\label{eq:gcv}
\end{equation}
Written on the log scale, GCV takes the form of equation \eqref{eq:GICsigUnknown} with $g(x) = \log(1-x/n)$ and
$C_n = -2/n$.

%\attn{This paragraph is not very clear to me. I think it's trying
%  to be general without misconstruing the literature (``considered known''
%  vs. ``known''; $\sigma^2$ vs. $\hat{\sigma}^2$) so it fails to make
%  the point well. I'm not sure what to do at the moment.}

%%%% Not Sure where this should go %%%%
%\begin{table}[t]
%  \centering
%  \begin{tabular}{llll}
%    \hline\hline
%    & $C_n$ & $g(x)$ & Common name \\
%    \hline
%    $\textrm{info}(C_n,g)$\\
%    \hline
%    \citealp{buhlmann2011statistics} & $\log (n)/n$ & $x$ & BIC \\
%    \citealp{wang2007tuning} & $\log (n)/n$ & $x$ & BIC \\
%    \citealp{Tibshirani1996}    & $-1$ & $\log(n - x)$ & GCV \\
%    \citealp{FanLi2001}    & $-1$ & $\log(n - x)$ & GCV \\
%    \citealp{flynn2013efficiency} & various & various & various\\
%    \hline
%    $\hat{R}(\sigma^2,g)$\\
%    \hline
%    \citealp{tibshirani2012degrees} & $2$ & $x$ & AIC (or SURE) \\
%    \citealp{zhang2010regularization} & $\log(n)/n$ & $x$ & BIC\\
%    \citealp{FanTang2013} &$\log \log (n) \log (p)$ & $x$ & $\textrm{GIC}_{a_n}$\\
%    \hline\hline
%  \end{tabular}
%  \caption{Common specifications of the $C_n$ and $g$ parameters in the two forms 
%    of GIC defined in equations \eqref{eq:GICsigUnknown} and
%    \eqref{eq:ourGeneralProp} } 
%  \label{tab:commonnames}
%\end{table}

% The above cited literature carefully investigates 
% properties of GIC in different settings.

While GIC-based tuning parameter selection has enjoyed good theoretical and empirical success
in a broad range of applications,
classical asymptotic arguments underlying GIC apply only for $p$ fixed and
rely on maximum likelihood estimates (or Bayesian posteriors) for all
parameters including $\sigma^2$. 
More recent investigations have explored theoretical regimes in which $p$ is allowed to increase, but the constraint $p<n$ is still
enforced.  \cite{wang2009shrinkage} shows that
the correct model is selected asymptotically even if $p\rightarrow
\infty$ as long as $p/n\rightarrow 0$. Additionally,
\cite{FlynnHurvich2013} investigates a variety 
GIC-based methods under increasing $p$, but again restricted to the case $p<n$. 

Theoretical support for GIC breaks down in the high-dimensional setting.
The most serious issue is that $\textrm{info}(C_n,g)$ from equation \eqref{eq:GICsigUnknown}
is unusable without modification if $n<p$ because it is possible to achieve
$\train_\beta = 0$ and hence $\log(\train_\beta) =
-\infty$. Therefore, as $\lambda \rightarrow 0$,
$\textrm{info}(C_n,g)$ will approach $-\infty$ unless $g(\df)
\rightarrow \infty$ faster, and $\lambda=0$ will always be selected. Simply forcing $\lambda > \epsilon$ for
some small positive $\epsilon$ often fails to remedy this situation in the sense that
$\lambda=\epsilon$ is selected.
Nonetheless, $\textrm{info}(C_n,g)$ is  still
commonly for use with the lasso, even 
in high-dimensional situations \citep[e.g.][]{BuhlmannGeer2011}.  

To provide some intuition for this last claim, we provide the following
trivial example which explores the behavior of AIC,  BIC, and GCV for
selecting the tuning parameter in a simple situation. We illustrate
this problem with $ \hat{\beta}_{ridge}(\lambda)$, as these GIC then
have a closed form.

\begin{example}
\label{sec:firstExample}
% Before laying out the specifics of our proposed methodology, we think
% it is useful to demonstrate the pathological behavior of information
% criteria with a very simple example. 
Consider the following regression
data set:
\begin{align*}
  Y &=\frac{\sigma}{\sqrt{2}} \begin{bmatrix} 1 \\ 
    -1 \end{bmatrix}, &\textrm{and} && \X&=
                       \frac{1}{\sqrt{2}}\begin{bmatrix}1 & 1 & \sqrt{2}\\ 1& -1 &
    0 \end{bmatrix}.
\end{align*}
In this no noise case, $Y$ is a scalar multiple of 
a column of $X$. 
%This setup could easily be obtained with two
%independent draws from
%the population model $y_i = x_{i2}+\epsilon_i$ where $\epsilon_i \sim
%(0,\sigma^2)$ and subsequently normalizing $\X$ and centering $Y$.
% Then simply normalize and center the columns of $X$ and
%center $Y$. Note that treating $\X$
%and $Y$ in this manner by centering and normalizing is standard in
%software implementations in order to avoid penalizing an intercept and
%so that the equal scale of the penalty is appropriate to all the
%coefficients.
%  Repeating this experiment numerous times will lead to
% data sets which look like this one on average. 

For ridge regression, one can show that % (see \autoref{sec:algebraapp} for details)
\begin{align*}
  \df(\lambda) 
  &= %tr\left(D(D^2+\lambda I)^{-1} D\right) = 
    \frac{3\lambda+4}{(2+\lambda)(1+\lambda)},
\end{align*}
\begin{align*}
\train_\lambda 
  &= %\frac{1}{2} Y^\top (I-H)^2 Y= 
    \frac{\sigma^2\lambda^2}
    {4}\left(\frac{1}{(2+\lambda)^2}+\frac{1}{(1+\lambda)^2}\right),
\end{align*}
and so,
\begin{align*}
  \textrm{info}(C_n,g) 
  &= \log\left(\frac{\sigma^2\lambda^2}
                         {4}\left(\frac{1}{(2+\lambda)^2}+\frac{1}{(1+\lambda)^2}\right)\right)\\
  &\quad +
  C_n \; g\left( \frac{3\lambda+4}{(2+\lambda)(1+\lambda)}\right).
\end{align*}
%\begin{align*}
%  AIC &= \log(\sigma^2/2) + 2\log \lambda -2\log(\lambda+3) +
%        \frac{6}{\lambda+3}\\
%  BIC &= \log(\sigma^2/2) + 2\log \lambda -2\log(\lambda+3)
%        +\frac{3\log 2}{\lambda+3}\\
%  GCV &= \log(\sigma^2/2) + 2\log \lambda -2\log(\lambda+3)
%        -\log \frac{2(\lambda+3)+3}{\lambda+3}.
%\end{align*}
%If we send $\lambda \rightarrow 0$, then all three go to
%$-\infty$. If $\lambda=O(n^\alpha),\ \alpha>0$, then the information
%criteria are $O(\log n)$.
For $0<\lambda<1$, 
$\frac{13\sigma^2\lambda^2}{144}\leq \train_\lambda \leq
\frac{5\sigma^2\lambda^2}{16}$, so $\log(\train_\lambda) \rightarrow-\infty$ like
$\log(\lambda)$ as $\lambda\rightarrow 0$.  Hence, minimizing
$\textrm{info}(C_n,g)$ will choose $\lambda=0$ unless
the second term increases at least as fast as $-\log(\lambda)$, that
is we require constants $c$ and $C$ such that
$g\left(\frac{3\lambda+4}{(2+\lambda)(1+\lambda)} \right)\geq C \log(1/\lambda)$ for all
$\lambda<c$. We see immediately 
that AIC and BIC, which both have $g(x) \equiv x$, will always
select $\lambda=0$. This corresponds to reporting the unregularized, least
squares solution. 

For GCV, the issue is a bit  
more subtle.  In this example, as $\textrm{rank}(\X) = n=2$, $-\log(1
- \df/n) \rightarrow \infty$ 
and hence the rate that $-\log(1-\df/n)$ goes to $\infty$,  
along with magnitude of the constants  involved, determines which 
trivial solution, $\lambda=0$ or $\lambda\rightarrow\infty$, is returned. In
particular, 
\begin{align*}
\log\left( \frac{5\sigma^2}{9}\right)
  &\geq \textrm{GCV} =
    \log\left(\frac{\sigma^2(2\lambda^2+6\lambda+5)}{(2\lambda+3)^2}\right)\\
  &\geq \lim_{\lambda\rightarrow \infty}\textrm{GCV}
    = \log\left(\frac{\sigma^2}{2}\right)
\end{align*}
which means GCV will select $\lambda\rightarrow\infty$ and
$\hat{\beta}\rightarrow 0$.

%$\dfHat \rightarrow n$ as $\lambda \rightarrow 0$, and hence $\log(n-\dfHat) \rightarrow -\infty$,
%BLAH while GCV can
%potentially avoid this pathology.  To see this, note
%\[
%\log(n - x)
%\]

In \autoref{fig:smallex}, we plot AIC and BIC for $\lambda \in
[1\times 10^{-5}, 1]$ (left plot) and GCV (right plot) for ridge
regression on this dataset. Using AIC would have us report the unregularized model; that is 
 using a least squares solution. 
% Of
%course had we simply used one column alone with a standard linear
%model, we would have both fit the data perfectly and precisely estimated the
%single parameter. 
We will illustrate how the lasso behaves with
$\textrm{info}(C_n,g)$ in greater detail below.
Finally, we note that the behavior of GCV in this example is the opposite of what
happens in the simulations we report below. There, the penalty term is unable to outweigh
the training error term, and hence, the unregularized, $\lambda = 0$, solution is usually returned.

\begin{figure*}[t!]
  \centering
  \includegraphics[width=\linewidth]{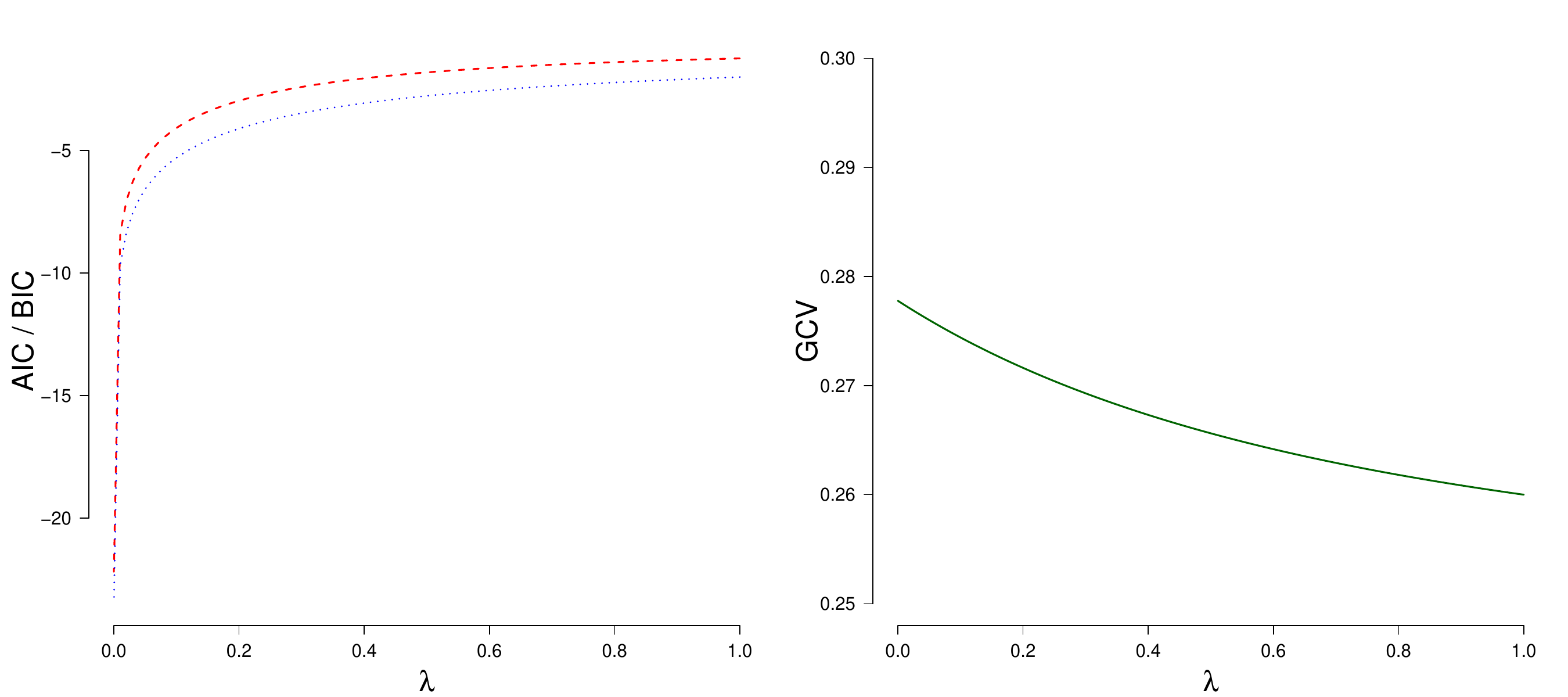}
  \caption{The left plot shows AIC (red, dashed) and BIC (blue, dotted) as we vary $\lambda$ from
    $1\times 10^{-5}$ to 1 for the small numerical example. The right
    plot shows the same setup but GCV instead. Notice that, using AIC
    or BIC, we
    would always choose the unregularized, $\lambda = 0$ model while GCV
    leads us to select $\lambda=\infty$.}
  \label{fig:smallex}
\end{figure*}

\end{example}

\section{Our procedure for tuning parameter selection via plug-in estimation}
\label{sec:sure}

To remedy the pathological behavior of $\textrm{info}(C_n,g)$ from equation \eqref{eq:GICsigUnknown}
in the high-dimensional case,
we propose to select $\lambda$ in the lasso problem via unbiased risk estimation.
Under the model in equation \eqref{eq:modelStatement}, 
the squared $\ell_2$ prediction risk of a coefficient vector $\beta$ can be written 
\begin{align*}
\risk_\beta & = n^{-1}\E\norm{\X\beta - \X\bstar}_2^2 \\
& = 
n^{-1}\E \norm{\X\beta - Y}_2^2  - \sigma^2 + 2n^{-1}\sum_{i=1}^n \textrm{Cov}(\hat{Y}_i,Y_i), \\
& =
n^{-1}\E \norm{\X\beta - Y}_2^2  - \sigma^2 + 2n^{-1}\sigma^2 \df.
\end{align*}
Therefore, a suite of sensible estimators of the squared $\ell_2$ prediction risk is produced via
\begin{equation}
\hat R_{\beta}(\hat\sigma^2,C_n) =   n^{-1}\norm{\X\beta - Y}_2^2  - \hat\sigma^2+ C_n\hat\sigma^2 \dfHat,
\label{eq:SURE}
\end{equation}
where $C_n$ is a
sequence of constants depending 
on $n$,
$\hat\sigma^2$ is an estimator of $\sigma^2$, and $\dfHat$ is an estimator of $\df$ for the procedure
under consideration. This general expression is commonly referred to
as Stein's unbiased risk estimator \citep[SURE,][]{Stein1981}.  For
simplicity, we will omit any arguments to $\hat R$ that aren't directly relevant to the discussion at hand and write
$\hat{R}_\lambda \equiv \hat{R}_{\beta(\lambda)}$ when $\beta$ is
indexed by the tuning parameter $\lambda$.

If
$\E[\hat\sigma^2\dfHat] = \sigma^2\df$  and $\E[\hat\sigma^2] = \sigma^2$ 
then $\hat R_{\beta}(\hat\sigma^2,C_n = 2n^{-1})$ 
is an unbiased estimator of 
$\risk_\beta$. For example, suppose that $n > p$, $\hat\beta(0)$ is a least squares solution, and $\hat\sigma^2 = (n-p)^{-1}\norm{Y - \X\hat\beta(0)}_2^2$ is the least squares estimator of $\sigma^2$.
Then $\E[\hat\sigma^2\dfHat] = \sigma^2\df$
and  $\hat{R}_{\hat\beta(0)}(\hat\sigma^2,C_n = 2n^{-1})$ is the classical Mallow's Cp 
\citep{Mallows1973}.  This follows as $\hat\beta(0)$ is linear in $Y$ and hence $\df = \dfHat = tr(H) =  \textrm{rank}(\X)$, 
where $H$ is such that $\X \hat\beta(0) = H Y$.  

As the lasso is not linear $Y$, we must use
an estimate of $\df$. \cite{ZouHastie2007,TibshiraniTaylor2012} show that for
the lasso, the degrees of freedom of $\hat{Y} = \X\lassoL$ is equal to $\E[\textrm{rank}(\X_{\S(\lambda)})]$,
suggesting the natural unbiased estimator $\dfHat=\dfHat(\lambda) =
\textrm{rank}(\X_{\S(\lambda)})$.  This is the degrees of freedom estimator we use
for both GIC and $\hat{R}_\lambda$.

Though  SURE is not in itself a new approach
to selecting tuning parameters in the lasso problem, the literature at this point
contains a major omission.  When  $\textrm{rank}(\X) = n \leq p$, the
choice of an estimator of the noise variance $\sigma^2$ is far from
straightforward.   For example, the lasso path algorithm in the {\tt
  R} package {\tt lars} avoids this issue. If $p<n$, it
provides a Cp-like score, which is superficially similar to equation
\eqref{eq:SURE}, with the least-squares variance estimator for the
largest possible model as $\hat{\sigma}^2$.  
Hence, it is unusable (and not produced) if $p>n$. 

In the recent theoretical
literature, results for high-dimensional
tuning parameter selection assume $\sigma^2$ is known to get around the difficult
task of high-dimensional variance estimation
\citep{chen2012extended,ZhangShen2010,kim2012consistent,FanTang2013}. However,
it is crucial to estimate $\sigma^2$ for $\hat R_{\lambda}$ to work
effectively in practice.  To demonstrate this necessity, we perform a second small simulation to 
illustrate the poor behavior of $\hat{R}(\sigma^2)$ when $\sigma^2$ is erroneously assumed known.

\begin{example}
\label{sec:secondExample}

We generate draws according to the model in equation \eqref{eq:modelStatement},
such that $n = 30$, $p = 150$, and  $\bstar$ has one nonzero coefficient drawn from the
standard Laplace distribution.
In \autoref{fig:introex}, we explore four methods for choosing $\lambda$ for the lasso. 
Clockwise from top left these methods are
 $\hat{R}_\lambda(\sigma^2  = 1)$, 
 $\hat{R}_\lambda(\hat\sigma_{\textrm{CV}}^2)$, $\hat{R}_\lambda(\hat\sigma_{\textrm{RCV}}^2)$
(see \autoref{sec:variance-estimation} for definitions of these variance estimators),
and lastly  $\textrm{info}(C_n = 2/n,g(x)=x)$, which corresponds to AIC.

\begin{figure*}[h!]
  \centering
  \includegraphics[width=.45\linewidth]{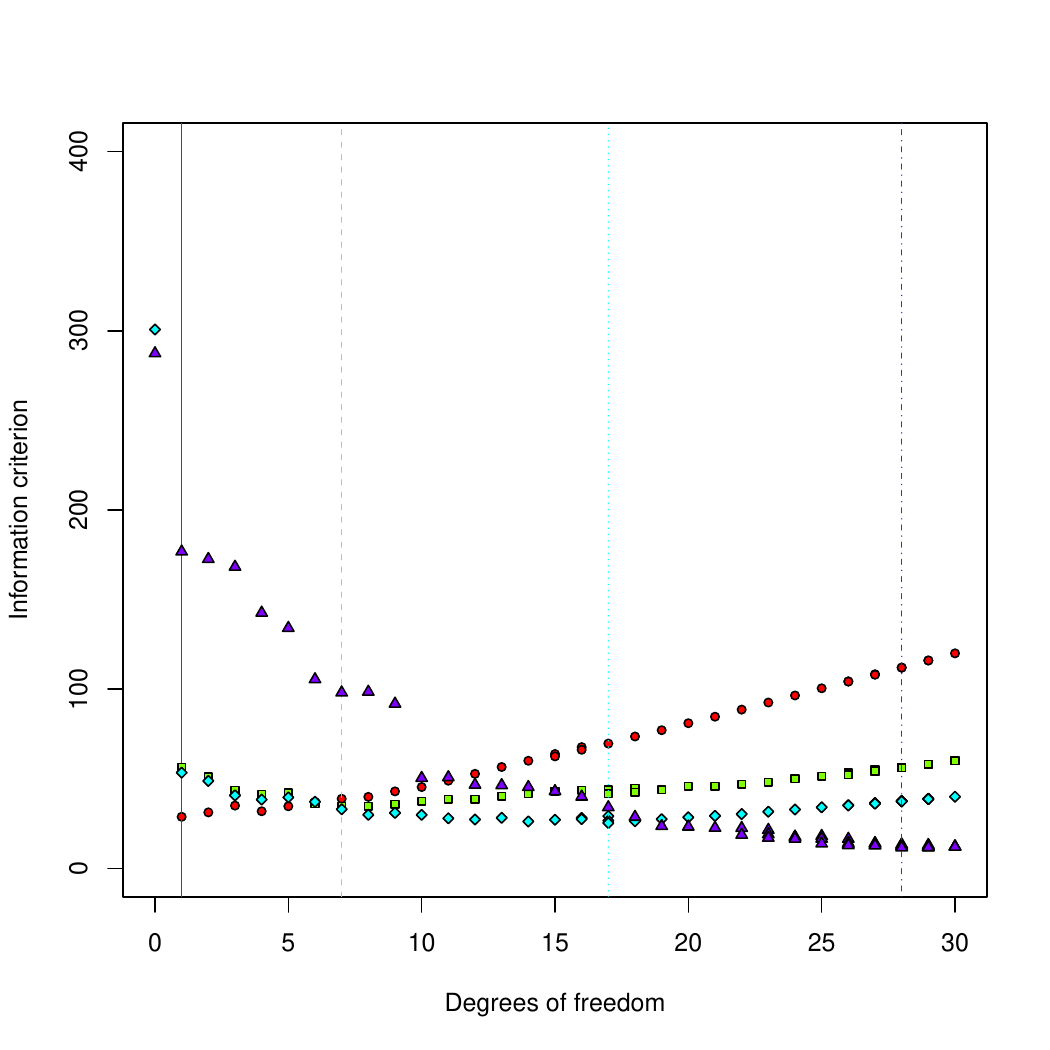}
  \includegraphics[width=.45\linewidth]{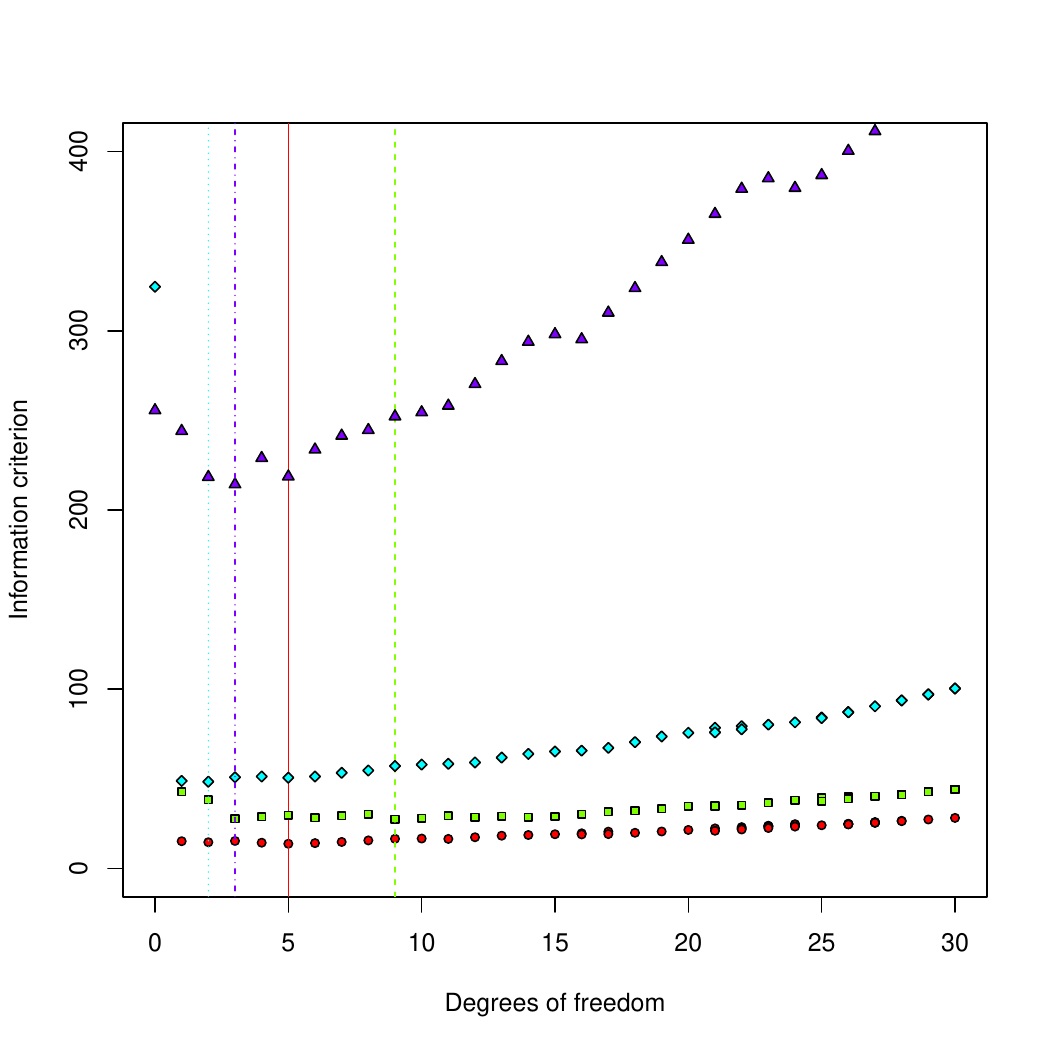} \\
  \includegraphics[width=.45\linewidth]{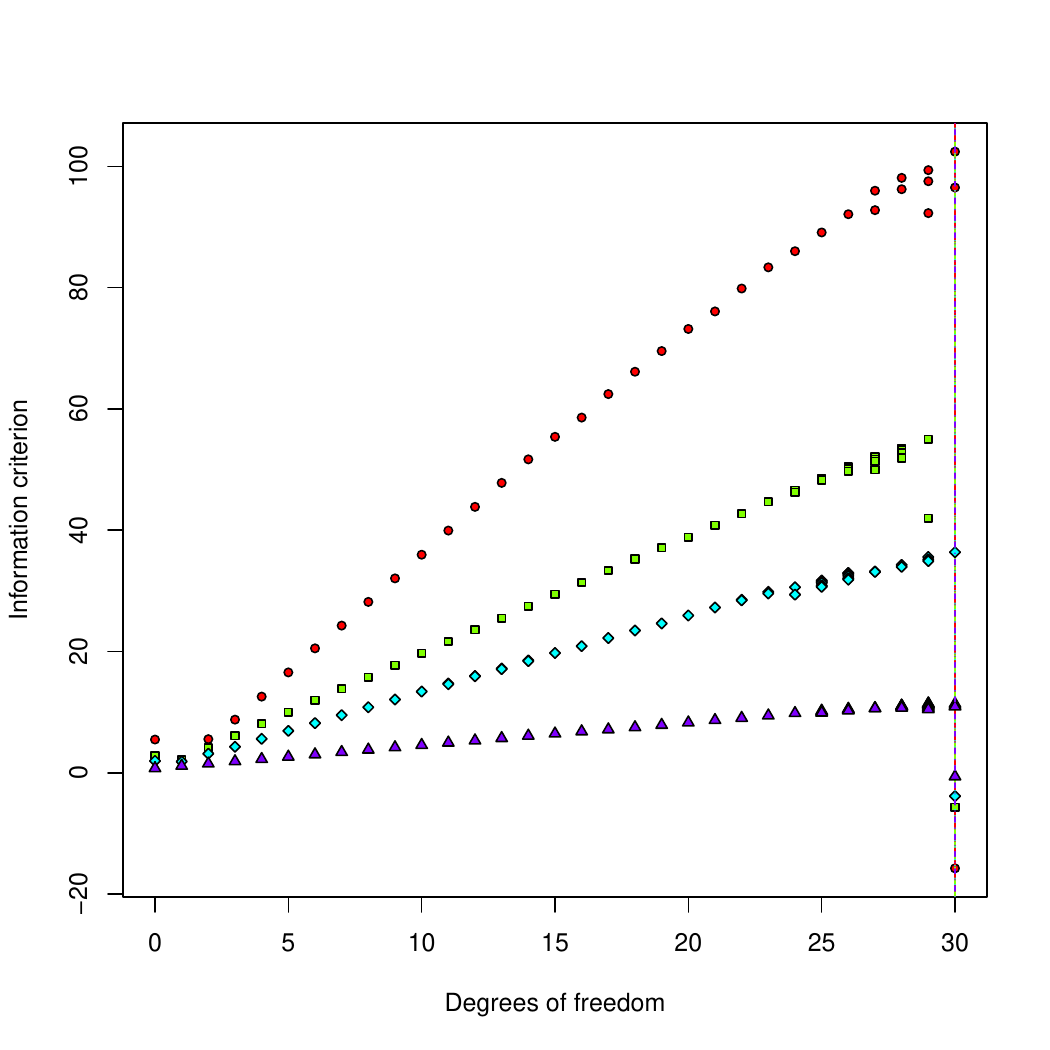}
  \includegraphics[width=.45\linewidth]{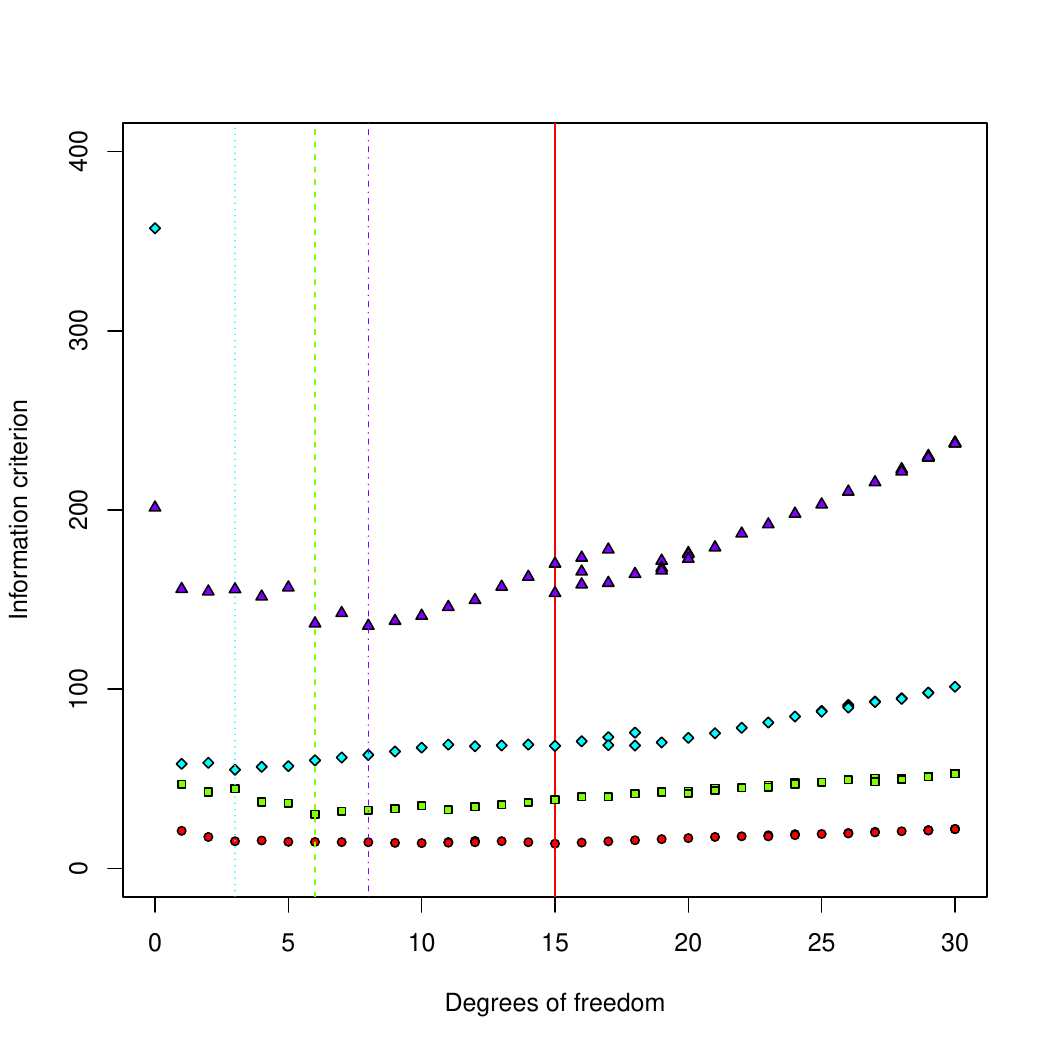}
%  \caption{Information criteria tend to overfit in high-dimensional
%    problems. We use four different values for $\sigma$: $\sigma=0.5$
%    (red, solid, circles), $\sigma=1$ (green, dashed, squares),
%    $\sigma=1.5$ (cyan, dotted, diamonds), $\sigma=5$ (violet,
%    dash-dot, triangles). Model selection methods, clockwise from top left:
%    Modified Cross-Validation; ``oracle'' AIC; cross validated variance estimator;
%    traditional AIC; and refitted cross-validation. Notice that
%    traditional AIC always selects the saturated model and modified CV
%    depends significantly on $\sigma$.}
      \caption{We use four different values for $\sigma$: $\sigma=0.5$
    (red, solid, circles), $\sigma=1$ (green, dashed, squares),
    $\sigma=1.5$ (cyan, dotted, diamonds), $\sigma=5$ (violet,
    dash-dot, triangles). A vertical line is drawn at the minimizer. Risk estimation methods, clockwise from top left:
    $\hat{R}(\sigma^2  = 1)$,  $\hat{R}(\hat\sigma_{\textrm{CV}}^2)$, 
    $\hat{R}(\hat\sigma_{\textrm{RCV}}^2)$, and $\textrm{info}(C_n = 2/n,g(x)=x)$. Notice that
    $\textrm{info}(C_n = 2/n,g(x)=x)$ always selects the unregularized model and  $\hat{R}(\sigma^2  = 1)$
    depends significantly on $\sigma$.}
  \label{fig:introex}
\end{figure*}

As expected, $\hat{R}_\lambda(\sigma^2  = 1)$ performs quite poorly
when $\sigma$ is far from 1.  In this case, the selected models have
widely varying degrees of freedom, choosing highly non-sparse models
despite there being only 1 non-zero true coefficient. Also,
$\textrm{info}(C_n = 2,g(x)=x)$ continues to choose the unregularized
solution, as predicted by the previous example, unless we arbitrarily
constrain $\df$ to be some value less than 30.   The other two, 
$\hat{R}(\hat\sigma_{\textrm{CV}}^2)$ and $\hat{R}(\hat\sigma_{\textrm{RCV}}^2)$,
perform much 
better.  We now discuss both of these estimators.  Occasionally in practice,
researchers may not compute $\textrm{info}(C_n = 2,g(x)=x)$ for all
$\lambda$. Instead, it is calculated from the most sparse to the
least sparse solutions, and then cut off when $\textrm{info}(C_n = 2,g(x)=x)$ does not
decrease. However, this procedure may not always work. In particular,
for $\sigma=5$, $\textrm{info}(C_n = 2,g(x)=x)$ is monotonically
increasing, except for $\df=30$. In other cases, $\textrm{info}(C_n =
2,g(x)=x)$ is not guaranteed to be convex, and this procedure will
result in possibly ignoring better solutions.

\end{example}

\subsection{High-dimensional variance estimation}
\label{sec:variance-estimation}

The literature on variance
estimation in high dimensions is a quickly growing
field. 
%\cite{spokoiny2002variance} focuses on variance
%estimation with local nonparametric estimates of the mean function. In
%particular, he argues that $O(n^{-1/2})$ rates are only achievable
%with sufficient smoothness and
%high first-order accuracy, demonstrating rates of $O(n^{-4/p})$ for
%the general case.
We use three high-dimensional variance estimators in our proposed risk
estimator. A comprehensive evaluation of these estimators (and some
others) is given by \cite{ReidTibshirani2016}, but we note that the
goal here is different: we do not wish to estimate $\sigma^2$ itself but rather
wish to
use it as an input to $\hat R_{\beta}$, which can then be used to select tuning parameters or
 estimate $\risk_\beta$.  It is not necessarily true that a good estimator of $\sigma^2$ leads to a 
 good estimator of $\risk_\beta$.

The first two approaches start by finding $\hat\beta(\hat\lambda_{CV})$ by
minimizing a $K$-fold cross-validation 
estimator of the risk to produce $\hat\lambda_{CV}$ (see 
\autoref{sec:cross-validation} for the details of cross validation) and 
finding a minimizer of equation \eqref{eq:optim} after inserting $\hat\lambda_{CV}$.
With this coefficient estimate, the squared $\ell_2$-norm of the
residuals can be used as a variance estimate, that is 
\begin{equation}
\label{eq:sigmaCV}
  \hat\sigma_{CV}^2 = \frac{1}{n- \dfHat(\hat\lambda_{CV})} \norm{ Y - \X \hat\beta(\hat\lambda_{CV})}_2^2.
\end{equation}
Alternatively, a restricted maximum likelihood-type
method can be formed by examining the orthogonal complement 
of the projection onto 
the column space of $\X_{\S(\hat\lambda_{CV})}$:
$H_{CV}^{\perp}$. Using this projection we define
\begin{equation*}
  \hat\sigma_{RMLE}^2 = \frac{1}{tr(H_{CV}^{\perp})} \norm{
    H_{CV}^{\perp}Y}_2^2 = \frac{1}{n - \dfHat(\hat\lambda_{CV})}\norm{
    H_{CV}^{\perp}Y}_2^2.
\end{equation*}
The second equality follows because, for the lasso,\\
$\textrm{trace}(H_{CV}^{\perp}) = \textrm{trace}(I - H_{CV}) = n - \textrm{rank}\left(\X_{\S(\hat\lambda_{CV})}\right)$,
which implies $\textrm{trace}(H_{CV}^{\perp}) = n - \dfHat(\hat\lambda_{CV})$.
Hence these two variance estimators differ only in the size of the residuals.
In fact,  due to the nature of projections, 
\[
\norm{ H_{CV}^{\perp}Y}_2^2 \leq \norm{ Y - \X \hat\beta(\hat\lambda_{CV})}_2^2.
\]
Thus, it must hold that $\hat\sigma_{RMLE}^2 \leq  \hat\sigma_{CV}^2$
and $\hat R(\hat\sigma_{RMLE}^2)$ penalizes model complexity less than
 $\hat R(\hat\sigma_{CV}^2)$.  In \autoref{sec:simulationConditions}, our simulations show that, when choosing $C_n = 2/n$,
 $\hat R(\hat\sigma_{CV}^2)$  results in lower prediction risk, better
 estimation consistency, and higher precision, while 
$\hat R(\hat\sigma_{RMLE}^2)$ has better recall.

The third variance estimation method we consider is known as refitted cross-validation \citep[RCV,][]{fan2012variance}.
After randomly splitting the data in half,  $\X_{\S(\hat\lambda_{CV})}$ is
formed on the first half and $\hat\sigma_1^2$ is formed via equation \eqref{eq:sigmaCV}, using 
the $Y$ and $\X$ values from the second half.  The procedure is
then repeated, exchanging the roles of the halves, producing $\hat\sigma_2^2$.  A final estimate is formed via
$\hat\sigma_{RCV}^2 = (\hat{\sigma}_1^2 + \hat{\sigma}_2^2)/2$.

In a comprehensive simulation study, \cite{ReidTibshirani2016} finds that $\hat\sigma_{CV}^2$
is the most reliable estimator for $\sigma^2$ out of those cited above, although,
as pointed out by \cite{fan2012variance}, it appears to have a
negative bias whereas $\hat\sigma_{RCV}^2$ does not. However, this  
doesn't mean that any of the above methods will necessarily produce
superior performance as a plug-in variance estimator for risk estimation or
tuning parameter selection.

Armed with any of the above high-dimensional variance estimators,
we can form an estimator of $\bstar$ via $\hat\beta(\hat\lambda)$, where
\begin{equation}
\label{eq:lambda-hat-definition}
\hat\lambda = \argmin_\lambda \hat R_{\lambda}(\hat\sigma,C_n).
\end{equation}

As discussed above, tuning parameter selection procedures based on
SURE or information criteria have no theoretical justification when
the variance is unknown and $p>n$. In the next section, we present a 
comprehensive empirical investigation of the performance of the lasso
with tuning parameter selected by the aforementioned methods.  Additionally, we 
include comparisons to other modified lasso-type methods for completeness.

\section{Empirical evaluation}
\label{sec:simulationConditions}

In the remainder of this paper, we evaluate our proposed risk
estimation methods for the purposes of choosing the tuning parameter
$\lambda$ for lasso.  We consider only the high-dimensional setting and
evaluate success  using several
criteria such as prediction risk and model selection. We first perform a comprehensive simulation and
then present results from a real-world application involving survival times as a function of gene expression data.

\subsection{Simulation parameters}
For our simulations, we consider a wide range of possible
conditions by varying the correlation in the design, $\rho$; the
number of measurements, $p$; the sparsity, $\alpha$; and the
signal-to-noise ratio, SNR.  In all cases, we let $n = 200$ (similar
results hold for $n=100$).

The design matrices, $\X \in \mathbb{R}^{n\times p}$,
 are produced by concatenating independent and identically distributed
 rows with mean zero and correlations introduced
by an autoregressive model:  $\textrm{Cov} (X_{ij},X_{ik}) = 
\rho^{|j - k|}$. For these simulations,
we consider correlations $\rho = 0.1,$ 0.5, and 0.8.

For sparsity, we define $\sstar = \lfloor n^{\alpha} \rfloor$ and generate
the $\sstar$ non-zero elements of $\bstar$ from a Laplace distribution
with parameter 1, which matches a Bayesian interpretation of the lasso.
  We let $\alpha$ be $0.4$ or $0.7$, which corresponds to
 $8$ or $40$ non-zero elements, respectively. We vary $\sigma^2$ so 
 that the signal-to-noise ratio, defined to be 
$\textrm{SNR} = n^{-1}\bstar^{\top} \E [\X^{\top}\X] \bstar/\sigma^2$, is 
$0.1$, $1$, or $10$.  Note that as
$\textrm{SNR}$ increases the observations go from a high-noise and
low-signal regime to a low-noise and high-signal one.  We let $p=400$ or
$p = 1500$. 

Lastly, we consider two different noise distributions, $\epsilon_i \sim
N(0,1)$ and $\epsilon_i \sim 3^{-1/2} t(3)$.  Here $t(3)$ indicates a
$t$ distribution with 3 degrees of freedom and the $3^{-1/2}$ term
makes the variance equal to 1 and the $\epsilon_i$ are independent.
As the results for these noise distributions  
are quite similar, we only present the Gaussian
simulations. Furthermore, while we have
simulated all combinations of these parameters and distributions, we 
include only a subset here for brevity. 

% A comprehensive
%record of the simulation results appears in the supplement. 
%\textbf{In the next section, we include plots for
%$p=50,250,1000$; $\rho=0.2,0.5,0.95,0.99$; $\textrm{SNR} = 0.5,5,20$;
%and Gaussian noise.}

\subsection{Modified lasso-type methods}
\label{sec:modif-lasso-crit}

For a more complete comparison, we include in our simulations
some variations on the lasso estimator that have been proposed.

First, \cite{SunZhang2012} develops `scaled sparse
regression' (SSR), which uses the fact that the optimal choice of $\lambda$ for lasso
is asymptotically proportional to
$\sigma$. By recasting the lasso problem as 
\begin{equation*}
\label{eq:ssr}
  \hat{\beta}_{SSR} = \argmin_{\beta,\sigma} \frac{1}{2n\sigma}
  \norm{Y-\X\beta}_2^2 + \frac{(1-a)\sigma}{2} + \EPconst \norm{\beta}_1,
\end{equation*}
and fixing $\EPconst$ and $a$, the authors develop theory for ``tuning parameter free''
lasso with simultaneous variance estimation. Though this is a
promising approach, the
objective function is not convex, hence the variance and the lasso solution are
iteratively computed and 
the solutions tend
to depend on the starting values. 
Nonetheless, SSR enjoys attractive
theoretical properties.

Alternatively, \cite{BelloniChernozhukov2011} suggests the \SQRT{}, or ``square root lasso,'' as a modification of the lasso problem
\begin{equation}
\label{eq:squareRootLasso}
\hat{\beta}_{\sqrt{\textrm{lasso}}} = \argmin_{\beta} \frac{1}{\sqrt{n}}
  \norm{Y-\X\beta}_2 + \frac{\lambda_n}{n} \norm{\beta}_1.
\end{equation}
Appealing to asymptotic arguments, they show that the 
minimizer of equation \eqref{eq:squareRootLasso} achieves near oracle
performance if $\lambda_n = c \sqrt{n}\Phi^{-1}(1- \alpha/(2p))$, which does not
depend on $\sigma$. Here, $\Phi^{-1}$ is the quantile function for the
standard Gaussian distribution.

We also consider the Smoothly
Clipped Absolute Deviation Penalty \citep{FanLi2001}:
\begin{align*}
\hat\beta_{SCAD} &= \argmin_\beta \frac{1}{2n}\norm{Y-\X\beta}_2^2 +
\sum_{j=1}^p g_\lambda(|\beta_j|),\\
\intertext{where}
g'_\lambda(\theta) &= \lambda\left[\indicator(\theta\leq\lambda) +
                     \frac{(a\lambda-\theta)_+}{(a-1)\lambda}\indicator(\theta
                     \geq \lambda)\right],
\end{align*}
for some $a>2$ and $\theta>0$. 

Lastly, our experiments show that GCV tends to dramatically under regularize in the lasso 
problem.  
Likewise, setting $C_n = \log(n)/n$ in $\hat{R}_\beta(\hat\sigma^2,C_n)$ 
tends to over regularize.  Hence, we investigate a two-stage method whereby 
an intial screening is performed by selecting $\hat{\lambda}_{GCV}$ and 
forming $\S_{\hat\lambda_{GCV}}$.  This often selects a very large model, 
typically with $|\S_{\hat\lambda_{GCV}}| = n$.
For the second stage, we use only the columns of $\X$ with indices in 
$\S_{\hat\lambda_{GCV}}$ to compute 
$\hat{R}_{\beta}(\hat{\sigma}^2,C_n = \log(n)/n)$, which is minimized 
over $\lambda$ 
to produce $\hat\lambda$.  Then, the output of this two-stage method is 
$\hat{\beta}(\hat\lambda)$. We refer to this procedure as ``2-stage''
and do not report results for GCV alone as it is uniformly poor. This
procedure is shown in \autoref{alg:2-stage}.
\begin{algorithm}[t]
  \KwIn{Design matrix $\X$, response $Y$, sequence of $\lambda$}
  Solve equation \eqref{eq:optim} for each $\lambda$\;
  Find $\hat{\lambda}_{GCV}$ by minimizing equation \eqref{eq:gcv}\;
  Set $S_{\hat{\lambda}_{GCV}}$ to be the non-zero elements of
  $\hat{\beta}(\hat{\lambda}_{GCV})$\;
  Compute $\hat{R}_{\beta}(\hat{\sigma}^2,C_n = \log(n)/n)$ using
  only the columns of $\X$ in ${S_{\hat{\lambda}_{GCV}}}$ for each 
  $\lambda$\;
  Select $\hat{\lambda}_{\textrm{2-stage}}$ by minimizing
  $\hat{R}_{\beta}(\hat{\sigma}^2,C_n = \log(n)/n)$\;
  \KwOut{Coefficient estimates $\hat{\beta}(\hat{\lambda}_{\textrm{2-stage}})$}    
  \caption{2-stage method for tuning parameter selection   \label{alg:2-stage}}
\end{algorithm}

GCV's behavior is intimately connected to the rate at which the
numerator, given by the training error, and the denominator, given by
$(1-\df/n)^2$, go to zero as $\lambda \rightarrow 0$.  In our
simulations, the numerator goes to zero at a faster rate than the
denominator and hence GCV tends to dramatically under-regularize.
Additionally, by noting that $1/(1-x)^2 \approx 1 + 2x$, GCV is
approximately the same as AIC. However, this approximation is only accurate for
$x$ near zero, which happens when $\df$ is forced to be small relative
to $n$.  In the classical case where $n\gg p$, this approximation is
quite accurate, but in the 
high-dimensional problem, relatively larger $\df$
may explain some of the underperformance of GCV as a tuning parameter
selection method. 

In the next section, we give more details about the numerical implementation
of the methods considered in this paper to aid in reproducibility.

\subsection{Implementation of methods and notation}
\label{sec:simulationMethods}
For ease of reference, \autoref{tab:listOfMethods} displays all of the
methods for which we present simulations.
\begin{table}[bt]
\centering
\caption{
  List of methods and abbreviations used in our empirical
    study}
%\resizebox{\linewidth}{!}{
  \begin{tabular}{@{}ll@{}}
    \toprule
    Abbreviation & Method\\
    \midrule
    CV-10-Fold & $10$-fold cross validation\\
    MCV & Modified Cross Validation\\
    R-Oracle-2 & $\hat{R}_\beta(\sigma^2,\ C_n=2/n)$\\
    \AICcv & $\hat{R}_\beta(\hat\sigma_{CV}^2,\ C_n=2/n)$\\
    \AICrmle & $\hat{R}_\beta(\hat\sigma_{RMLE}^2,\ C_n=2/n)$\\
    \AICrcv & $\hat{R}_\beta(\hat\sigma_{RCV}^2,\ C_n=2/n)$\\
    R-Oracle-logn & $\hat{R}_\beta(\sigma^2,\ C_n=\log(n)/n)$\\
    \BICcv & $\hat{R}_\beta(\hat\sigma_{CV}^2,\ C_n=\log(n)/n)$\\
    2-stage & Two-stage method using GCV then \BICcv\\
    SCAD & Smoothly clipped absolute deviation\\
    SSR & Scaled sparse regression\\
    SQRT & \SQRT{}\\
    SQRT refitted & OLS estimation on the model selected with \SQRT{}\\
    \bottomrule
  \end{tabular}
%}
  \label{tab:listOfMethods}
\end{table}
Since all of these methods rely on numerical optimization routines,
it is important to discuss the particular 
implementation of the solvers used to generate $\hat\beta(\lambda)$.  

Two widely 
used implementations for lasso are {\tt glmnet} \citep{FriedmanHastie2010},
which uses coordinate 
descent and a grid of $\lambda$ values, and 
{\tt lars}, which leverages the piece-wise linearity of the lasso
solution path.  The package 
{\tt glmnet} is much faster than {\tt lars},  however, {\tt glmnet} 
only examines a grid of $\lambda$ values and returns
an approximate solution at each $\lambda$ (due to the
iterative nature of the algorithm). 
Additionally, {\tt glmnet} suffers from numerical stability issues for
small $\lambda$ values when $p > n$.

Because the {\tt lars} path
will necessarily change for different cross-validation folds, the grid-based nature
of {\tt glmnet} is more suited for use with cross-validation. For this reason, we use {\tt glmnet} for CV-10-Fold and to find 
$\hat{\sigma}^2_{CV}$, $\hat{\sigma}^2_{RCV}$, and
$\hat{\sigma}^2_{RMLE}$.  

With any high dimensional variance estimator $\hat\sigma^2$,  
we need to compute $\hat\lambda = \argmin\hat{R}_\lambda(\hat\sigma^2)$.
We use {\tt lars} to find the entire lasso solution path on all of the data 
to compute $\hat{R}_\lambda(\hat\sigma^2)$ and then report the minimizer
$\hat\lambda$ and $\hat\beta(\hat\lambda)$.

To optimize the modified lasso problems' objective functions, we use the {\tt R} package
{\tt scalreg} to fit SSR and the {\tt R} package
{\tt flare} to fit the $\sqrt{\textrm{lasso}}$. 
For {\tt scalreg}, we choose the starting point for the iteration via the quantile method \citep{SunZhang2013}.
For {\tt flare}, we set the tuning parameter to $\lambda = c
\sqrt{n}\Phi^{-1}(1- \alpha/(2p))$ with 
$c=1.1$ and $\alpha = 0.05$, as suggested by
\cite{BelloniChernozhukov2011}. As \SQRT{} 
tended to pick the correct model but with overly regularized
coefficient estimates, we will additionally examine a refitted version of \SQRT{} in
which the unregularized least squares solution of $Y$ on
$\X_{\S(\hat{\beta}_{\sqrt{\textrm{lasso}}})}$ is reported. In an
attempt to get as close as possible to the global optimum, we decrease 
the {\tt prec} (precision) option to $1 \times 10^{-10}$ 
and increase {\tt max.ite} (maximum iterations) to $1 \times
10^{7}$.

To fit SCAD, we use the package {\tt ncvreg}~\citep{BrehenyHuang2011} with
default settings ($a=3.7$) and choose
$\lambda$ via the built in CV function. We note that
\cite{FanLi2001} suggests using either CV or an approximation to GCV
which uses the trace of the projection
matrix from the final iteration to form an estimate $\dfHat$. However, this matrix is a function of $Y$, so the
calculated $\df$ is not unbiased. We therefore only report the default cross-validation-based method, and we
note that subsequent work~\citep{zhang2010regularization,WangLi2007}
has carefully investigated information criteria using SCAD.

%%%%%%%%move the modified-lasso section
%In contrast to \SQRT (and the closed-form equation in Example 1), our experiments show that GCV tends to dramatically under regularize.  
%Likewise, setting $C_n = n^{-1}\log(n)$ in $\hat{R}_\beta(\hat\sigma^2,C_n)$ 
%tends to over regularize.  Hence, we investigate a two-stage method whereby 
%an intial screening is performed by selecting $\hat{\lambda}_{GCV}$ and 
%forming $\S_{\hat\lambda_{GCV}}$.  This often selects a very large model, 
%typically the number of elements in  $\S_{\hat\lambda_{GCV}}$ is equal to $n$.
%For the second stage, we use only the columns of $\X$ with indices in 
%$\S_{\hat\lambda_{GCV}}$ to compute 
%$\hat{R}_{\beta}(\hat{\sigma}^2,C_n = \log(n)/n)$, which is minimized 
%over $\lambda$ 
%to produce $\hat\lambda$ and the output of the two-stage method is 
%$\hat{\beta}(\hat\lambda)$. We refer to this procedure as ``2-stage''
%and do not report results for GCV alone as it is uniformly poor.

The ideal, or oracle, version of our method in equation \eqref{eq:SURE} would use
the known variance. We refer to this as the oracle risk estimator and
note that it is unbiased. Obviously this is not a viable
estimator in practice, but it is useful for normalizing comparisons in
our simulation study. We provide two versions of this oracle estimator:
$\hat{R}_{\beta}(\sigma^2,C_n = 2/n)$ and 
$\hat{R}_{\beta}(\sigma^2,C_n = \log(n)/n)$. 

\subsection{Simulation results}
\label{sec:simulationResults}

We present results for four different metrics based on
different data analysis objectives.  If the risk estimation methods are used
to select tuning parameters, then the data analysts could be interested in the
prediction risk, which evaluates how well we can predict a new $Y$ given a new $X$; consistency, which measures how
far the procedure $\hat\beta$ is from $\bstar$; or F-score, which
considers how well a method does at model selection.  Alternatively, 
when evaluating the success of a method, or when comparing it to another method,
the risk estimate itself is of interest. We evaluate these four
criteria in the following subsections. \autoref{tab:listOfMethods} shows
the correspondence between the mathematical notation we have used so
far, and the arabic letters used in the figures. When describing each
figure, we will refer to different methods with the arabic letters for clarity.
% We evaluate the quality of these risk estimates
% in a neighborhood around the true coefficient vector $\bstar$.

\subsubsection{Prediction risk}
\begin{figure*}[t!]
\centering
\includegraphics[width=\textwidth]{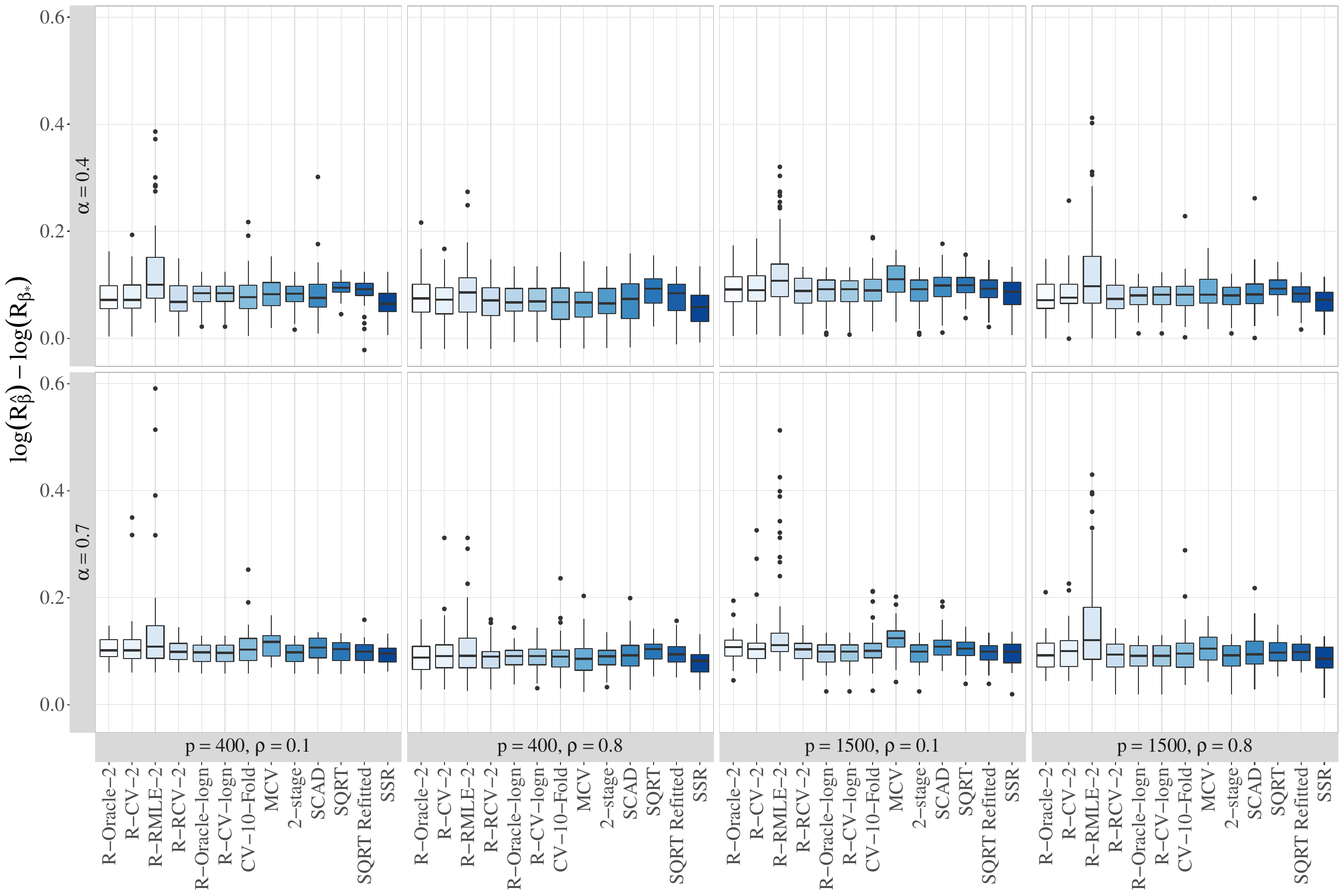}
\caption{ Comparison of log prediction risk for $SNR = 0.1$. Top row:
  $\alpha = 0.4$. Bottom row: $\alpha = 0.7$.  }
 % Top row:
 %  $\alpha = 0.4$. Bottom row: $\alpha = 0.7$.  \attn{I included this
 %    as an illustration. I've also made similar figures for all the
 %    other simulations, so we can talk about what you like. I did
 %    PredRisk on the log-scale to deal with the y-axis issue.}}
 \label{fig:predSNR01}
\end{figure*}

\begin{figure*}[t!]
\centering
\includegraphics[width=\textwidth]{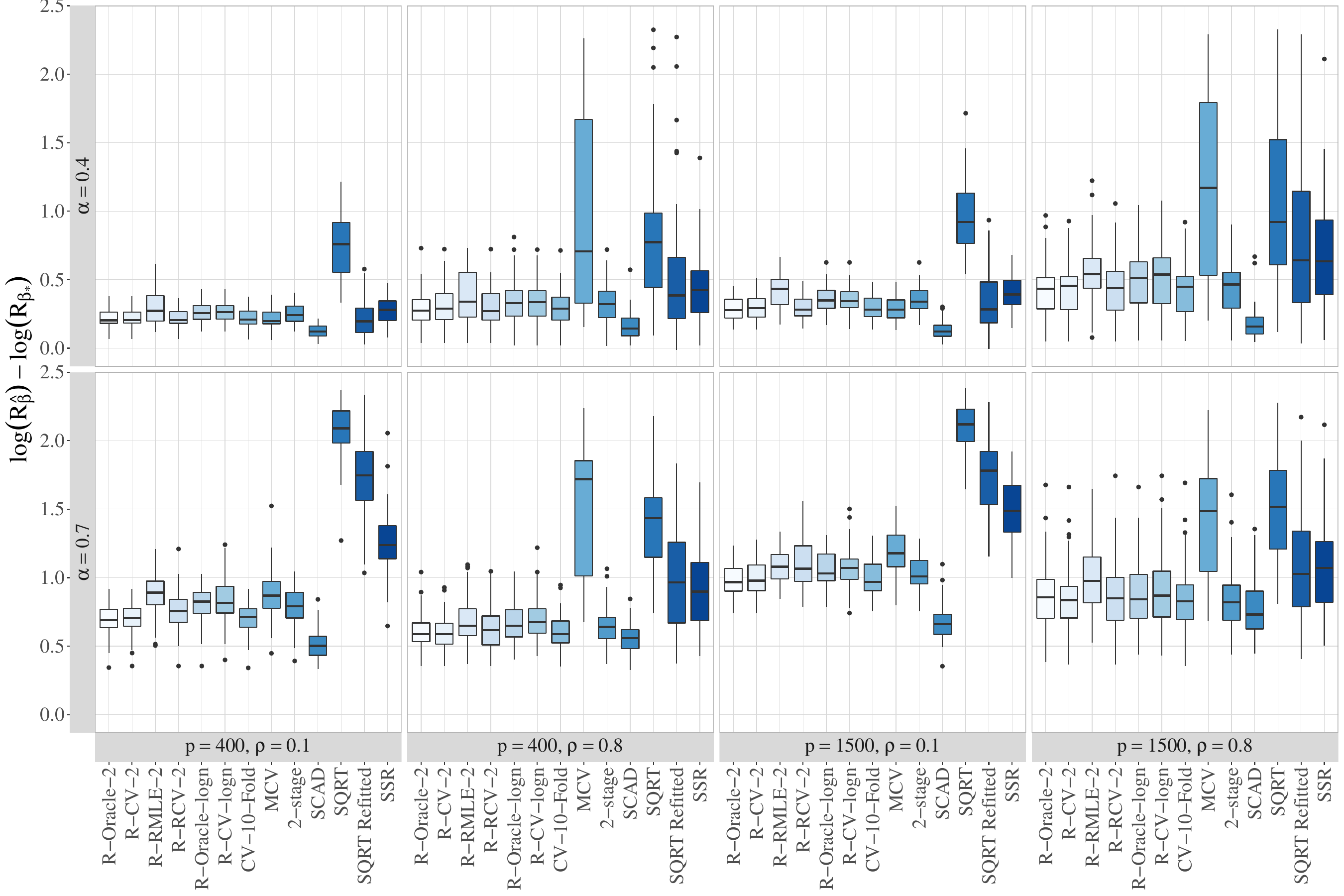}
\caption{ Comparison of log prediction risk for $SNR = 10$. Top row:
  $\alpha = 0.4$. Bottom row: $\alpha = 0.7$.  }
\label{fig:predSNR10}
\end{figure*}

Prediction risk is an important criterion as it is often a major goal in modern data analysis applications.
For these simulations, we approximate $\textrm{R}_\beta$ in equation \eqref{eq:risk} with the
average squared error over 5000 test observations and normalize it by subtracting $\sigma^2$, but continue to
denote it $\risk_\beta$. 
We present boxplots for the log of the prediction risk of the selected models in
\autoref{fig:predSNR01} and \autoref{fig:predSNR10} for SNR 0.1 and 10
respectively. 

% \begin{figure}[t!]
% \centering
%  \resizebox{\linewidth}{!}{
%  \begin{tabular}{cccc}
%  \includegraphics[width=1.45in,trim=30 0 100 70,clip]{predRisk/predRisk__n200p400_NoiseTypeGaussian_Alpha4e-01_Snr1e-01_Rho1e-01_Sigma1e+00} & 
%  \includegraphics[width=1.45in,trim=30 0 100 70,clip]{predRisk/predRisk__n200p400_NoiseTypeGaussian_Alpha4e-01_Snr1e-01_Rho8e-01_Sigma1e+00}  &
%  \includegraphics[width=1.45in,trim=30 0 100 70,clip]{predRisk/predRisk__n200p1500_NoiseTypeGaussian_Alpha4e-01_Snr1e-01_Rho1e-01_Sigma1e+00} &
%  \includegraphics[width=1.45in,trim=30 0 100 70,clip]{predRisk/predRisk__n200p1500_NoiseTypeGaussian_Alpha4e-01_Snr1e-01_Rho8e-01_Sigma1e+00} \\
%  \includegraphics[width=1.45in,trim=30 0 100 70,clip]{predRisk/predRisk__n200p400_NoiseTypeGaussian_Alpha7e-01_Snr1e-01_Rho1e-01_Sigma1e+00} &
%  \includegraphics[width=1.45in,trim=30 0 100 70,clip]{predRisk/predRisk__n200p400_NoiseTypeGaussian_Alpha7e-01_Snr1e-01_Rho8e-01_Sigma1e+00} &
%  \includegraphics[width=1.45in,trim=30 0 100 70,clip]{predRisk/predRisk__n200p1500_NoiseTypeGaussian_Alpha7e-01_Snr1e-01_Rho1e-01_Sigma1e+00} &
%  \includegraphics[width=1.45in,trim=30 0 100 70,clip]{predRisk/predRisk__n200p1500_NoiseTypeGaussian_Alpha7e-01_Snr1e-01_Rho8e-01_Sigma1e+00} \\
%  $p = 400, \rho=0.1$ & $p = 400, \rho=0.8$ & $p = 1500, \rho=0.1$ & $p = 1500, \rho=0.8$ 
%  \end{tabular}
%  }
% \caption{ Comparison of prediction risk for $SNR = 0.1$. Top row:
%   $\alpha = 0.4$. Bottom row: $\alpha = 0.7$.  }
% \label{fig:predSNR01}
% \end{figure}
For low SNR, MCV, R-RMLE-2, SQRT, and SQRT refitted all perform noticeably worse than
the competing methods.  For high SNR, SCAD performs best, especially
when $p = 1500$ and 
when the true vector $\bstar$ is non-sparse ($\alpha = 0.7$).  Also,
CV-10-Fold and R-CV-2  
both perform somewhat better than R-RCV and R-CV-logn.

%KEEP TO RECOVER OLD TRIM
%\includegraphics[width=1.45in,trim=70 0 100 100,clip]{predRisk/

% \begin{figure}[t!]
% \centering
%  \resizebox{\linewidth}{!}{
%  \begin{tabular}{cccc}
%  \includegraphics[width=1.45in,trim=30 0 100 70,clip]{predRisk/predRisk__n200p400_NoiseTypeGaussian_Alpha4e-01_Snr1e+01_Rho1e-01_Sigma1e+00} & 
%  \includegraphics[width=1.45in,trim=30 0 100 70,clip]{predRisk/predRisk__n200p400_NoiseTypeGaussian_Alpha4e-01_Snr1e+01_Rho8e-01_Sigma1e+00}  &
%  \includegraphics[width=1.45in,trim=30 0 100 70,clip]{predRisk/predRisk__n200p1500_NoiseTypeGaussian_Alpha4e-01_Snr1e+01_Rho1e-01_Sigma1e+00} &
%  \includegraphics[width=1.45in,trim=30 0 100 70,clip]{predRisk/predRisk__n200p1500_NoiseTypeGaussian_Alpha4e-01_Snr1e+01_Rho8e-01_Sigma1e+00} \\
%  \includegraphics[width=1.45in,trim=30 0 100 70,clip]{predRisk/predRisk__n200p400_NoiseTypeGaussian_Alpha7e-01_Snr1e+01_Rho1e-01_Sigma1e+00} &
%  \includegraphics[width=1.45in,trim=30 0 100 70,clip]{predRisk/predRisk__n200p400_NoiseTypeGaussian_Alpha7e-01_Snr1e+01_Rho8e-01_Sigma1e+00} &
%  \includegraphics[width=1.45in,trim=30 0 100 70,clip]{predRisk/predRisk__n200p1500_NoiseTypeGaussian_Alpha7e-01_Snr1e+01_Rho1e-01_Sigma1e+00} &
%  \includegraphics[width=1.45in,trim=30 0 100 70,clip]{predRisk/predRisk__n200p1500_NoiseTypeGaussian_Alpha7e-01_Snr1e+01_Rho8e-01_Sigma1e+00} \\
%  $p = 400, \rho=0.1$ & $p = 400, \rho=0.8$ & $p = 1500, \rho=0.1$ & $p = 1500, \rho=0.8$ 
%  \end{tabular}
%  }
% \caption{ Comparison of prediction risk for $SNR = 10$. Top row:
%   $\alpha = 0.4$. Bottom row: $\alpha = 0.7$.  }
% \label{fig:predSNR10}
% \end{figure}

\subsubsection{Consistency}

The second performance metric we use examines the ability of $\hat\beta(\hat\lambda)$
to produce accurate estimates of the true
parameter $\bstar$. We examine a normalized version of the deviation between
the estimated coefficients and the size of the parameter:
\[
C(\hat{\beta}) = \frac{\E\norm{\hat{\beta} - \bstar}_2^2}{\norm{\bstar}_2^2}.
\]
Thus, smaller values are better, and values near 1 often represent overly
sparse solutions as $\hat{\beta}\equiv 0 \Rightarrow
\E\norm{\hat{\beta}-\bstar}= \norm{\bstar}$.
\begin{figure*}[t!]
\centering
\includegraphics[width=\textwidth]{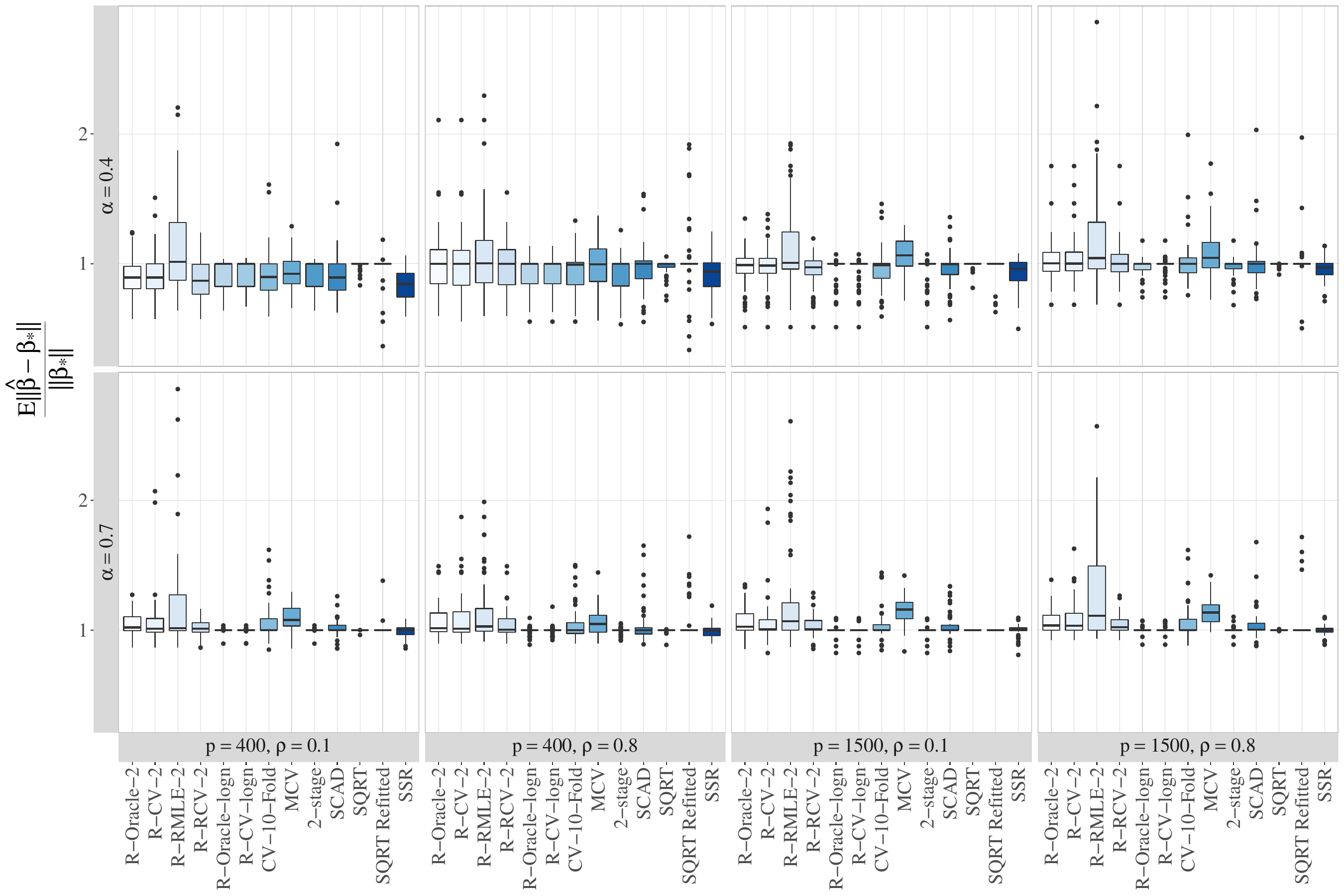}
\caption{ Comparison of consistency for $SNR = 0.1$. Top row:
  $\alpha = 0.4$. Bottom row: $\alpha = 0.7$.  }
\label{fig:consSNR01}
\end{figure*}

\begin{figure*}[t!]
\centering
\includegraphics[width=\textwidth]{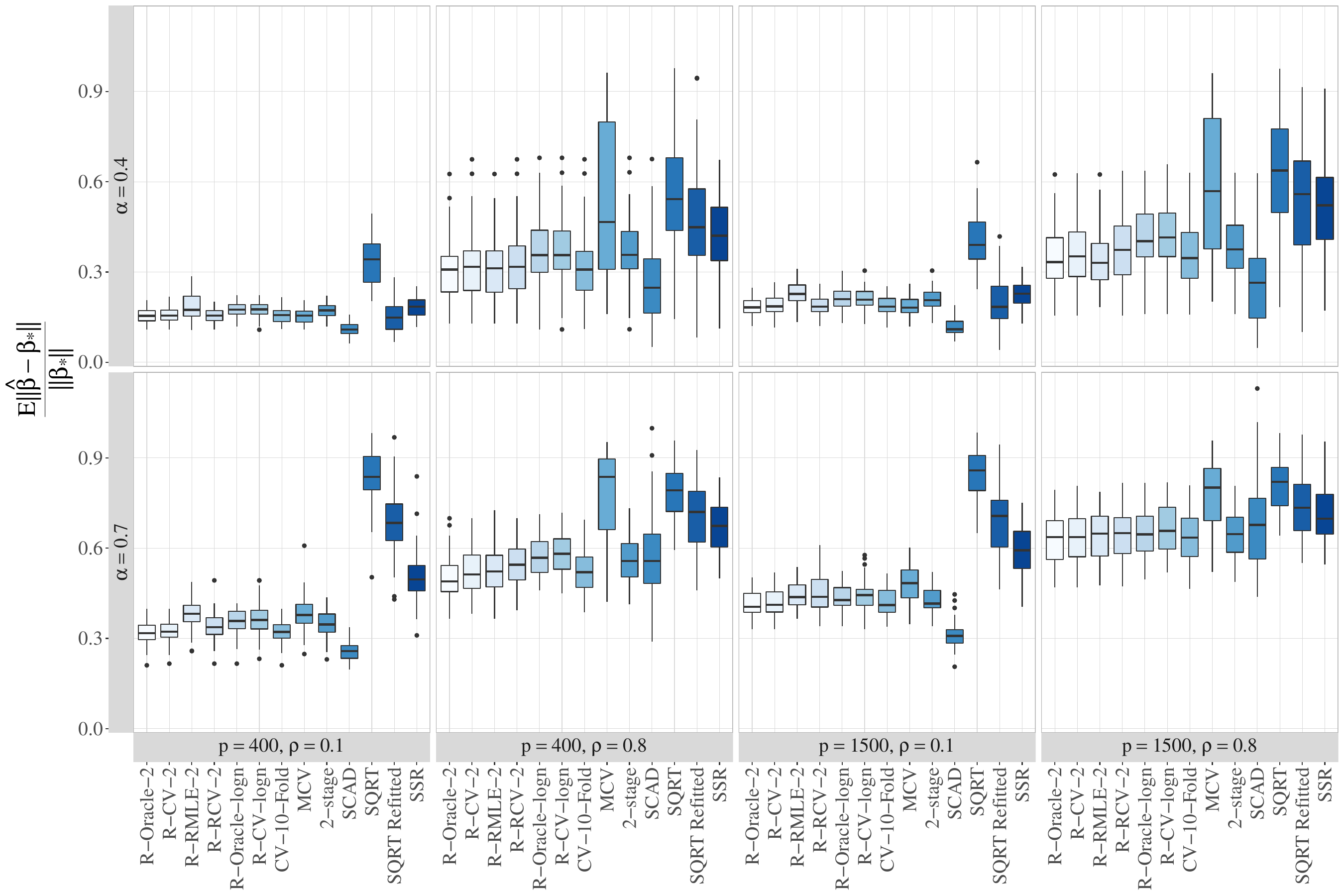}
\caption{ Comparison of consistency for $SNR = 10$. Top row:
  $\alpha = 0.4$. Bottom row: $\alpha = 0.7$.  }
\label{fig:consSNR10}
\end{figure*}

In the low-SNR regime (\autoref{fig:consSNR01}), no procedure
performs particularly well, as one would expect. The R-CV-logn, 2-stage, SQRT,
and SQRT refitted nearly always select $\hat{\beta}\equiv 0$ with occasional exceptions.  This results
in slightly better $C(\hat{\beta})$ than the other methods.  For the
high-SNR regime (\autoref{fig:consSNR10}), SCAD performs
best, particularly in the sparse scenario $(\alpha = 0.4)$ or when $p
= 1500$ and $\rho=0.1$.   
CV-10-Fold and R-CV-2 perform similarly to each other and are slightly
better than the other methods. 
MCV, R-RMLE-2, SQRT, and SQRT refitted all perform rather poorly.

\subsubsection{F-score}
To examine the ability of these procedures to perform model selection
directly, we define the precision and recall for a particular 
$\beta$ to be respectively (recalling that $\S = \{j: |\beta_j| > 0\}$
and $|\S|$ is the number of elements in $\S$) 
\begin{align*}
  P(\S) &= \frac{|\S \cap \Sstar|}{|\S|}&\textrm{and}&&
  R(\S) &= \frac{|\S \cap \Sstar|}{|\Sstar|}.
\end{align*}
To parsimoniously represent both precision and recall
at the same time, we use the  $F$-score (sometimes 
referred to as the $F1$-score), which is the harmonic mean of the
precision and recall: 
 \begin{equation*}
   F(\S) = \frac{2 R(\S) P(\S)}{R(\S) + P(\S)} = \frac{2}{\frac{1}{R(\S)} + \frac{1}{P(\S)}}.
 \end{equation*}
 Observe that $F(\S)$ is equal to one if and only if $R(\S)$ and
 $P(\S)$ are both equal to one and equal to zero if either $R(\S)$ or
 $P(\S)$ are equal to zero. Thus, higher values represent better performance.  As an aside,
 the SQRT and SQRT refitted methods will have the same F-score (as they select the same model).
 We nonetheless plot both of the methods to maintain easier comparability to other figures.

\begin{figure*}[t!]
\centering
\includegraphics[width=\textwidth]{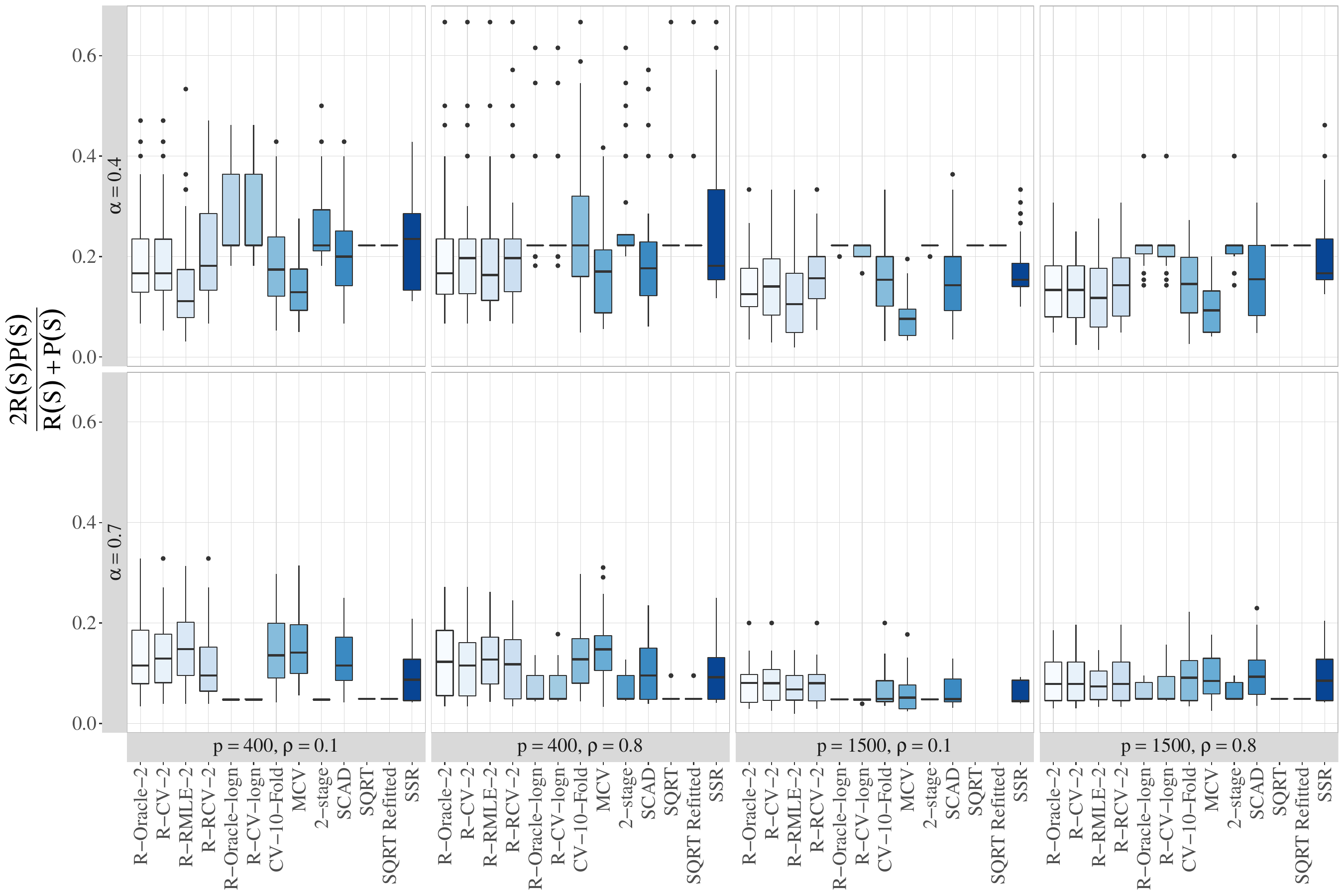}
\caption{ Comparison of F-score for $SNR = 0.1$. Top row:
  $\alpha = 0.4$. Bottom row: $\alpha = 0.7$.  }
\label{fig:FscoreLowSNR}
\end{figure*}

\begin{figure*}[h!]
\centering
\includegraphics[width=\textwidth]{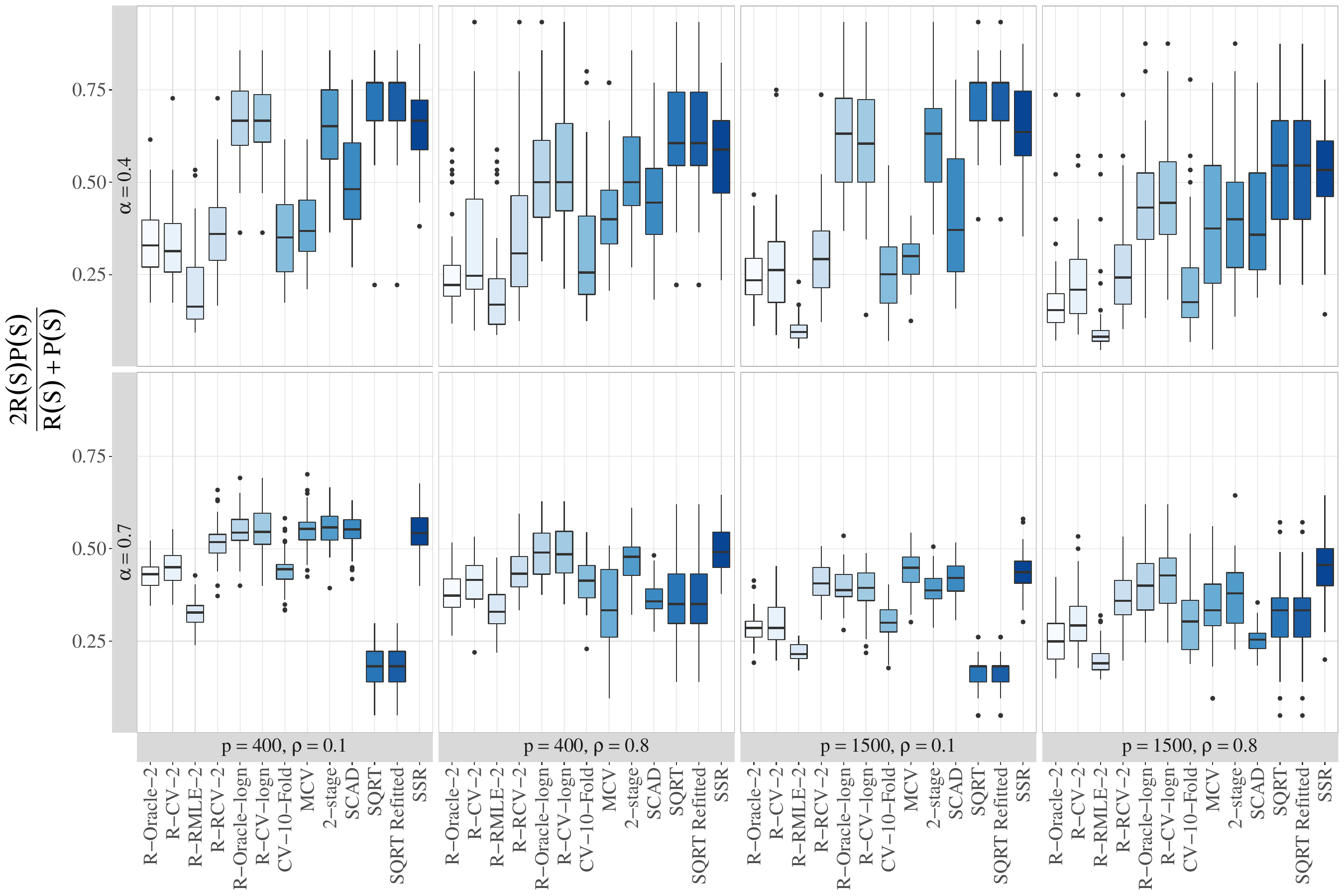}
\caption{ Comparison of F-score for $SNR = 10$. Top row:
  $\alpha = 0.4$. Bottom row: $\alpha = 0.7$.  }
\label{fig:FscoreHighSNR}
\end{figure*}

For the low SNR case (\autoref{fig:FscoreLowSNR}), no methods are consistently good.  For the high SNR case
(\autoref{fig:FscoreHighSNR}), the 2-stage method, SSR, and R-CV-logn work well across
all settings of $\alpha$ and $\rho$.  When $\bstar$ is sparse ($\alpha = 0.4$),
SQRT has good F-score performance, but it is one of the worst
when $\alpha$ is large. The performance of SCAD has similar
discrepancies:
it one of the best performers when $\rho = 0.1$ and one of the worst when $\rho = 0.8$.  
This is potentially useful because $\rho$ can be  estimated by
the data analyst before fitting the regression (as compared to the SNR
or sparsity which cannot). Thus, one could use SCAD in the
uncorrelated setting but avoid it when the design is highly correlated.
It is notable that for F-score 
in the high SNR case only, R-CV-logn and 2-stage outperform CV-10-Fold, R-CV-2, 
and R-RCV-2.

% \begin{figure}[t!]
% \centering
%  \resizebox{\linewidth}{!}{
%  \begin{tabular}{cccc}
%  \includegraphics[width=1.45in,trim=70 0 100 100,clip]{F1/F1__n200p400_NoiseTypeGaussian_Alpha4e-01_Snr1e-01_Rho1e-01_Sigma1e+00} & 
%  \includegraphics[width=1.45in,trim=70 0 100 100,clip]{F1/F1__n200p400_NoiseTypeGaussian_Alpha4e-01_Snr1e-01_Rho8e-01_Sigma1e+00}  &
%  \includegraphics[width=1.45in,trim=70 0 100 100,clip]{F1/F1__n200p1500_NoiseTypeGaussian_Alpha4e-01_Snr1e-01_Rho1e-01_Sigma1e+00} &
%  \includegraphics[width=1.45in,trim=70 0 100 100,clip]{F1/F1__n200p1500_NoiseTypeGaussian_Alpha4e-01_Snr1e-01_Rho8e-01_Sigma1e+00} \\
%  \includegraphics[width=1.45in,trim=70 0 100 100,clip]{F1/F1__n200p400_NoiseTypeGaussian_Alpha7e-01_Snr1e-01_Rho1e-01_Sigma1e+00} &
%  \includegraphics[width=1.45in,trim=70 0 100 100,clip]{F1/F1__n200p400_NoiseTypeGaussian_Alpha7e-01_Snr1e-01_Rho8e-01_Sigma1e+00} &
%  \includegraphics[width=1.45in,trim=70 0 100 100,clip]{F1/F1__n200p1500_NoiseTypeGaussian_Alpha7e-01_Snr1e-01_Rho1e-01_Sigma1e+00} &
%  \includegraphics[width=1.45in,trim=70 0 100 100,clip]{F1/F1__n200p1500_NoiseTypeGaussian_Alpha7e-01_Snr1e-01_Rho8e-01_Sigma1e+00} \\
%  $p = 400, \rho=0.1$ & $p = 400, \rho=0.8$ & $p = 1500, \rho=0.1$ & $p = 1500, \rho=0.8$ 
%  \end{tabular}
%  }
% \caption{ Comparison of F-score for $SNR = 0.1$. Top row: $\alpha =
%   0.4$. Bottom row: $\alpha = 0.7$.  }
% \label{fig:FscoreLowSNR}
% \end{figure}

% \begin{figure}[t!]
% \centering
%  \resizebox{\linewidth}{!}{
%  \begin{tabular}{cccc}
%  \includegraphics[width=1.45in,trim=70 0 100 100,clip]{F1/F1__n200p400_NoiseTypeGaussian_Alpha4e-01_Snr1e+01_Rho1e-01_Sigma1e+00} & 
%  \includegraphics[width=1.45in,trim=70 0 100 100,clip]{F1/F1__n200p400_NoiseTypeGaussian_Alpha4e-01_Snr1e+01_Rho8e-01_Sigma1e+00}  &
%  \includegraphics[width=1.45in,trim=70 0 100 100,clip]{F1/F1__n200p1500_NoiseTypeGaussian_Alpha4e-01_Snr1e+01_Rho1e-01_Sigma1e+00} &
%  \includegraphics[width=1.45in,trim=70 0 100 100,clip]{F1/F1__n200p1500_NoiseTypeGaussian_Alpha4e-01_Snr1e+01_Rho8e-01_Sigma1e+00} \\
%  \includegraphics[width=1.45in,trim=70 0 100 100,clip]{F1/F1__n200p400_NoiseTypeGaussian_Alpha7e-01_Snr1e+01_Rho1e-01_Sigma1e+00} &
%  \includegraphics[width=1.45in,trim=70 0 100 100,clip]{F1/F1__n200p400_NoiseTypeGaussian_Alpha7e-01_Snr1e+01_Rho8e-01_Sigma1e+00} &
%  \includegraphics[width=1.45in,trim=70 0 100 100,clip]{F1/F1__n200p1500_NoiseTypeGaussian_Alpha7e-01_Snr1e+01_Rho1e-01_Sigma1e+00} &
%  \includegraphics[width=1.45in,trim=70 0 100 100,clip]{F1/F1__n200p1500_NoiseTypeGaussian_Alpha7e-01_Snr1e+01_Rho8e-01_Sigma1e+00} \\
%  $p = 400, \rho=0.1$ & $p = 400, \rho=0.8$ & $p = 1500, \rho=0.1$ & $p = 1500, \rho=0.8$ 
%  \end{tabular}
%  }
% \caption{ Comparison of F-score for $SNR = 10$. Top row: $\alpha = 0.4$. Bottom row: $\alpha = 0.7$.  }
% \label{fig:FscoreHighSNR}
% \end{figure}

\subsubsection{Estimating the risk of the oracle linear model} 

Instead of using a risk estimate as a tool to empirically choose tuning parameters, sometimes it is important to 
directly estimate the risk of a
procedure to evaluate or compare its performance. In this subsection, we
investigate the risk estimation property of  
both $K$-fold CV and $\hat{R}(\hat\sigma,C_n)$ for a few choices of
$K$ and $\hat\sigma^2$.  As MCV, SSR, 2-stage, SCAD, and SQRT
are model selection/estimation procedures and not risk estimators, we leave them out of this comparison. 
The goal here is to determine whether 
equation~\eqref{eq:SURE} can yield good risk estimates in the
high-dimensional setting the same way that unbiased risk estimation
can in the low-dimensional setting. 
Hence, we set $C_n = 2/n$ as this would be the unbiased choice if
either $\sigma^2$ is known and $\dfHat$ is 
unbiased or $\hat\sigma^2$ is unbiased and $\dfHat$ doesn't depend on $Y$.

Using, $K$-fold CV or $\hat{R}$ to both choose $\hat\lambda$ and evaluate the risk $\hat\beta(\hat\lambda)$
conflates $\hat{R}$'s performance at tuning parameter selection and risk estimation.
Hence, for this evaluation only, 
%of Rather than using $\hat{R}$ to select the tuning
%parameter, then estimating $\beta$ and $\hat\sigma^2$, and then
%plugging the results into 
%equation \eqref{eq:ourGeneralProp}, 
we use as a $\bstar$-estimation procedure the oracle least squares estimator.
%abstract away from the model
%selection step and focus on risk estimation by using the oracle
%least squares estimator to calculate $\train_\beta$; 
That is, we set 
\[
\hat\beta_O = \argmin_\beta \norm{Y - \X_{\Sstar}\beta}_2^2
\]
and then calculate $\hat{R}_{\hat\beta_O}(\hat\sigma^2_{CV},2/n)$,
$\hat{R}_{\hat\beta_O}(\hat\sigma^2_{RCV},2/n)$, and
$\hat{R}_{\hat\beta_O}(\hat\sigma^2_{RMLE},2/n)$ where $\hat\sigma^2$
is estimated with the relevant high-dimensional variance estimator. We
also include 
2-Fold CV and 10-Fold CV.
This choice of $\bstar$ estimation procedure 
is still a function of the data, and hence is random, but it does not
require the selection of a tuning parameter. It should, however, be in a neighborhood of $\bstar$.

\begin{figure*}[t!]
\centering
\includegraphics[width=\textwidth]{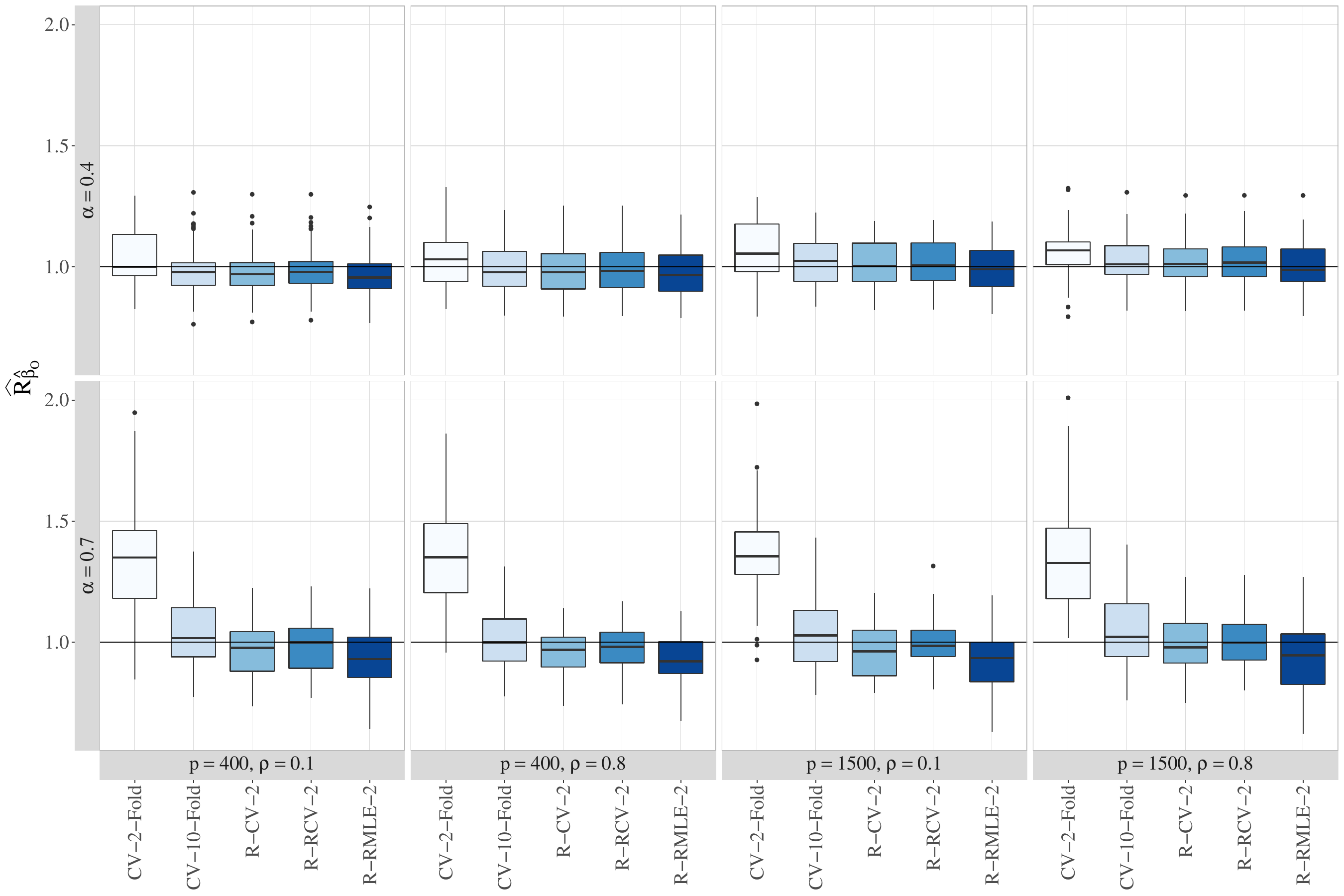}
\caption{ Comparison of risk estimation for $SNR = 0.1$. Top row:
  $\alpha = 0.4$. Bottom row: $\alpha = 0.7$.  }
\label{fig:riskSNR01}
\end{figure*}

\begin{figure*}[t!]
\centering
\includegraphics[width=\textwidth]{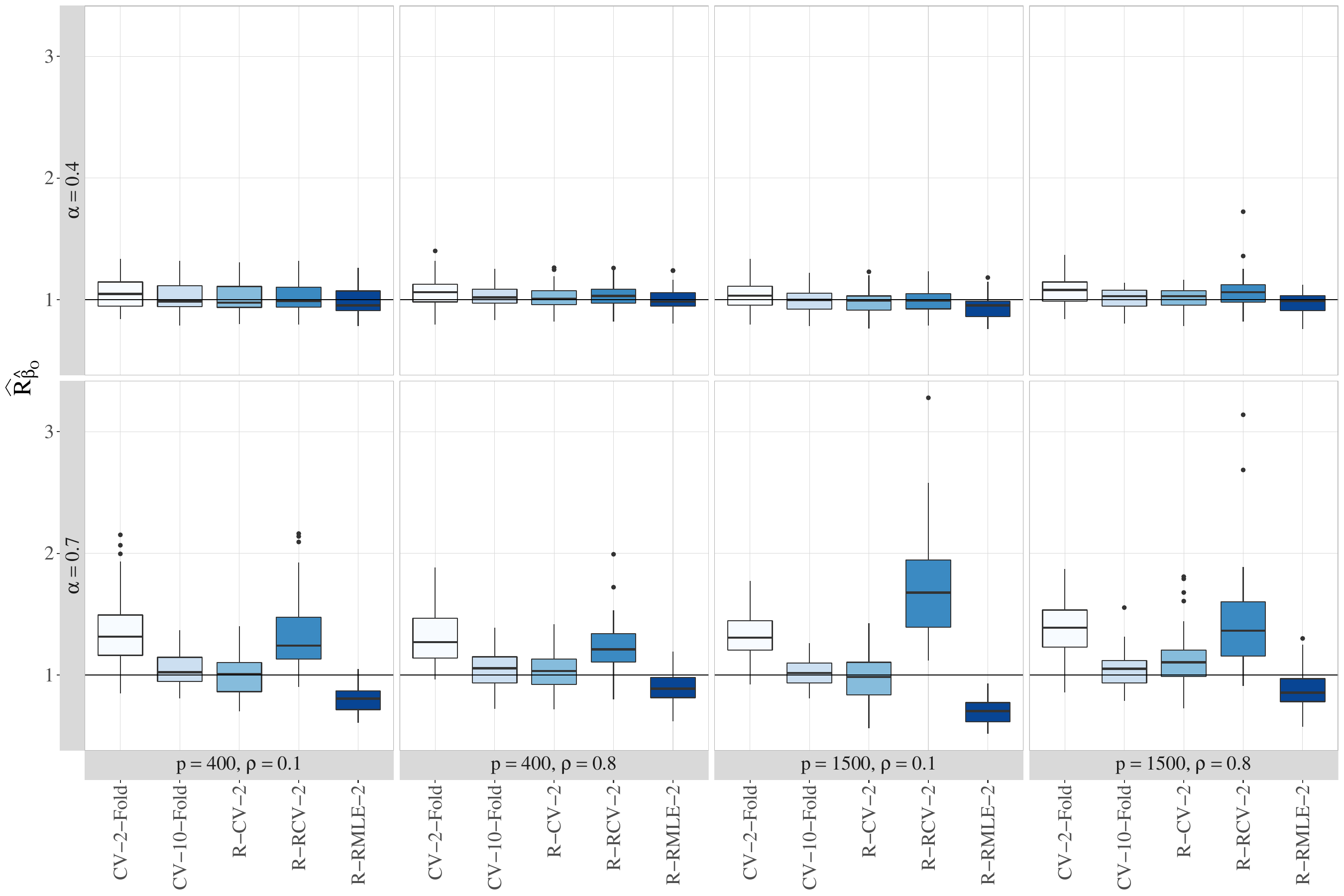}
\caption{ Comparison of risk estimation for $SNR = 10$. Top row:
  $\alpha = 0.4$. Bottom row: $\alpha = 0.7$.  }
\label{fig:riskSNR10}
\end{figure*}

We find that for sparse models (\autoref{fig:riskSNR01} and
\autoref{fig:riskSNR10}, top rows), there is very little difference
between these five procedures: all are unbiased on median, though
2-Fold  CV has slightly larger variance. However, with less sparse
models,
2-Fold CV greatly overestimates the risk, while
10-Fold CV is quite accurate. For high SNR and low sparsity, R-RCV-2
has a large upward bias, though it is otherwise quite accurate. For
another take, \autoref{tab:riskEstMSE} shows the squared
difference between the risk estimate and the true risk ($\sigma^2$ in all
cases), averaged across the simulation runs---the risk of the risk
estimator. Looking 
down the table for low SNR, \AICrcv\ is
the best method according to this metric, although for sparse models, 10-Fold CV and \AICcv\ are close behind in
terms of MSE. This is because the small negative bias of
\AICrcv\ is outweighed by the smaller variance it has relative to
10-fold CV and \AICcv, which are relatively unbiased. With
high SNR and dense models, \AICrcv\ is terrible with high positive bias and huge
variance, worse than even 2-Fold CV.  Note that \AICrcv\ uses a
version of 2-Fold CV to estimate $\sigma^2$.
Here, 10-Fold CV is easily the best, \AICcv\ has low bias, but relatively
large variance, while \AICrmle\ has a pronounced downward bias with
small variance.

\begin{table}[tb]
\caption{The root mean squared error of all five risk estimators. Bold
  values indicate the best method(s) (those within .005 of the
  minimum) in each case.}
% \resizebox{\linewidth}{!}{
\centering
\begin{tabular}{@{}ccccccccc@{}}
  \toprule
 snr & alpha & p & rho & CV-2-Fold & CV-10-Fold & R-CV-2 & R-RCV-2 & R-RMLE-2 \\ 
  \midrule
  0.1 & 0.4 & 400 & 0.1 & 0.125 & {\bf 0.108} & {\bf 0.104} & {\bf 0.105} & {\bf 0.104} \\ 
  0.1 & 0.4 & 400 & 0.8 & 0.118 & {\bf 0.105} & {\bf 0.105} & {\bf 0.104} &{\bf  0.106} \\ 
  0.1 & 0.4 & 1500 & 0.1 & 0.145 & {\bf 0.107} & {\bf 0.106} & {\bf 0.104} & {\bf 0.106} \\ 
  0.1 & 0.4 & 1500 & 0.8 & 0.128 & {\bf 0.104} & {\bf 0.102} & 0{\bf .104} & {\bf 0.103} \\ 
  0.1 & 0.7 & 400 & 0.1 & 0.447 & 0.147 & 0.131 & {\bf 0.119} & 0.158 \\ 
  0.1 & 0.7 & 400 & 0.8 & 0.405 & 0.122 & {\bf 0.112} & {\bf 0.107} & 0.132 \\ 
  0.1 & 0.7 & 1500 & 0.1 & 0.426 & 0.137 & 0.114 & {\bf 0.100} & 0.146 \\ 
  0.1 & 0.7 & 1500 & 0.8 & 0.413 & 0.150 & 0.118 & {\bf 0.109} & 0.156 \\ 
  10 &0.4 & 400 & 0.1 & 0.142 & 0.115 & {\bf 0.112} & 0.115 & {\bf 0.107} \\ 
  10 & 0.4 & 400 & 0.8 & 0.142 & 0.099 &{\bf  0.097} & {\bf 0.099} & {\bf 0.093} \\ 
  10 & 0.4 & 1500 & 0.1 & 0.117 & {\bf 0.101} & {\bf 0.103} & {\bf 0.099} & 0.111 \\ 
  10 & 0.4 & 1500 & 0.8 & 0.139 & {\bf 0.087} & {\bf 0.092} & 0.163 & {\bf 0.089} \\ 
  10 & 0.7 & 400 & 0.1 & 0.463 & {\bf 0.147} & 0.160 & 0.465 & 0.227 \\ 
  10 & 0.7 & 400 & 0.8 & 0.393 & {\bf 0.156} & 0.161 & 0.308 & 0.165 \\ 
  10 & 0.7 & 1500 & 0.1 & 0.371 & {\bf 0.110} & 0.192 & 0.831 & 0.313 \\ 
  10 & 0.7 & 1500 & 0.8 & 0.440 & {\bf 0.158} & 0.272 & 0.588 & 0.198 \\ 
\bottomrule
\end{tabular}

\label{tab:riskEstMSE}
\end{table}

The poor performance of CV-2-Fold and R-RCV-2 (for dense, high SNR conditions) deserves additional
comment. According to \cite[Figure 9]{ReidTibshirani2016},
the ability of  $\hat\sigma_{RCV}^2$ to estimate the variance
deteriorates with increasing SNR, which is in line with our simulations.
%(see \autoref{fig:varCompare}, for example).  
This is an area for
further investigation as neither we nor \cite{ReidTibshirani2016} can
provide a careful explanation for this phenomenon.  One possibility is
that splitting the data in 
half provides insufficient training data for accurate estimation and
one or two additional splits may be sufficient to remedy the issue. 
% \begin{figure}
% \includegraphics[width=1.1in]{./reviews/varCompare_n200p1500_NoiseTypeGaussian_Alpha4e-01_Snr1e-01_Rho1e-01_Sigma1e+00}
% \includegraphics[width=1.1in]{./reviews/varCompare_n200p1500_NoiseTypeGaussian_Alpha7e-01_Snr1e-01_Rho1e-01_Sigma1e+00}
% \includegraphics[width=1.1in]{./reviews/varCompare_n200p1500_NoiseTypeGaussian_Alpha4e-01_Snr1e+01_Rho1e-01_Sigma1e+00}
% \includegraphics[width=1.1in]{./reviews/varCompare_n200p1500_NoiseTypeGaussian_Alpha7e-01_Snr1e+01_Rho1e-01_Sigma1e+00}
% \caption{\attn{Fix me.  Perhaps we should just remove this and leave the sentence as ``.. in line with our simulations.''} Left to right: low alpha/low snr, high alpha/low snr, low alpha/high snr, and high alpha/high snr. }
% \label{fig:varCompare}
% \end{figure}

Another possible area for further investigation is the construction of
a confidence interval for the risk estimator. 
As cross-validation averages over $K$ folds in the training data, the variation 
of the prediction error on each fold can be used to form an informal confidence
interval for the risk.  This confidence interval can be useful in practice, for 
example when using the so-called ``one standard error rule'' 
\citep{FriedmanHastie2010}. The risk estimator in equation 
\eqref{eq:SURE} does not rely directly on subsampling and 
hence does not by default produce a confidence interval.  If the data 
 analyst desires such  an uncertainty estimate, a sensible, though
  computationally expensive, approach would be via the bootstrap.

\subsection{Data example: survival times for leukemia patients}

% \attn{Darren: Can you redo this over 10 replications with different
%   training test splits?}

We examine a microarray data set consisting of diffuse large B-cell lymphoma
(DLBCL) patients \citep{rosenwald2002use,bair2004semi}. This data set consists 
of measurements of 7399 genes made on 160 training patients and 80 test 
patients, matching the training and test split used by \cite{bair2004semi}.  
The  response, $Y$, is the survival time for each patient which we
transform as $\log(Y + 1)$ due to skewness.

\begin{figure*}
\centering
\includegraphics[width=2.25in]{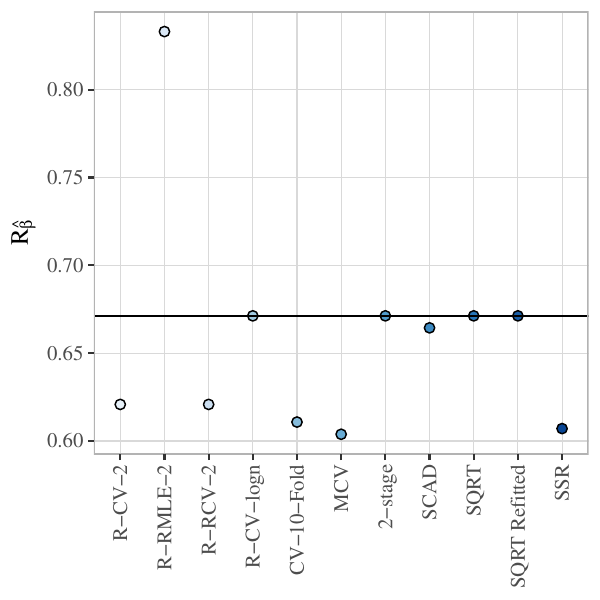}
\includegraphics[width=2.25in]{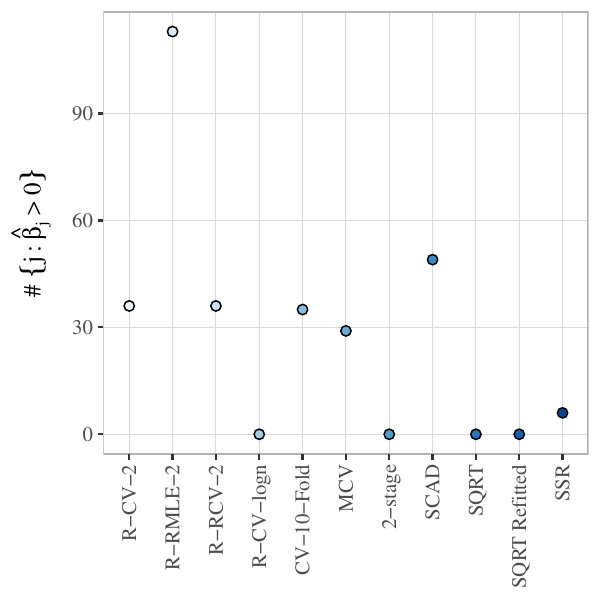}
\caption{Analysis of leukemia patient survival times.  Left plot: prediction risk on test data.  The horizontal line indicates the risk of
the identically zero estimator $\hat\beta(\lambda) \equiv 0$.  Right plot: number of selected genes
for methods that report $\hat\beta(\lambda) \neq 0$.}
\label{fig:realData}
\end{figure*}

Our results, which can be found in \autoref{fig:realData} (left plot), are that many of
the tuning parameter selection methods choose $\hat\lambda$ such
that $\hat\beta(\hat\lambda) \equiv 0$; that is, the identically zero vector.  
CV-10-Fold, MCV, R-CV-2, R-RCV-2, SSR, and SCAD produce non-trivial coefficient estimates
that improve on the risk of the zero estimator while R-RMLE-2 produces 
a nontrivial coefficient estimate that is much worse than the zero estimator.
For reference, the variance estimators $\hat\sigma^2$ are approximately
0.23, 0.68, and 0.69 for $\hat\sigma_{RMLE}^2$, $\hat\sigma_{CV}^2$, 
and $\hat\sigma_{RCV}^2$, respectively.  Additionally, each method suggests
dramatically different numbers of selected genes (\autoref{fig:realData}, right plot), ranging from 6 for SSR
to 116 for R-RMLE-2.  The intersection of the selected models for
those methods which produce nontrivial
coefficient estimates are genes 3822 and 4131, which may be reasonable candidates for
further investigation.

\section{Theoretical analysis}
\label{sec:theoretical-results}

In this section, we provide a result demonstrating that, under a
number of standard conditions, our risk estimator will produce a
predictor whose performance is comparable to that of the true
model. For convenience, we define  $x_j$ to be the
$j^{th}$ column of $\X$  
and $X_{ij}$ to be the $i,j$ entry of $\X$.  Also,
let $\S \subseteq \{1,2,\ldots,p\}$ be an index set with $|\S|$ elements
and define $\S^c$ to be complement: $\S^c := \{1,2,\ldots,p\} \setminus \S$.

We define the following conditions.  
\begin{enumerate}[align=left,label=\textbf{Condition~\arabic*.},ref=Condition~\arabic*]
\item \label{cond:linear-model} Assume that $Y = \X\bstar + \epsilon$
\item \label{cond:sub-Gaussian-noise} The $\epsilon_i$ are distributed i.i.d 
  sub-Gaussian with variance $\sigma^2$. That is $\forall t\in\R$,
  $\E[\exp(t\epsilon_i)]\leq \exp\{\sigma^2t^2/2\}$. 
\item \label{cond:design} The design matrix $\X$ satisfies
 $C_X:= \max_{1\leq j \leq p}\norm{x_j}_2^2/n =O(1)$ for all $p$. 
% \attn{We
%    need something here so that as $p$ increases, this doesn't blow
%    up. Imposing this condition is a
%    bit weaker, but simplifies the discussion relative to $s_*$ and
%    $\lambda_{\max}$. (Also makes the ugly a bit less so.) Otherwise,
%    we'd have to include some talk about how $C_X(n,p)$ can behave with
%    $p$.}
% \item \label{cond:model-size} Suppose $\sstar = o(\sqrt{n/\log
% (p)})$.
\item \label{cond:comp-const}  If $\beta \in \R^p$ is such that 
 $\norm{\beta_{\S^c}}_1 \leq
  L\norm{\beta_{\S}}_1$,  for some $L \geq 0$, then

  \[
  \norm{\beta_{\S}}_1^2 \leq \frac{|\S|}{\phi^2n}\beta^\top\X^\top\X\beta,
  \]
  where $\phi \equiv \phi(L) > 0$ is known as a compatibility constant.
\end{enumerate}

  %  \attn{And possibly
  % the additional condition that $\dfHat(\lambda_{\min}) \leq a_n$}.

These conditions are well-known and appear frequently in lasso-related theoretical
results.  We assume that the data is actually generated by a linear
model, as was the case in our simulated analysis. We assume
homoscedastic noise which has reasonable tails. Gaussian distributions
satisfy \ref{cond:sub-Gaussian-noise} as well as bounded distributions
and other standard ``light-tailed'' distributions. Our goal will be to
consider the standard high dimensional setting where $p\gg n$ and both
approach infinity. Because of this, we need to ensure that as we add
columns to the design matrix, larger and larger entries do not come to
dominate the solution. \ref{cond:design} says that the maximum column norm
grows like its length but not with $p$. This condition can be
eliminated without any difficulty, but it allows for easier
interpretation of the result. Finally, we assume that the design
matrix satisfies the so-called ``compatibility
condition''~\citep{GeerBuhlmann2009}. This allows us to relate the
$\ell_1$-norm of the coefficient vector with the $L_2$-norm of the
predicted values for a collection of sufficiently sparse coefficient
vectors. This condition is also related to the
restricted eigenvalue
condition~\citep{BickelRitov2009,BuneaTsybakov2007} which is an alternative.

We state the core result, showing an upper bound on the prediction
loss of the lasso with tuning parameter chosen by $\hat{R}$ versus the
true coefficient vector $\beta_*$.    Set $\Lambda = [\lambda_{\min},\ \lambda_{\max}]$
to be the optimization grid for the tuning parameter $\lambda$.
\begin{theorem}

%$\lambda_{\min} \leq \EPconst$ and assume
%  \[
%  \lambda_n=4\sigma\max_{1\leq j \leq
%    p}\norm{x_j}_2\sqrt{\frac{t^2+2\log p}{n}}\geq 2\EPconst.
%  \]
Assume  \ref{cond:linear-model}--\ref{cond:comp-const}.  Let $\delta >
0$ and $\Lambda =[\lambda_{\min},\ \lambda_{\max}]$.
Set $\lambda_{\min}=2\sigma \sqrt{ \frac{2C_X \left(
    \log(p) + \delta\right)}{n}}$.
Then, with probability at least $1-2e^{-\delta}$,
  \begin{align*}
  \frac{1}{n}\norm{\X\bstar - \X\hat\beta(\hat\lambda)}_2^2 
  & \leq
    \left(  \frac{2\sstar}{\phi^2}\right) \left(9\lambda_{\max}^2 +\frac{8\sigma^2 C_X \left(
    \log(p) + \delta\right)}{n}\right)\\
    &\quad+ \rho\left(4\sigma \sqrt{ \frac{2C_X \left(
    \log(p) + \delta\right)}{n}}\right).
  % \left(  \frac{\sstar}{\phi^2}\right) \left(
  %   \lambda\left(\frac{8\max_j\norm{x_j}_2^2\sigma^2}{n^2} \left(
  %   \log(2p) - \log(\delta)\right) \right)^{-1/2} \right) \cdot \\ 
  % & \qquad \left( 2\lambda_{\max}^2 +
  %   \left(\frac{8\max_j\norm{x_j}_2^2\sigma^2}{n^2} \left( \log(2p) -
  %   \log(\delta)\right) \right) \right) + \rho(\lambda). 
%  \frac{4s_*}{\phi^2} \left( \lambda_{\max} \left(\frac{8\norm{x_j}_2^2\sigma^2}{n^2} \left( \log(2p) - \log(\delta)\right) \right)^{1/2}+\lambda^2 \right) + \rho(\lambda).
  \end{align*}
  \label{thm:main}
\end{theorem}
%\left(\frac{\sstar}{2\phi^2}\right) \left(\frac{   \lambda }{M} \right)\left(4\lambda_{\max}^2 + 2\EPconst^2 \right)
%+ \rho(\lambda).

The first part of the upper bound depends on $\lambda_{\max}$. The
second part depends on the penalty $\rho$.
Results for the lasso with oracle tuning parameter deal only with an
upper bound that looks like
\[
 \frac{2\sstar}{\phi^2} \frac{8\sigma^2 C_X \left(
    \log(p) + \delta\right)}{n}.
\]
Therefore, for convergence, they examine the case where $s_*$ goes to
infinity as fast as possible.  Thus, the lasso ``works'' as long as
$s_*=o(n/\log(p))$. For our bound to be meaningful when $s_*$ grows
this quickly, we must have
$\lambda_{\max} =O(\sqrt{\log(p)/n})$, the same order as
$\lambda_{\min}$.  That is,  if $s_*$ grows as fast as
possible, we get a trivial $\Lambda$ interval with the upper
and lower bounds having the same order (though they can differ by an
arbitrary constant). If, instead,
$s_*$ is constant, hence growing as slowly as possible, then
we simply need $\lambda_{\max}=o(1)$.

Finally, we require $\rho\left(4\sigma \sqrt{ \frac{2C_X \left(
    \log(p) + \delta\right)}{n}}\right)$ to go to zero at a similar
rate. This of course depends on the penalty selected. For AIC, we require
\[
\rho\left(4\sigma \sqrt{ \frac{2C_X \left(
    \log(p) + \delta\right)}{n}}\right) = \frac{2}{n}\hat{\sigma}^2\hat{\df}\left(4\sigma \sqrt{ \frac{2C_X \left(
    \log(p) + \delta\right)}{n}}\right).
\]
Thus, if $\hat{\sigma}^2=O(1)$, then 
\[
\hat{\df}\left(4\sigma \sqrt{ \frac{2C_X \left(
    \log(p) + \delta\right)}{n}}\right) = o(n)
\]
is sufficient. In particular, for $\hat\df = O(s_*)$ gives convergence.

% \attn{old corollary}
% \begin{corollary}
%   Set $\Lambda_n = [c_0(\log(p)/n)^{-1/2},\ c_1(\log(p)/n)^{-1/4}]$ for
%   some constants $c_0$ and $c_1$ such that $c_0(\log(p)/n)^{-1/2}\geq
%   2\EPconst$. Then under
%   \ref{cond:linear-model}--\ref{cond:model-size},
%   \[
%   \frac{1}{n}\norm{\X\bstar - \X\hat\beta(\hat\lambda)}_2^2 = o_{\mathbb{P}}(1).
%   \]
% \end{corollary}

%\begin{corollary}
%  Set $\Lambda_n = [c_0(\log(p)/n)^{-1/2},\ c_1(\log(p)/n)^{-1/4}]$ for
%  some constants $c_0$ and $c_1$ such that $c_0(\log(p)/n)^{-1/2}\geq
%  2\EPconst$. Then under
%  \ref{cond:linear-model}--\ref{cond:model-size},
%  \[
%  \frac{1}{n}\norm{\X\bstar - \X\hat\beta(\hat\lambda)}_2^2 = o_{\mathbb{P}}(1).
%  \]
%\end{corollary}

\section{Discussion}
\label{sec:discussion}
In this paper, we investigate
a large number of procedures for selecting $\lambda$ 
in high-dimensional lasso problems. Our results supplement and elaborate
upon those of \cite{FlynnHurvich2013} which apply to the low-dimensional
setting ($p<n$). In general, the
unbiased-risk-estimation methods we present perform consistently well across
conditions. They exhibit many of the familiar properties from the
AIC-vs.-BIC debate (BIC selects smaller models, AIC is better for
prediction) as well as some variation across variance estimators due
to estimation bias. Our simulations lead us to suggest a novel
two-stage method (see \autoref{sec:modif-lasso-crit} and \autoref{alg:2-stage})
that also performs consistently well and warrants further theoretical investigations. 

Substantial theory exists for the optimal choice of the tuning parameter for the lasso and related
methods.  These results, however,
depend both on unknown properties of the data generating process and unknown constants. 
Though there are many data-dependent methods for 
choosing the tuning parameters, there is a distinct lack of guidance in the literature about 
which method to use.  This uncertainty is even more pronounced when faced with high-dimensional
data where $p\gg n$.

We give examples that show that one commonly advocated approach,
a generalized information criterion which has desirable theoretical
properties in low dimensions, would necessarily choose the unregularized
model with $\lambda = 0$
when $p>n$.  Therefore, we propose a risk
estimator motivated by Stein's unbiased risk estimation.  This estimator
requires three ingredients: an estimate of the degrees of freedom ($\dfHat$), a
constant that may depend on $n$ ($C_n$), and an estimator of the variance ($\hat\sigma^2$).
While the degrees of freedom for the lasso problem is well understood, the other two
choices are much less so.  In particular, high-dimensional variance estimation is a difficult problem 
in its own right.

\subsection{Overall recommendations}

In general, CV-10-Fold performs similarly to R-CV-2, which tends to
outperform both R-RCV-2 and R-CV-logn.  A notable exception 
is that R-CV-logn dramatically outperforms for model selection when in
the high SNR regime.  In all other cases, 
both CV-10-Fold and R-CV-2 should perform satisfactorily in practice
relative to the other methods we examine. 

For the oracle risk estimation methods, R-oracle-2 and R-oracle-logn,
$\hat\sigma_{CV}^2$ is a good estimator of $\sigma^2$  in practice  
and hence R-CV-2 and R-CV-logn behave very similarly to R-oracle-2 and R-oracle-logn, respectively.  
However, the variance estimator $\hat\sigma_{RMLE}^2$ tends to dramatically underestimate $\sigma^2$
and hence R-RMLE-2 tends to under-regularize.  Also,
though MCV performs the best on the genetics data set, it
performed very poorly in the simulations.  Hence,  R-RMLE-2 and MCV
should be avoided in practice.

SCAD performs well for both prediction risk and consistency, particularly when $p$ is large and the
true model is not sparse.  On the other hand, 
SQRT refitted performs substantially better than SQRT and hence should be used as an additional step to 
SQRT in practice.  However, SQRT refitted tends to underperform the other methods in our simulations.

In general, the SURE-based methods we develop perform
quite well across different simulation conditions and evaluation metrics. The
2-stage method described in \autoref{sec:simulationMethods} also performs
well and warrants further investigation. Standard 10-fold
CV performs adequately while the behavior of  scaled-sparse regression,
\SQRT{} variants, and MCV depends strongly on the
simulation condition. In particular, these modern methods often
underperform the SURE-based methods presented in this paper.

\section*{Funding} 
Darren Homrighausen is supported by the National Science
Foundation under grant DMS--1407543 and the Institute for New
Economic Thinking; under grant INO14-00020. Daniel J.\ McDonald is
supported by the National Science Foundation under grant DMS--1407439 and the Institute for
New Economic Thinking under grant INO14-00020.

\section*{Biographical note}

Darren Homrighausen is Assistant Professor of Statistics at Colorado
State University. He has a bachelors degree from the University of
Colorado in economics and math, and a and Ph.D.\ from Carnegie Mellon
University in statistics. He has worked extensively on developing both
methods and theory for solving various problems in astronomy and
cosmology as well as investigating the prediction risk implications of
empirical tuning parameter selection for lasso-type methods. More
recently, he has become interested in examining the statistical
implications of computational approximations. \\

\noindent Daniel J. McDonald is Assistant Professor of Statistics and Adjunct
Assistant Professor of Computer Science at Indiana University,
Bloomington. His research interests involve the estimation and
quantification of prediction risk, especially developing methods for
evaluating the predictive abilities of complex dependent data. This
includes the application of statistical learning techniques to time
series prediction problems in the context of economic forcasting, as
well as investigations of cross-validation and the bootstrap for risk
estimation.

\bibliographystyle{mybibsty}
\bibliography{csdaRefs}

\appendix
\section{Proof of \autoref{thm:main} and supporting results}
\label{sec:appendix}

\begin{lemma}[Generalization of \citep{BuhlmannGeer2011}, Lemma 6.2]
  \label{lem:sub-gaussian}
  Define
  \[
  \mathcal{G} = \left\{ \max_{1\leq j\leq p} 2|\epsilon^\top x_j|
    / n < \EPconst\right\}.
  \]
   Suppose \ref{cond:sub-Gaussian-noise} holds. For any $\delta>0$, if 
  \[
  \EPconst := 2\sigma \sqrt{ \frac{2C_X \left(
    \log(p) + \delta\right)}{n} } 
  \]
  then
  \[
  \mathbb{P}(\mathcal{G}) \geq 1-2e^{-\delta}.
  \]
\end{lemma}
\begin{proof}
  Define $x_j$ to be the $j^{th}$ column of $\X$ and
  recalling that $X_{ij}$ is the $j^{th}$ entry of the $i^{th}$
  covariate vector.   Define \[
  Z_j := \frac{2\epsilon^\top x_j }{n}.
  \]
  Let $t \geq 0$ be given.  Then,
  under \ref{cond:sub-Gaussian-noise}, we have
  \begin{align*}
    \E\left[\exp(tZ_j)\right]
    &= \prod_{i=1}^n
      \E\left[\exp\left(\frac{2t\epsilon_iX_{ij}}{n}\right)\right]
    \leq \prod_{i=1}^n
      \exp\left(\frac{4t^2\sigma^2X^2_{ij}}{2n^2}\right)
    = \exp \left(\frac{2t^2\sigma^2}{n^2} \norm{x_j}_2^2\right).
  \end{align*}
  Therefore,
  \begin{align*}
  1 - \mathbb{P} \left(  \mathcal{G}\right)  
  &  =
    \mathbb{P}\left(\max_j |Z_j| \geq \EPconst\right)  \\
    & \leq \sum_j \mathbb{P}\left( |Z_j| \geq \EPconst\right) \\
    &   \leq p\max_j \mathbb{P}\left(|Z_j| \geq \EPconst\right)\\
    & \leq 2p\max_j\inf_t \exp(-t\EPconst) \exp \left(\frac{2t^2\sigma^2}{n^2}
      \norm{x_j}_2^2\right)\\ 
    & = 2p\inf_t \exp(-t\EPconst) \exp \left(\frac{2t^2\sigma^2}{n^2}
      \max_j\norm{x_j}_2^2\right)\\ 
    &= 2p\exp\left\{ -\frac{n^2\EPconst^2}{8 \sigma^2\max_j\norm{x_j}_2^2}\right\}.
  \end{align*}
  Thus, for any $\delta > 0$, if we set 
  \[
  \EPconst:= \sqrt{ \frac{8\max_j\norm{x_j}_2^2\sigma^2}{n^2} \left( \log(p) - \log(\delta)\right) }
  \]
  then
  \[
    \mathbb{P} \left(  \mathcal{G}\right)  \geq 1 - 2\delta.
  \]
  Redefine $\delta\rightarrow e^{-\delta}$ and use
  $C_X\geq n^{-1}\max_j\norm{x_j}_2^2$ to get the result.
\end{proof}

\begin{lemma}
  \label{lem:new-basic-inequality}
  Define $\hat\lambda$ as in equation~\eqref{eq:lambda-hat-definition}. 
  Set
  $\rho(\lambda)=C_n\hat{\sigma}^2\dfHat(\lambda)$.
  Then for any $\lambda \geq 0$,
  \[
  \frac{1}{n}\norm{\X\bstar - \X\hat\beta(\hat\lambda)}_2^2 + \lambda\norm{\hat\beta(\lambda)}_1
  \leq  \frac{2}{n}\epsilon^\top\X(\bstar-\hat\beta(\hat\lambda)) + \lambda\norm{\bstar}_1 + \rho(\lambda)
  \]
\end{lemma}
\begin{proof}
  \begin{align*}
     \frac{1}{n} \norm{Y - \X\hat\beta(\hat\lambda)}_2^2 + \lambda\norm{\hat\beta(\lambda)}_1
    & \leq
        \frac{1}{n}\norm{Y - \X\hat\beta(\hat\lambda)}_2^2 + \rho(\hat\lambda) +
      \lambda\norm{\hat\beta(\lambda)}_1 \\ 
    & \leq
        \frac{1}{n}\norm{Y - \X\hat\beta(\lambda)}_2^2 + \rho(\lambda) +
      \lambda\norm{\hat\beta(\lambda)}_1 \\ 
    & \leq
        \frac{1}{n}\norm{Y - \X\bstar}_2^2 + \rho(\lambda) +
      \lambda\norm{\bstar}_1.
  \end{align*}
  Here we have used the fact that $\hat\lambda$ minimized $n^{-1}\norm{Y -
    \X\hat\beta(\lambda)}_2^2 + \rho(\lambda)$ and
  $\hat\beta(\lambda)$ minimized       $n^{-1}\norm{Y - \X\beta}_2^2 + \lambda\norm{\beta}_1$.
  Using $Y=\X\bstar + \epsilon$ gives
  \begin{align*}
    &\norm{Y - \X\hat\beta(\hat\lambda)}_2^2 =  \norm{\X\bstar +\epsilon - \X\hat\beta(\hat\lambda)}_2^2  \\
    &= \norm{\epsilon}_2^2 + \norm{\X(\bstar-\hat\beta(\hat\lambda))}_2^2
      +2\epsilon^\top\X(\bstar-\hat\beta(\hat\lambda))
  \end{align*}
  while $\norm{Y - \X\bstar}_2^2 = \norm{\epsilon}_2^2$. Therefore,
\begin{align*}
    \frac{1}{n}\norm{\X\bstar - \X\hat\beta(\hat\lambda)}_2^2 +
  \lambda\norm{\hat\beta(\lambda)}_1  
  & \leq
   \frac{2}{n}\epsilon^\top\X(\bstar-\hat\beta(\hat\lambda))
    +\rho(\lambda) + \lambda\norm{\bstar}_1. 
\end{align*}
\end{proof}

\begin{lemma}[Generalization of \cite{BuhlmannGeer2011}, Theorem 6.1]
\label{lem:ell1normBound}
  Suppose \ref{cond:linear-model} and \ref{cond:comp-const} hold. Then on
  $\mathcal{G}$, for any 
  $\lambda >   \EPconst$,
  \[
\norm{\hat\beta(\lambda) - \bstar}_1 
\leq \frac{\sstar ( 3\lambda + \EPconst )^2}{4(\lambda -   \EPconst)\phi^2}.
  \]
\end{lemma}
\begin{proof}
  
  Note that $\bstar=0$ on $\Sstar^c$. Then, by the triangle
  inequality, we have,
  \begin{align}
    \norm{\hat\beta(\lambda)}_1 
    &\geq \norm{\hat\beta_{\Sstar^c}(\lambda)}_1 -
      \norm{\hat\beta_{\Sstar}(\lambda) - \bstar}_1 + \norm{\bstar}_1.\label{eq:reverse-triangle}
  \end{align}
  Therefore, on $\mathcal{G}$ for any $\lambda \geq 0$,
  \begin{align*}
    &\frac{1}{n}\norm{\X(\hat\beta(\lambda)-\bstar)}_2^2 + \lambda\left(
      \norm{\hat\beta_{\Sstar^c}(\lambda)}_1 - 
      \norm{\hat\beta_{\Sstar}(\lambda) - \bstar}_1 + \norm{\bstar}_1\right) \\
    &\leq \frac{1}{n}\norm{\X(\hat\beta(\lambda)-\bstar)}_2^2 +
      \lambda\norm{\hat\beta(\lambda)}_1,\\
    &\leq \frac{2}{n}
      \epsilon^\top\X(\hat\beta(\lambda)-\bstar) +
      \lambda\norm{\bstar}_1\\
    &\leq \EPconst \norm{\hat\beta(\lambda)-\bstar}_1  +
      \lambda\norm{\bstar}_1 \\
    &=  \EPconst \norm{\hat\beta_{\Sstar}(\lambda) - \bstar}_1 +
      M\norm{\hat\beta_{\Sstar^c}(\lambda)}_1 +  \lambda\norm{\bstar}_1,
  \end{align*}
  where the first inequality is due to equation \eqref{eq:reverse-triangle}
  and 
  the second and third follow from
  \autoref{lem:new-basic-inequality}. The final equality follows by
  noting that 
  \[
  \norm{\hat\beta(\lambda) - \bstar}_1 =
  \norm{\hat\beta_{\Sstar}(\lambda) - \bstar}_1 +
  \norm{\hat\beta_{\Sstar^c}(\lambda)}_1.
  \]
  Collecting terms shows that 
  \begin{equation}
    \frac{1}{n}\norm{\X(\hat\beta(\lambda)-\bstar)}_2^2 + (\lambda -
    \EPconst)\norm{\hat\beta_{\Sstar^c}}_1  
    \leq 
    (\lambda + \EPconst)\norm{\hat\beta_{\Sstar} - \beta_*}_1.
    \label{eq:important}
  \end{equation}
  By using the above inequality twice, we see that
  \begin{align*}
    &\frac{1}{n}\norm{\X(\hat\beta(\lambda)-\bstar)}_2^2 +
      (\lambda-\EPconst)\norm{\hat\beta(\lambda) - \bstar}_1\\
    &\leq \frac{1}{n}\norm{\X(\hat\beta(\lambda)-\bstar)}_2^2 + 
      (\lambda-\EPconst)\norm{\hat\beta_{\Sstar}(\lambda) - \bstar}_1 +
    (\lambda + \EPconst)\norm{\hat\beta_{\Sstar} - \beta_*}_1 \\
    & = \frac{1}{n}\norm{\X(\hat\beta(\lambda)-\bstar)}_2^2 + 
      2\lambda\norm{\hat\beta_{\Sstar}(\lambda) - \bstar}_1 \\
    &\leq
      (\lambda+\EPconst)\norm{\hat\beta_{\Sstar}(\lambda)- \bstar}_1 +
      2\lambda\norm{\hat\beta_{\Sstar}(\lambda) - \bstar}_1 \\
    &= (3\lambda + \EPconst) \norm{\hat\beta_{\Sstar}(\lambda)-
      \bstar}_1
    \end{align*}
By equation \eqref{eq:important}, 
$\norm{\hat\beta_{\Sstar^c}} \leq (\lambda + \EPconst)(\lambda - \EPconst)^{-1}\norm{\hat\beta_{\Sstar} - \beta_*}_1$
and hence \ref{cond:comp-const} with $L = (\lambda + \EPconst)(\lambda - \EPconst)^{-1}$ applies.  Also,
observe that $uv\leq u^2/4 + v^2$.  Therefore,
\begin{align*}
(3\lambda + \EPconst) \norm{\hat\beta_{\Sstar}(\lambda)- \bstar}_1
      & \leq
(3\lambda + \EPconst) \left(\frac{\sqrt{\sstar}}{\phi \sqrt{n}}\right)\norm{\X(\hat\beta(\lambda)- \bstar)}_2 \\
& \leq
\left(\frac{(3\lambda + \EPconst) ^2\sstar}{4\phi^2 }\right) + \frac{1}{n}\norm{\X(\hat\beta(\lambda)- \bstar)}_2^2.
\end{align*}
Rearranging produces the desired result as long as $\lambda> M$.
%\label{cond:comp-const}  If $\beta \in \R^p$ is such that 
% $\norm{\beta_{\S^c}}_1 \leq
%  L\norm{\beta_{\S}}_1$,  for some $L \geq 0$, then
%
%  \[
%  \norm{\beta_{\S}}_1^2 \leq 
%  \]
%  where $\phi \equiv \phi(L) > 0$ is known as a compatibility constant.
%

%    &\leq \sqrt{\sstar}(3\lambda + \EPconst)
%      \norm{\hat\beta_{\Sstar}(\lambda)- 
%      \bstar}_2\\
%    &\leq \sqrt{\frac{\sstar}{n}} \frac{3\lambda + \EPconst}{\phi_0}
%      \norm{\X( \hat\beta(\lambda) - \bstar)}_2\label{eq:comp}\\
%    &=\sqrt{\frac{2\sstar}{n}} \frac{3\lambda + \EPconst}{\phi_0}
%      \frac{\norm{\X( \hat\beta(\lambda) - \bstar)}_2}{\sqrt{2}}\\
%    &\leq \frac{\sstar(3\lambda+\EPconst)^2}{2\phi_0^2} +
%      \frac{1}{2n}\norm{\X( \hat\beta(\lambda) - \bstar)}^2_2\label{eq:quad-thing}.
%  \end{align}
%  Here equation \eqref{eq:prev-lemma} follows from \autoref{lem:basic-ineq-2},
%  equation \eqref{eq:comp} follows from \ref{cond:comp-const}, and
%  equation \eqref{eq:quad-thing} follows because $uv\leq u^2/4 +
%  v^2$. The result follows by subtracting $\frac{1}{2n}\norm{\X(
%    \hat\beta(\lambda) - \bstar)}_2$ from both sides and multiplying
%  by 2.
\end{proof}

\begin{proof}[Proof of \autoref{thm:main}]
  On the set $\mathcal{G}$, 
  \[
  \frac{2}{n}\epsilon^\top\X(\bstar-\hat\beta(\hat\lambda)) 
  <
  \EPconst\norm{\hat{\beta}(\hat{\lambda})-\bstar}_1.
  \]
By \autoref{lem:new-basic-inequality} and  \autoref{lem:ell1normBound} for any $\lambda > M$
  \begin{align*}
  \frac{1}{n}\norm{\X\bstar - \X\hat\beta(\hat\lambda)}_2^2
    & <
    \EPconst\norm{\hat{\beta}(\hat{\lambda})-\bstar}_1
      + \lambda\norm{\bstar}_1-\lambda\norm{\hat\beta(\lambda)}_1 +
      \rho(\lambda)\\
    &\leq \EPconst\sup_{\lambda' \in\Lambda}\norm{\hat{\beta}(\lambda')-\bstar}_1
      + \lambda\norm{\bstar-\hat\beta(\lambda)}_1 +
      \rho(\lambda)\\
    &\leq (\EPconst+\lambda)\sup_{\lambda' \in\Lambda}\norm{\hat{\beta}(\lambda')-\bstar}_1
      +
      \rho(\lambda)\\
    &\leq 2\lambda\sup_{\lambda' \in\Lambda}\norm{\hat{\beta}(\lambda')-\bstar}_1
      +
      \rho(\lambda)\\
          &\leq  
   2\lambda\sup_{\lambda' \in\Lambda}  \frac{\sstar ( 3\lambda' + \EPconst )^2}{4(\lambda' -   \EPconst)\phi^2}
+ \rho(\lambda) \\
& \leq \left(\frac{\sstar}{2\phi^2}\right) \left(\frac{   \lambda ( 3\lambda_{\max} + \EPconst )^2}{M}\right)
+ \rho(\lambda) \\
& \leq \left(\frac{\sstar}{\phi^2}\right) \left(\frac{   \lambda }{M} \right)\left(9\lambda_{\max}^2 + \EPconst^2 \right)
+ \rho(\lambda).
%    &\leq  
%    \EPconst\sup_{\lambda' \in\Lambda}  \frac{\sstar ( 2\lambda' + \EPconst )^2}{4(\lambda' -   \EPconst)\phi^2}
%+ \lambda\frac{\sstar ( 2\lambda + \EPconst )^2}{4(\lambda -   \EPconst)\phi^2}
%+ \rho(\lambda) \\
%& \leq  \frac{\sstar ( 2\lambda_{\max} + \EPconst )^2}{\phi^2}
%+ \lambda\frac{\sstar ( 2\lambda + \EPconst )^2}{4\EPconst\phi^2}
%+ \rho(\lambda).
  \end{align*}
  % \attn{I intentionally made this result weaker so that the theorem is
  %   simpler to state. Why not also take $\lambda=2M$ so that we worry
  %   about $\rho(2M)$ and dump the $\lambda/M$ term?}
Where for this last inequality we use that $\lambda_{\min} =
2M$. Finally, since this inequality holds for all $\lambda>M$ and
$\rho(\lambda)$ is decreasing in $\lambda$, we take $\lambda=2M$.
\end{proof}

\section{Supplementary graphics}

\hypertarget{prediction-risk-figures}{%
\subsection{Prediction risk figures}\label{prediction-risk-figures}}

\begin{center}
\includegraphics[width=6in,height=4in]{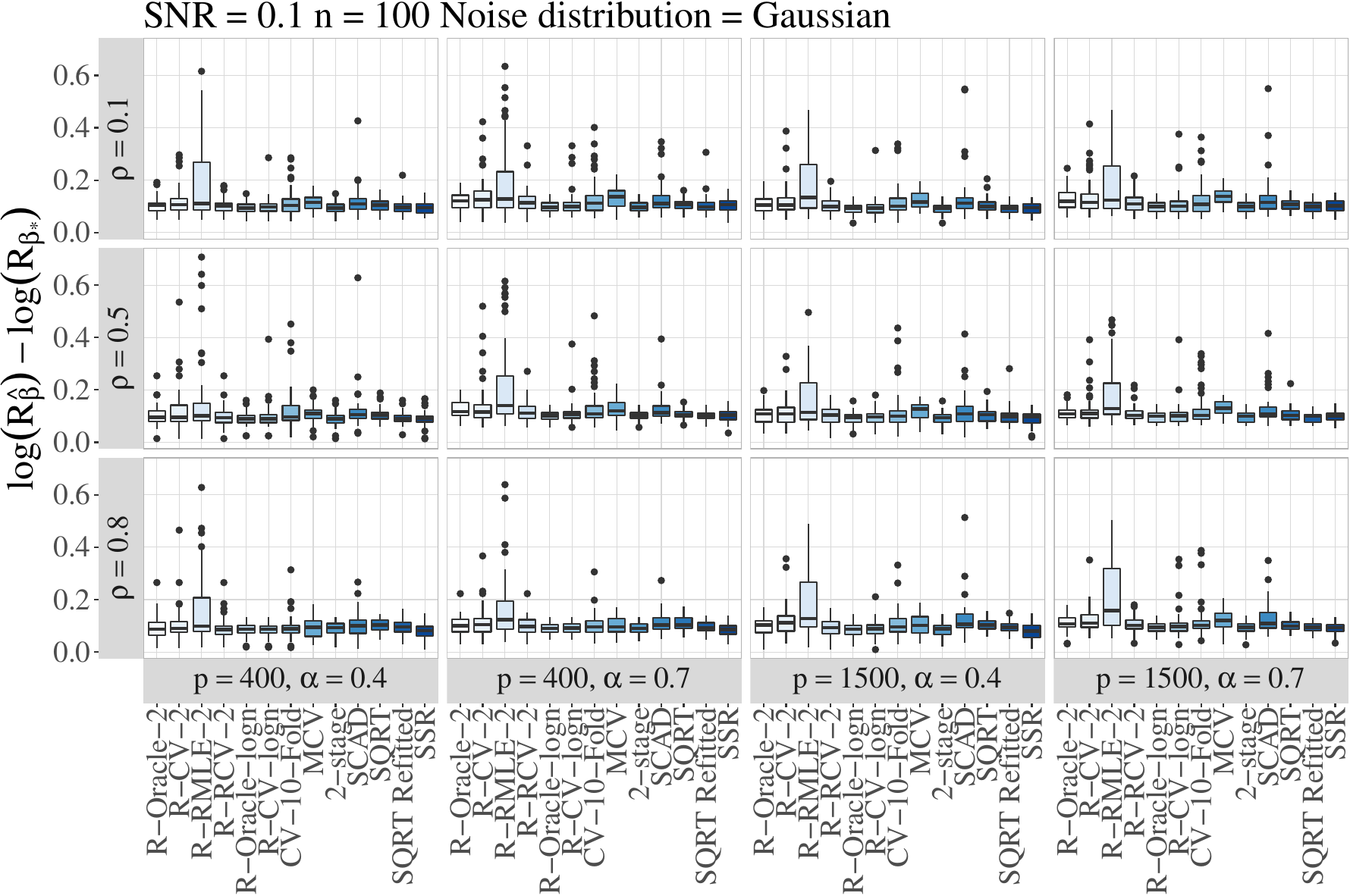}
\includegraphics[width=6in,height=4in]{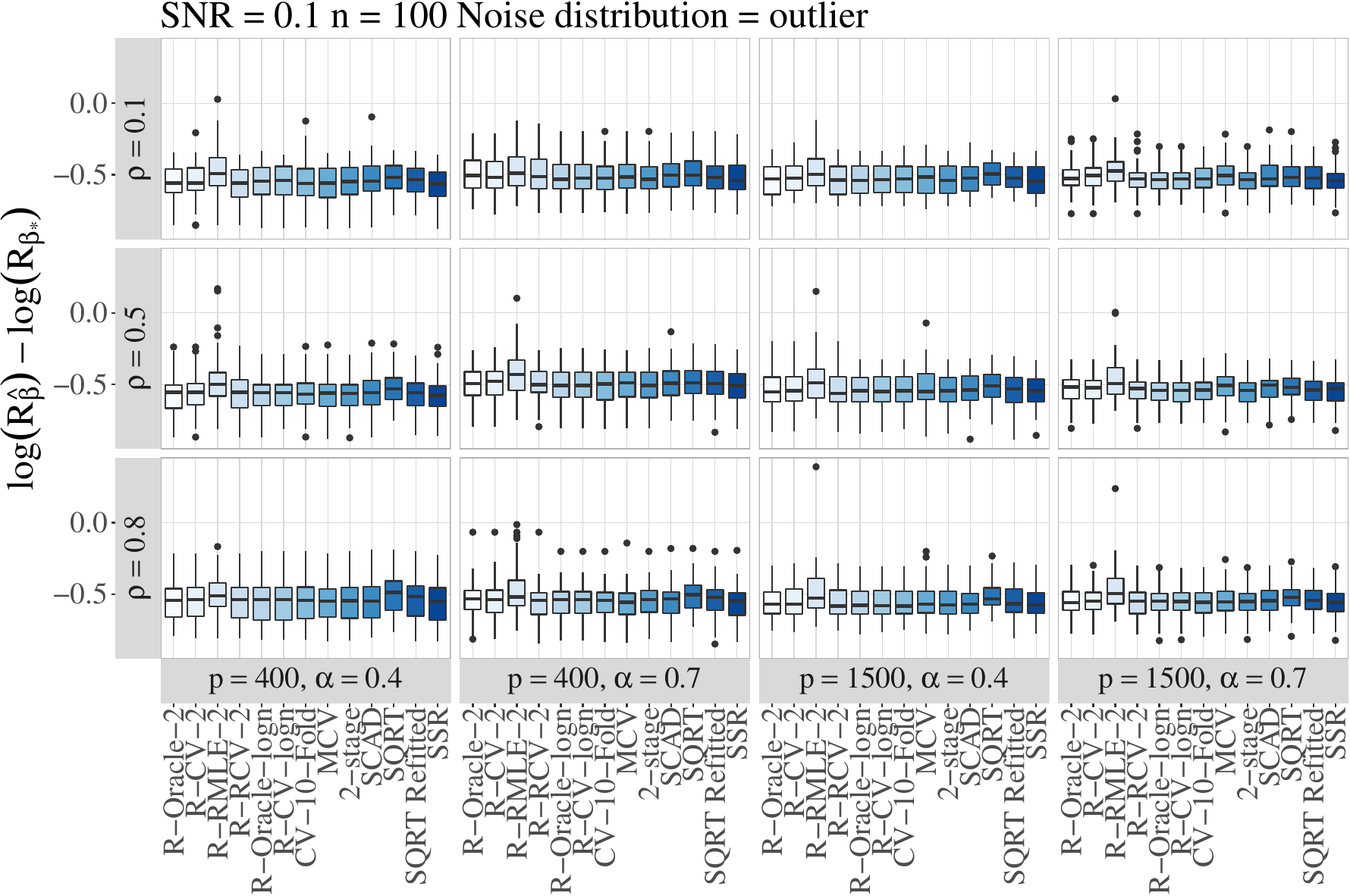}
\includegraphics[width=6in,height=4in]{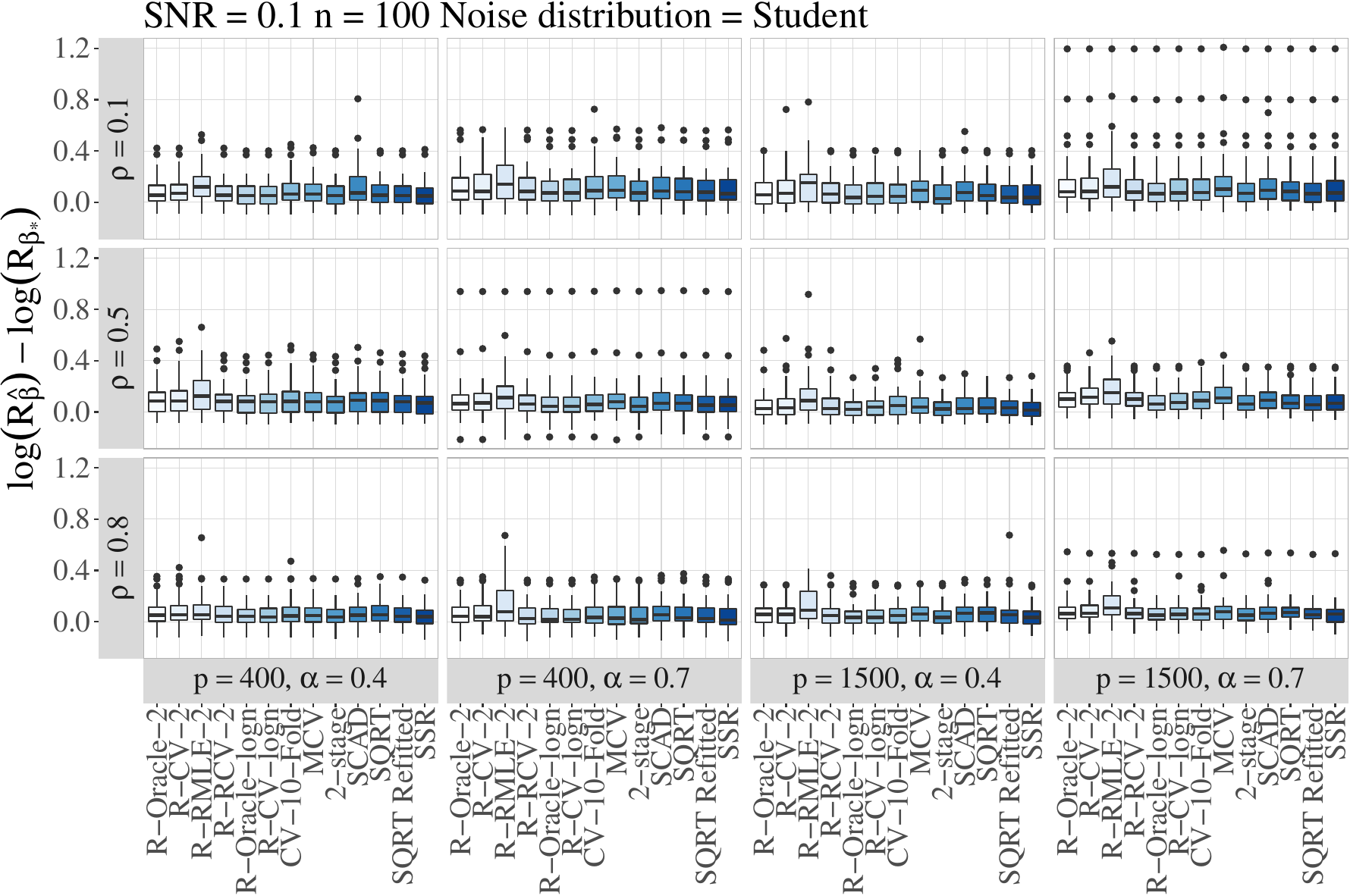}
\includegraphics[width=6in,height=4in]{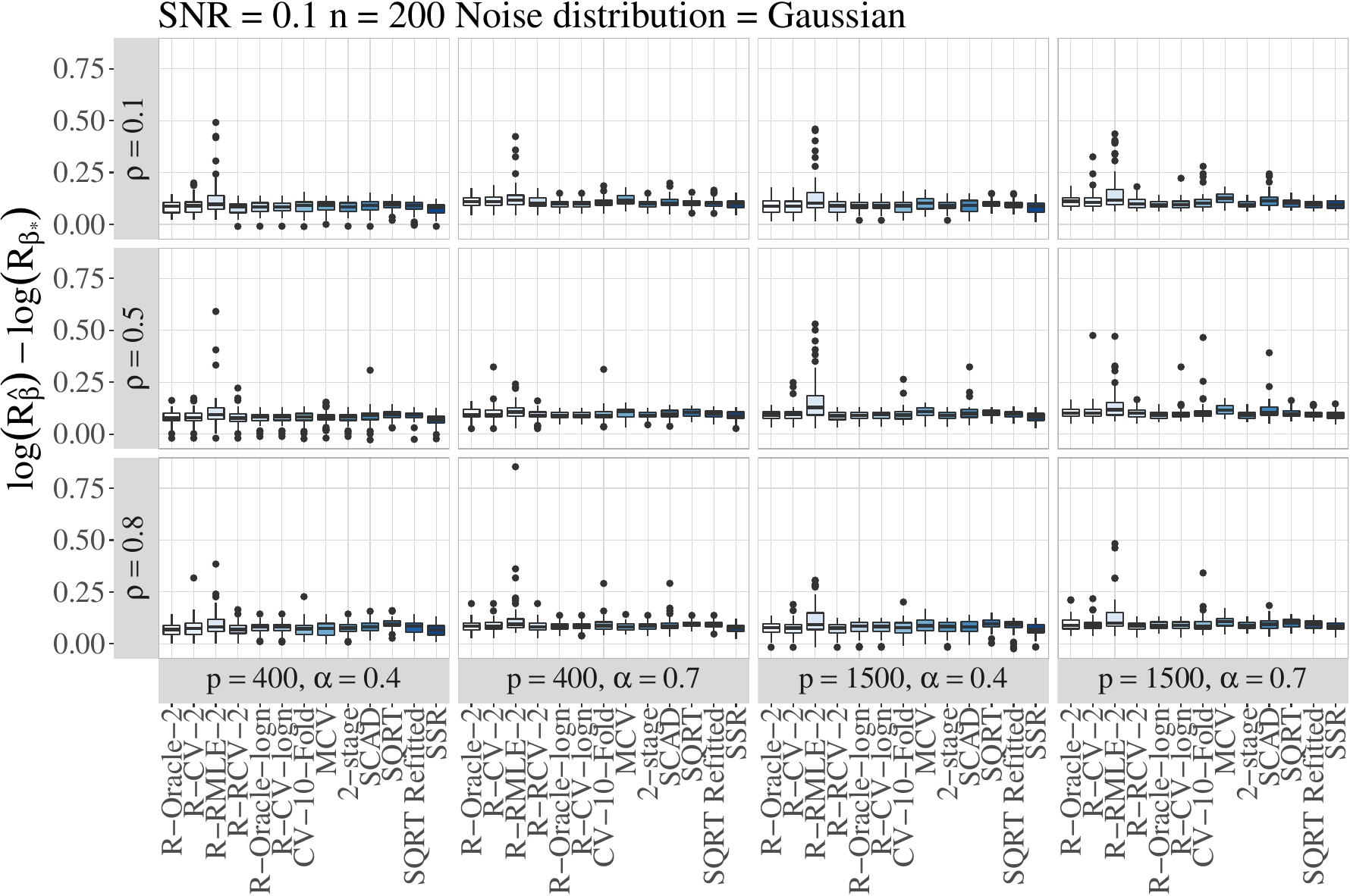}
\includegraphics[width=6in,height=4in]{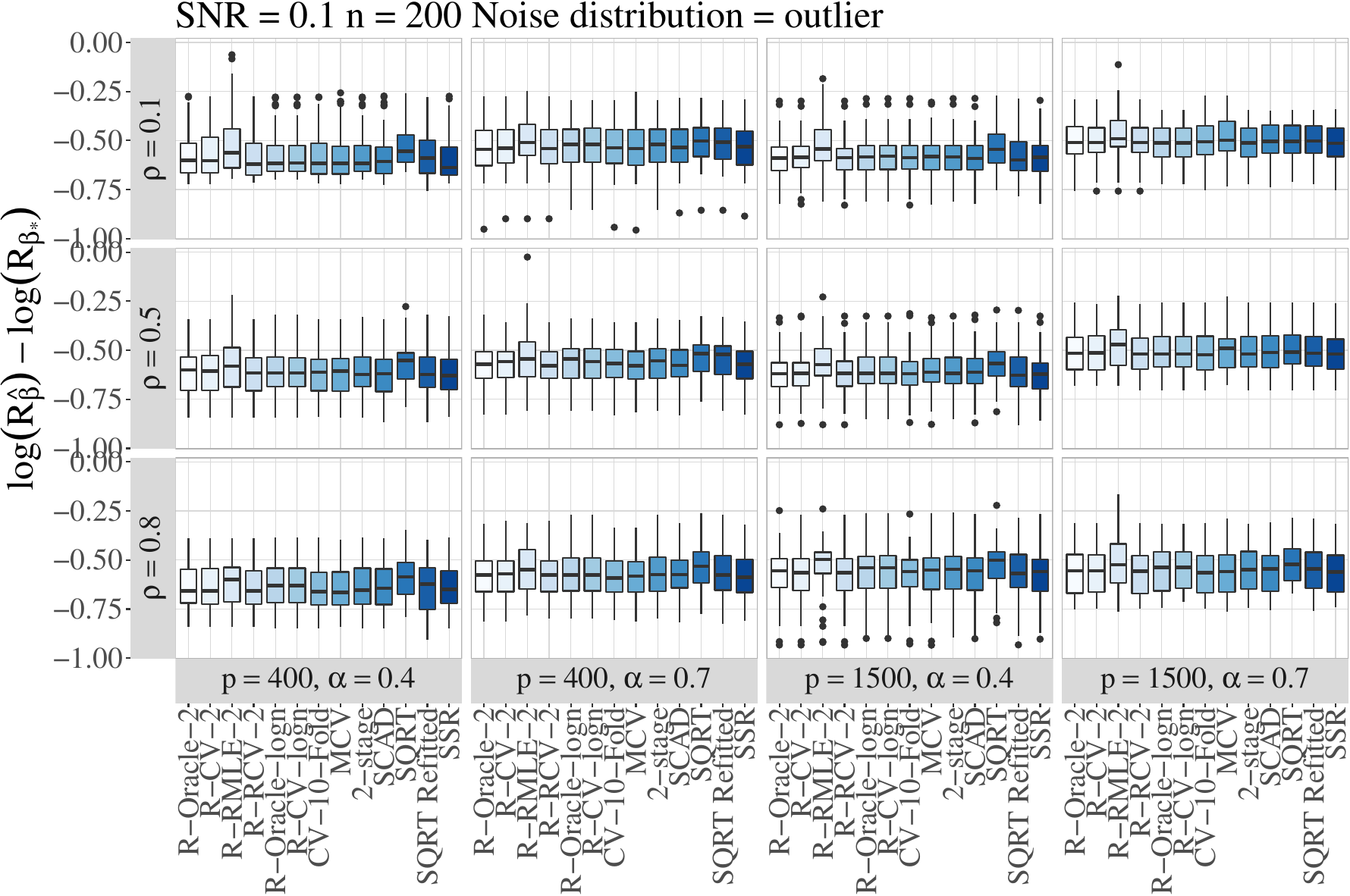}
\includegraphics[width=6in,height=4in]{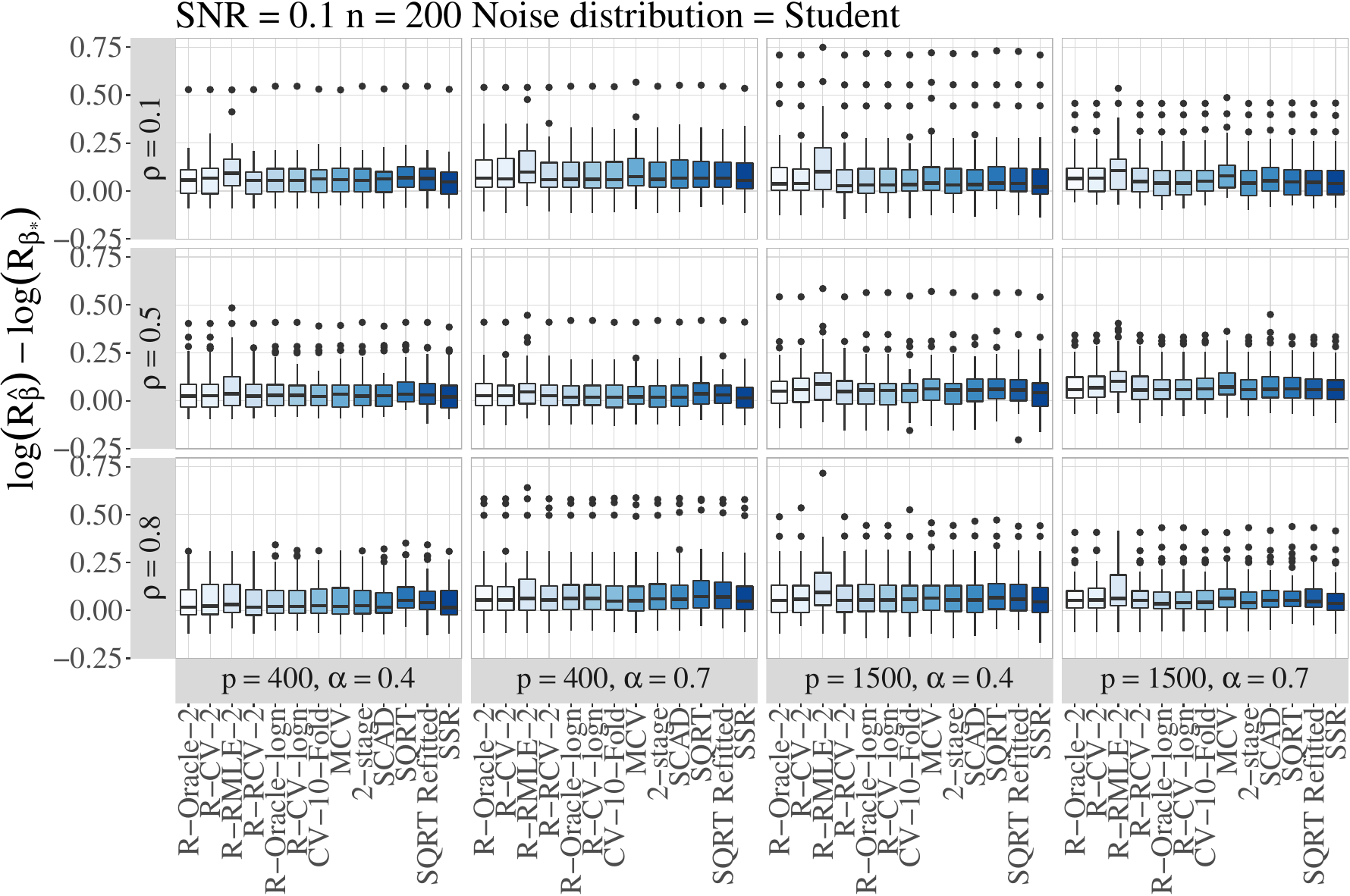}
\includegraphics[width=6in,height=4in]{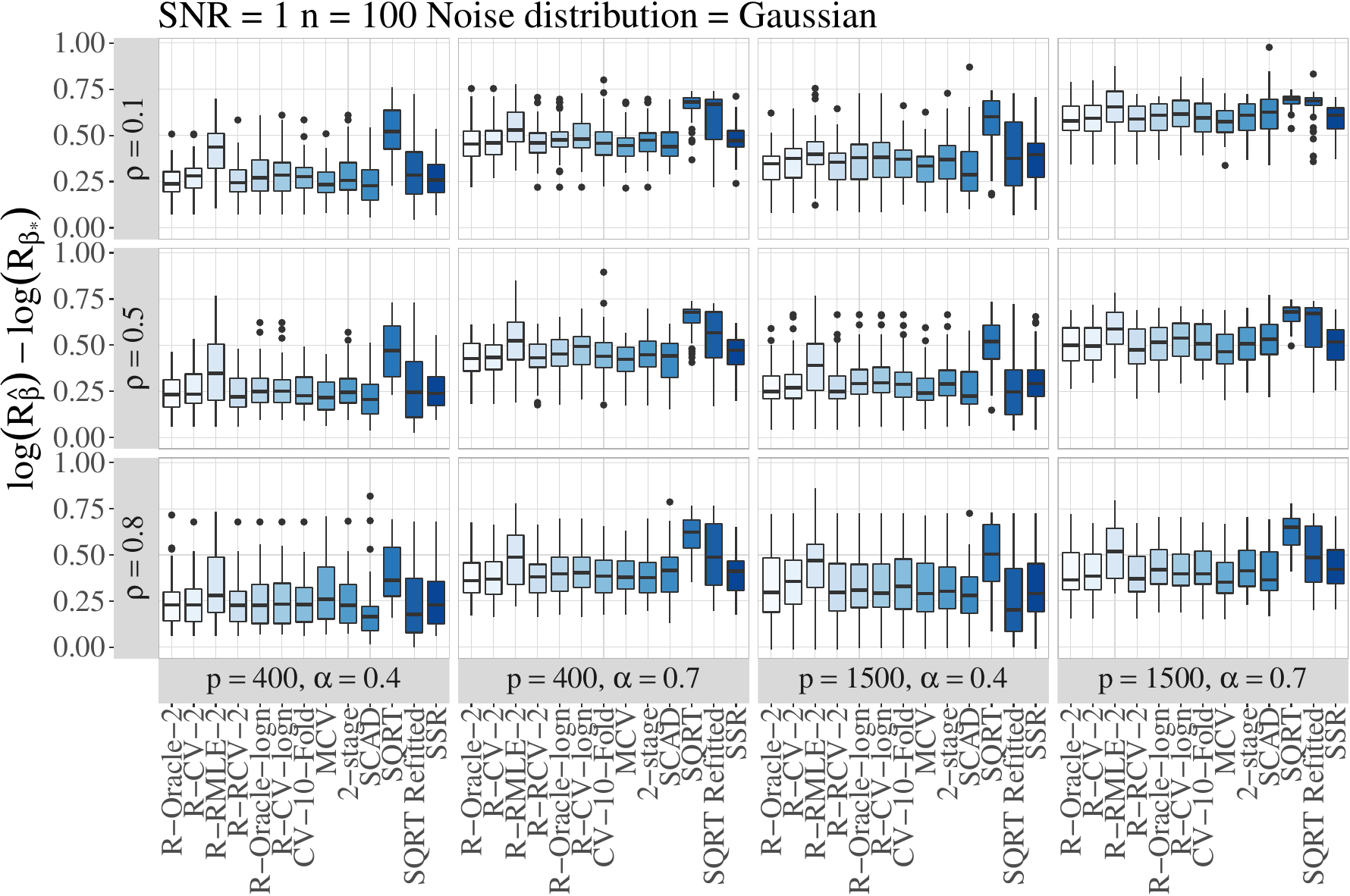}
\includegraphics[width=6in,height=4in]{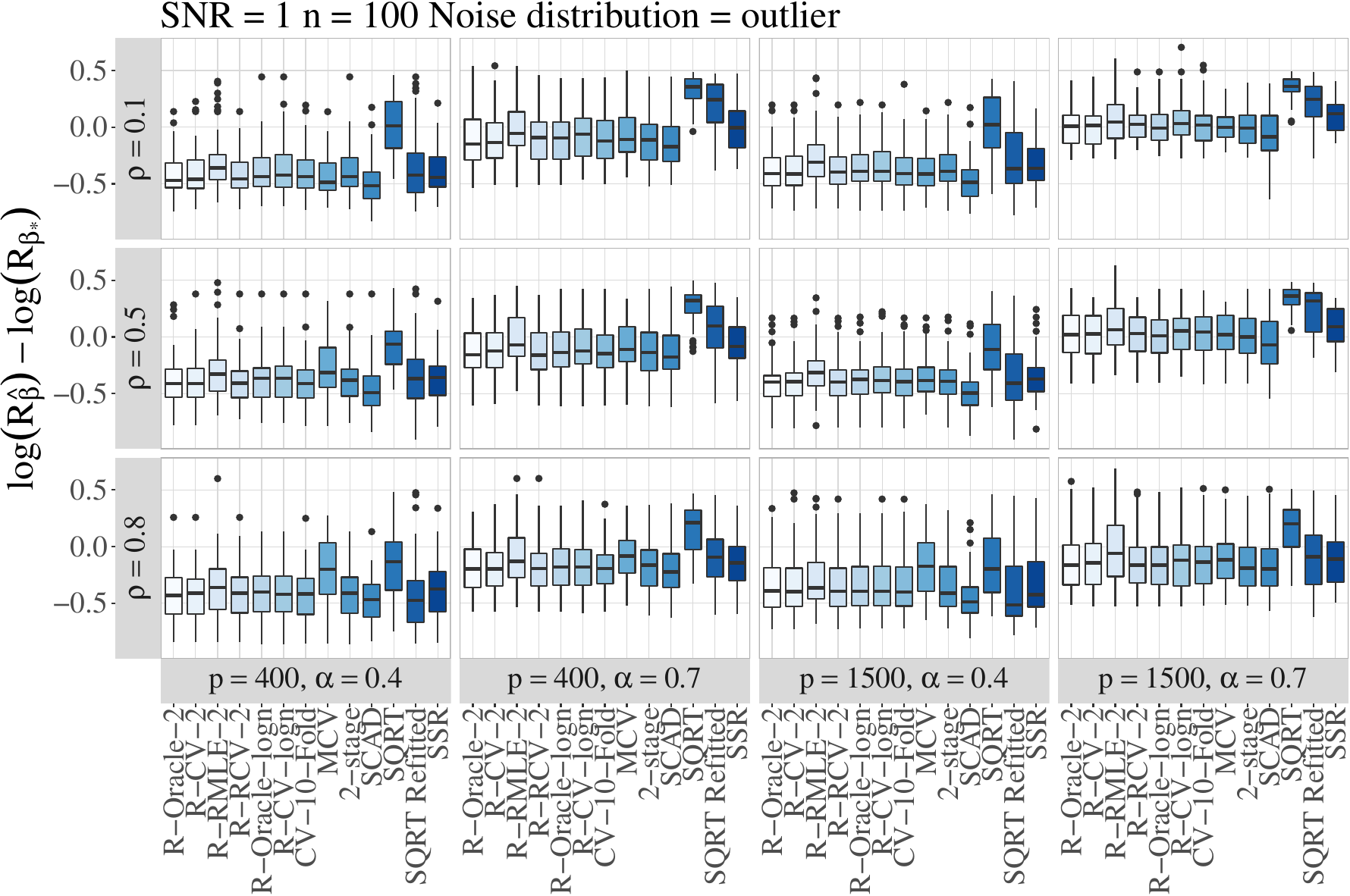}
\includegraphics[width=6in,height=4in]{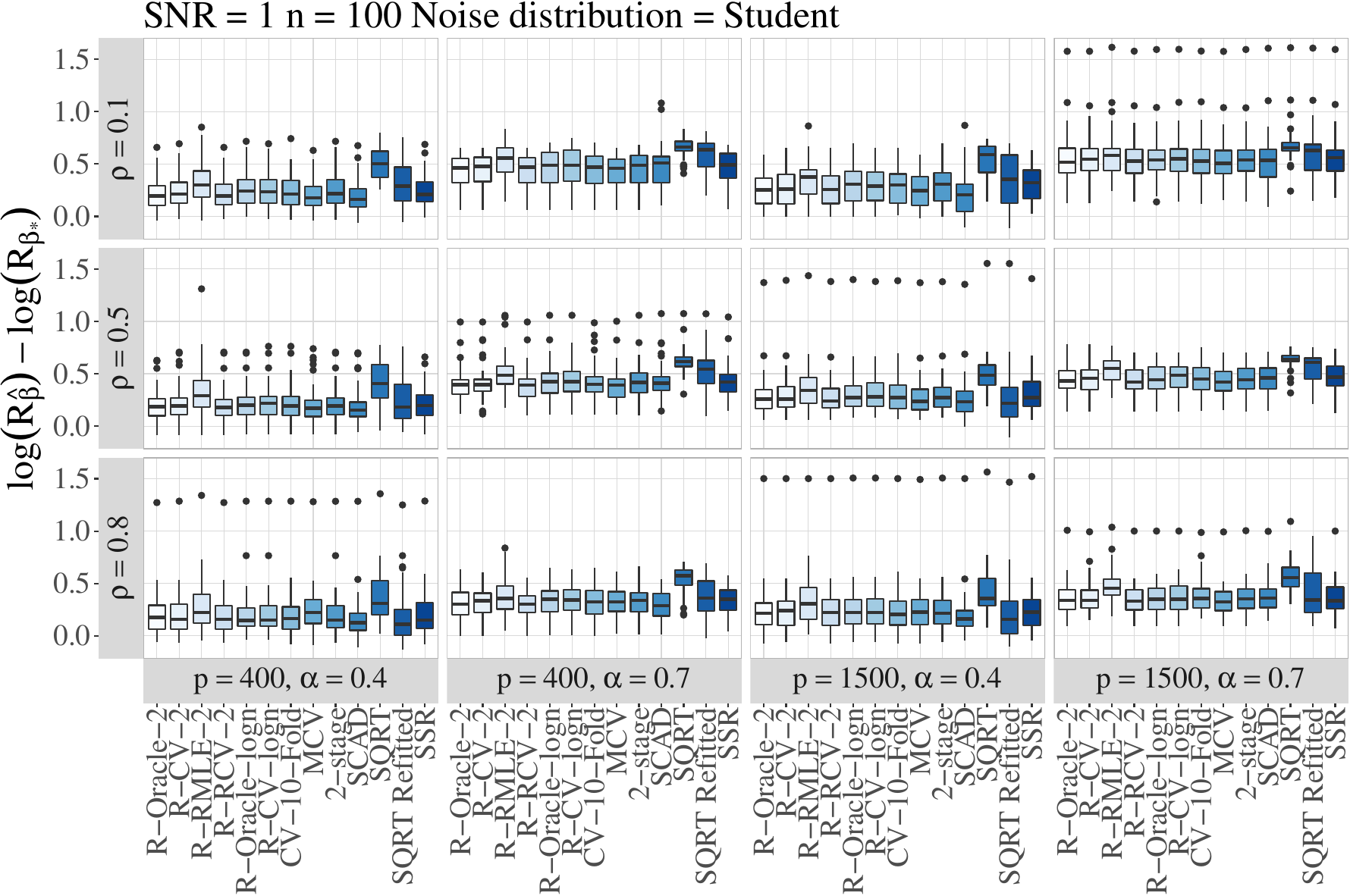}
\includegraphics[width=6in,height=4in]{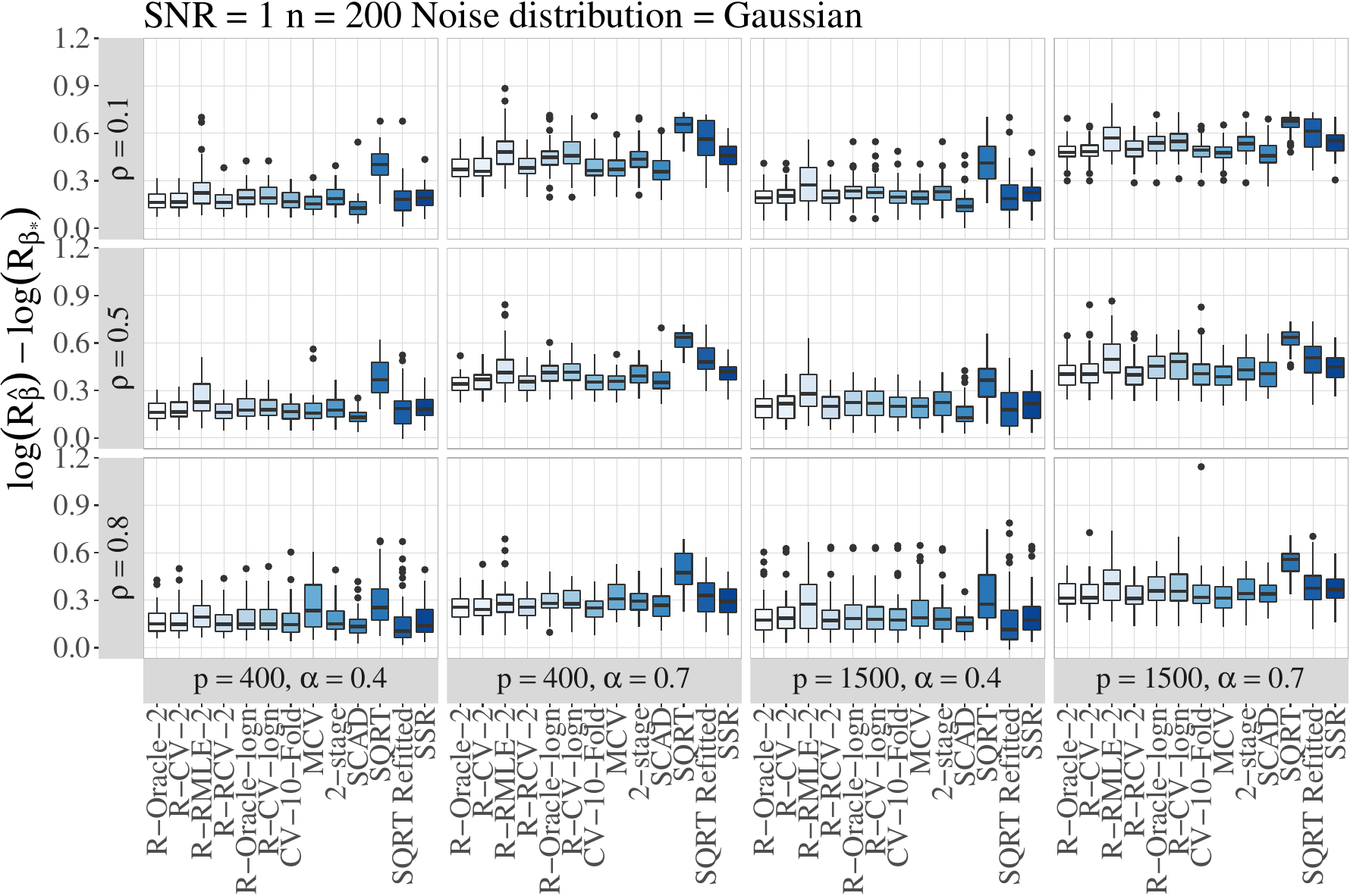}
\includegraphics[width=6in,height=4in]{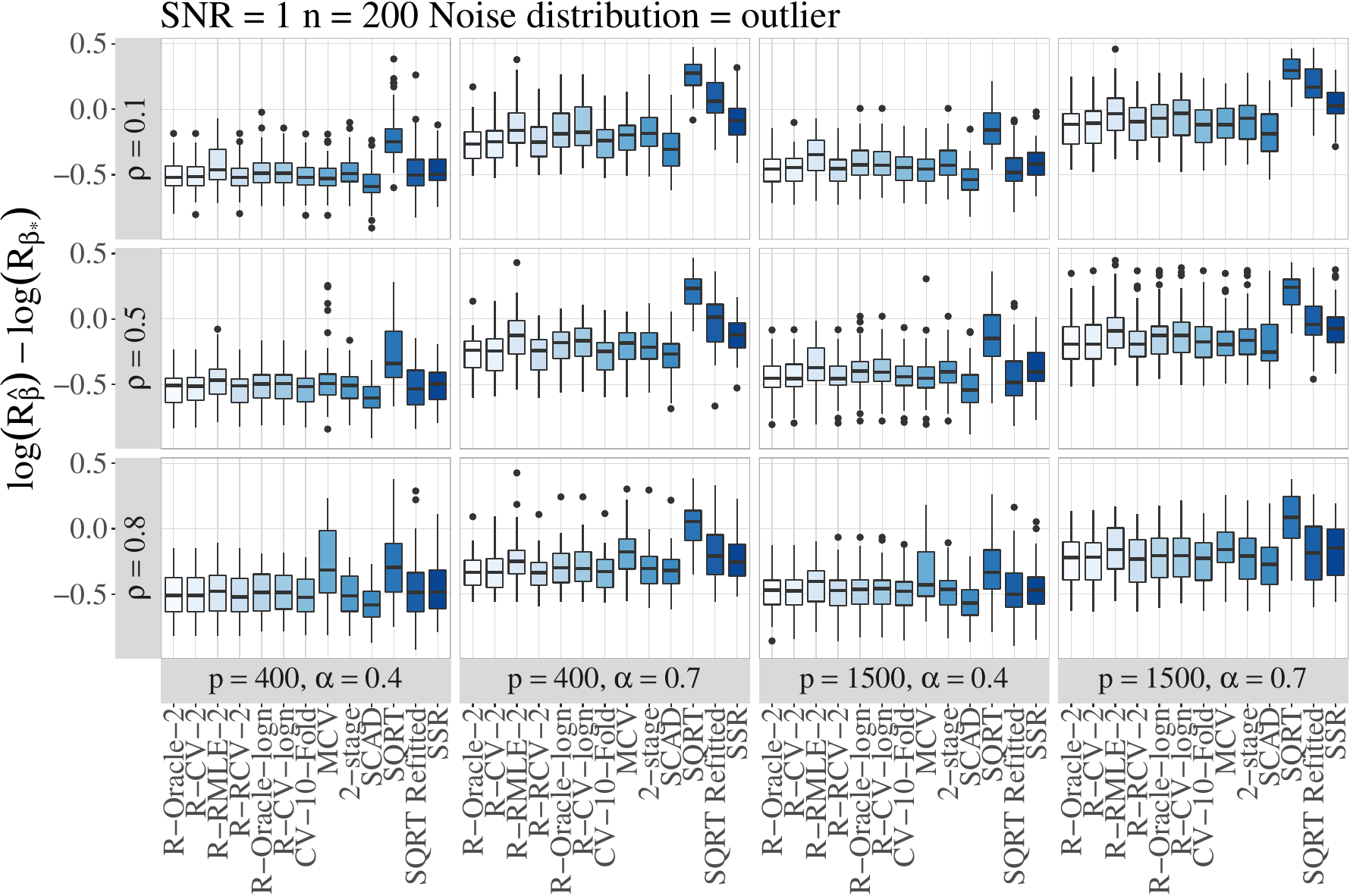}
\includegraphics[width=6in,height=4in]{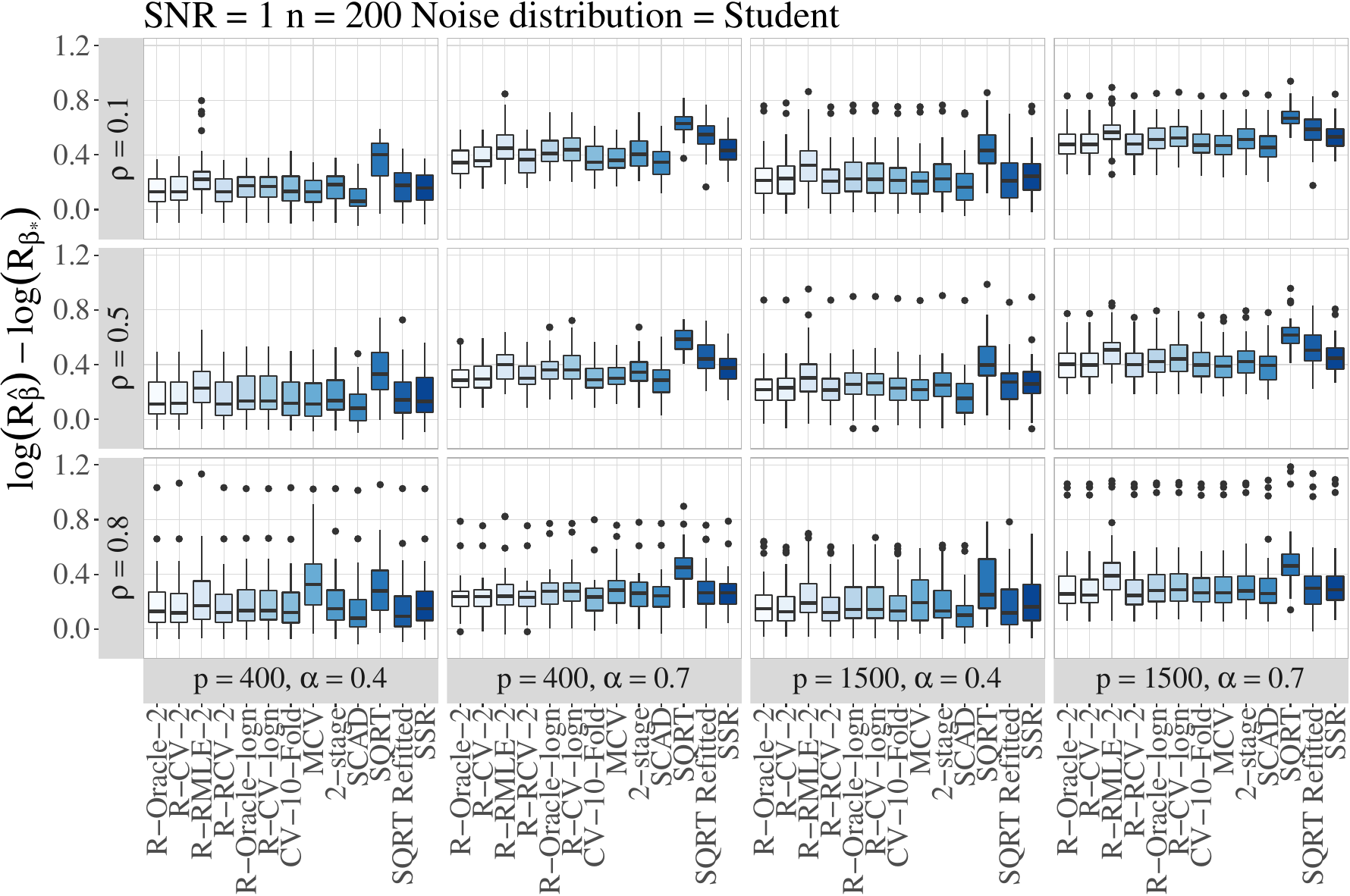}
\includegraphics[width=6in,height=4in]{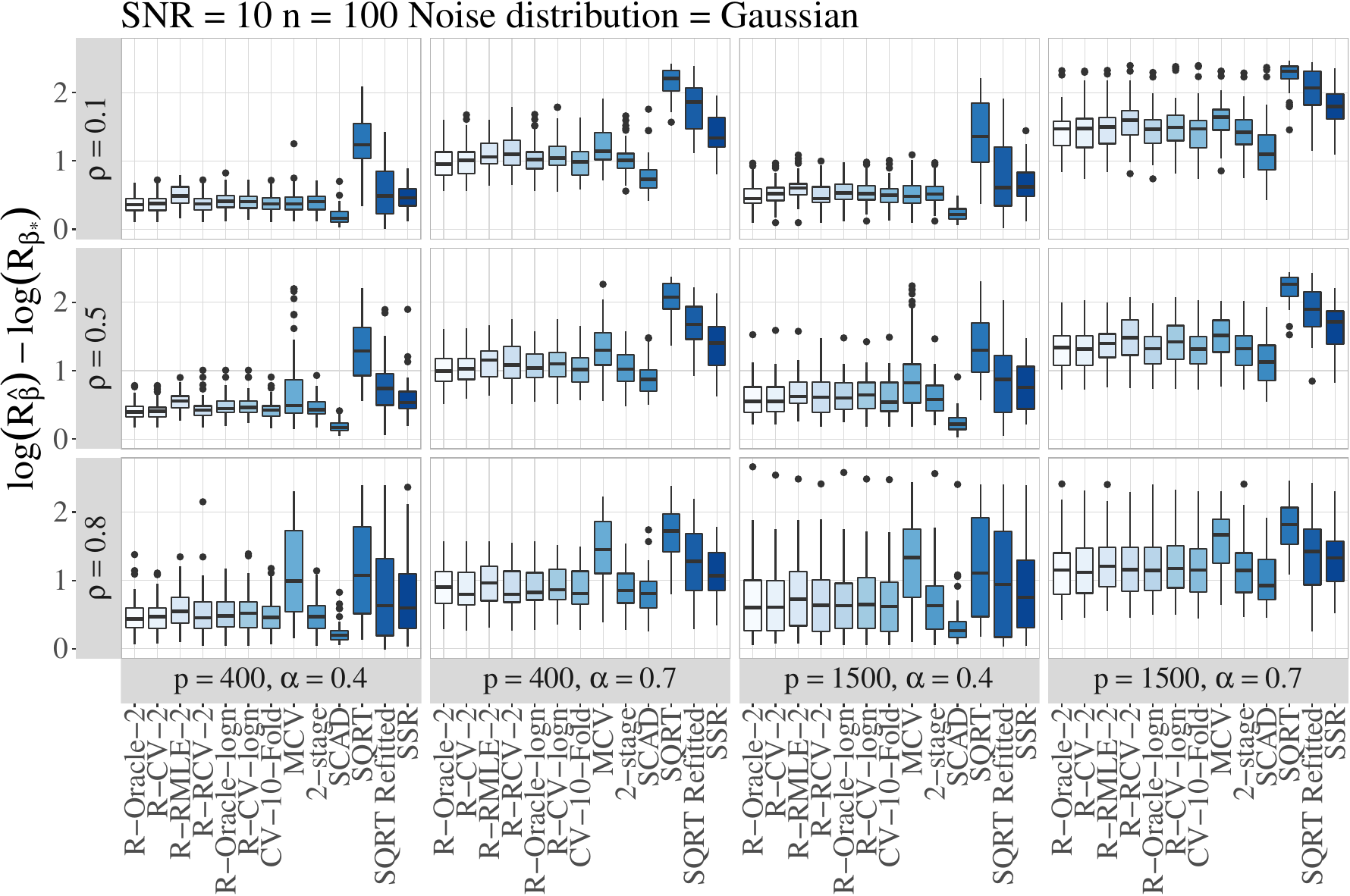}
\includegraphics[width=6in,height=4in]{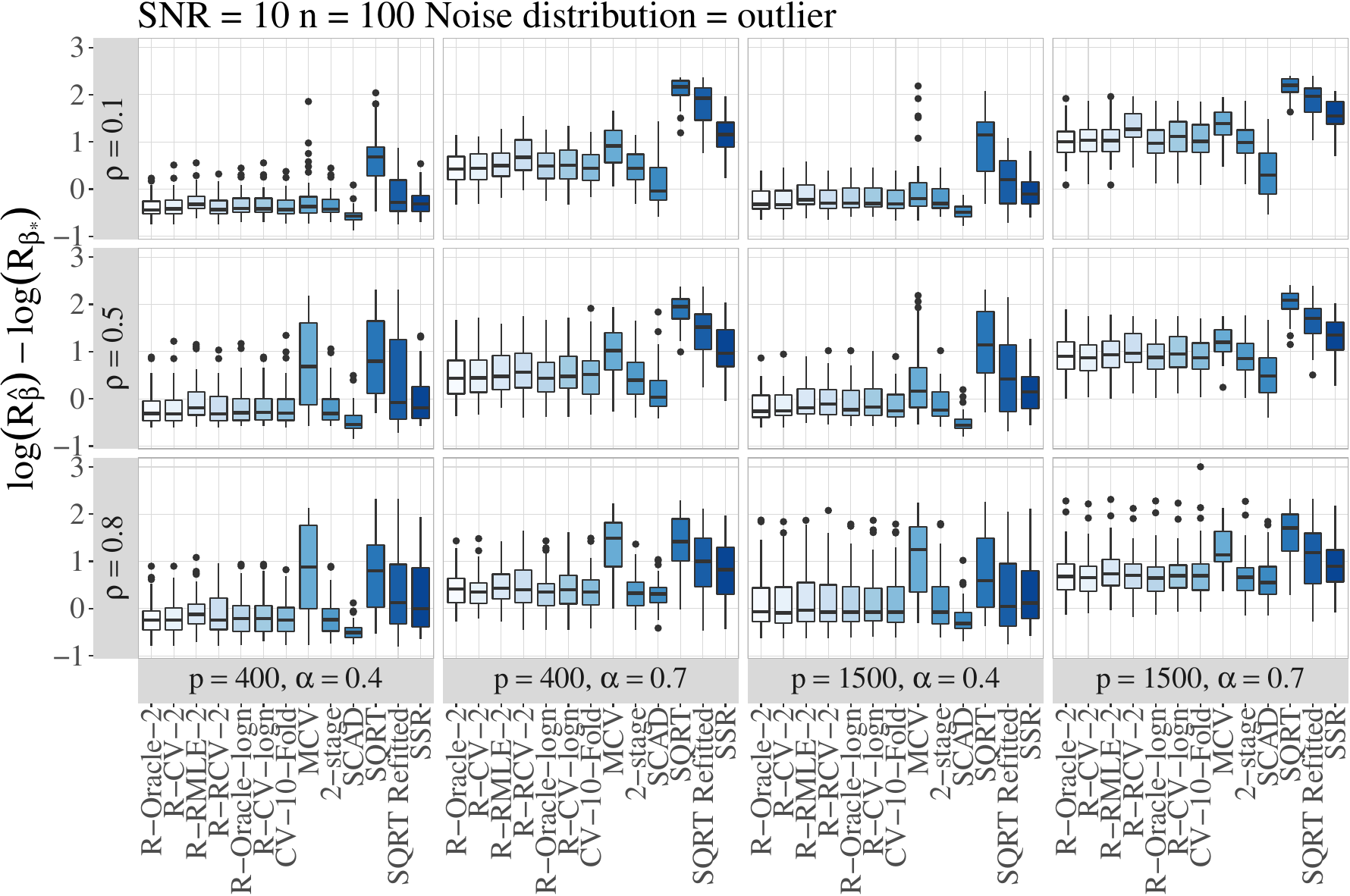}
\includegraphics[width=6in,height=4in]{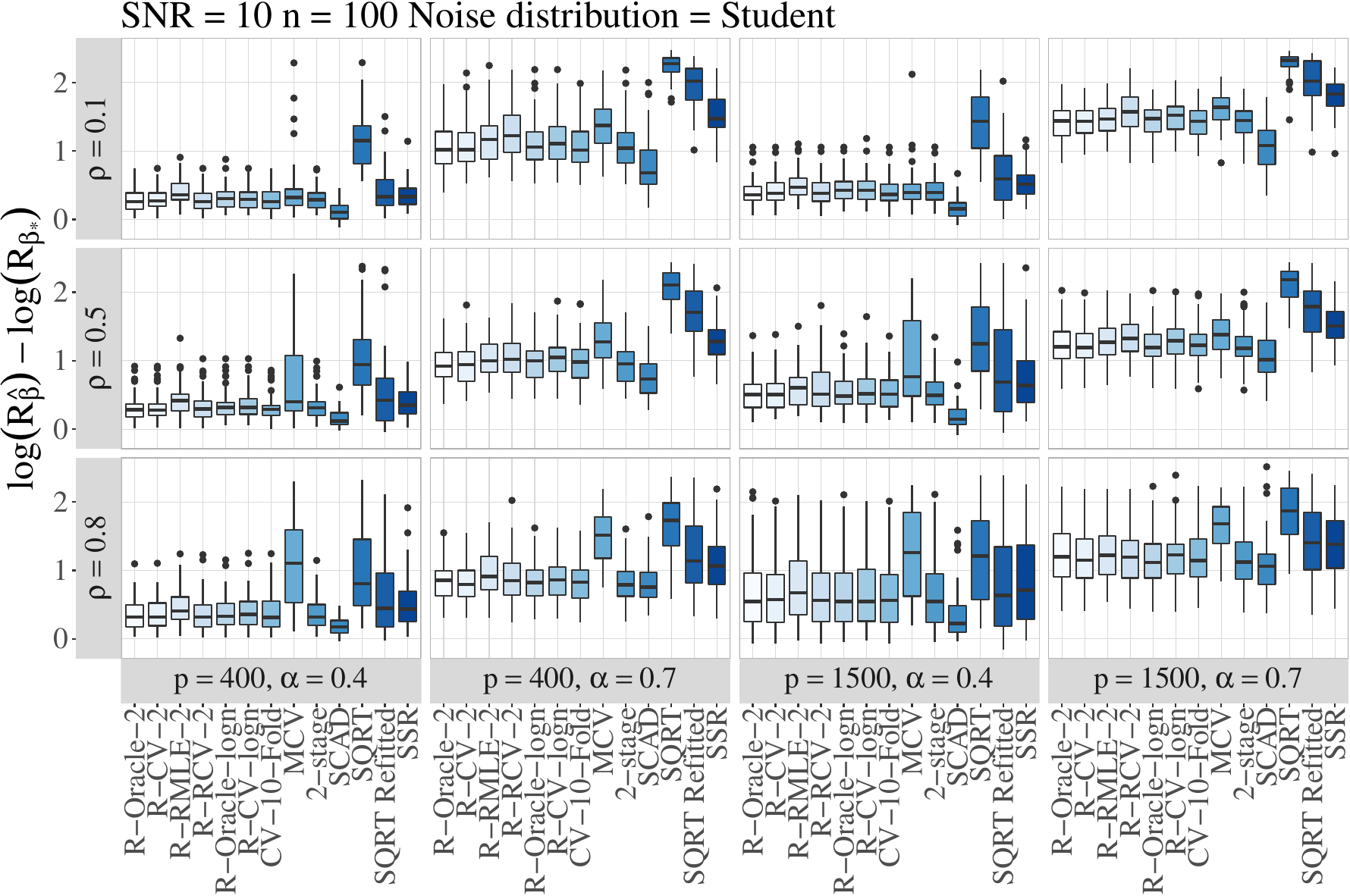}
\includegraphics[width=6in,height=4in]{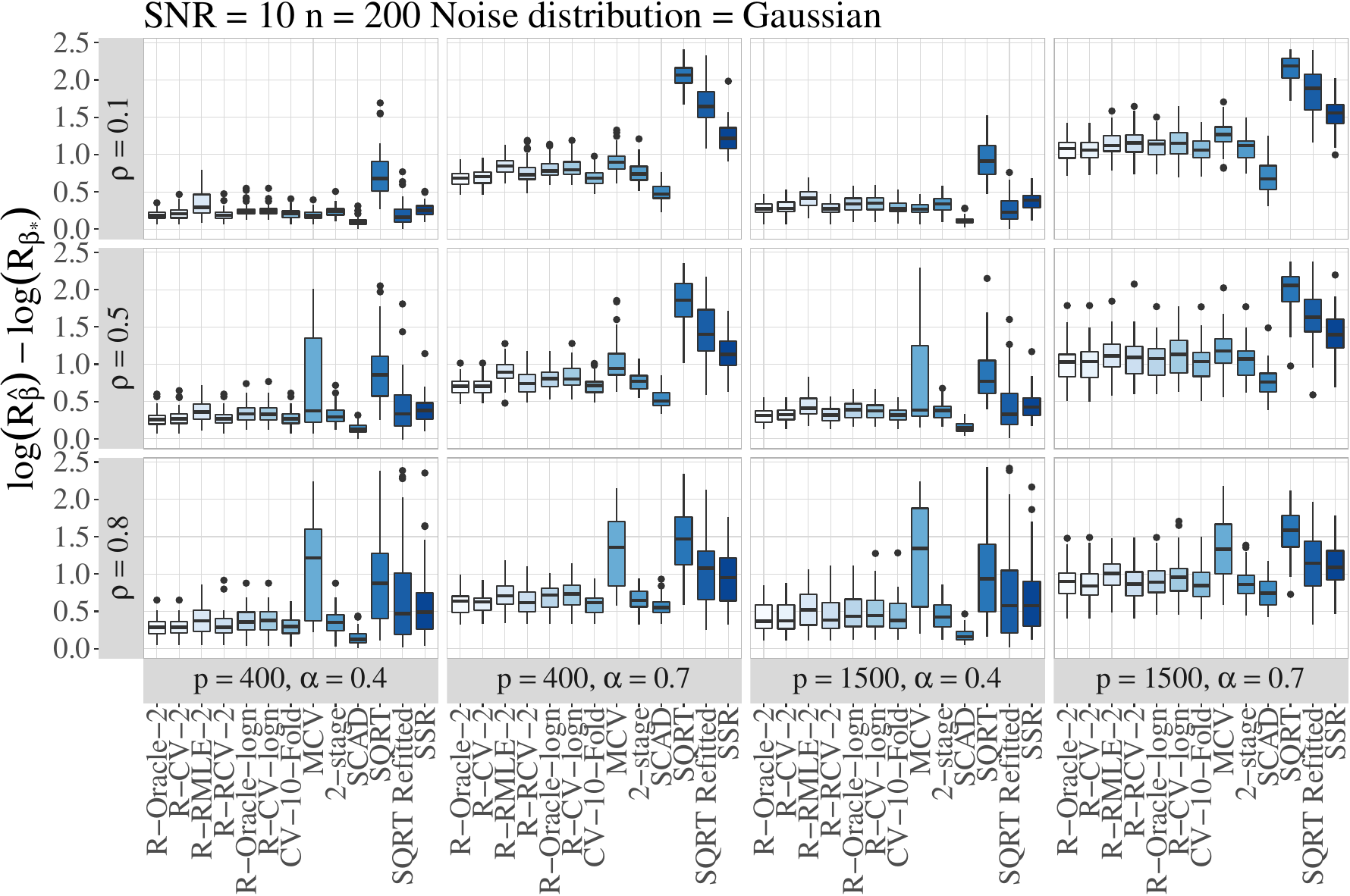}
\includegraphics[width=6in,height=4in]{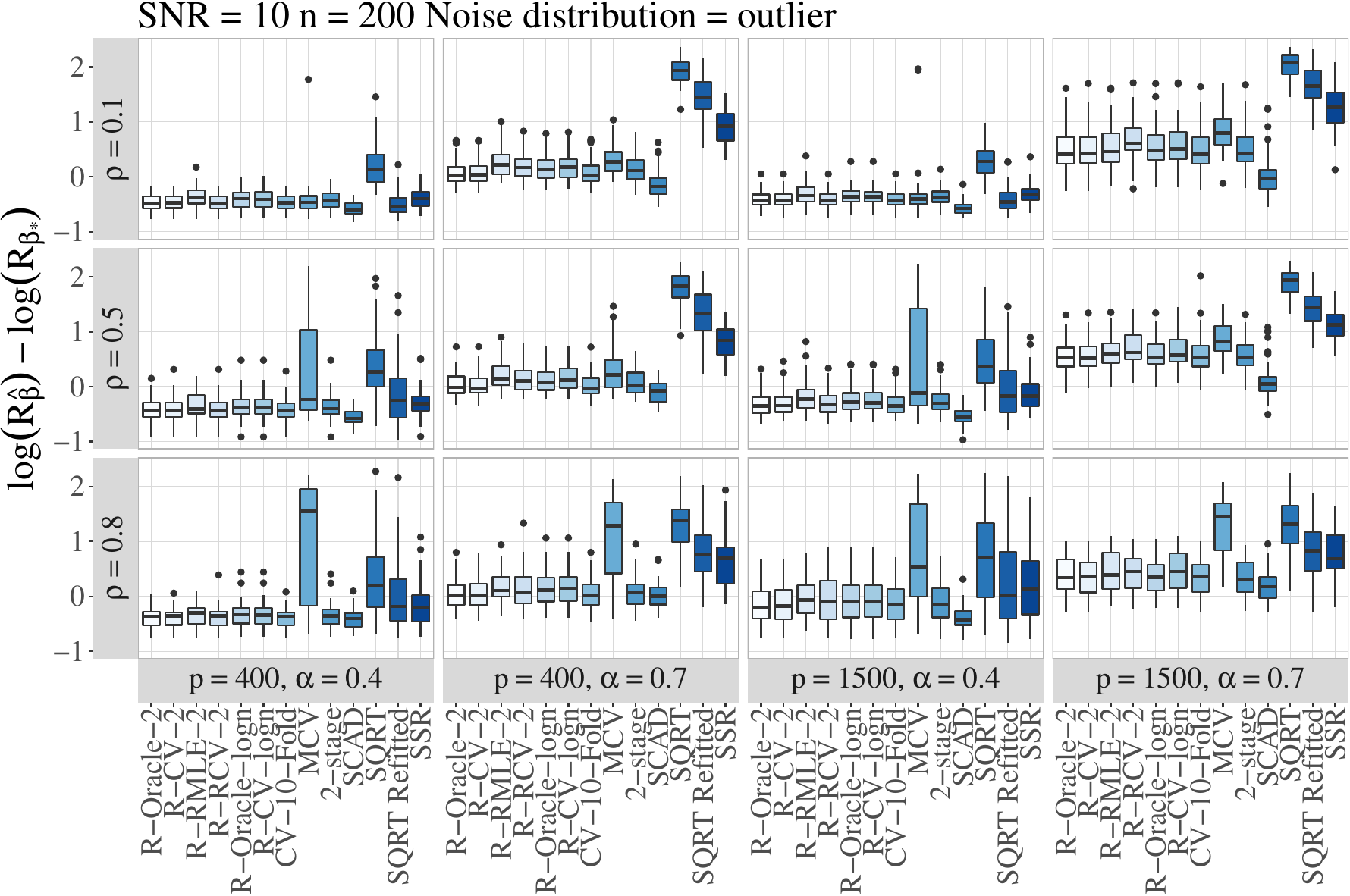}
\includegraphics[width=6in,height=4in]{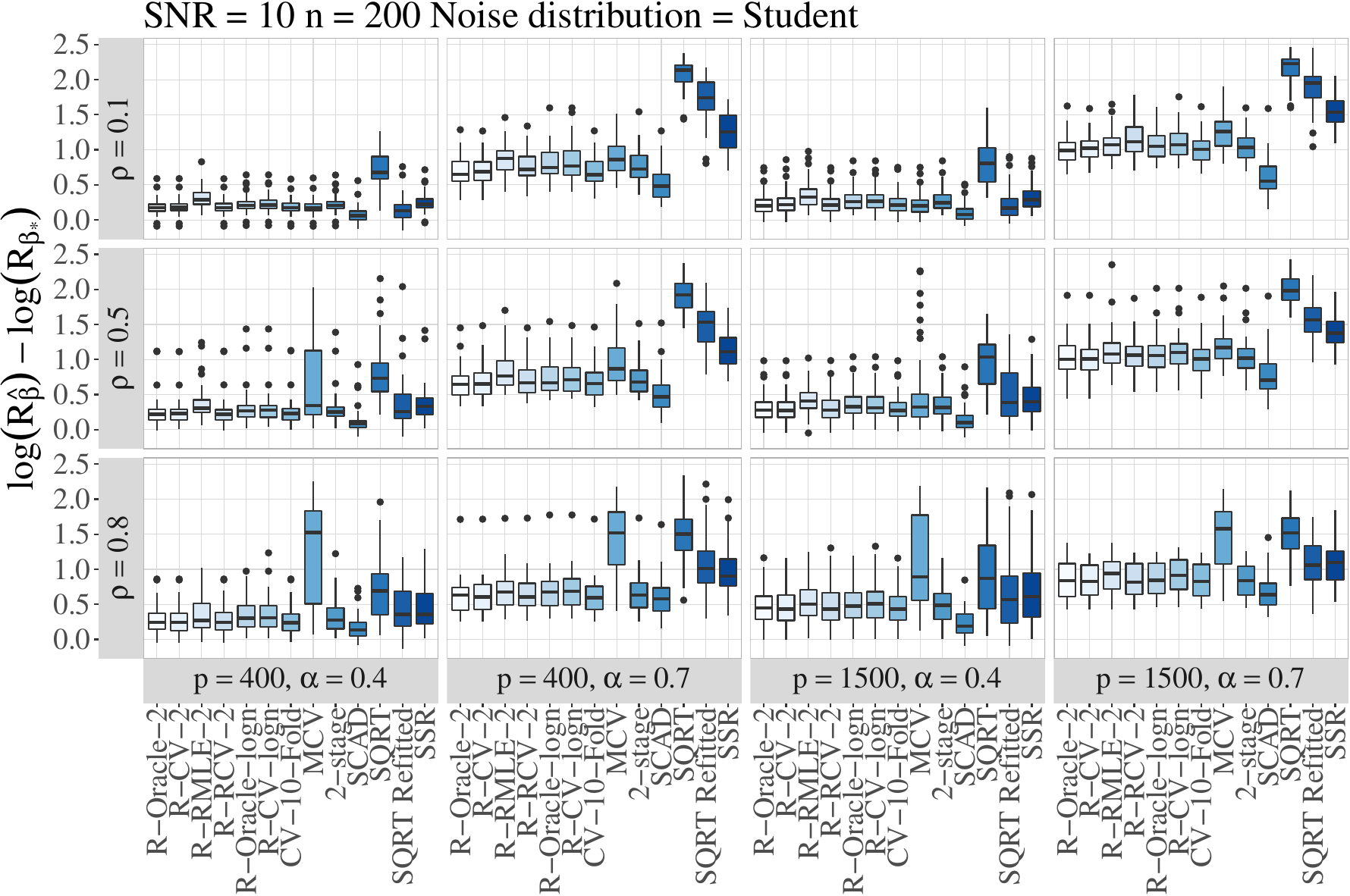}
\end{center}

\hypertarget{consistency-figures}{%
\subsection{Consistency figures}\label{consistency-figures}}

\begin{center}
\includegraphics[width=6in,height=4in]{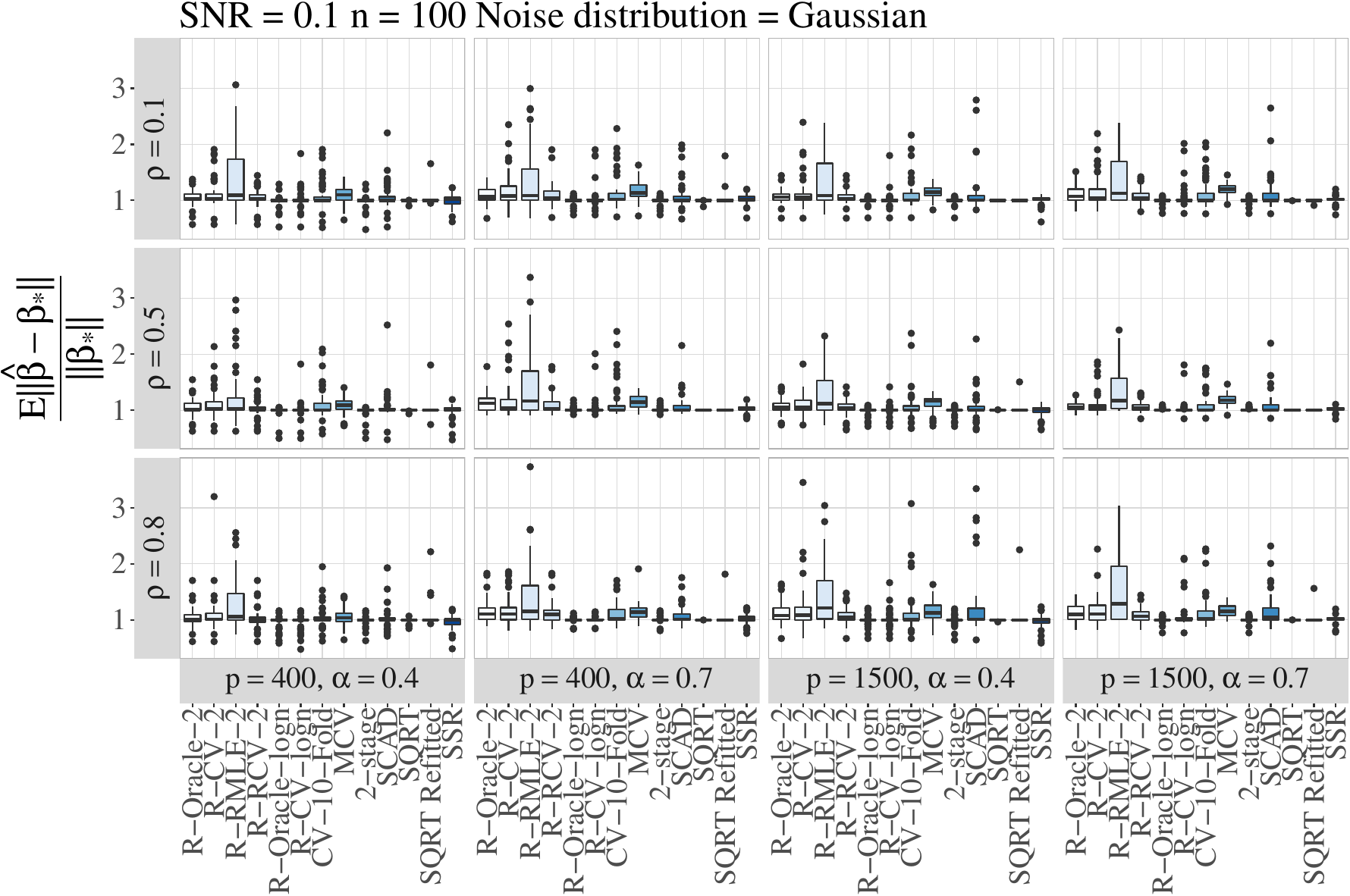}
\includegraphics[width=6in,height=4in]{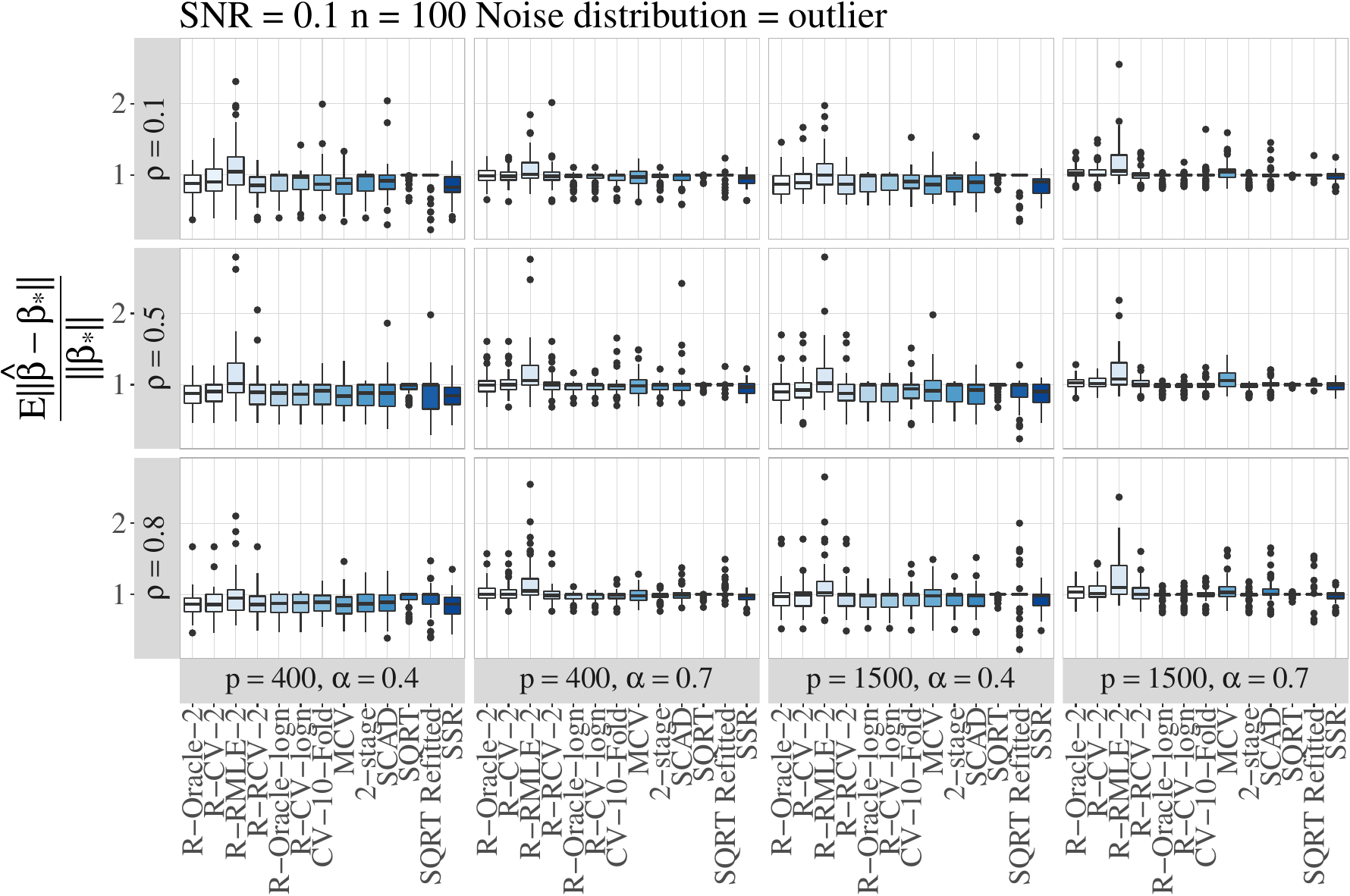}
\includegraphics[width=6in,height=4in]{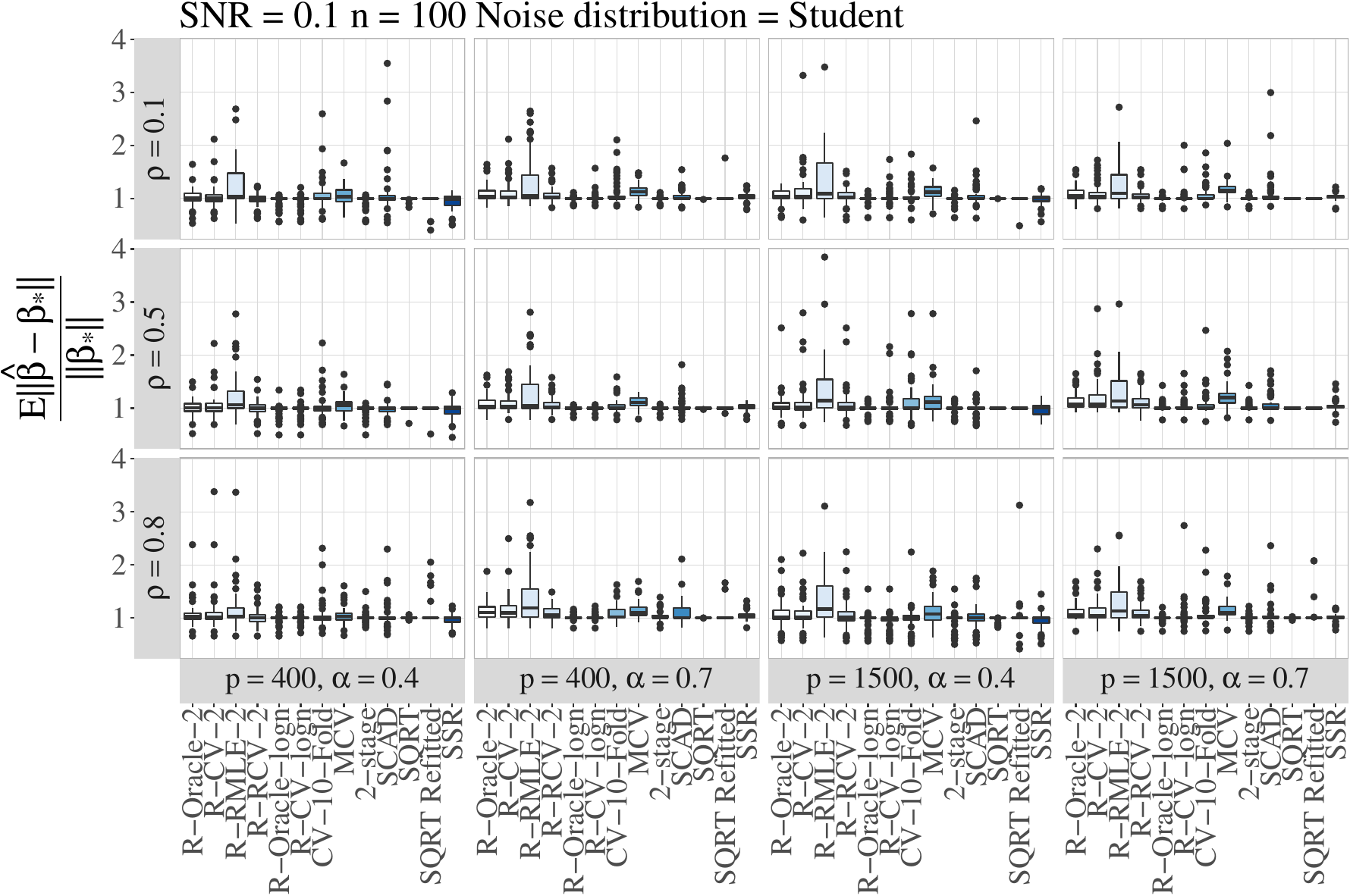}
\includegraphics[width=6in,height=4in]{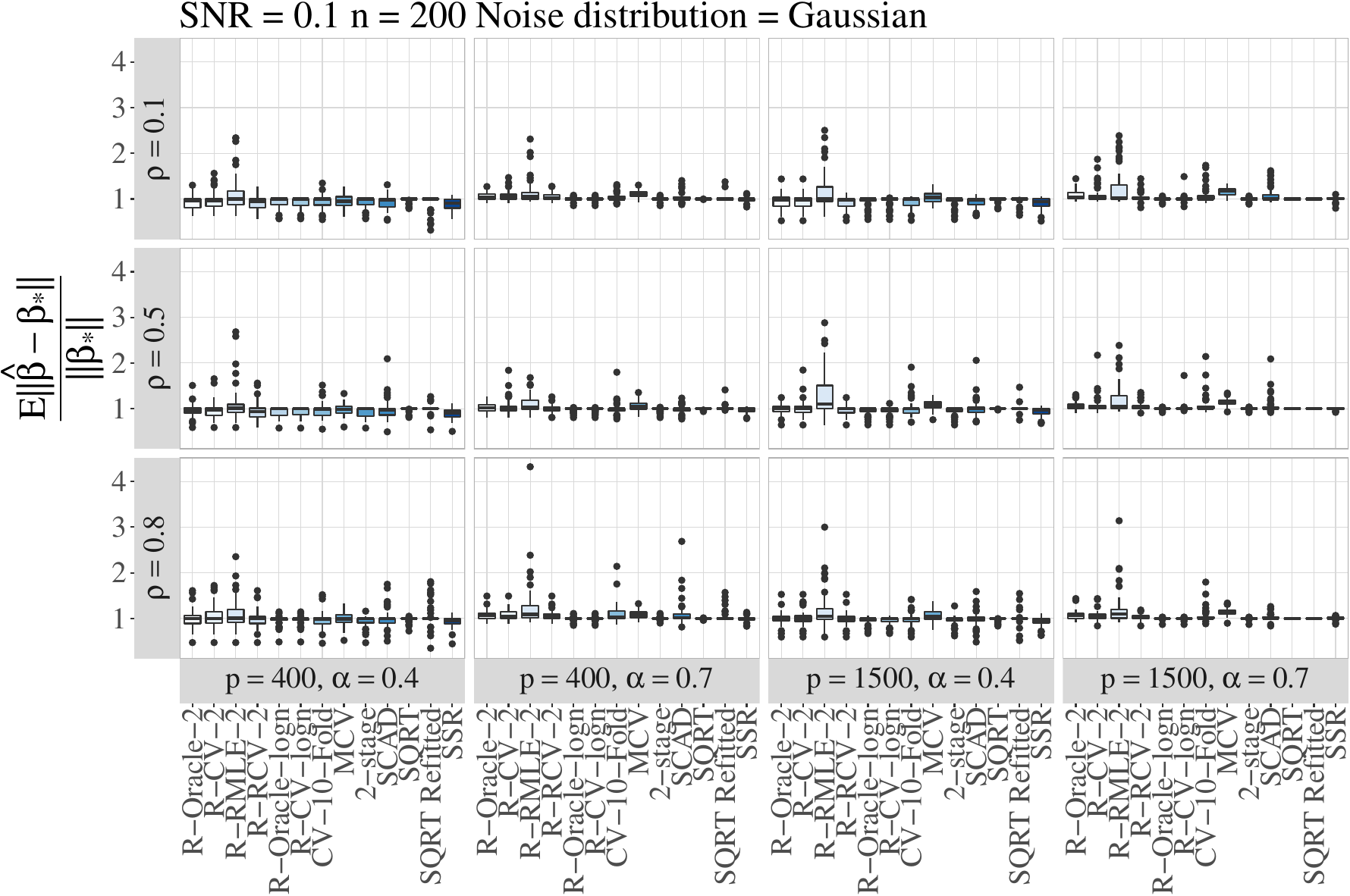}
\includegraphics[width=6in,height=4in]{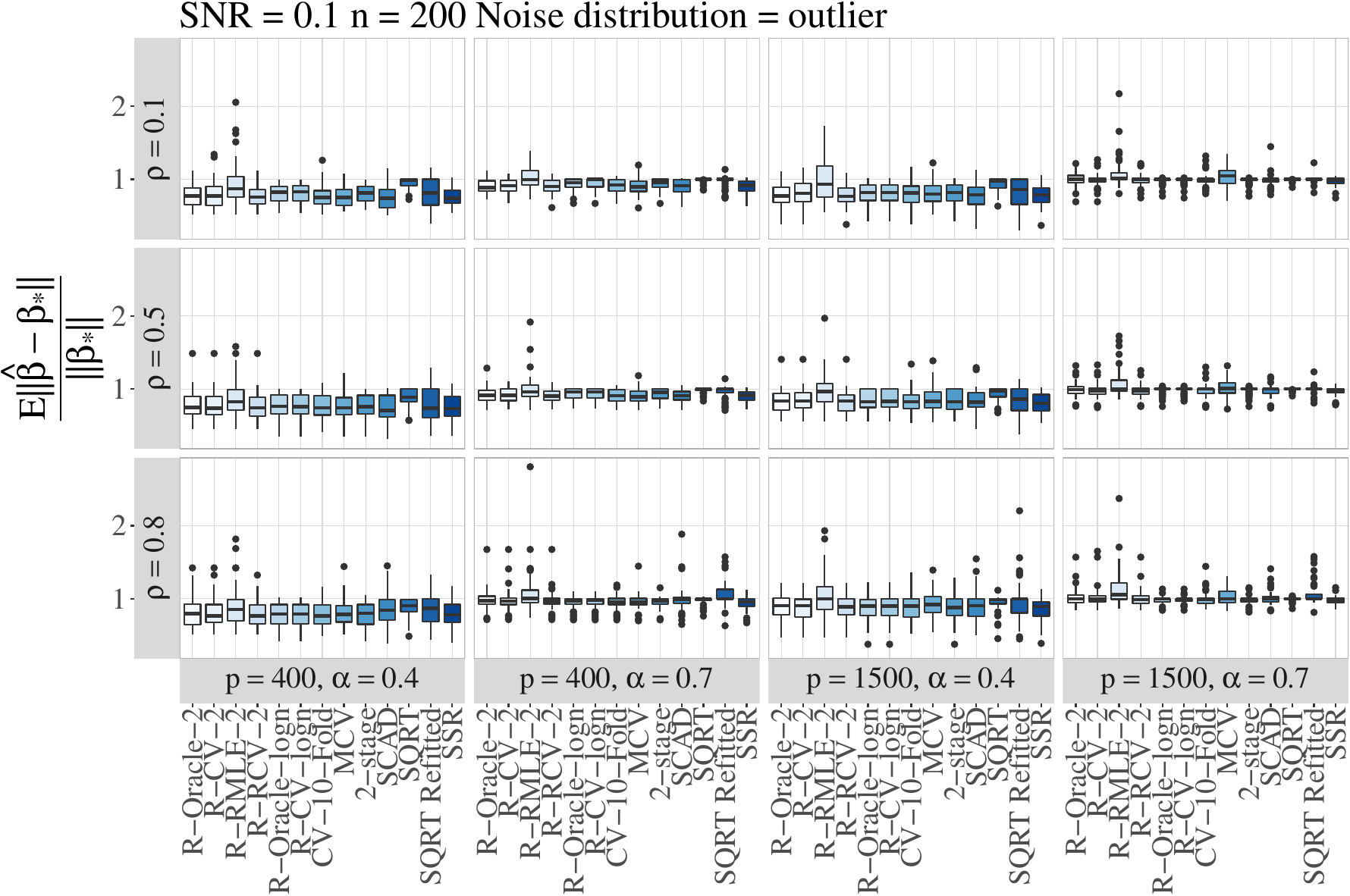}
\includegraphics[width=6in,height=4in]{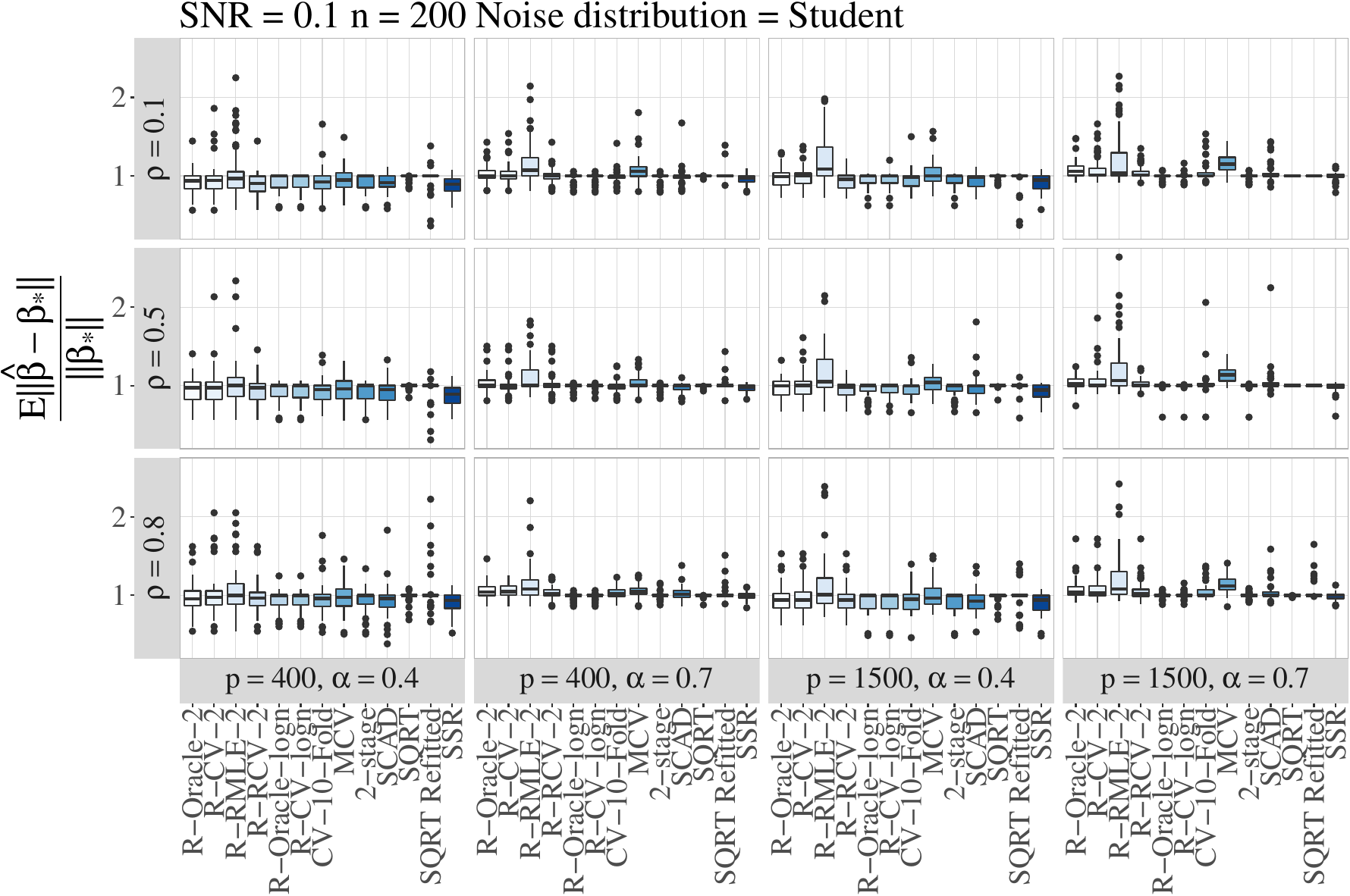}
\includegraphics[width=6in,height=4in]{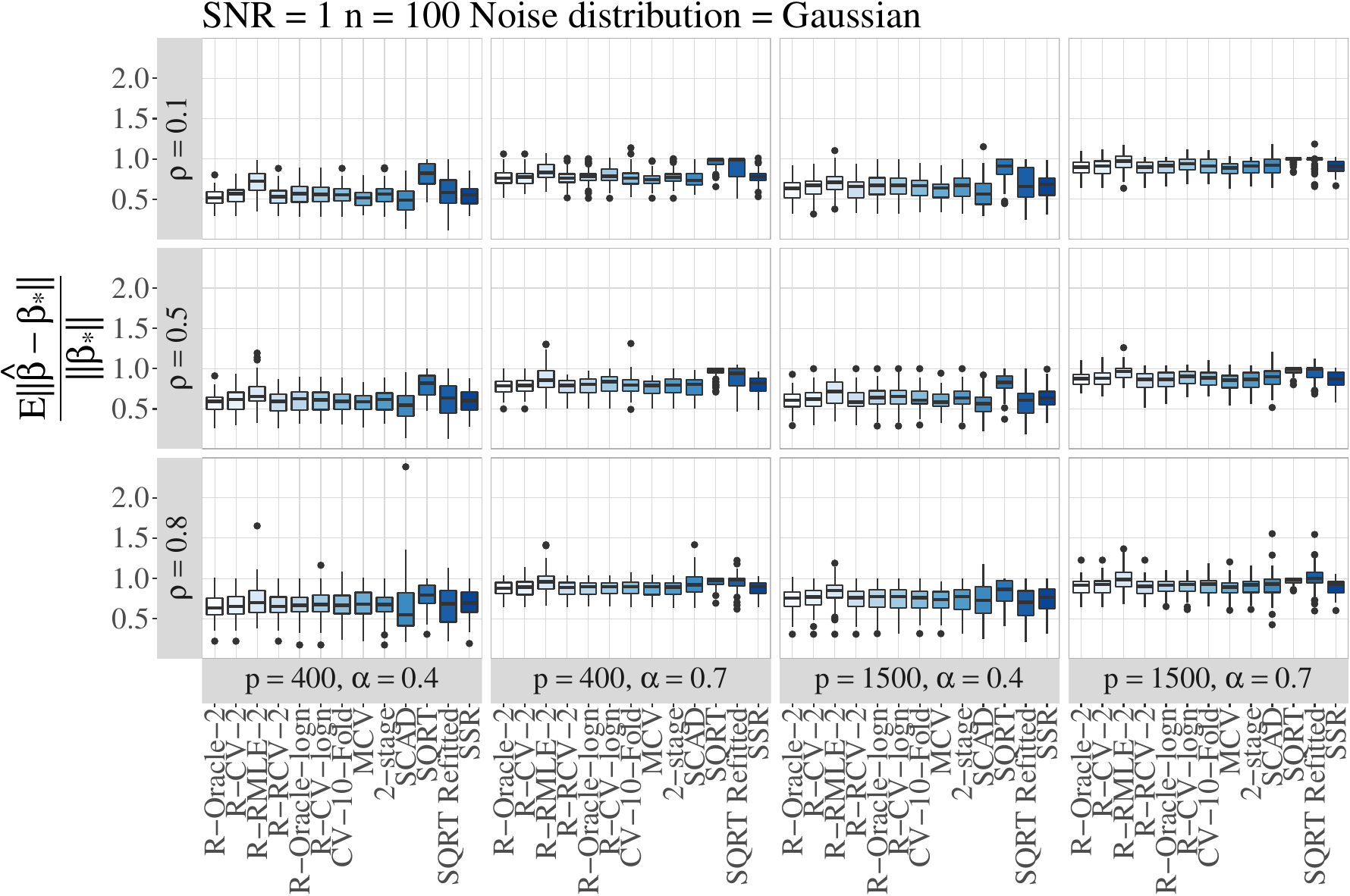}
\includegraphics[width=6in,height=4in]{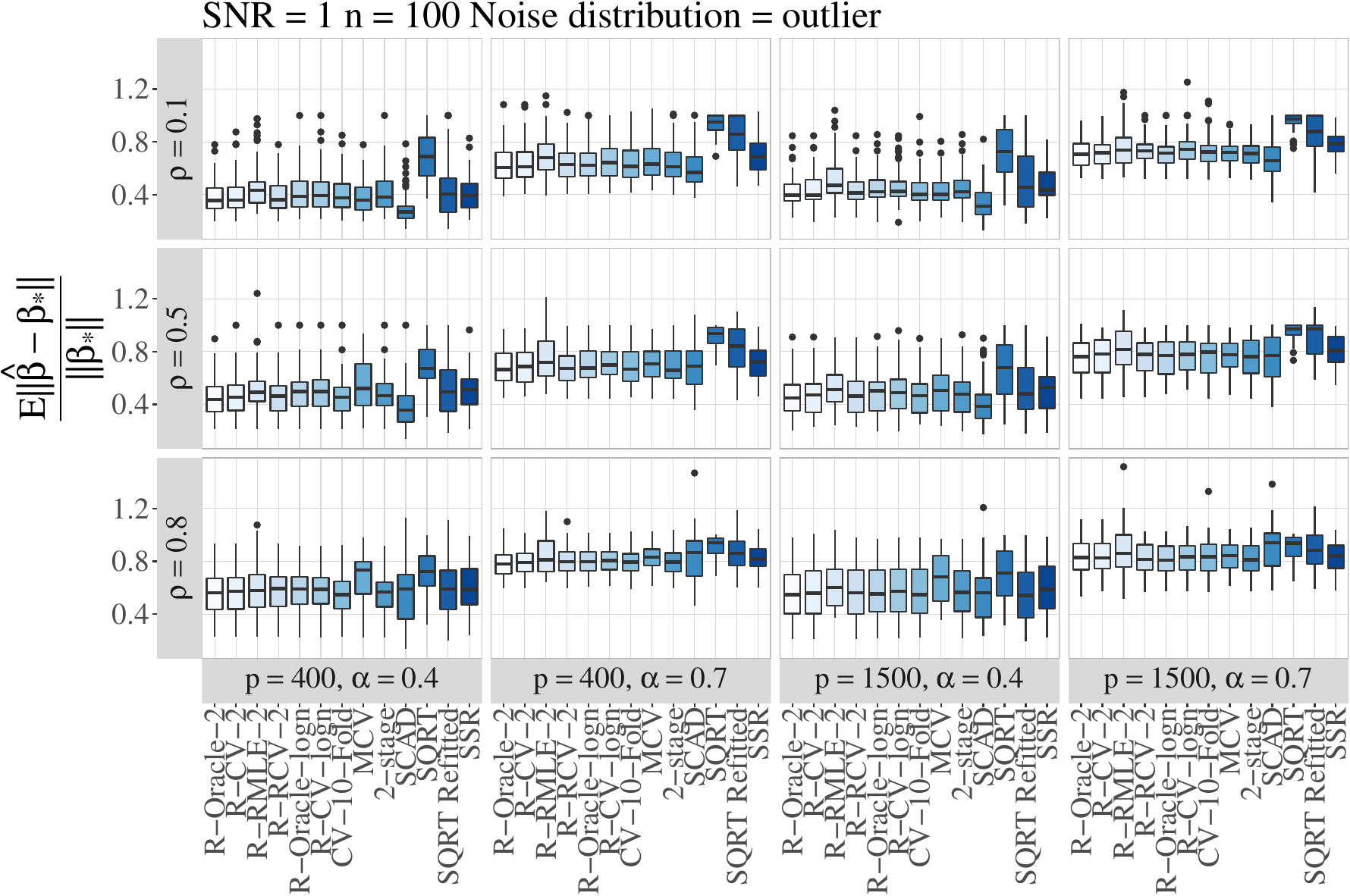}
\includegraphics[width=6in,height=4in]{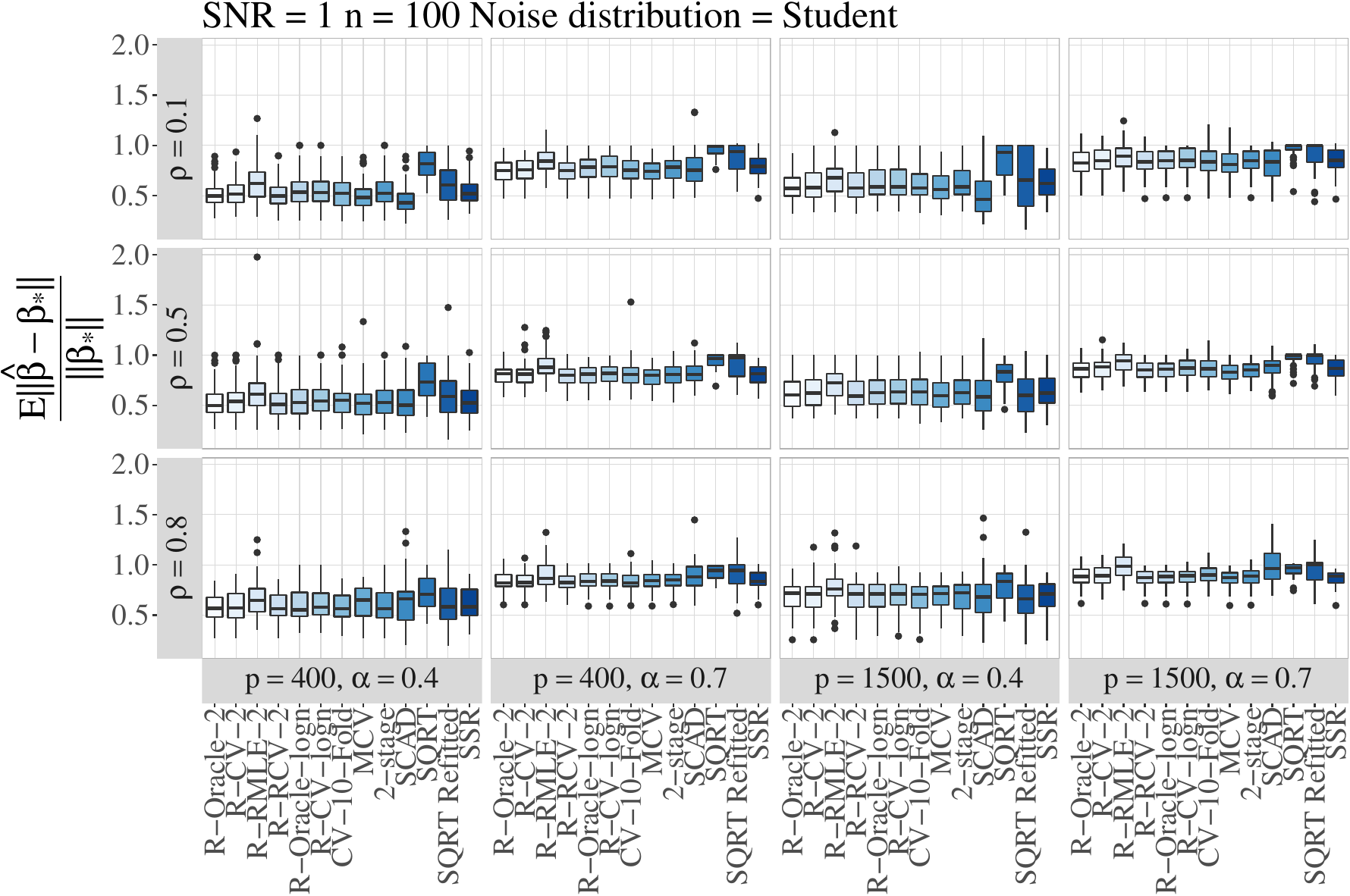}
\includegraphics[width=6in,height=4in]{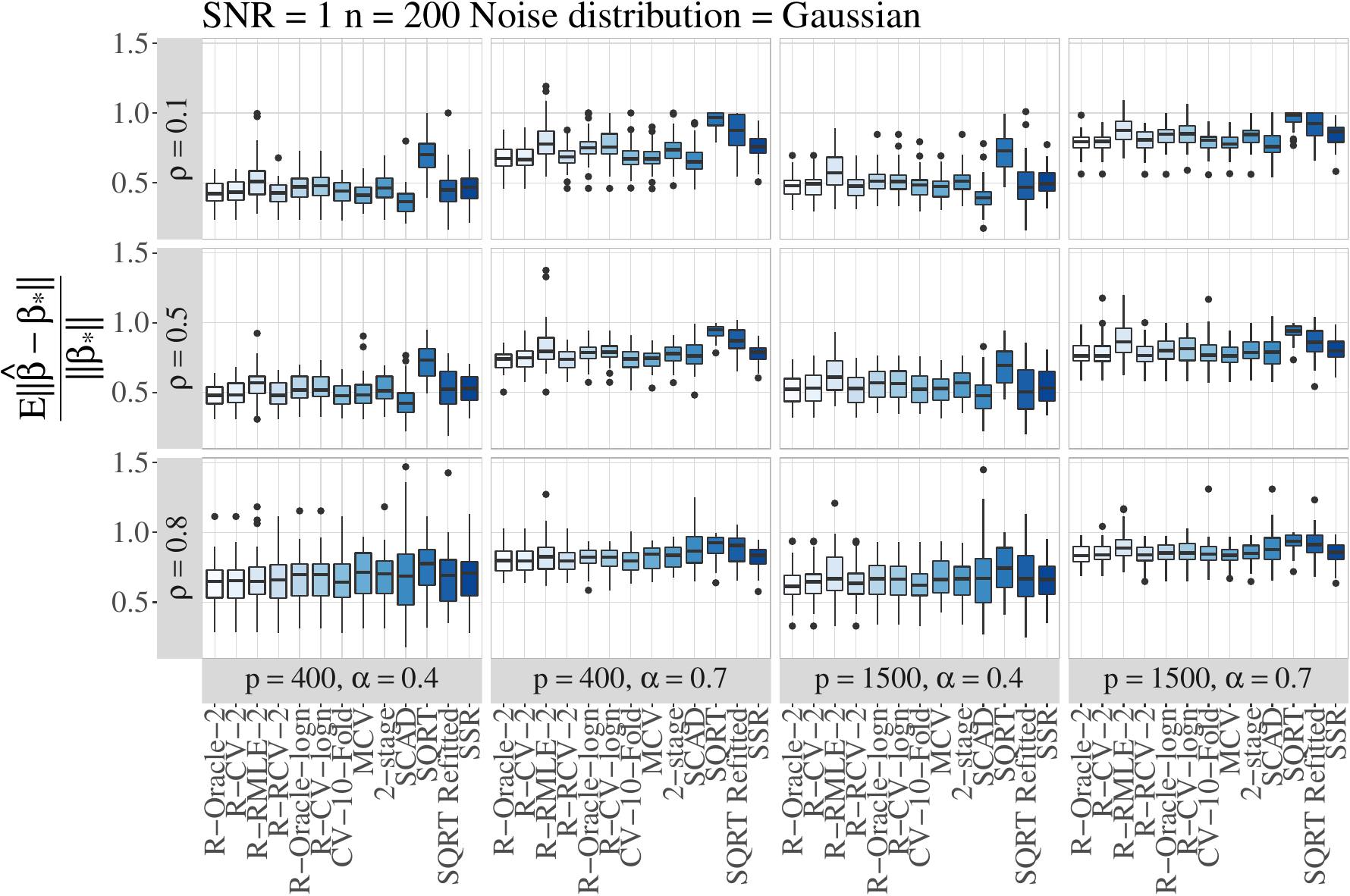}
\includegraphics[width=6in,height=4in]{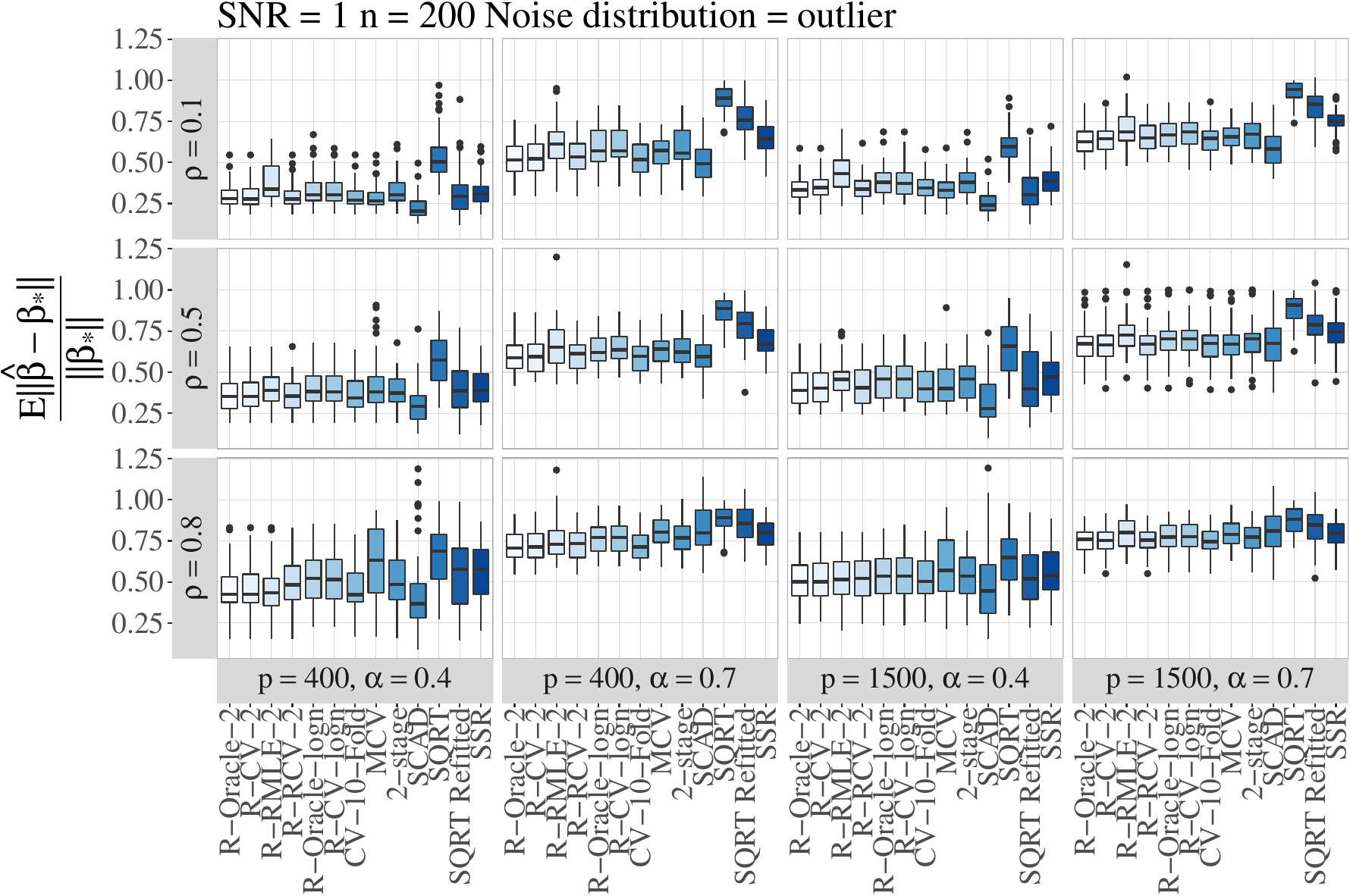}
\includegraphics[width=6in,height=4in]{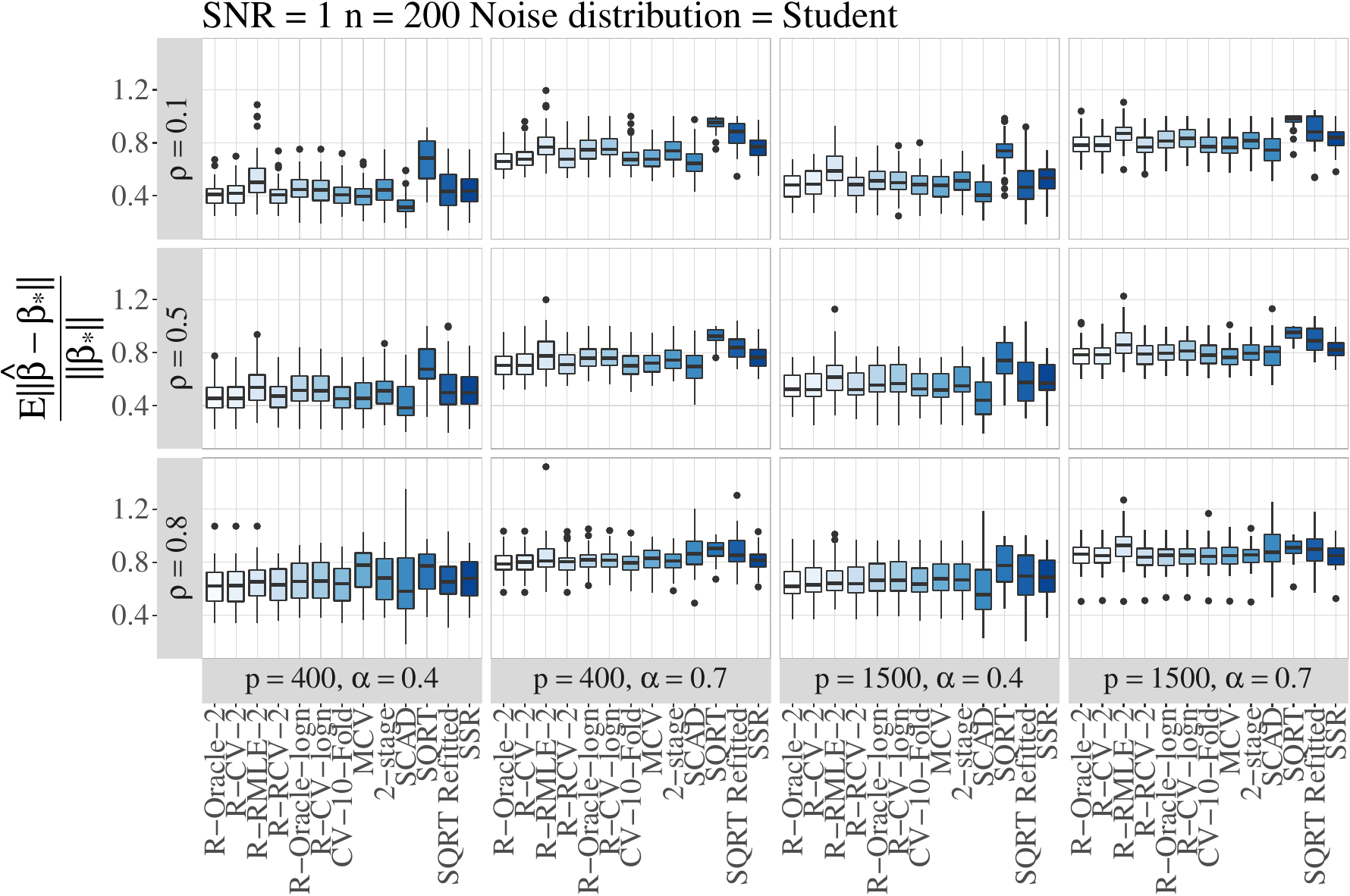}
\includegraphics[width=6in,height=4in]{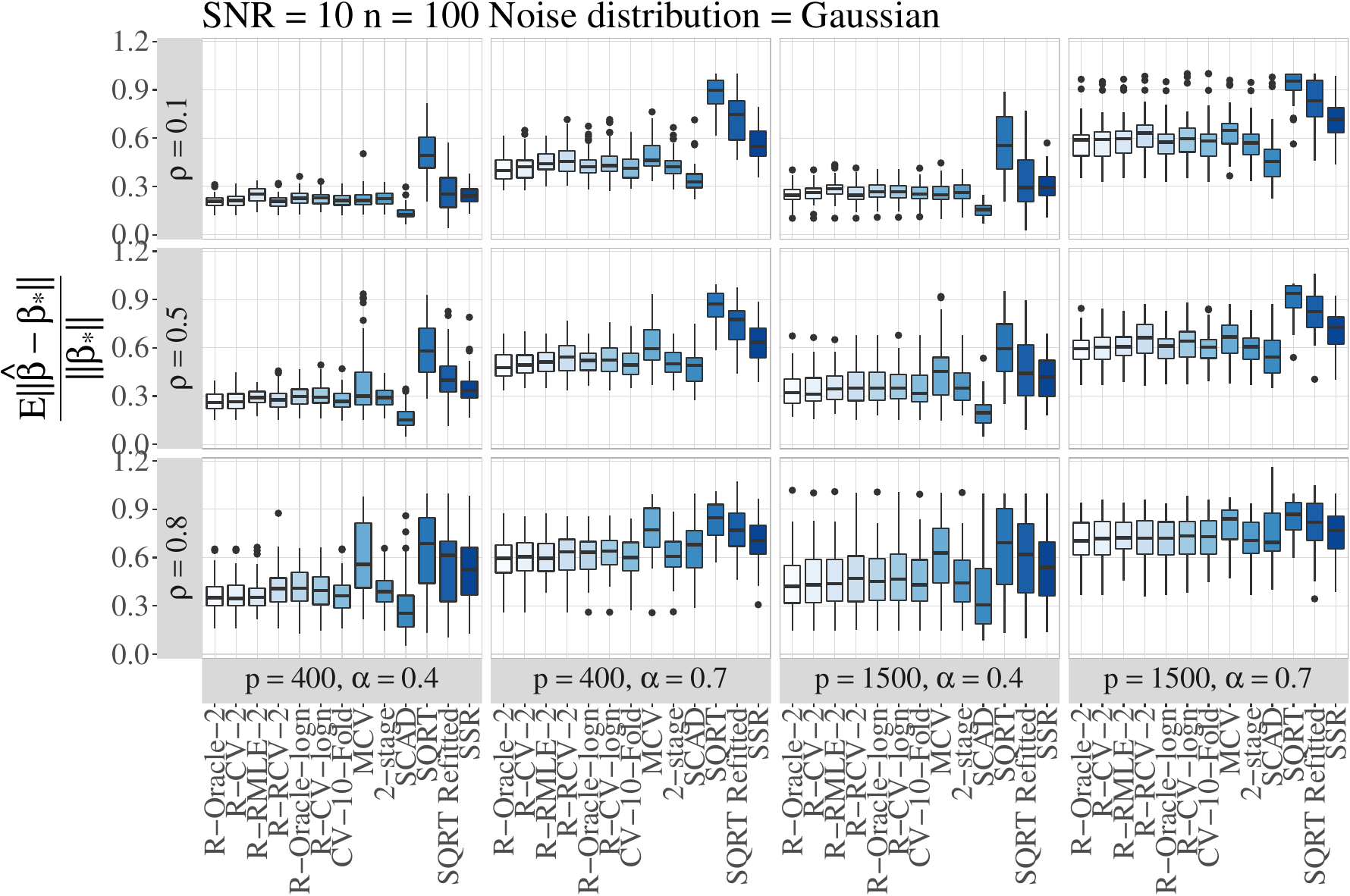}
\includegraphics[width=6in,height=4in]{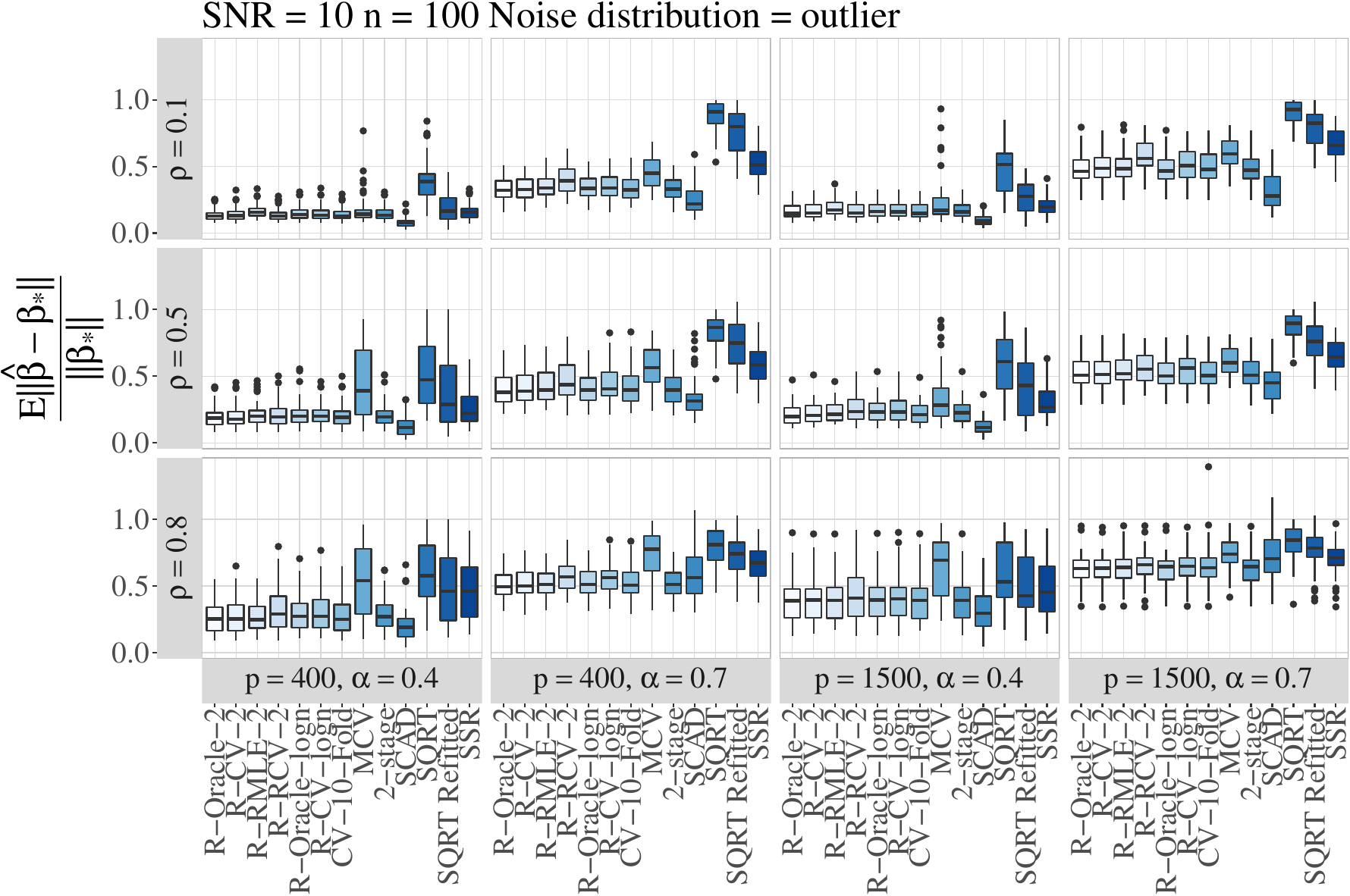}
\includegraphics[width=6in,height=4in]{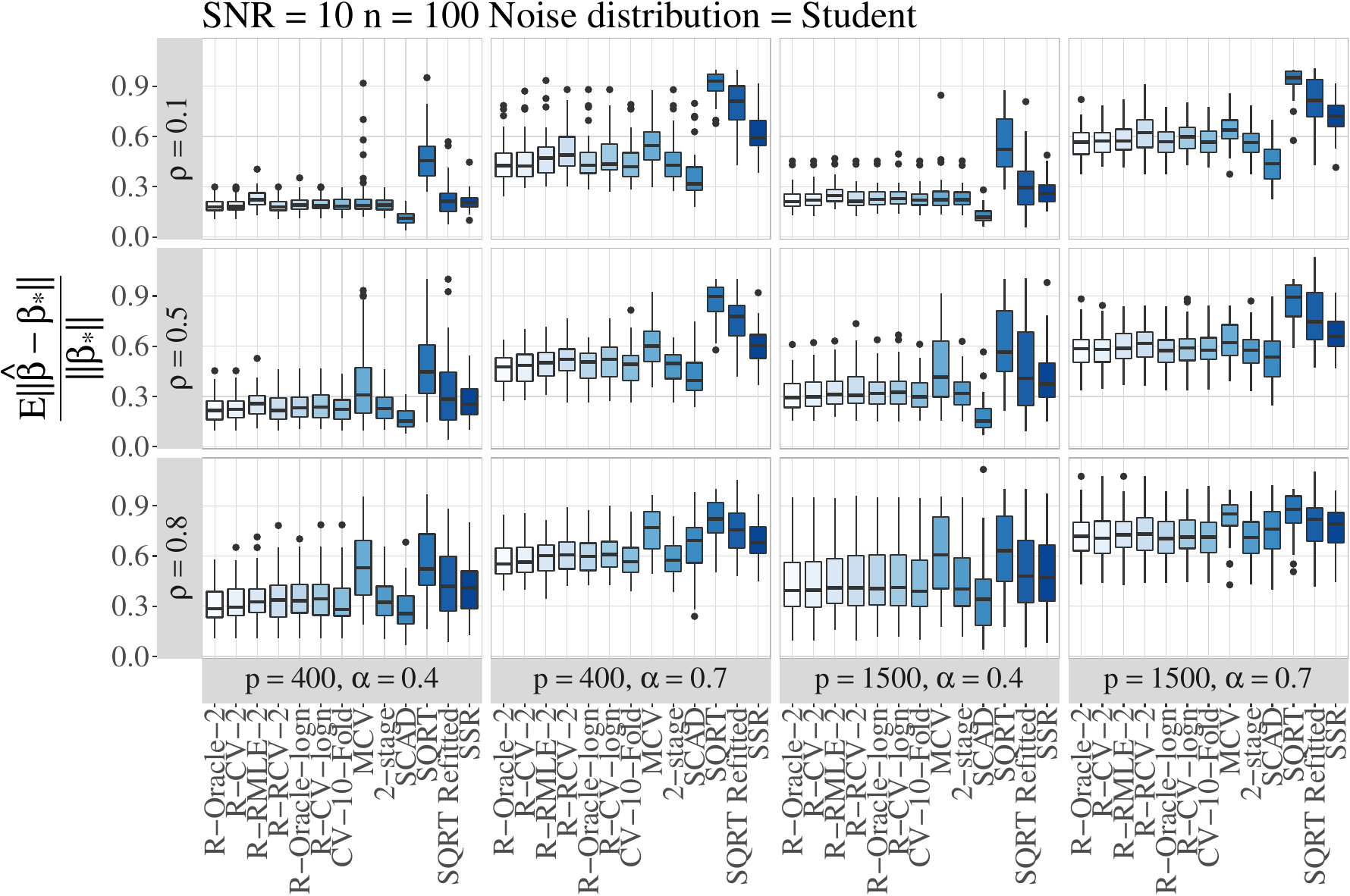}
\includegraphics[width=6in,height=4in]{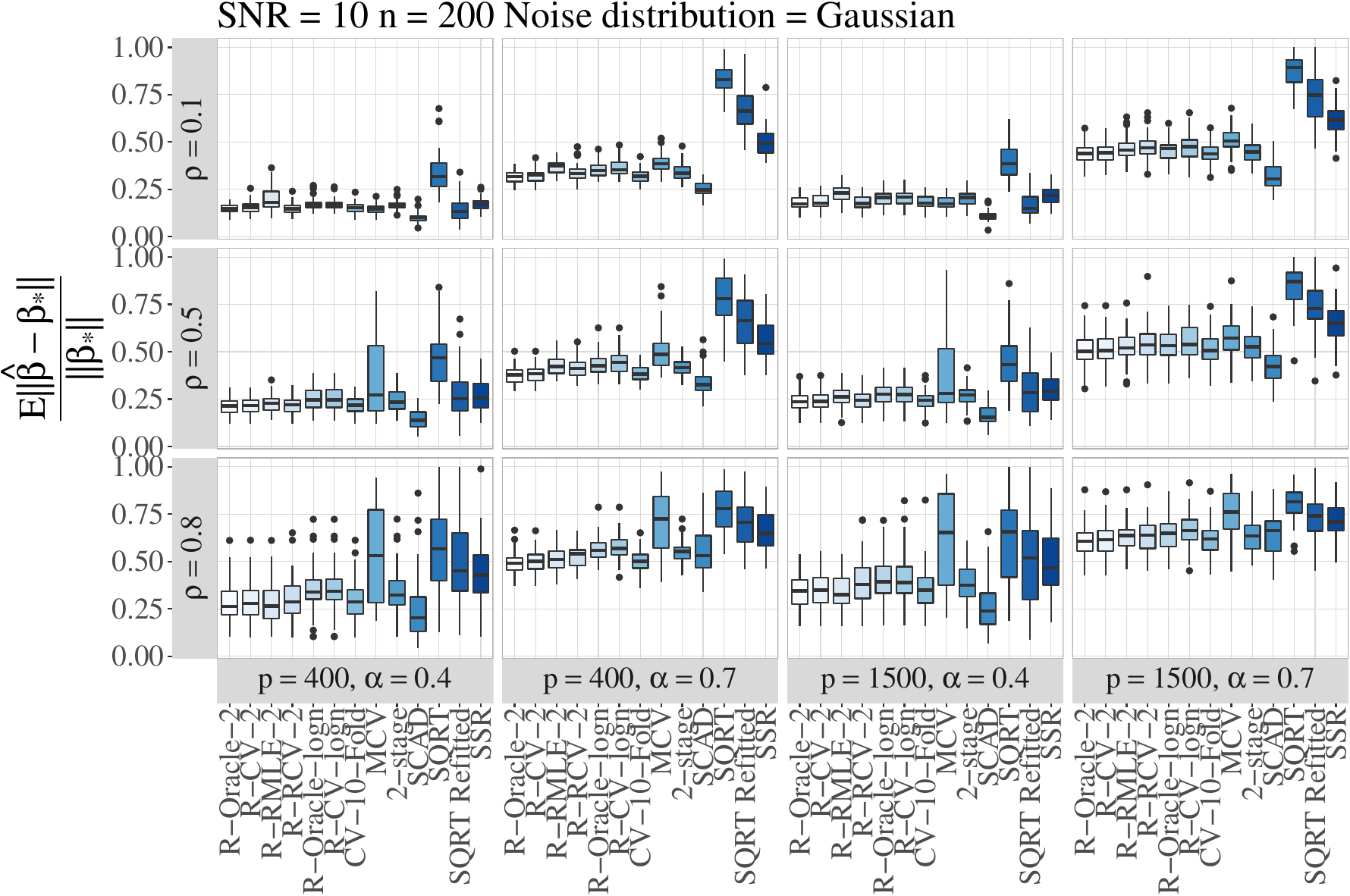}
\includegraphics[width=6in,height=4in]{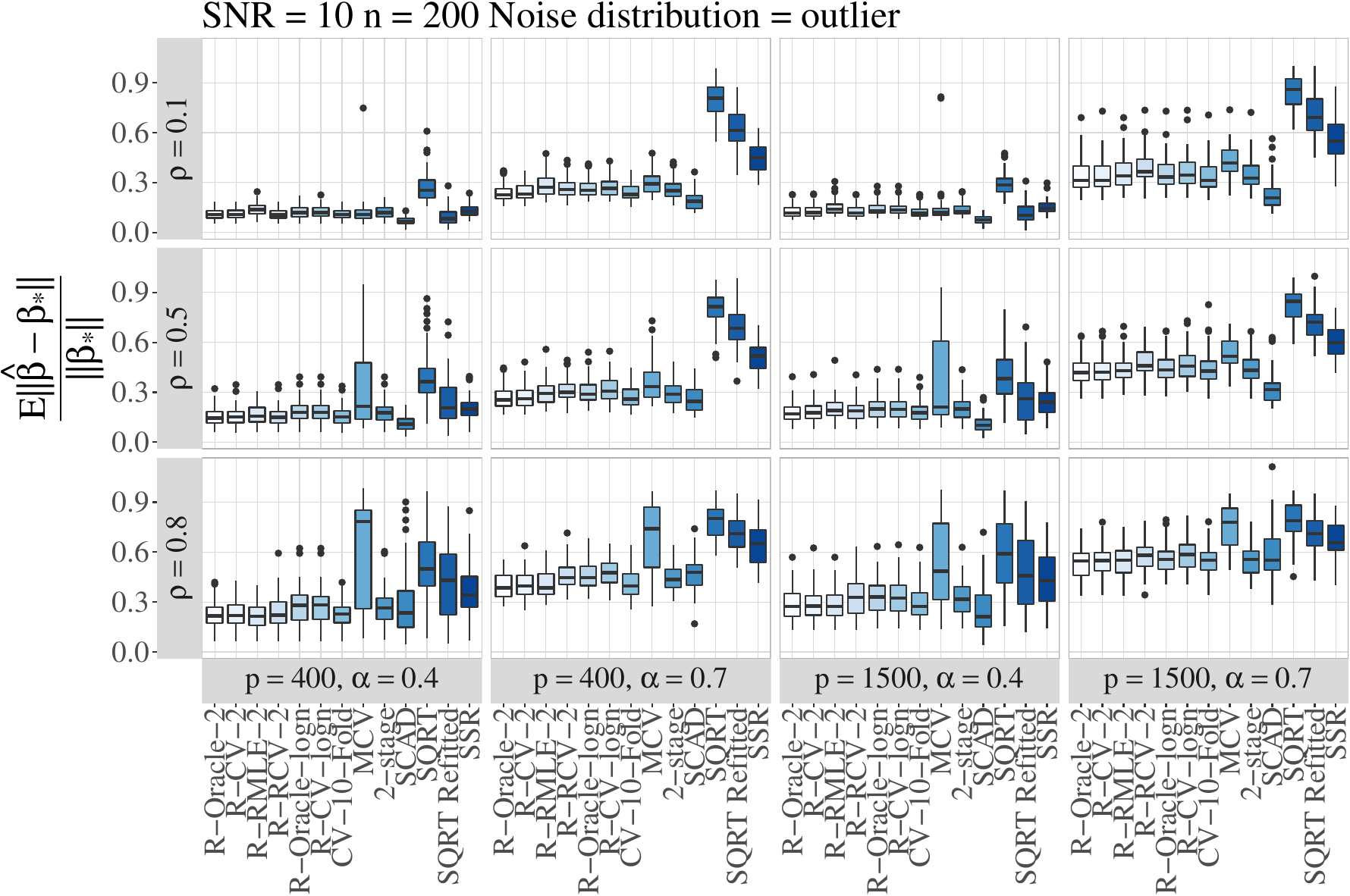}
\includegraphics[width=6in,height=4in]{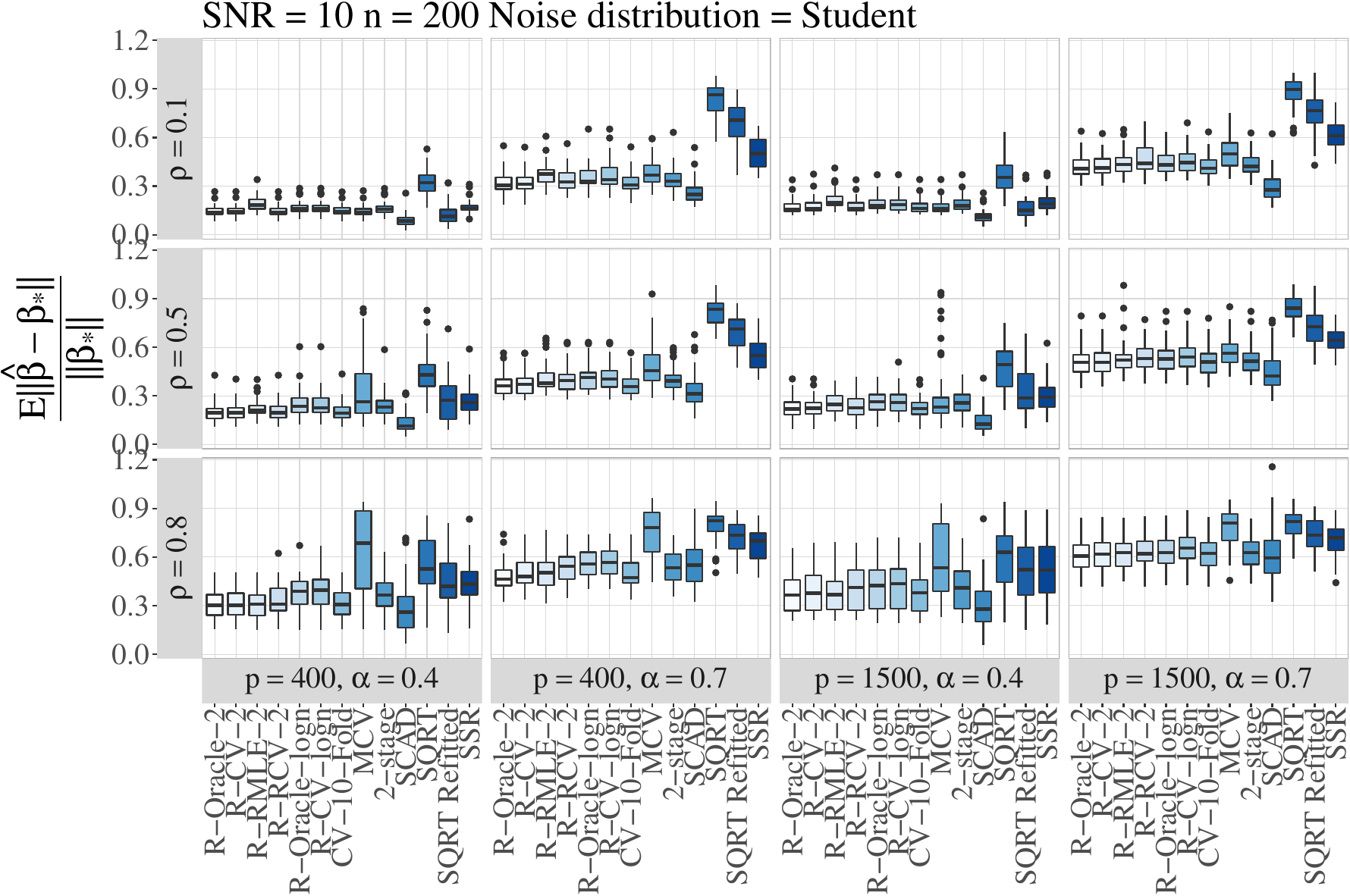}
\end{center}

\hypertarget{f-score}{%
\subsection{F score}\label{f-score}}

\begin{center}
\includegraphics[width=6in,height=4in]{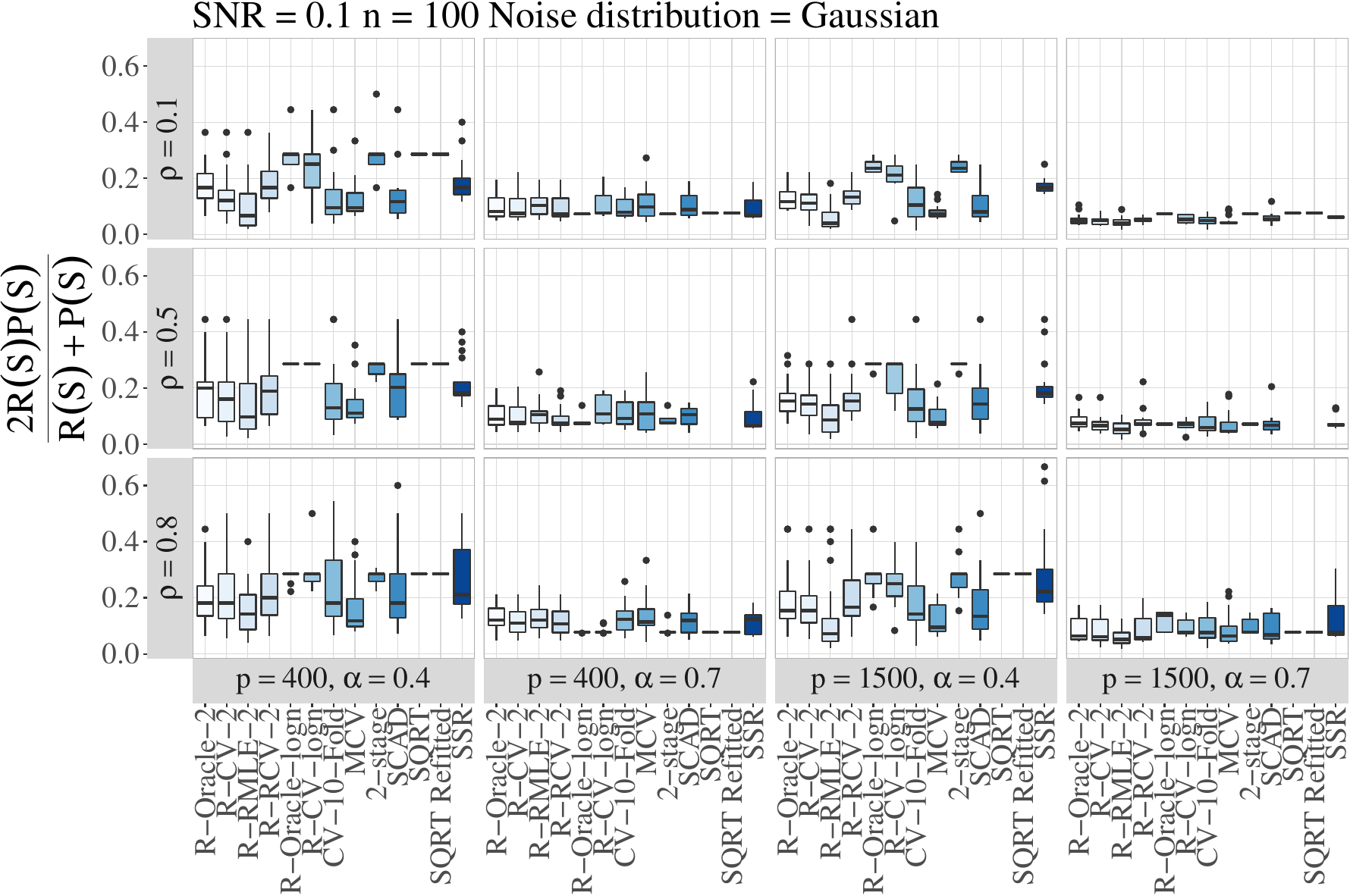}
\includegraphics[width=6in,height=4in]{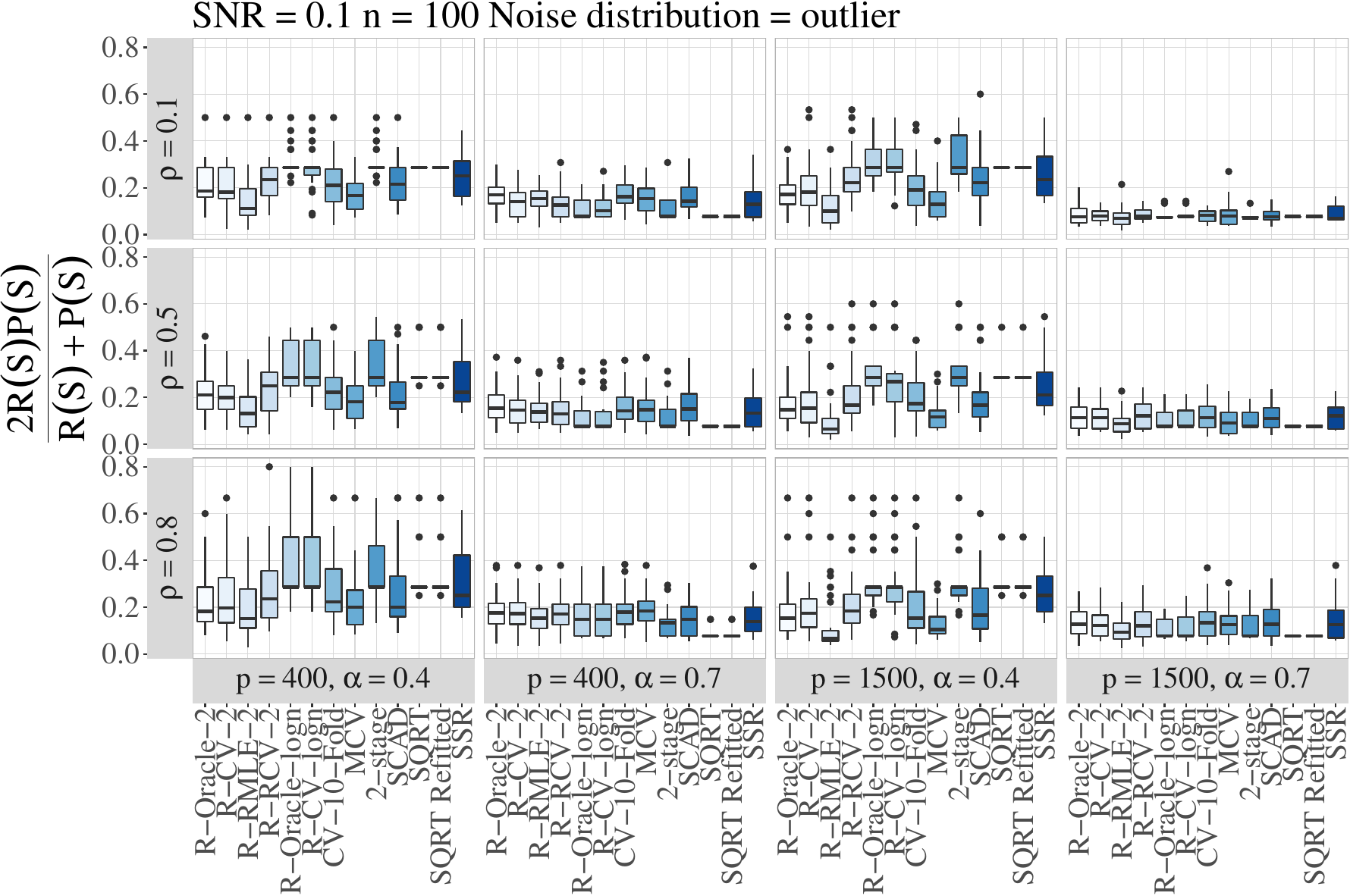}
\includegraphics[width=6in,height=4in]{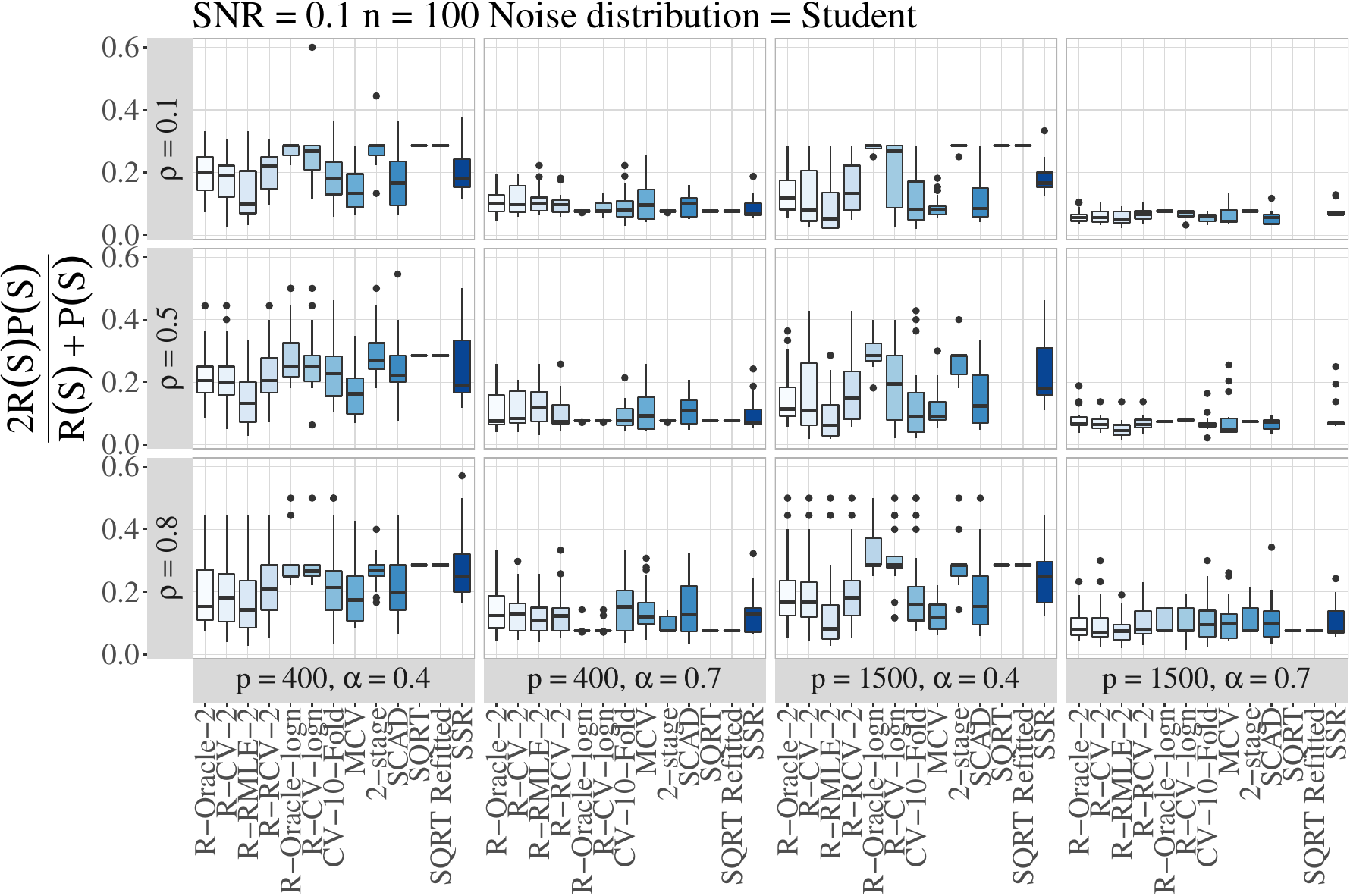}
\includegraphics[width=6in,height=4in]{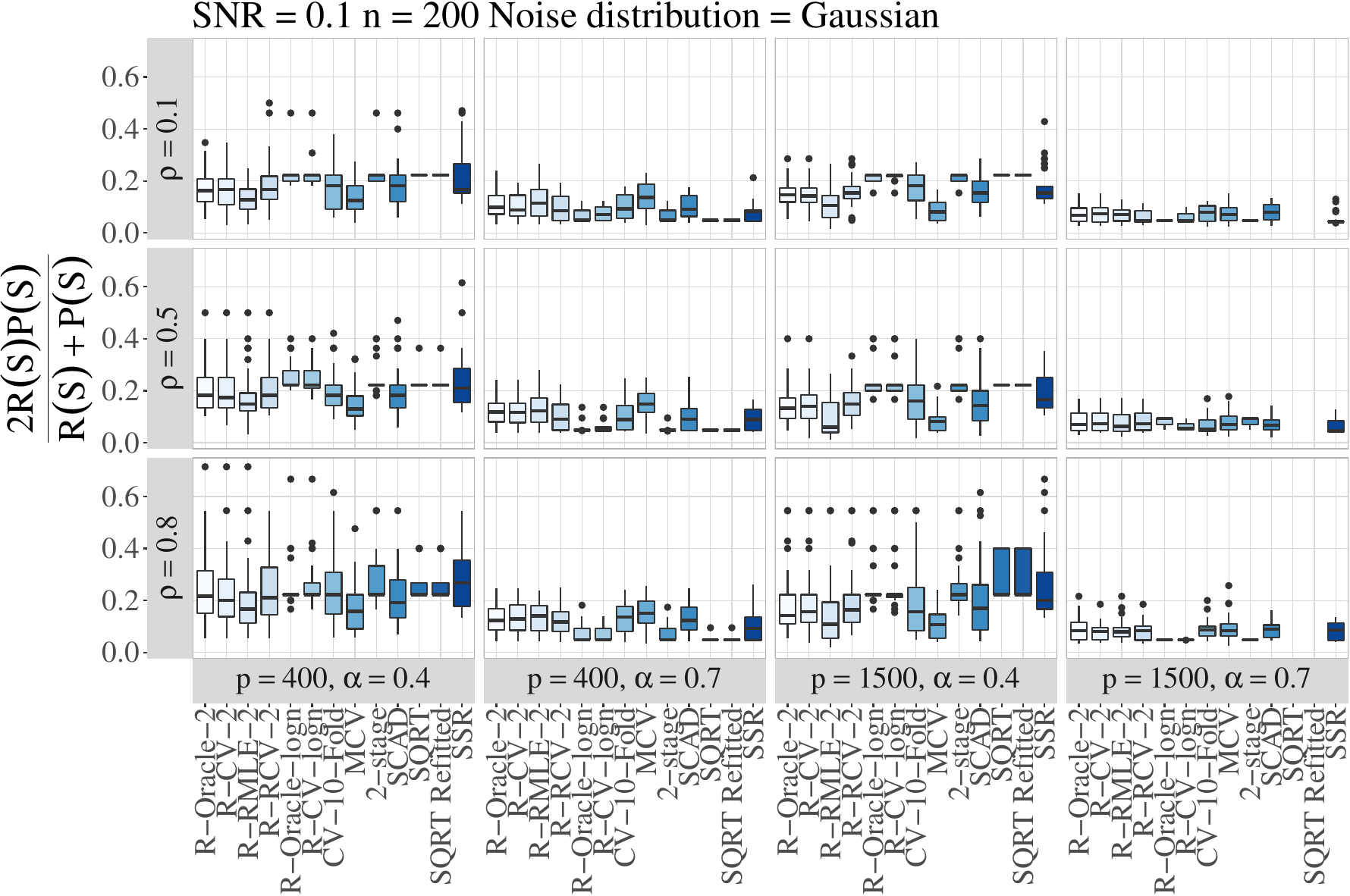}
\includegraphics[width=6in,height=4in]{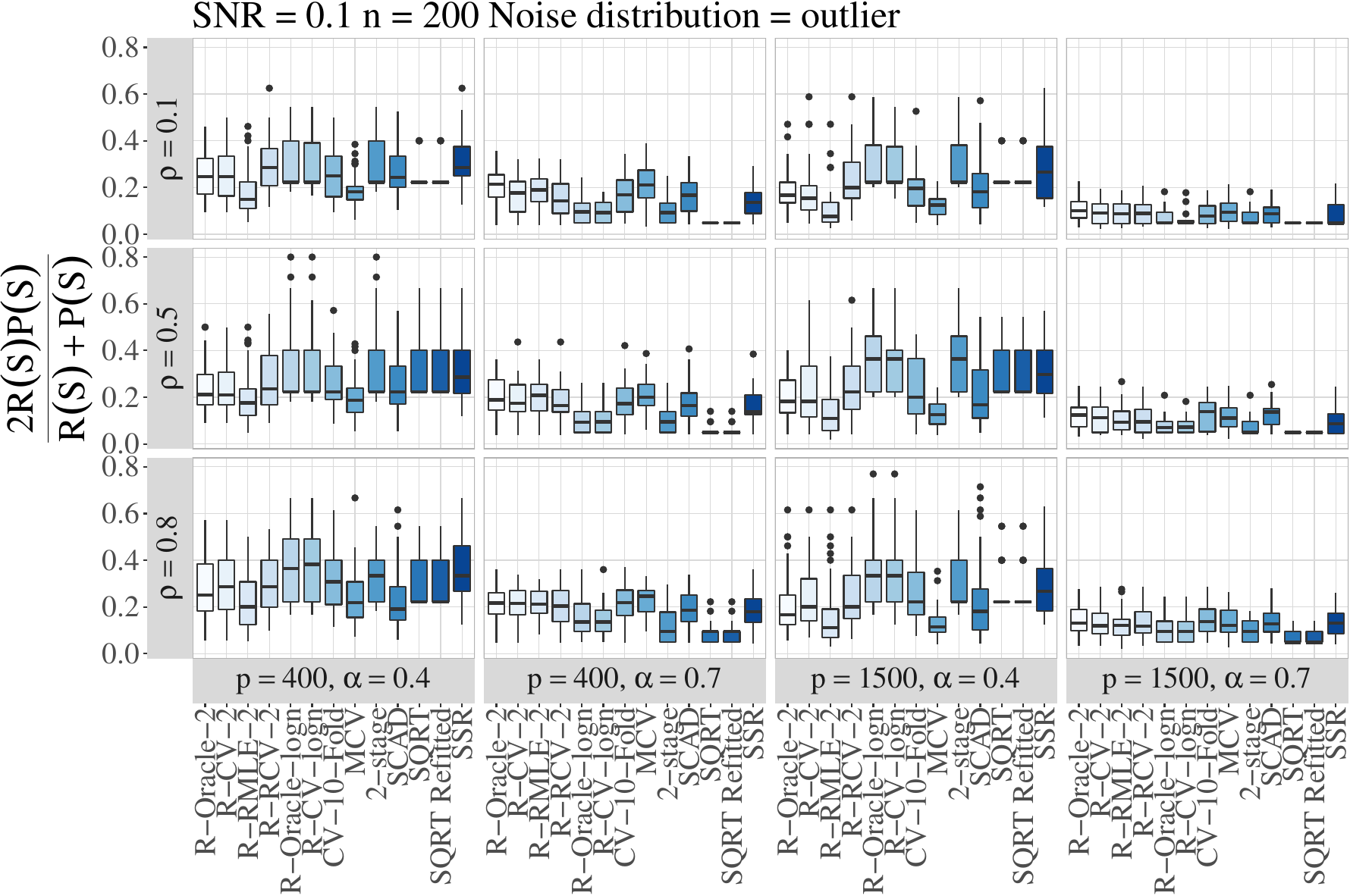}
\includegraphics[width=6in,height=4in]{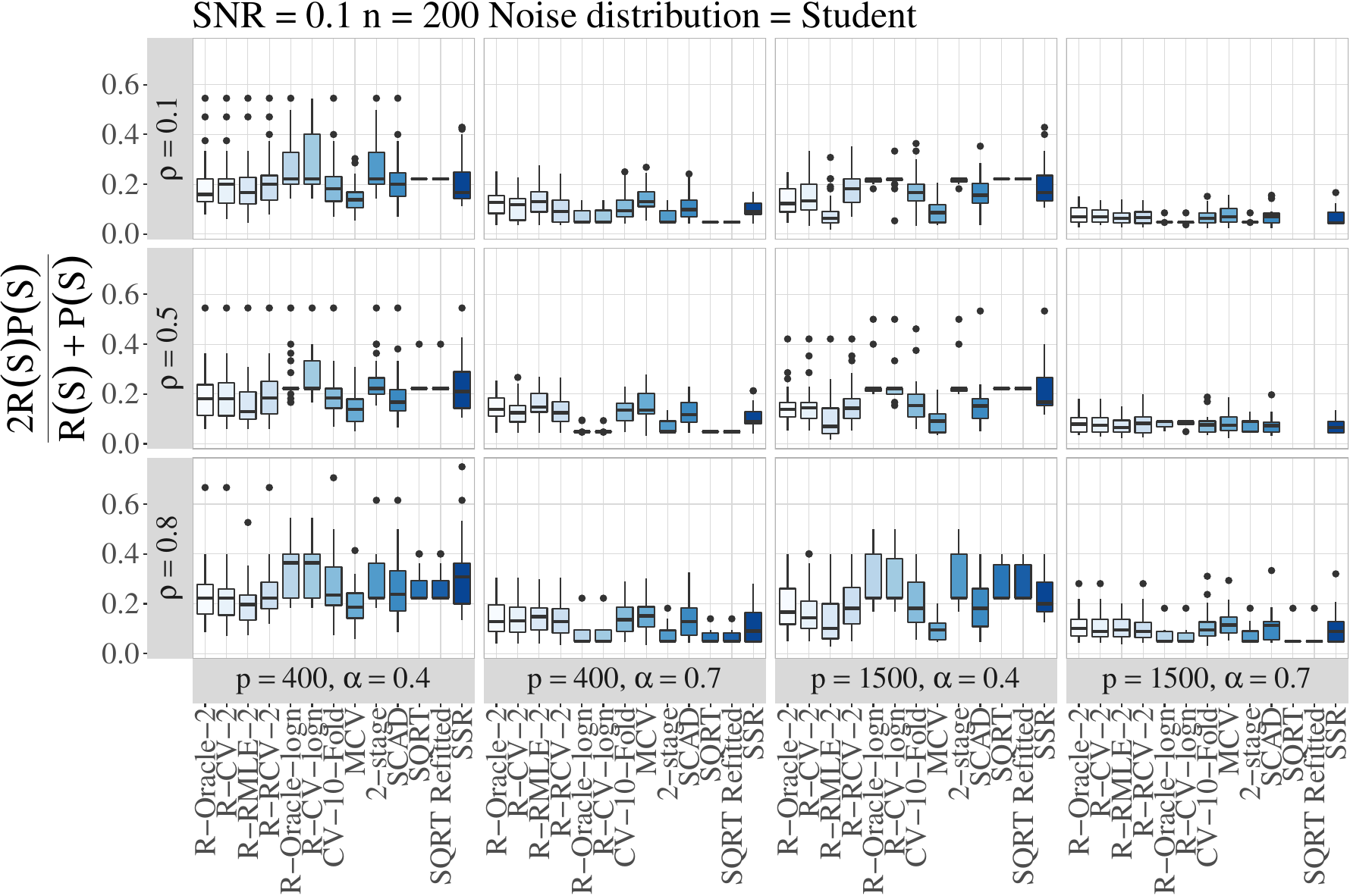}
\includegraphics[width=6in,height=4in]{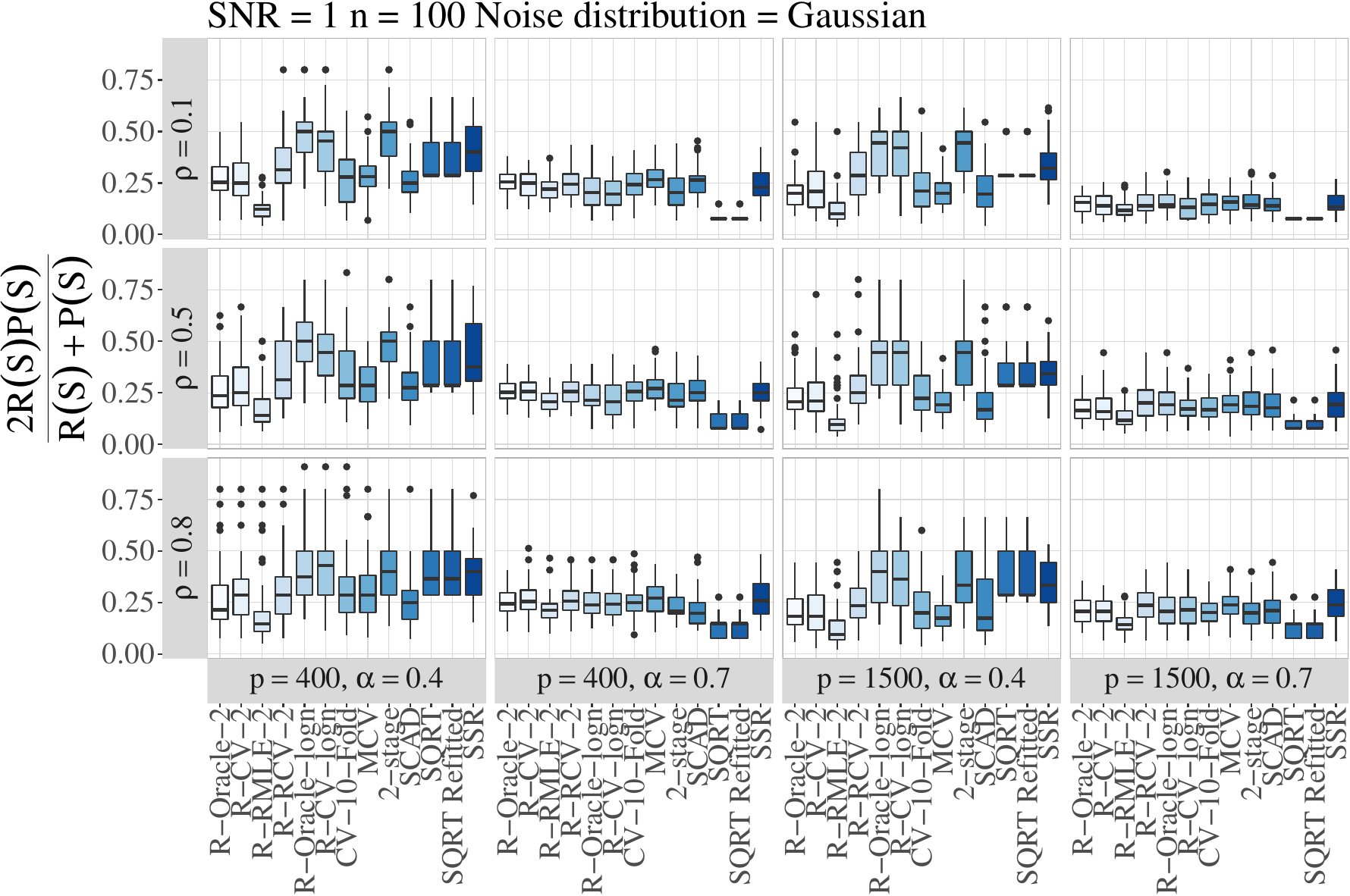}
\includegraphics[width=6in,height=4in]{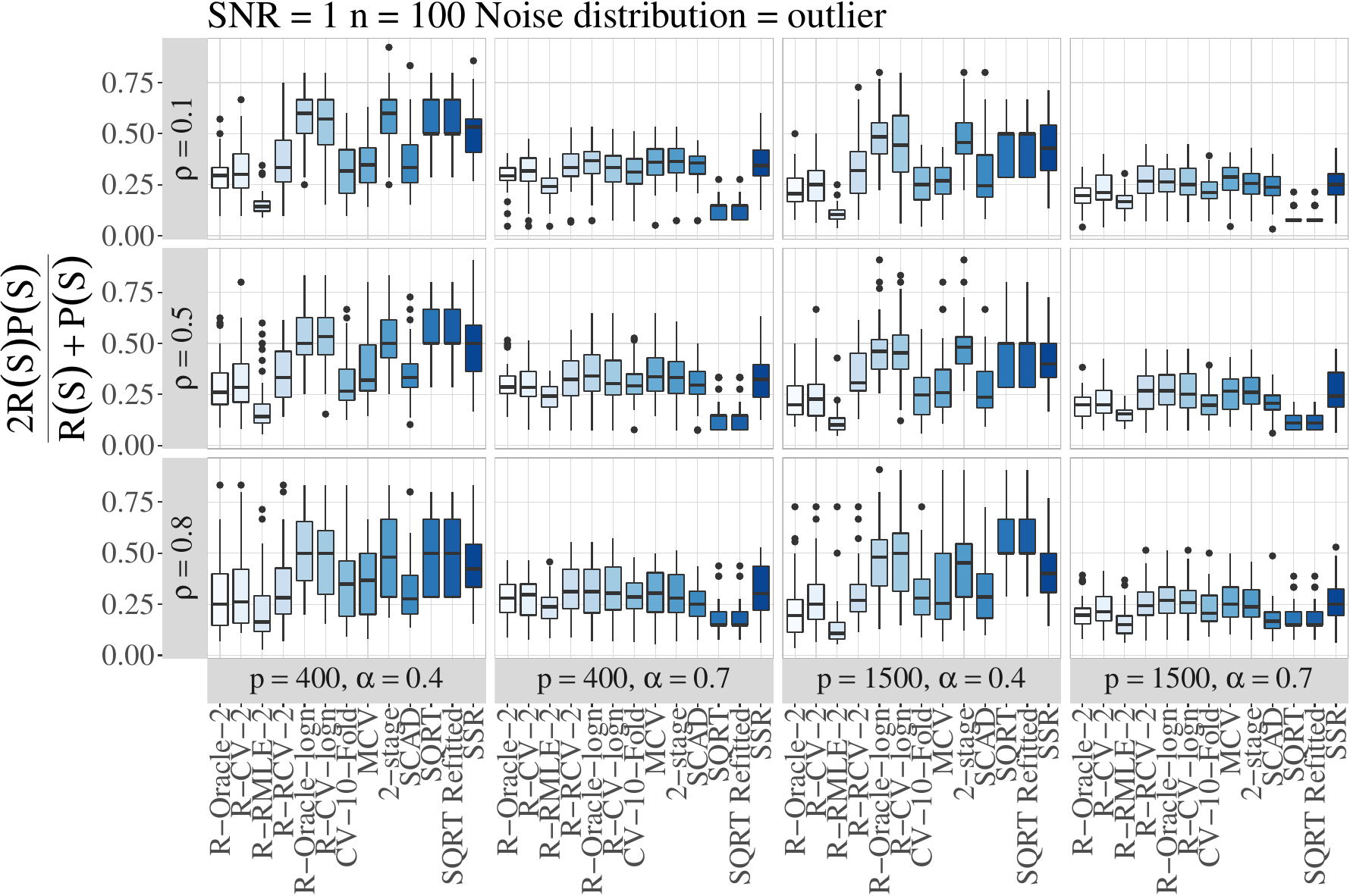}
\includegraphics[width=6in,height=4in]{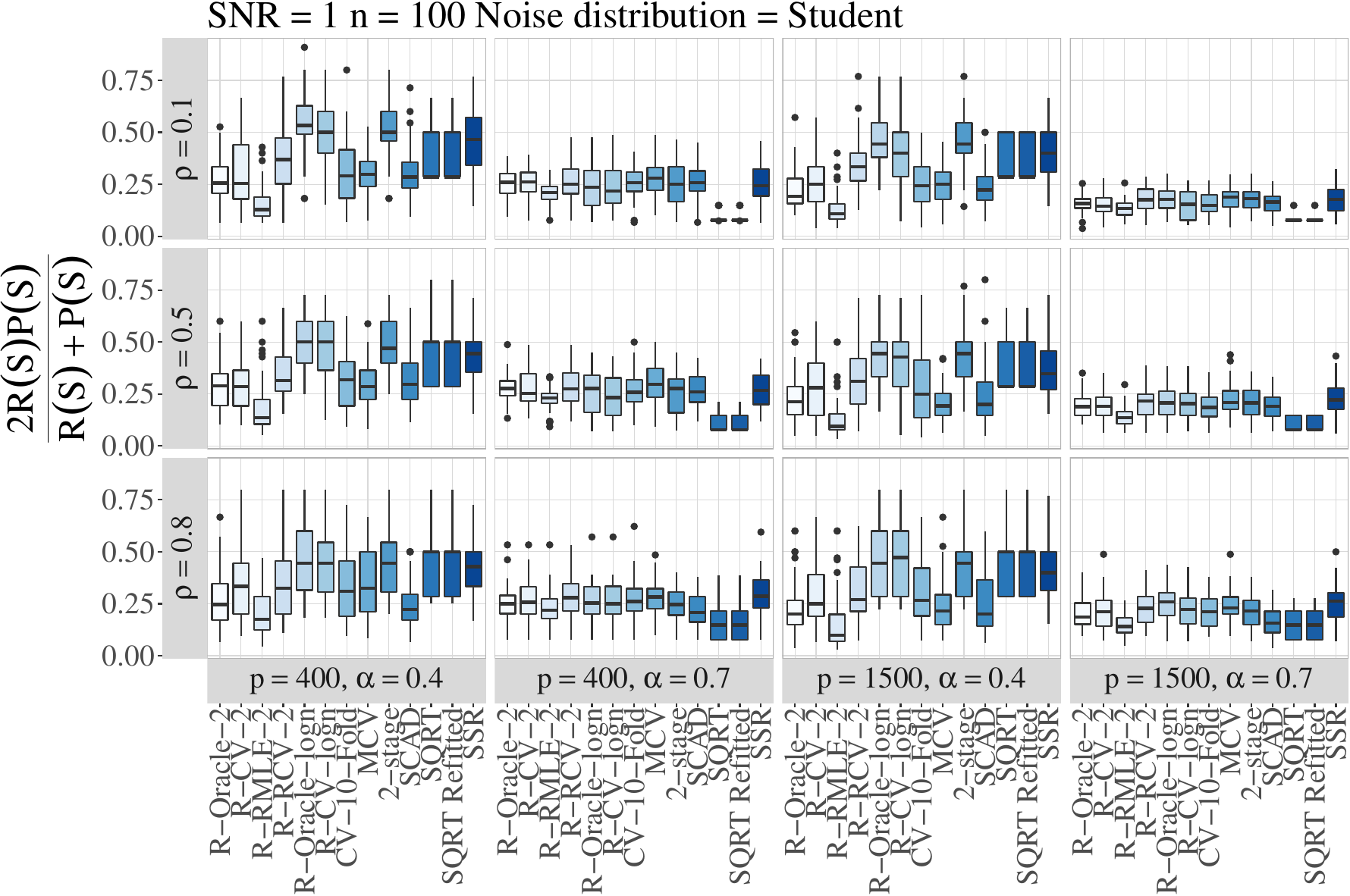}
\includegraphics[width=6in,height=4in]{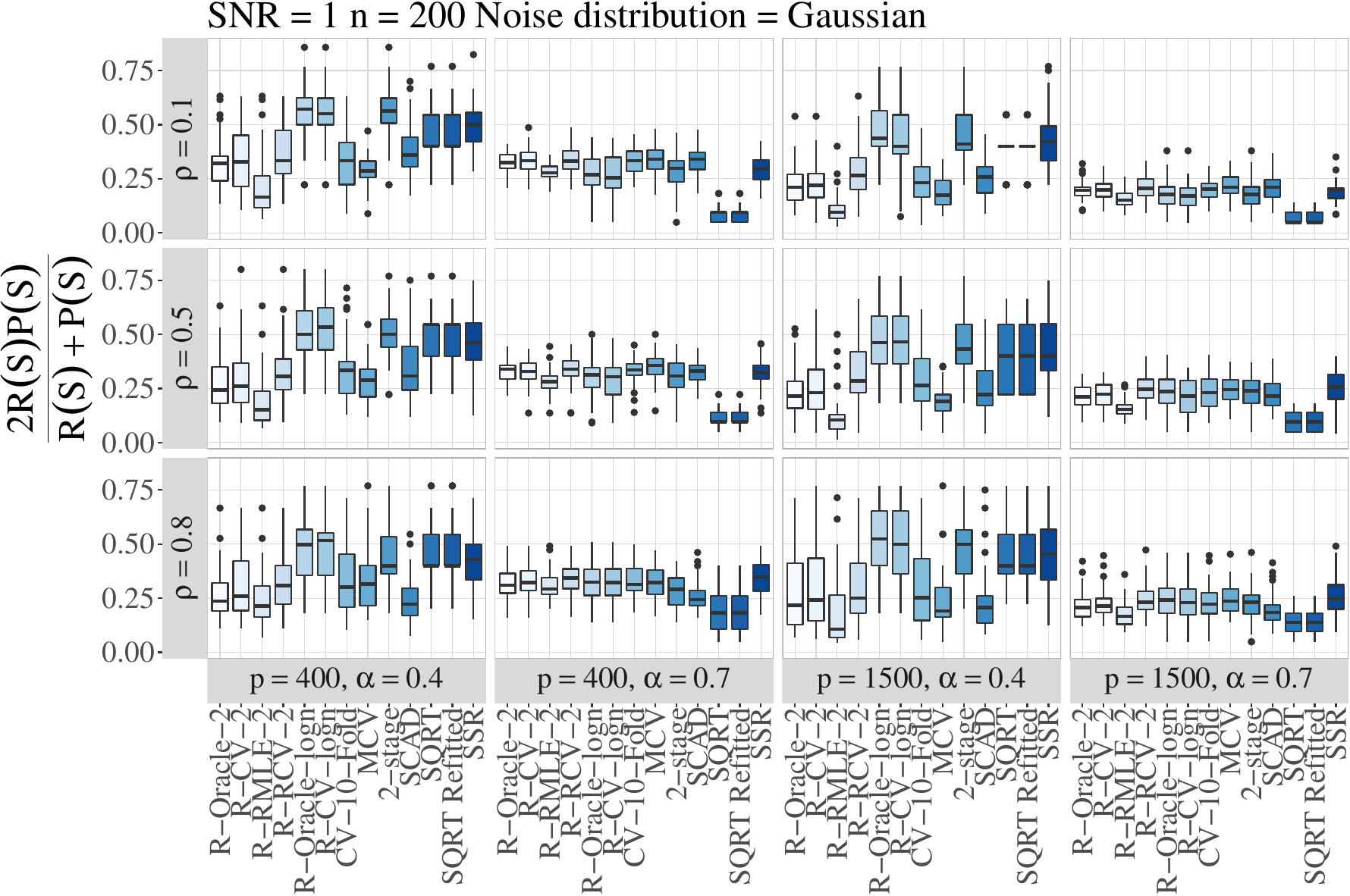}
\includegraphics[width=6in,height=4in]{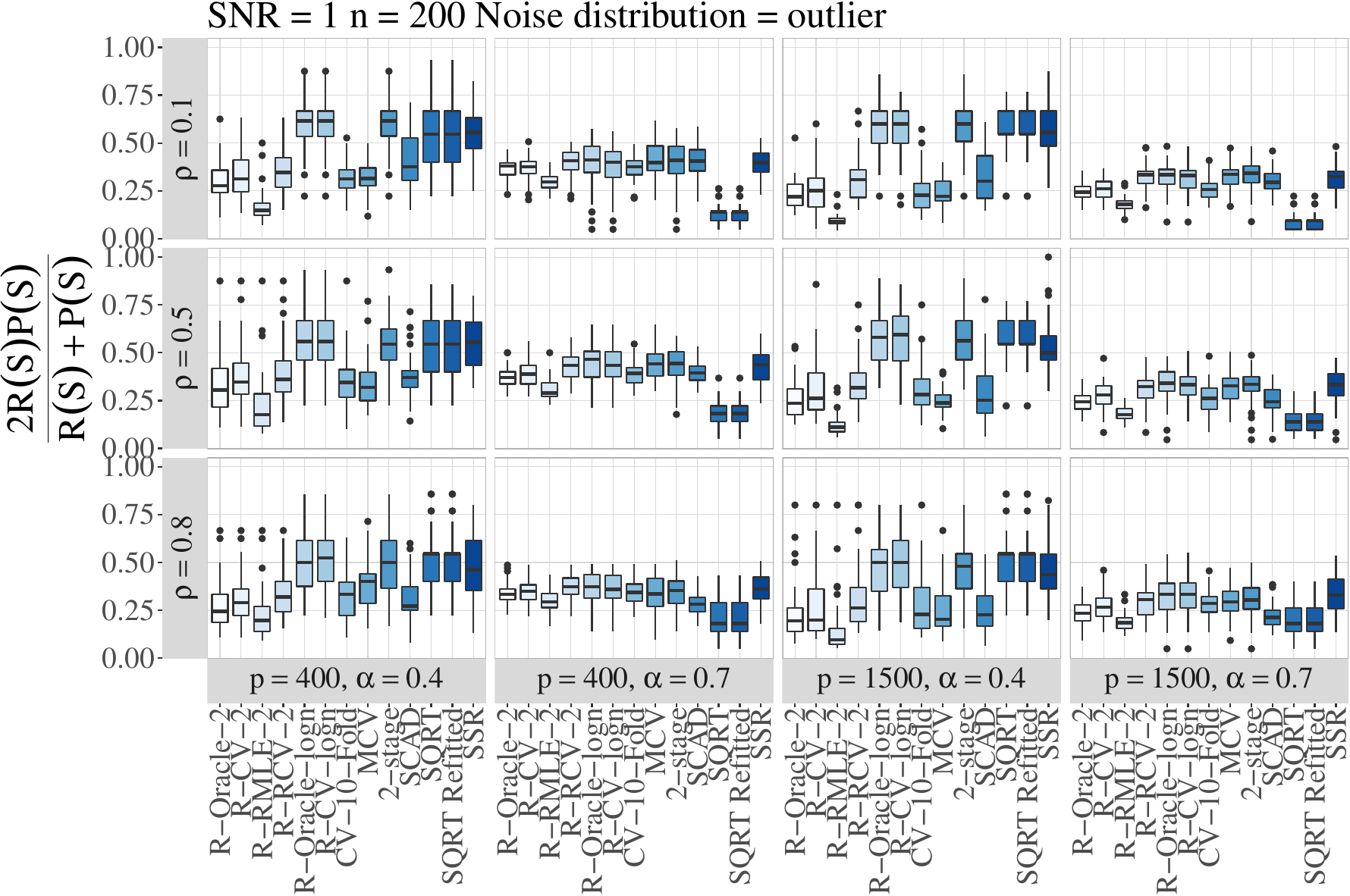}
\includegraphics[width=6in,height=4in]{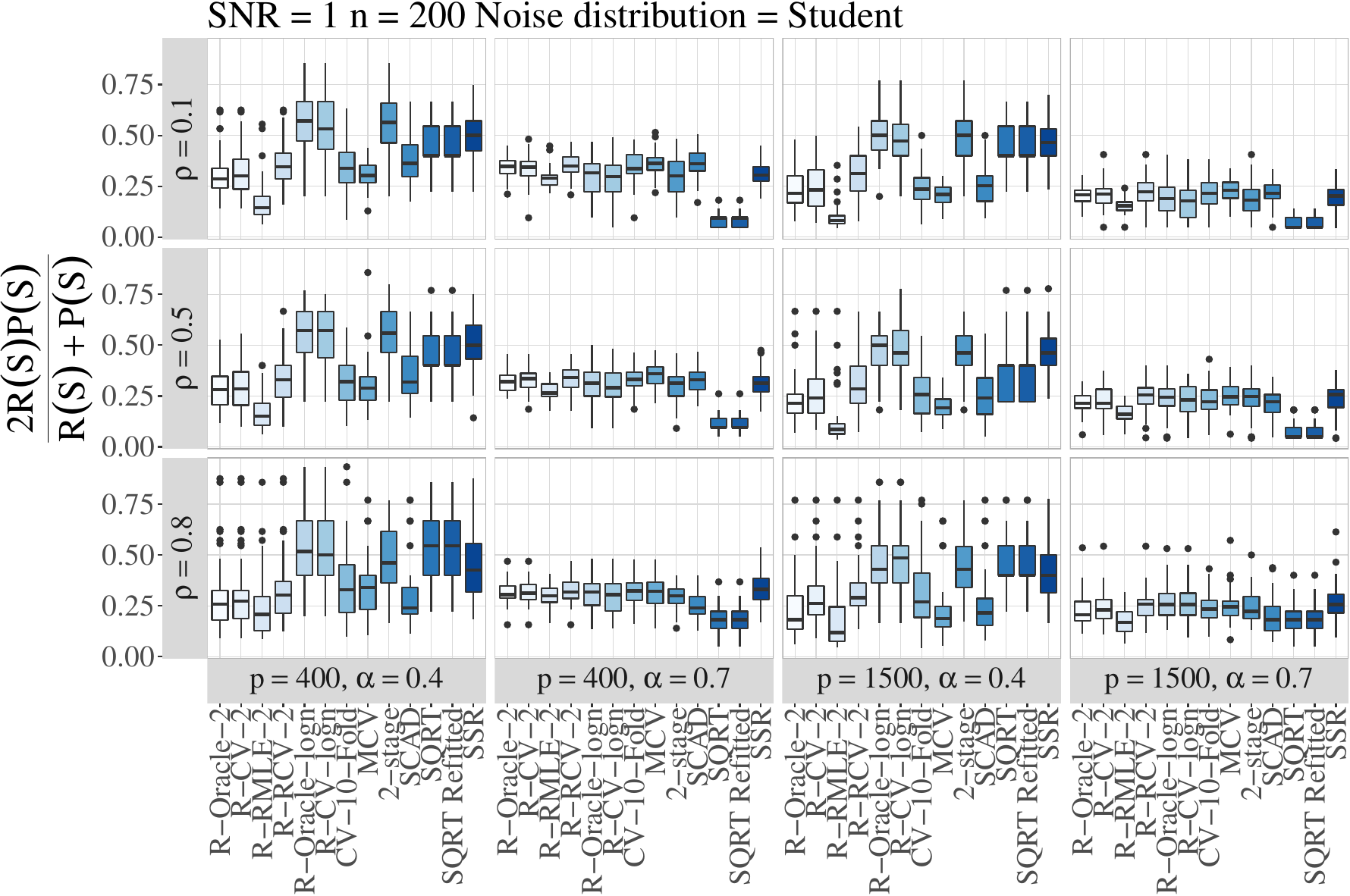}
\includegraphics[width=6in,height=4in]{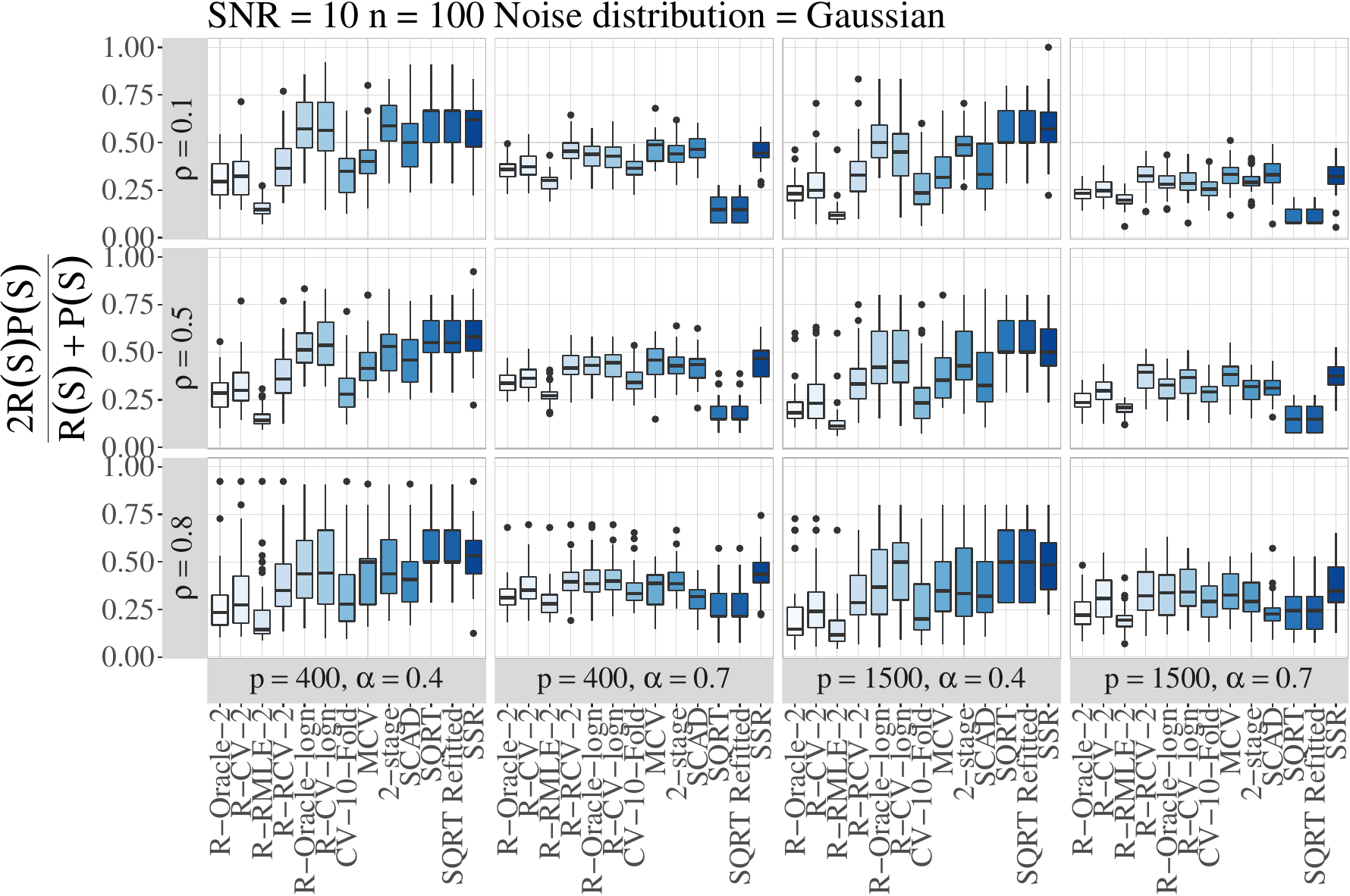}
\includegraphics[width=6in,height=4in]{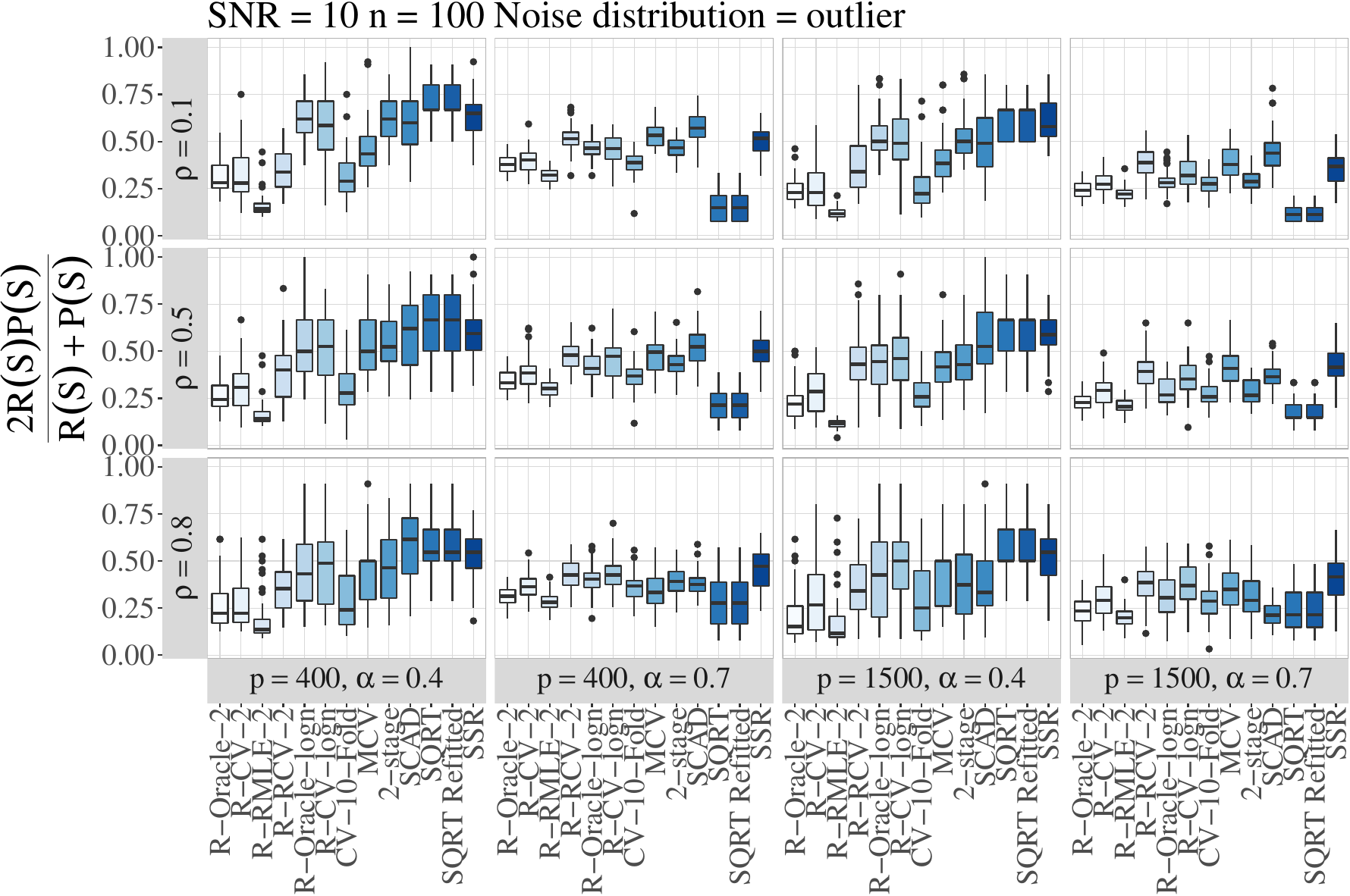}
\includegraphics[width=6in,height=4in]{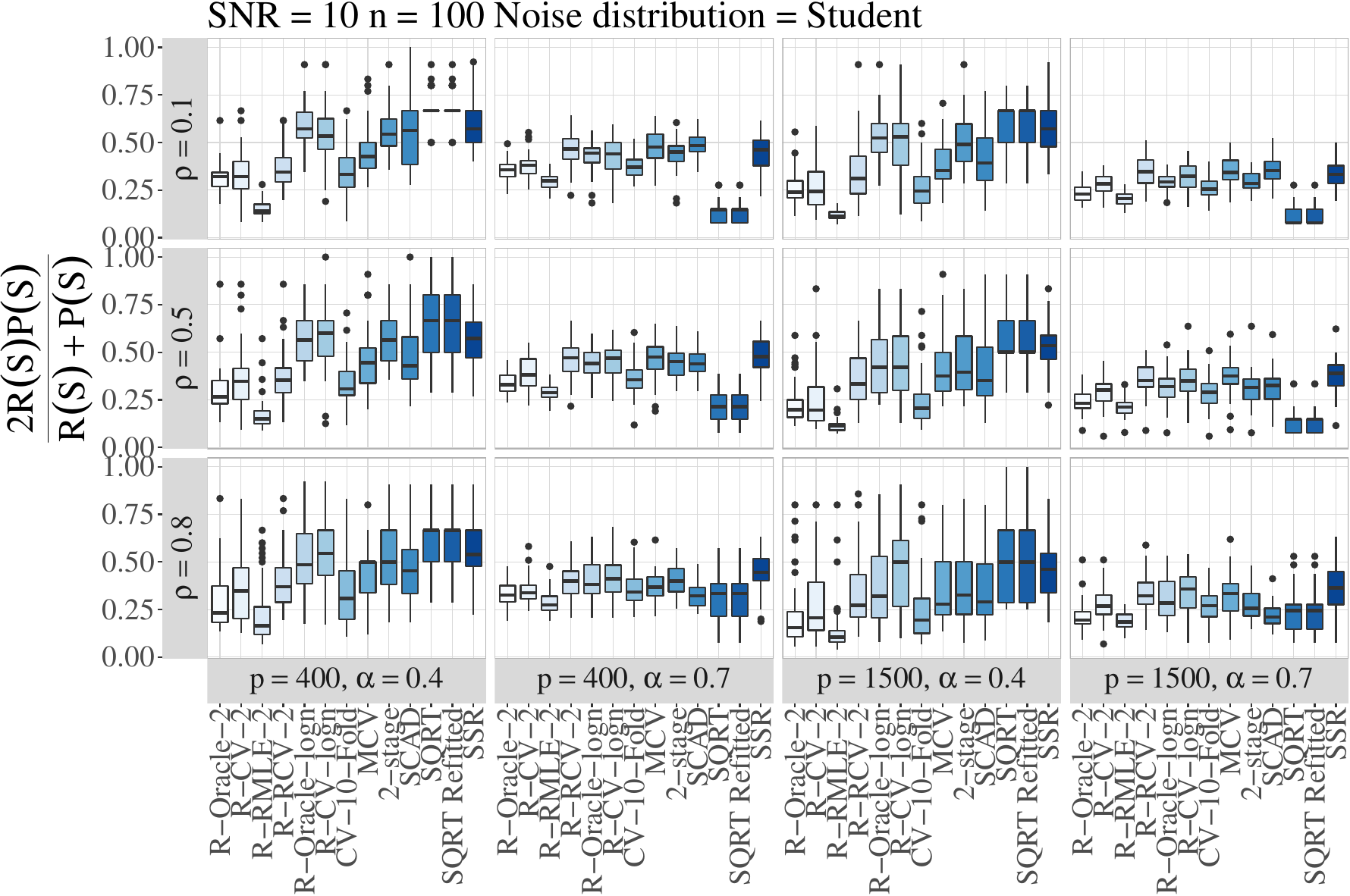}
\includegraphics[width=6in,height=4in]{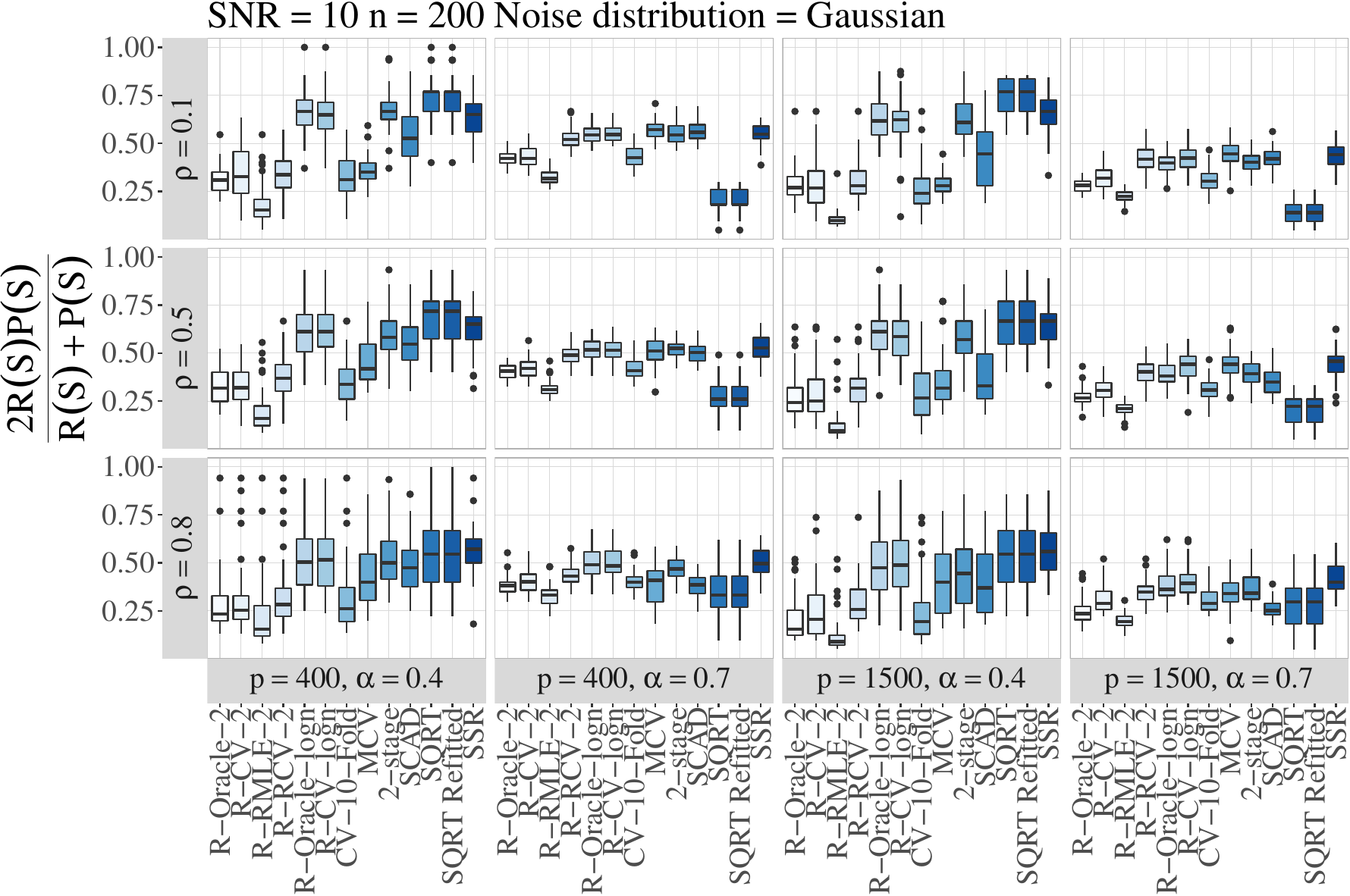}
\includegraphics[width=6in,height=4in]{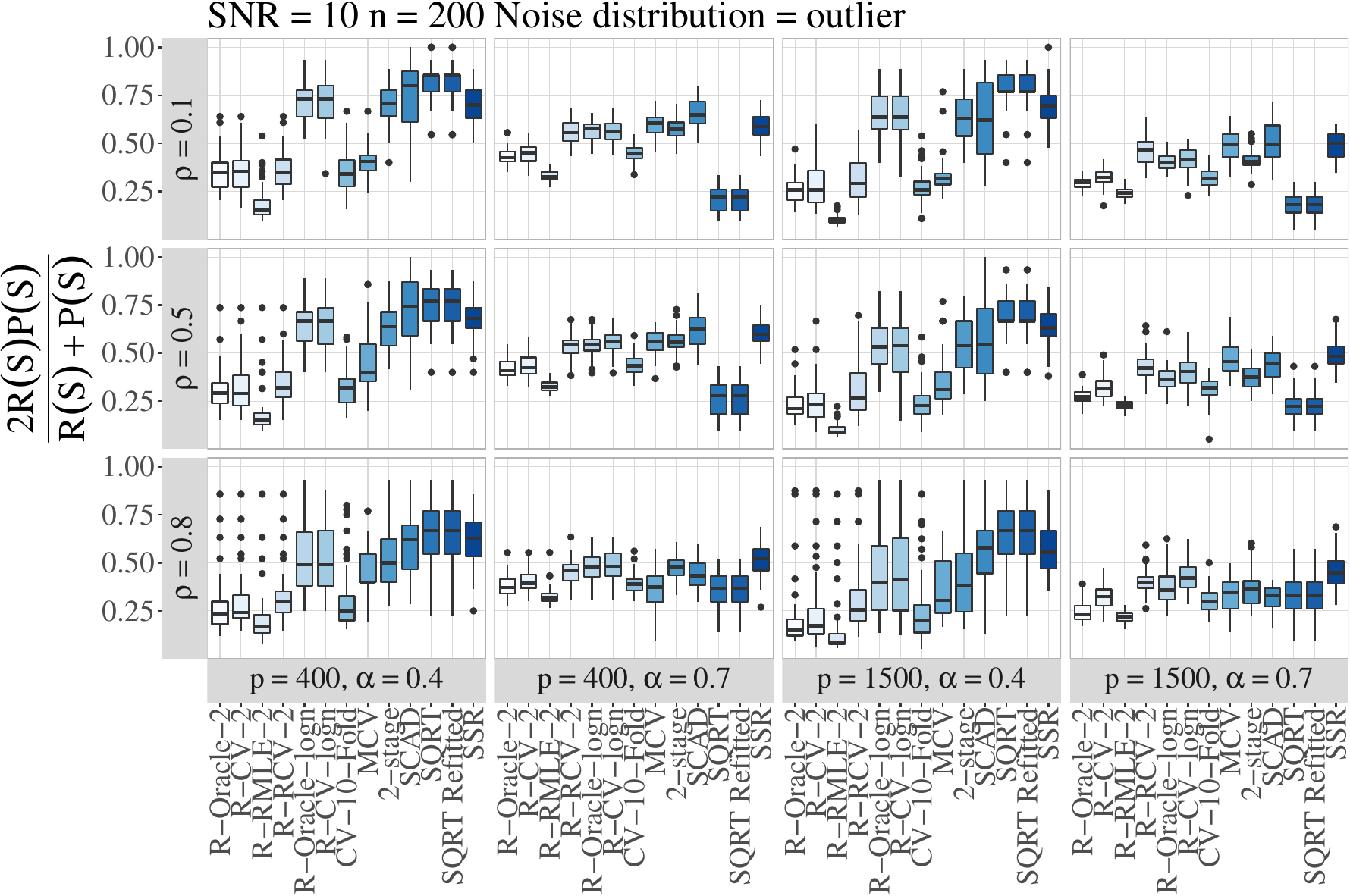}
\includegraphics[width=6in,height=4in]{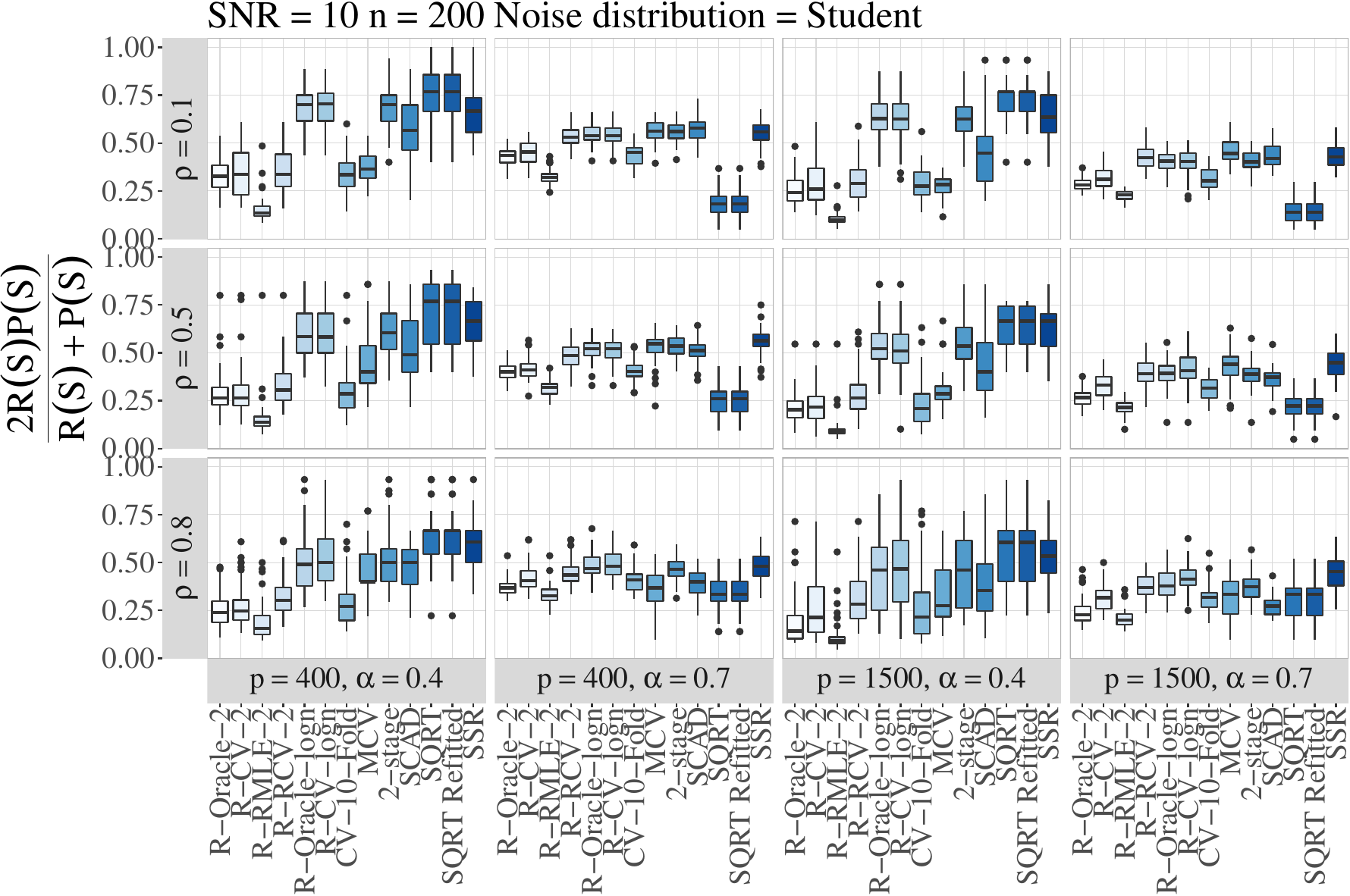}
\end{center}

\hypertarget{risk-estimation}{%
\subsection{Risk estimation}\label{risk-estimation}}

\begin{center}
\includegraphics[width=6in,height=4in]{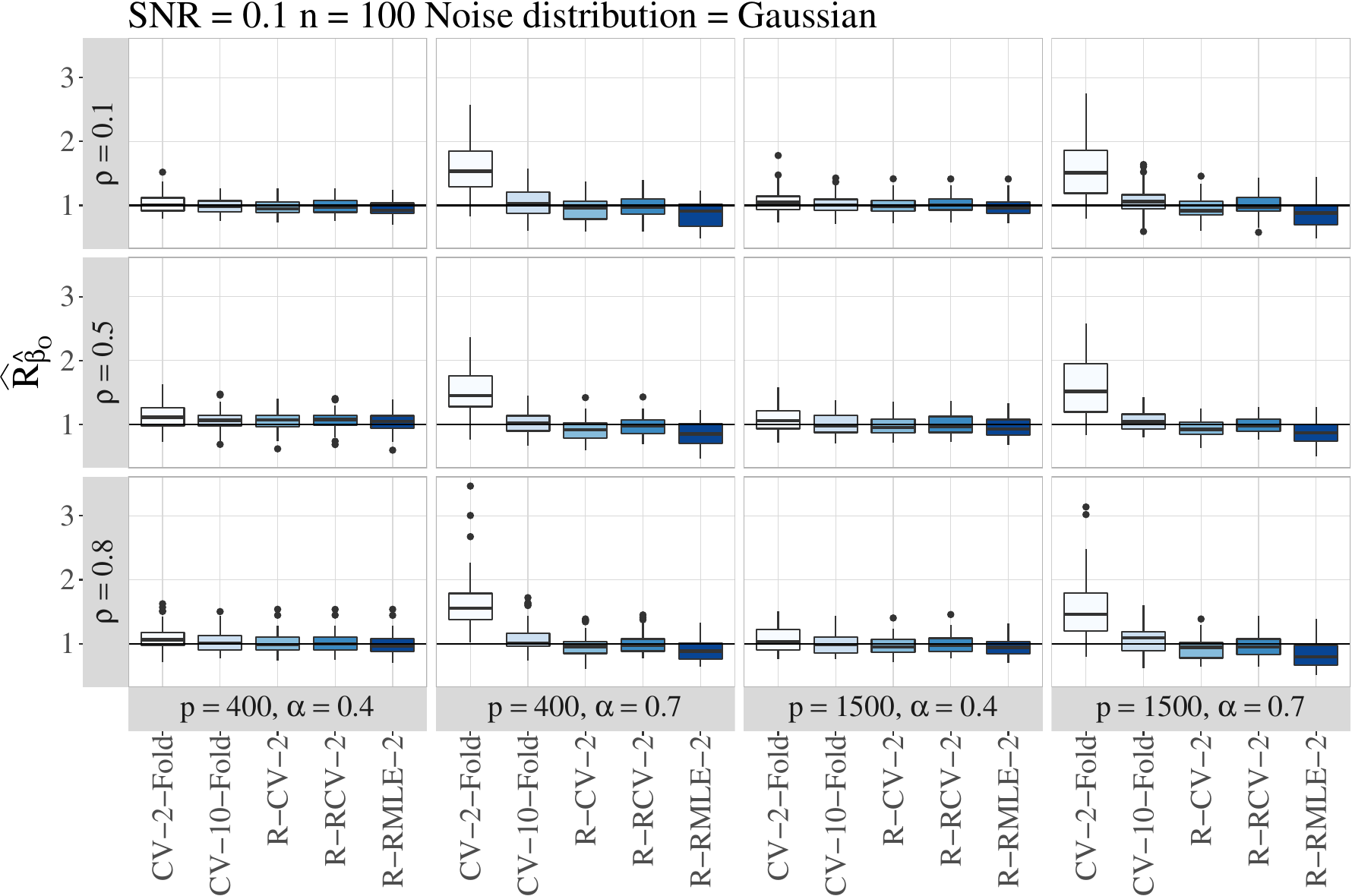}
\includegraphics[width=6in,height=4in]{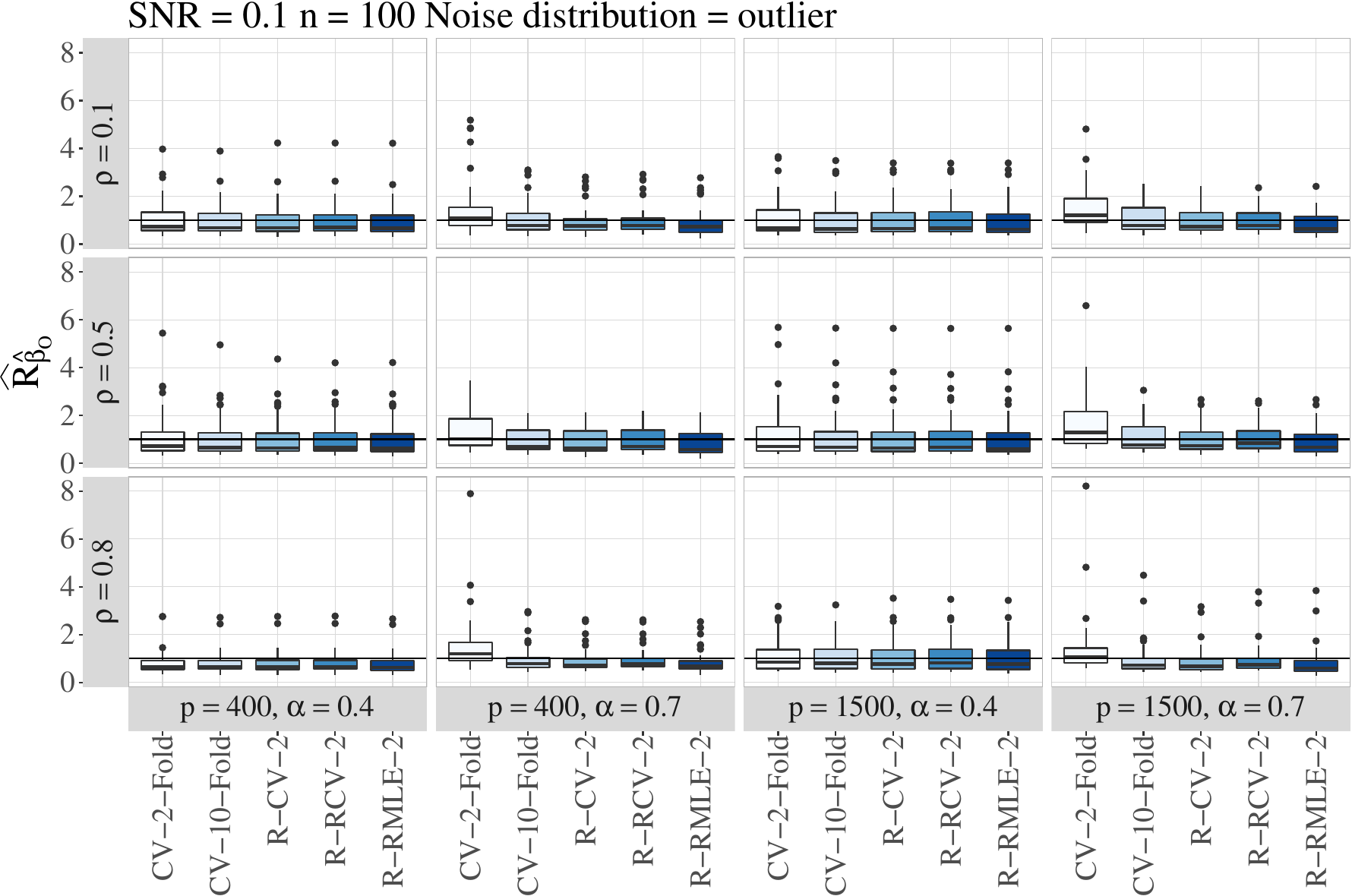}
\includegraphics[width=6in,height=4in]{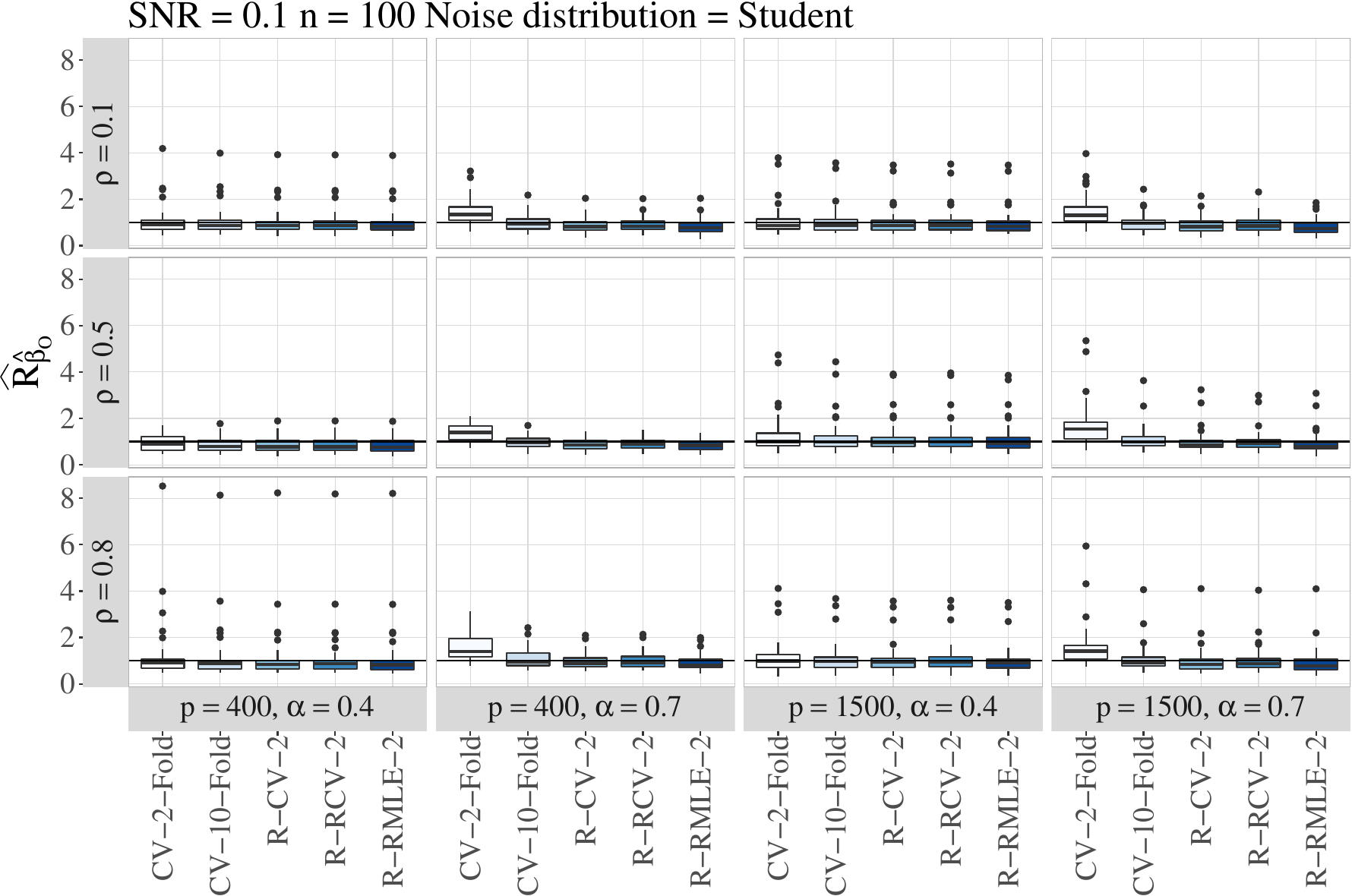}
\includegraphics[width=6in,height=4in]{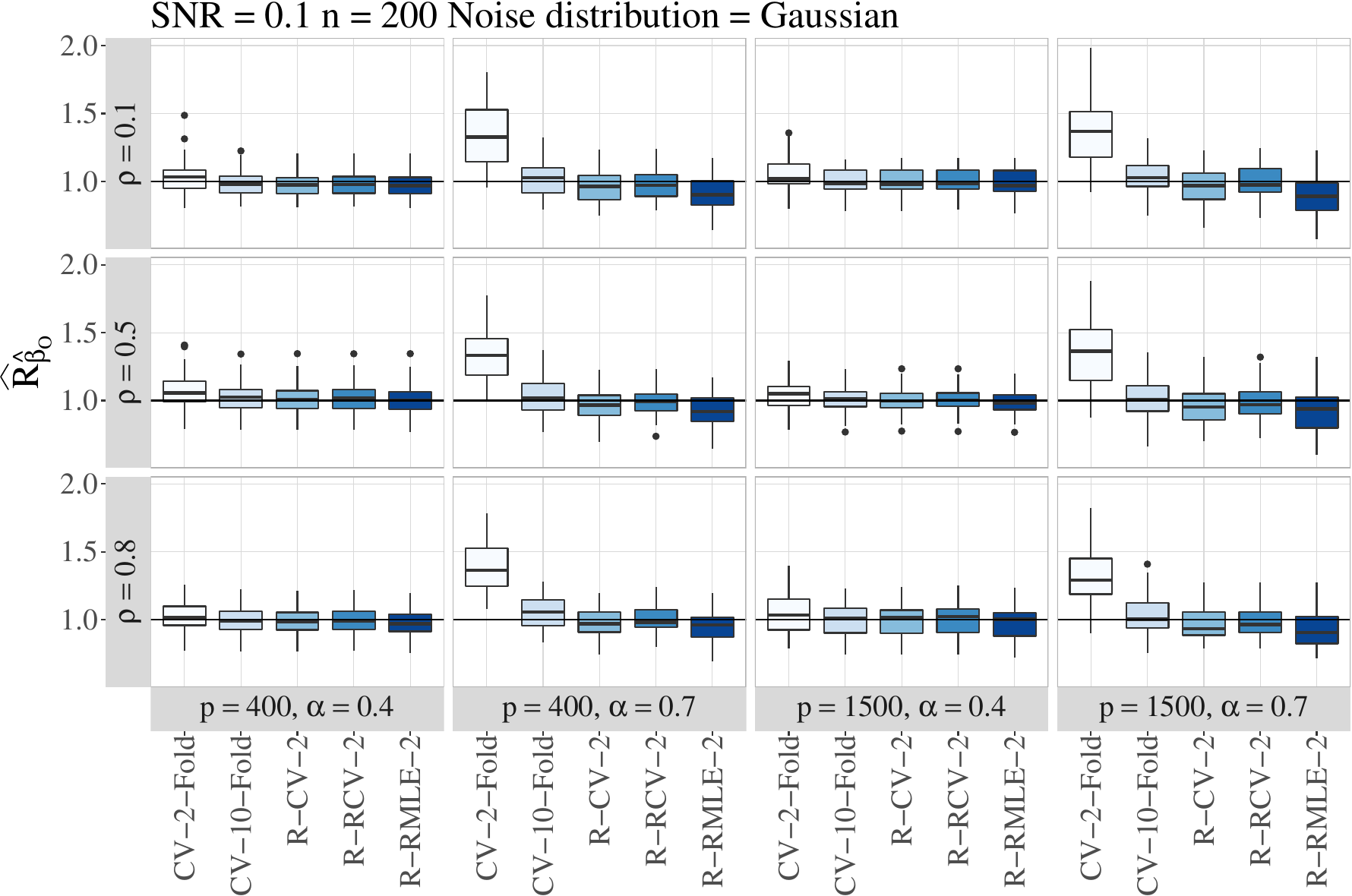}
\includegraphics[width=6in,height=4in]{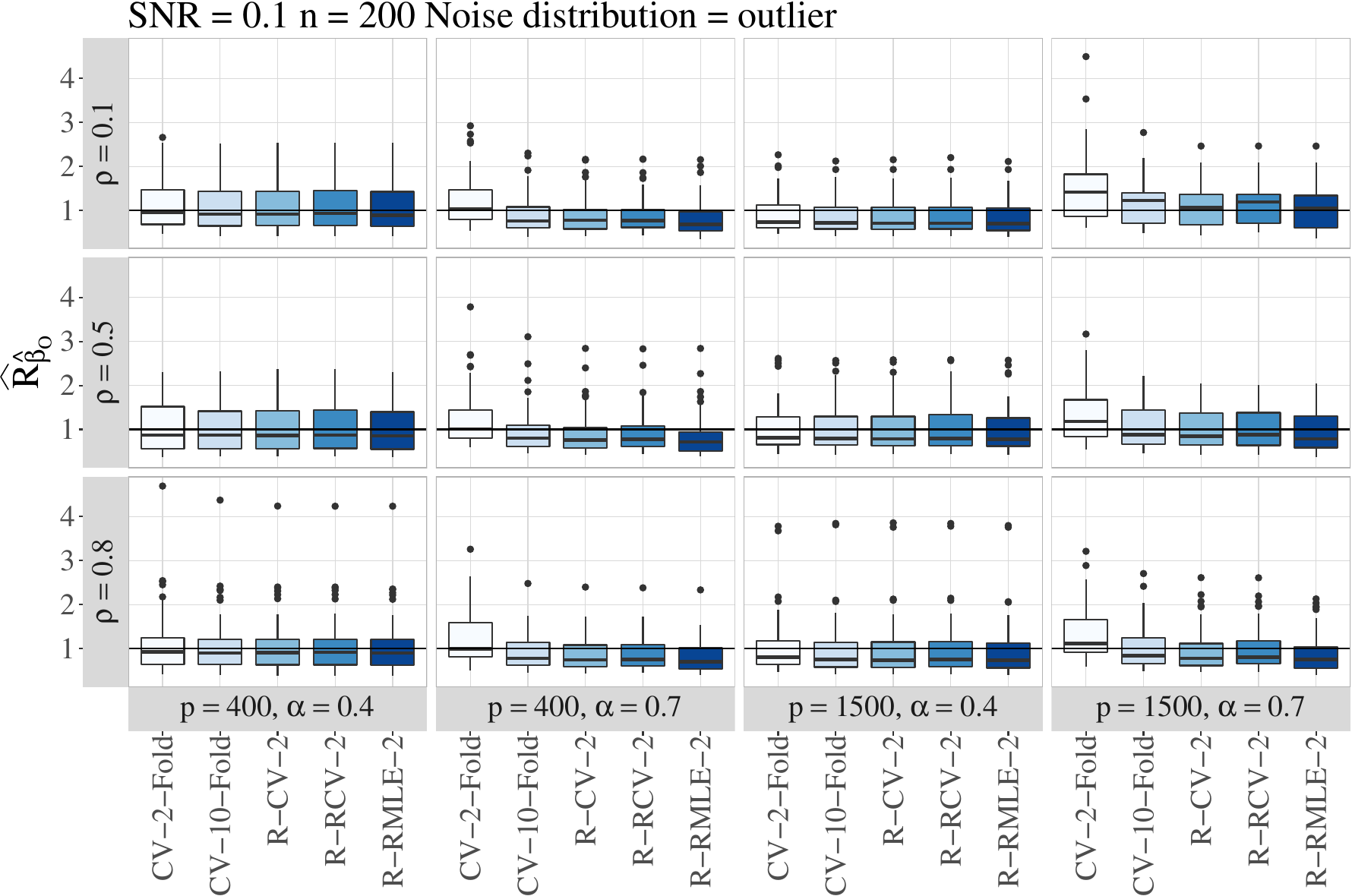}
\includegraphics[width=6in,height=4in]{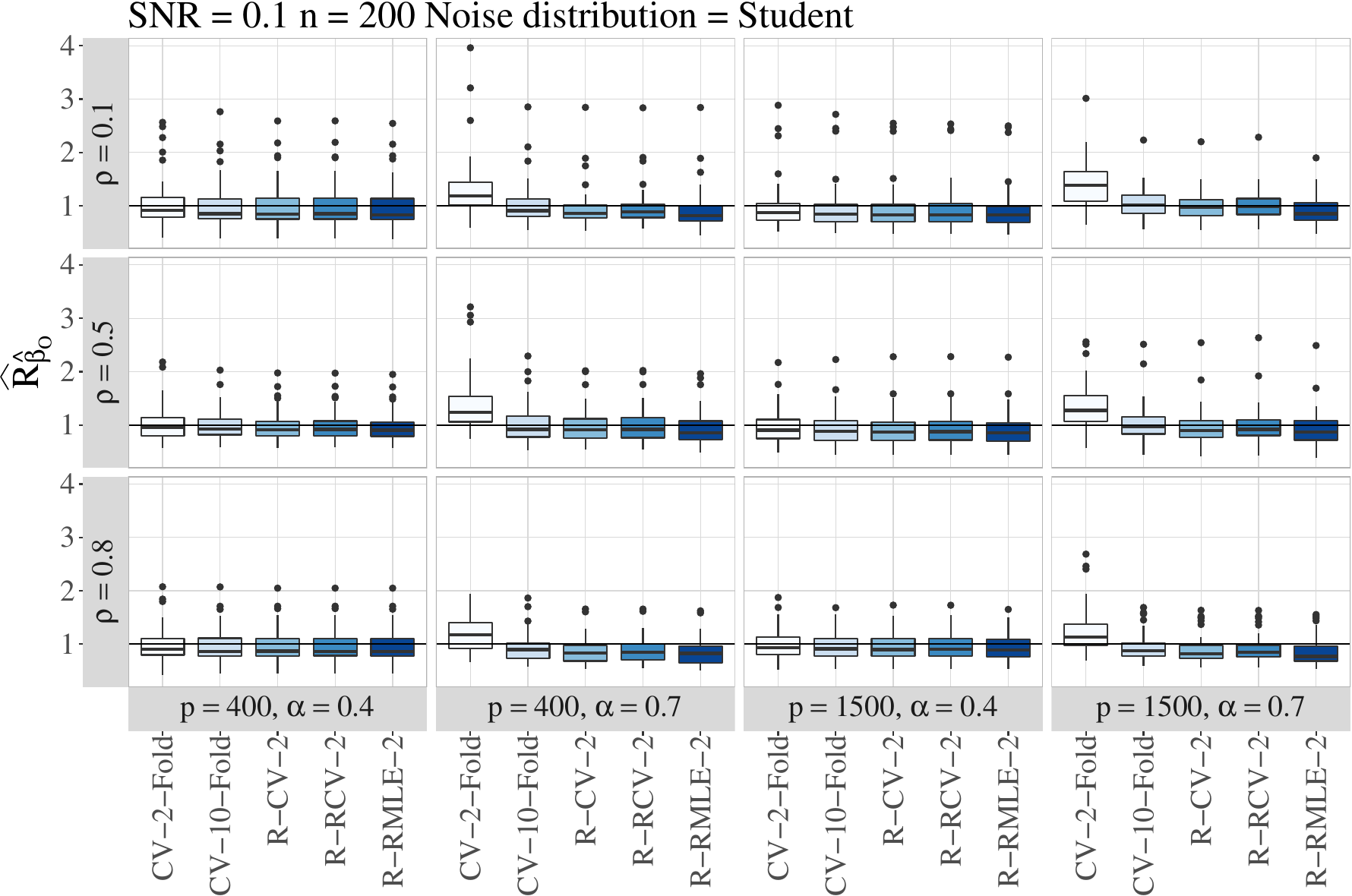}
\includegraphics[width=6in,height=4in]{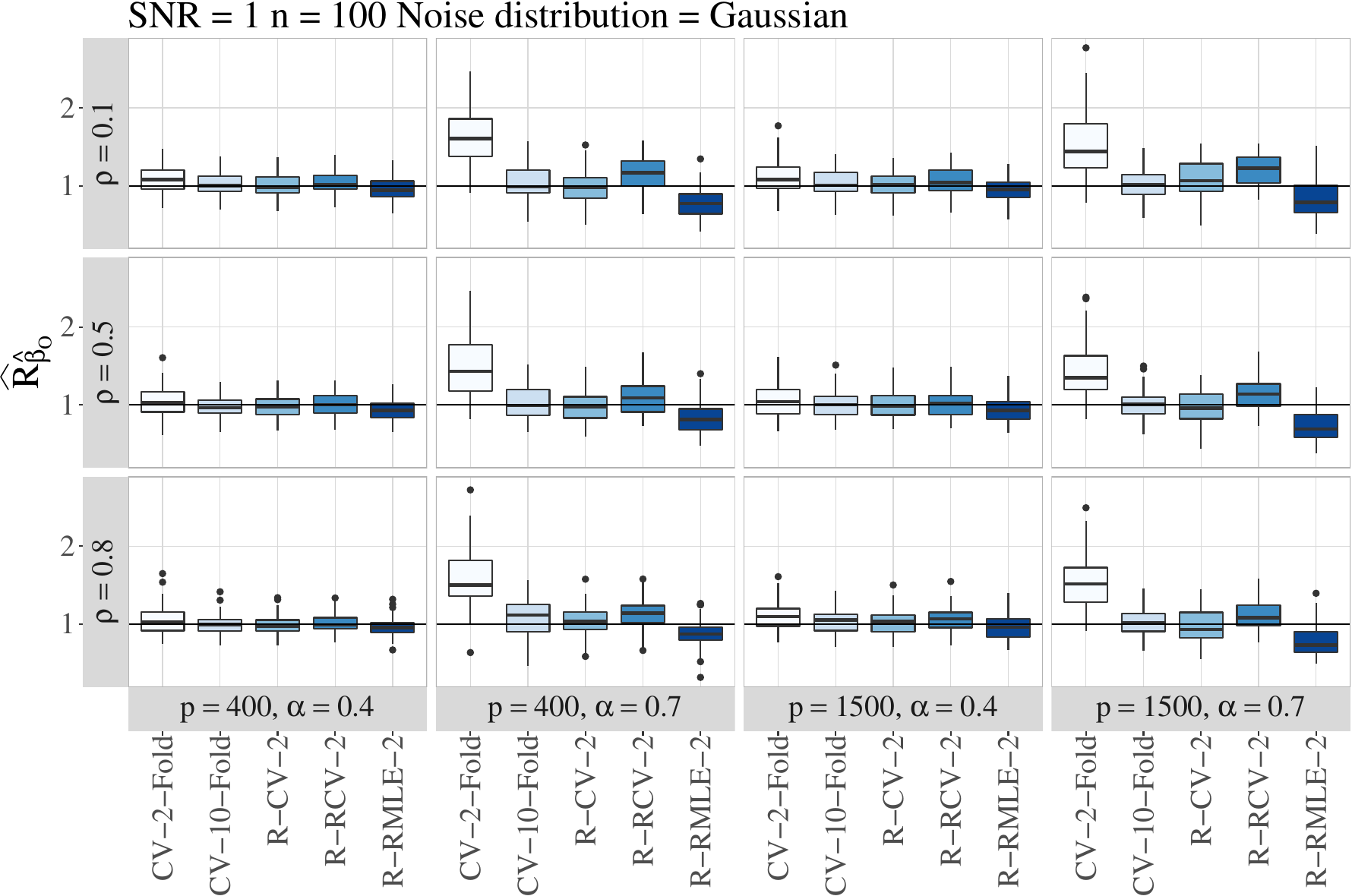}
\includegraphics[width=6in,height=4in]{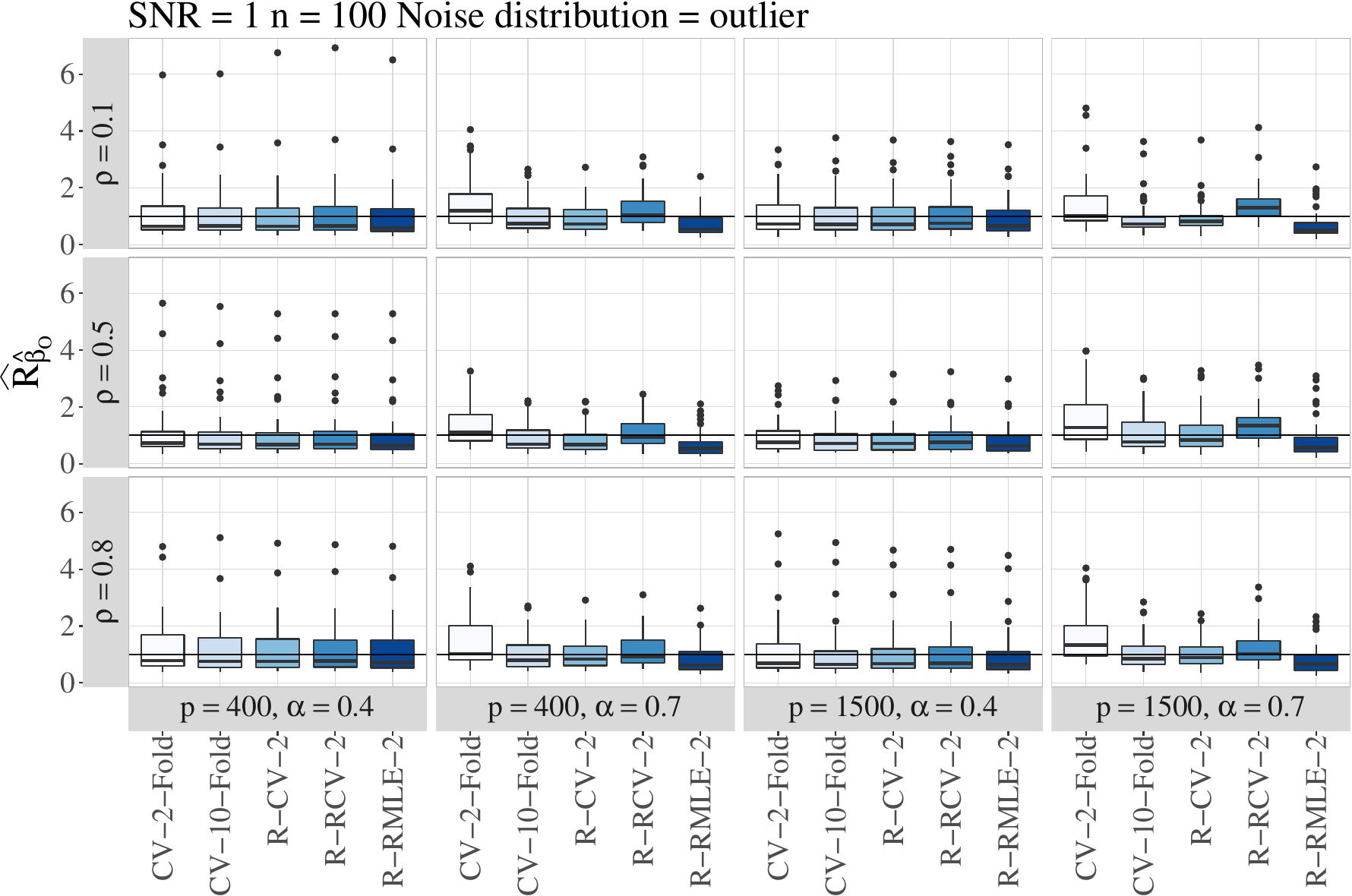}
\includegraphics[width=6in,height=4in]{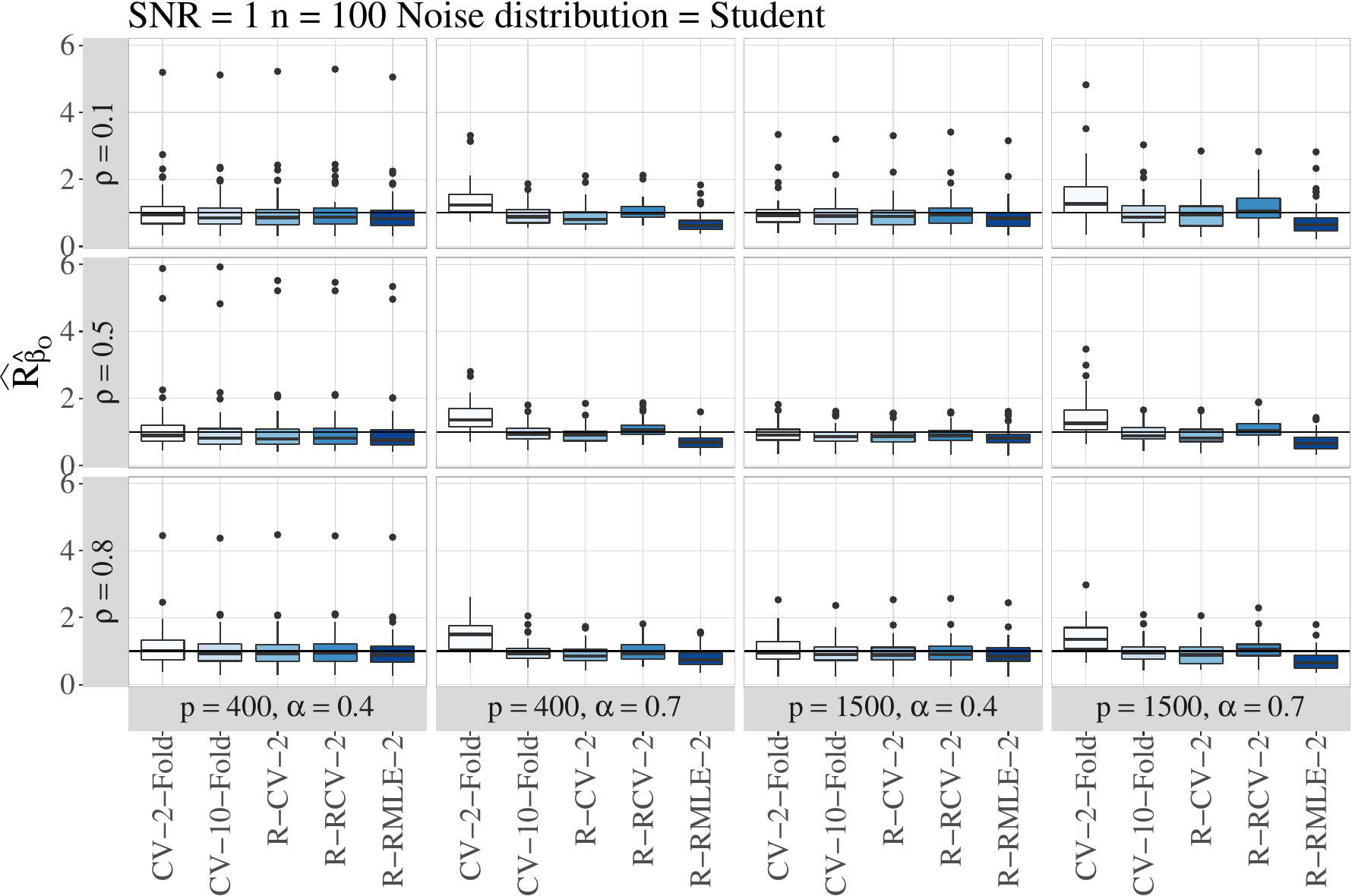}
\includegraphics[width=6in,height=4in]{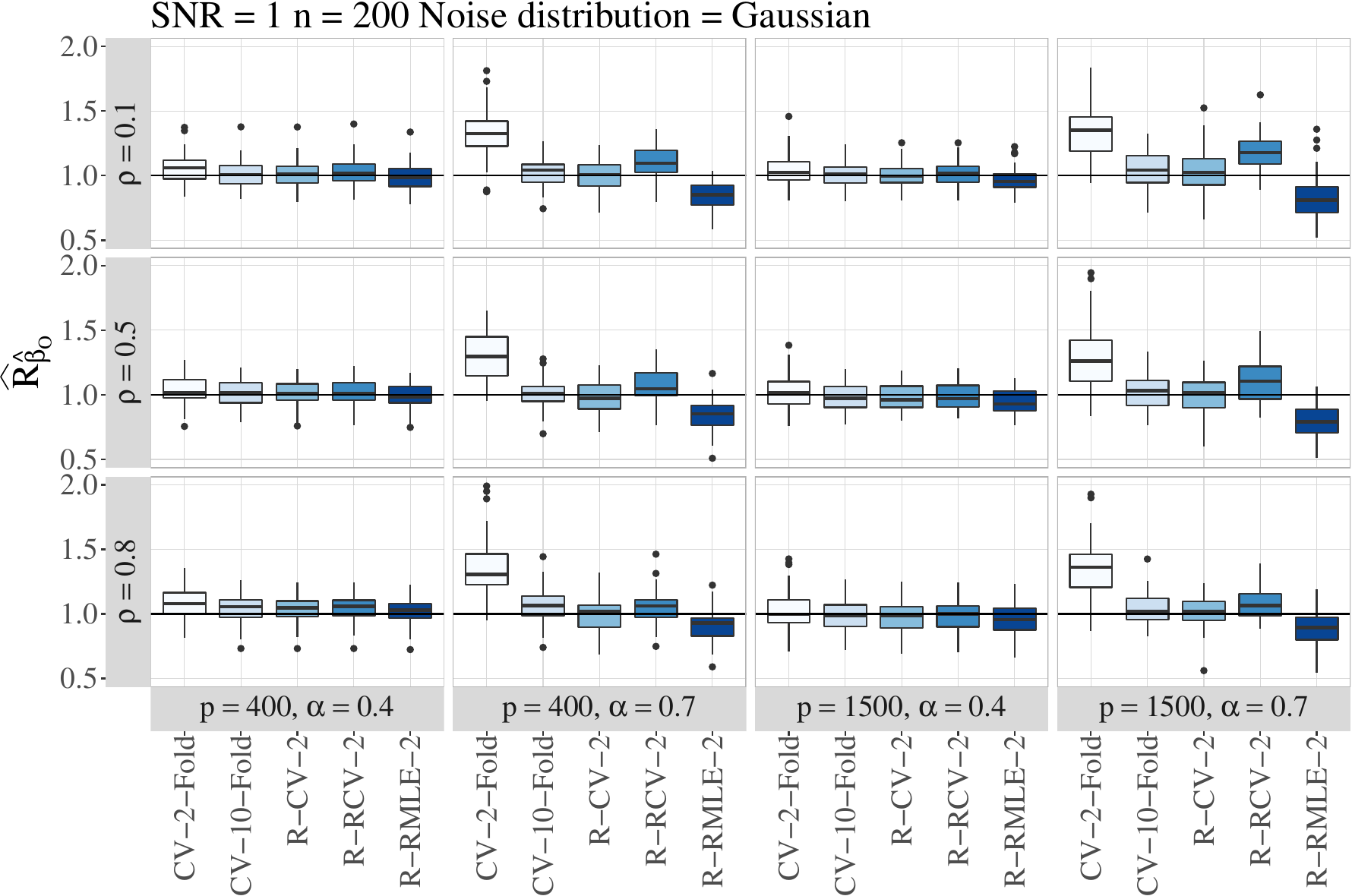}
\includegraphics[width=6in,height=4in]{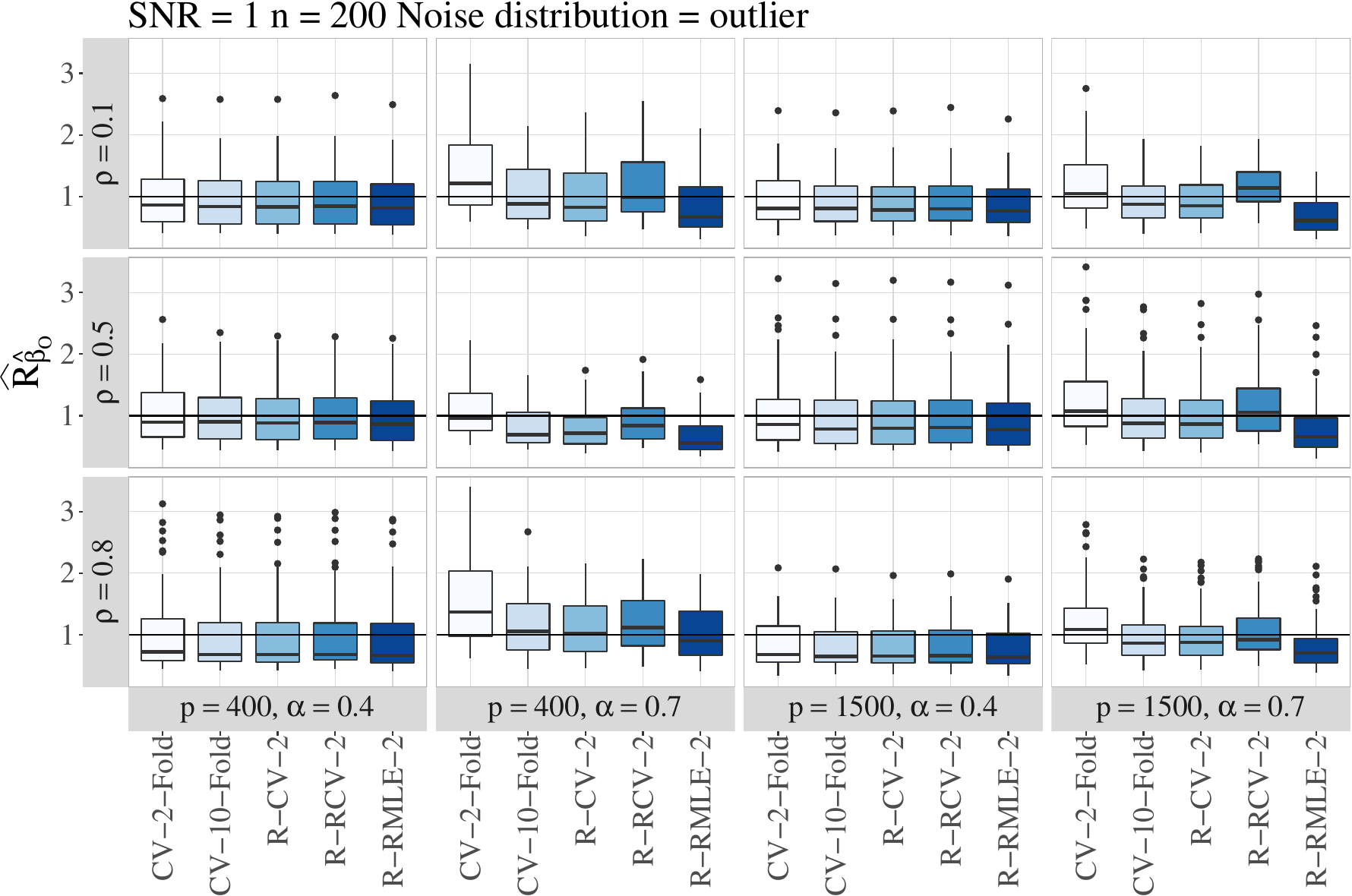}
\includegraphics[width=6in,height=4in]{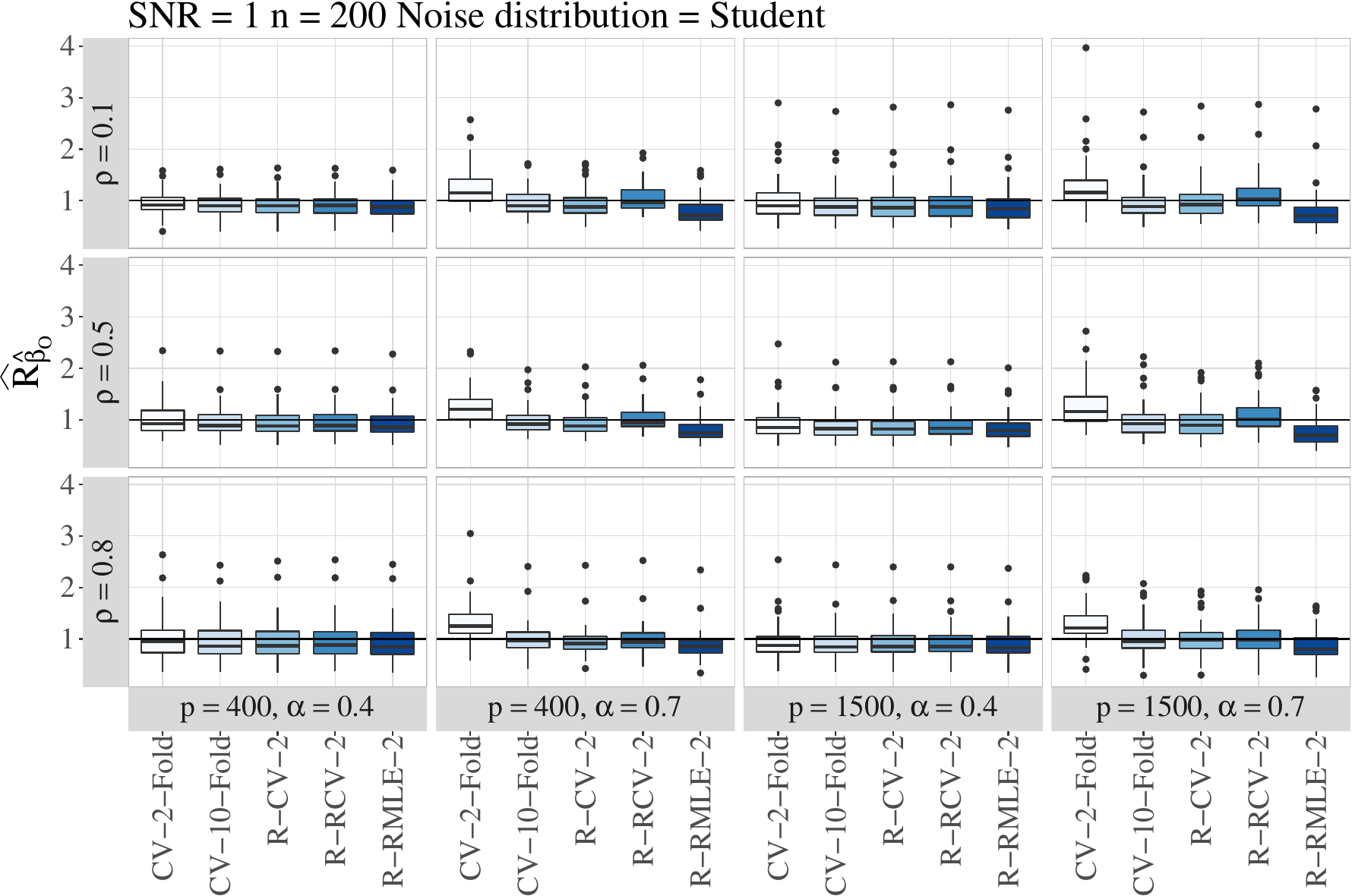}
\includegraphics[width=6in,height=4in]{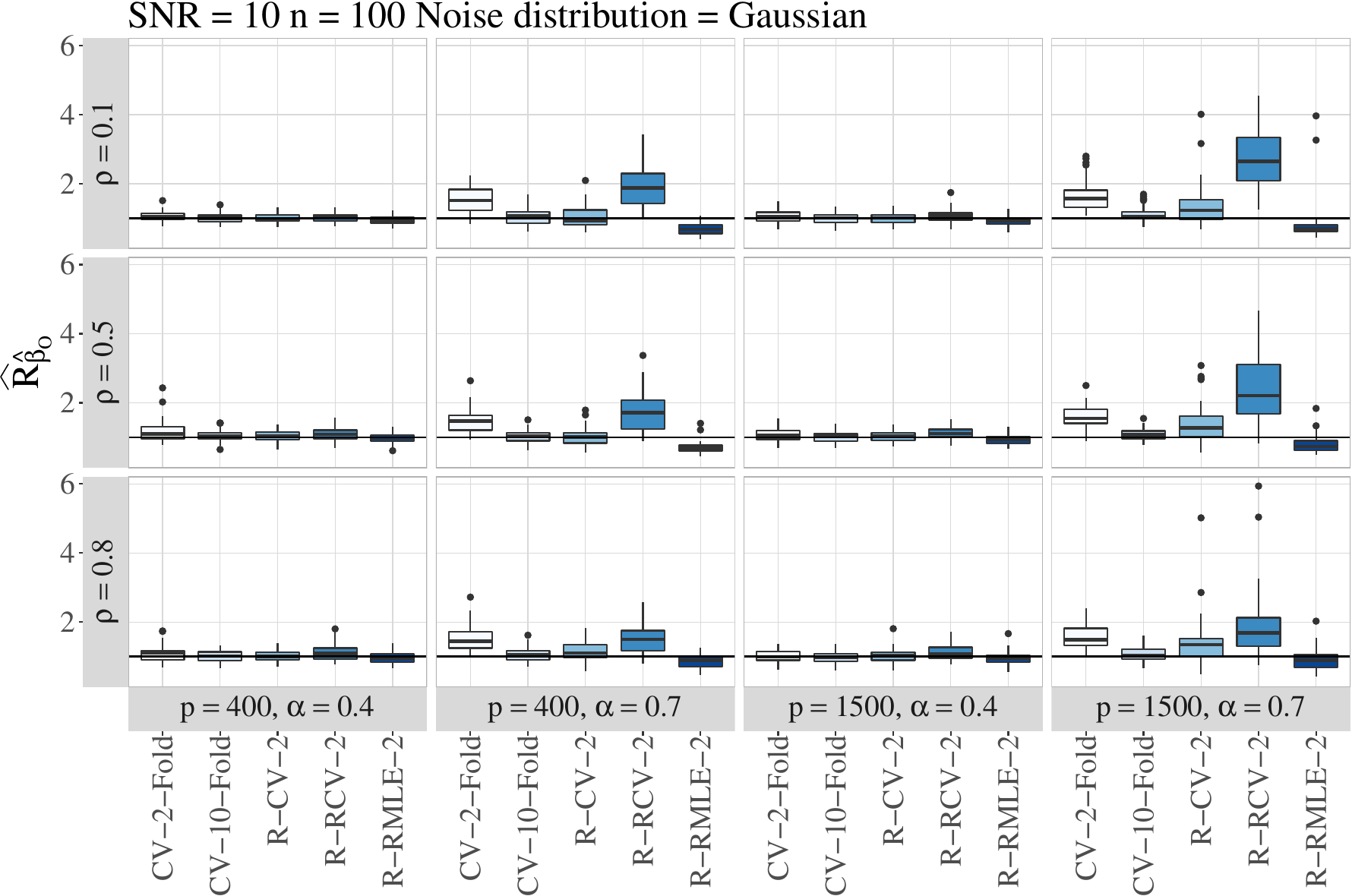}
\includegraphics[width=6in,height=4in]{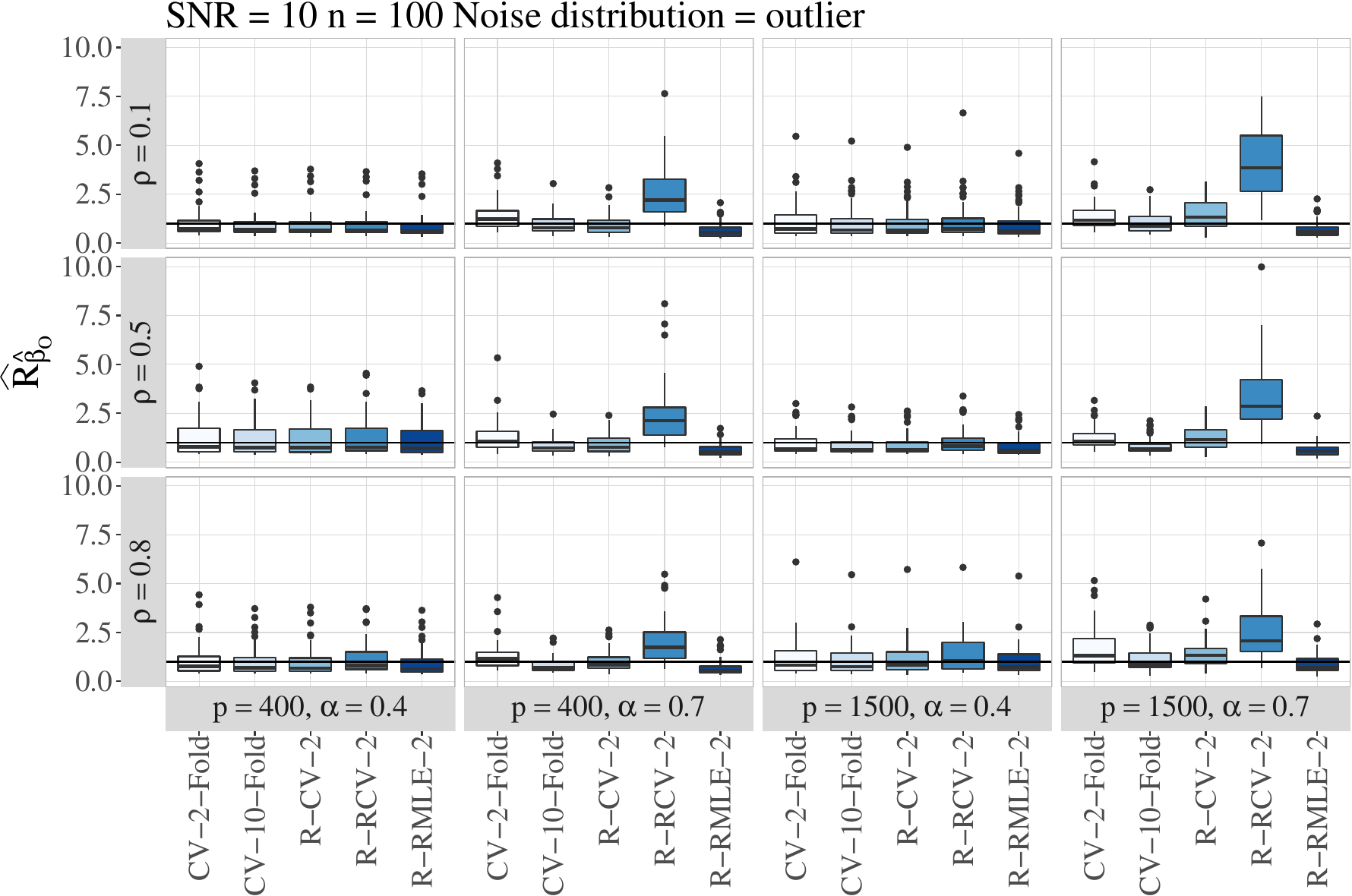}
\includegraphics[width=6in,height=4in]{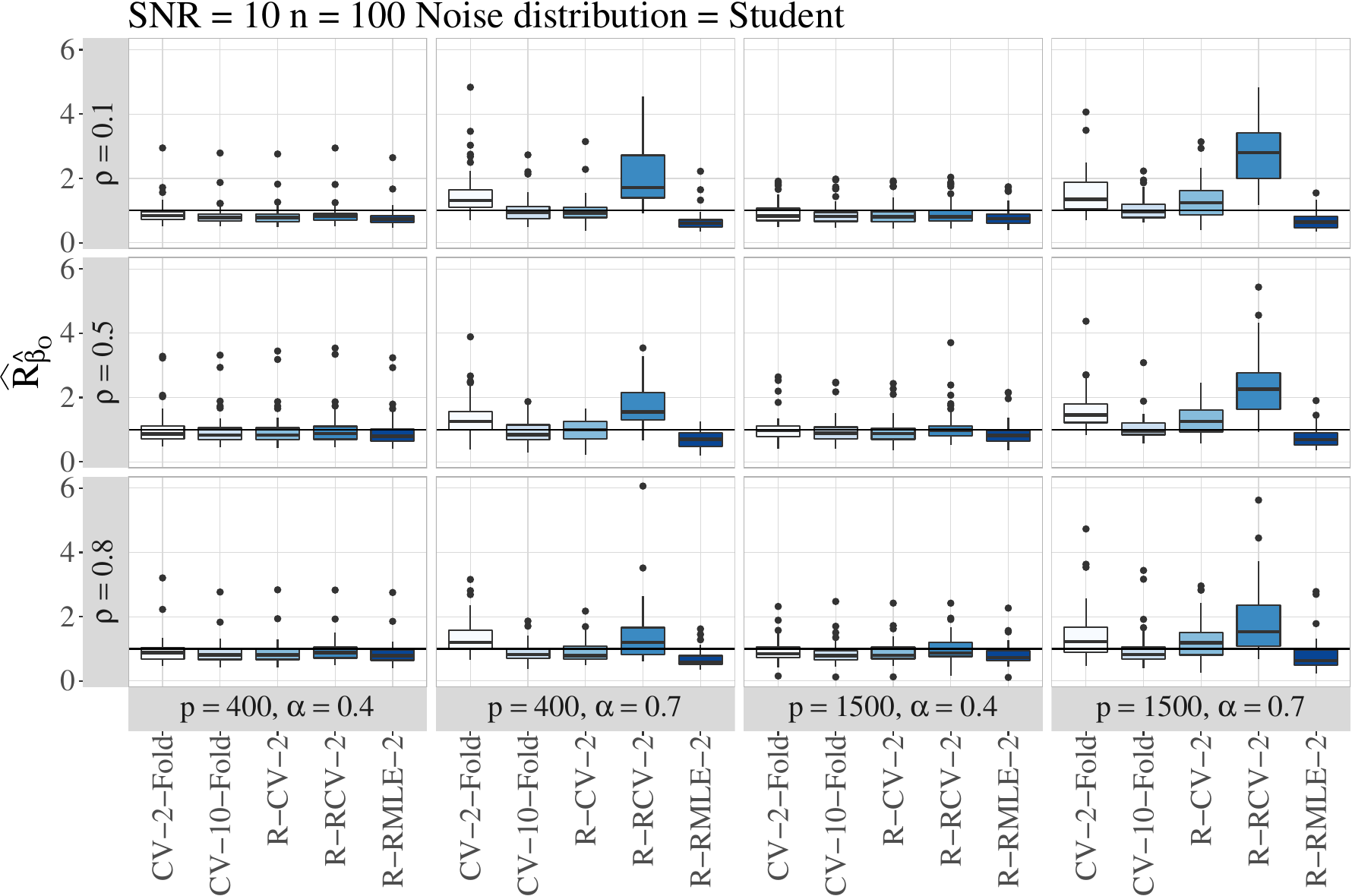}
\includegraphics[width=6in,height=4in]{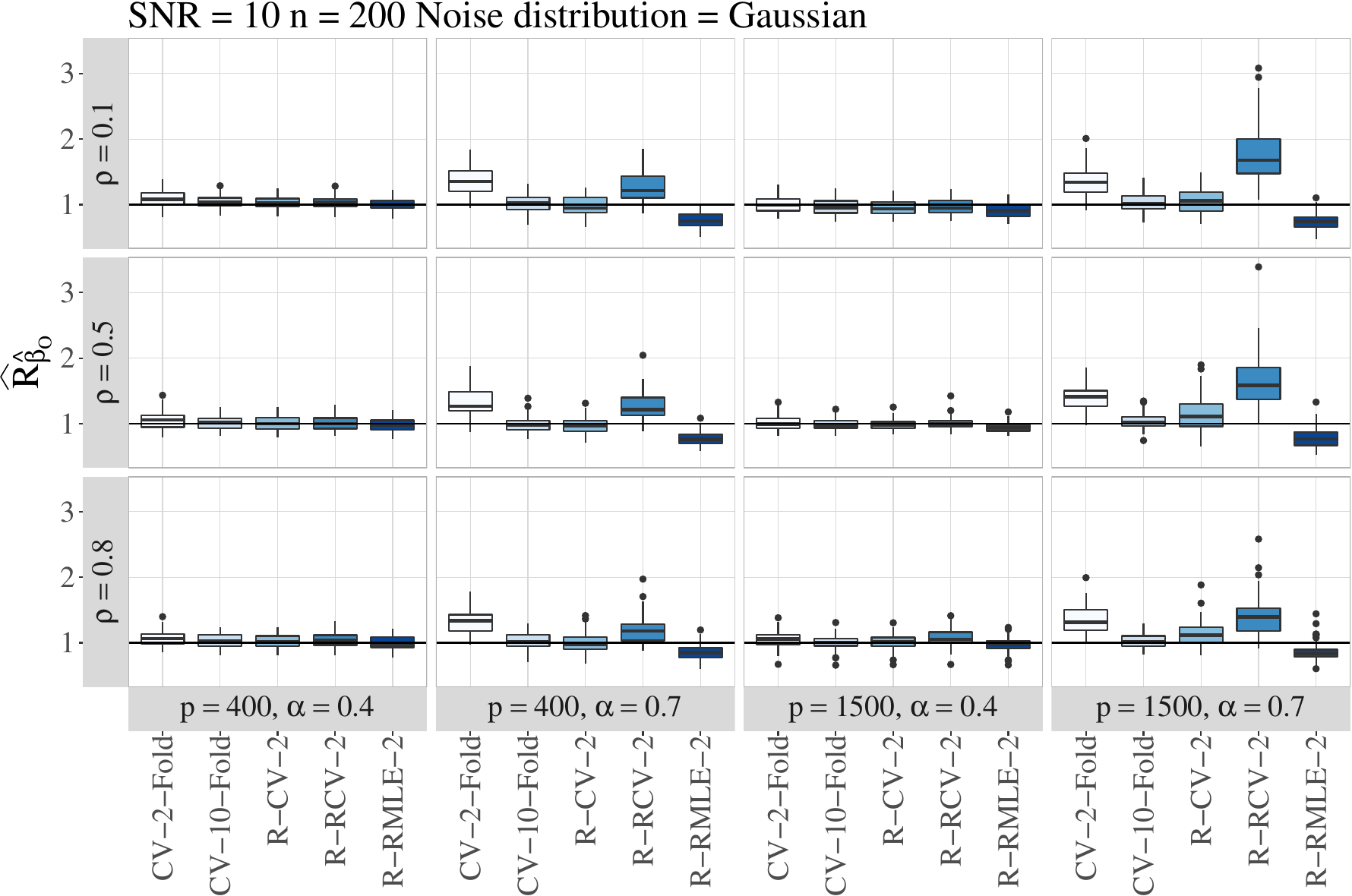}
\includegraphics[width=6in,height=4in]{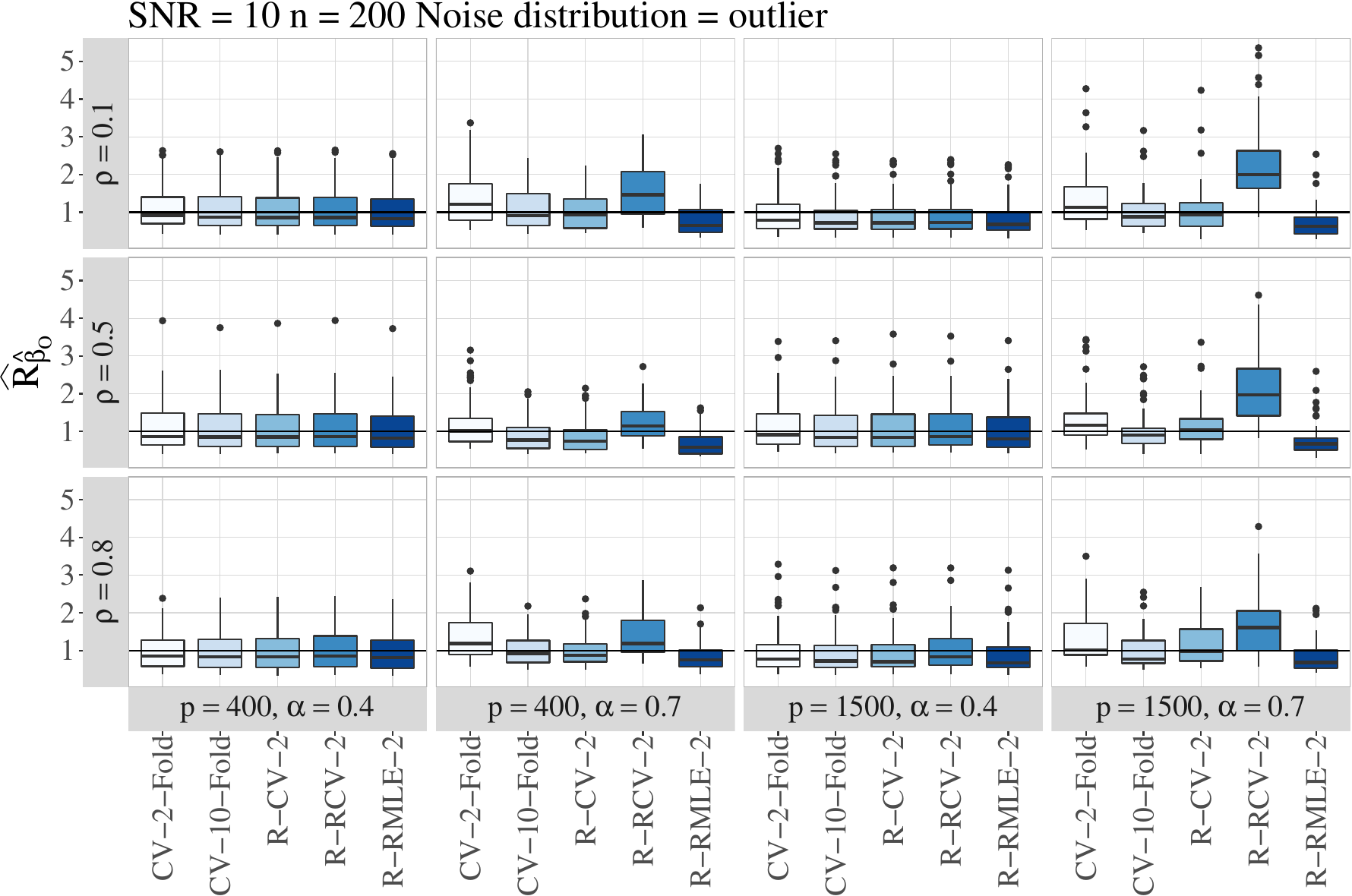}
\includegraphics[width=6in,height=4in]{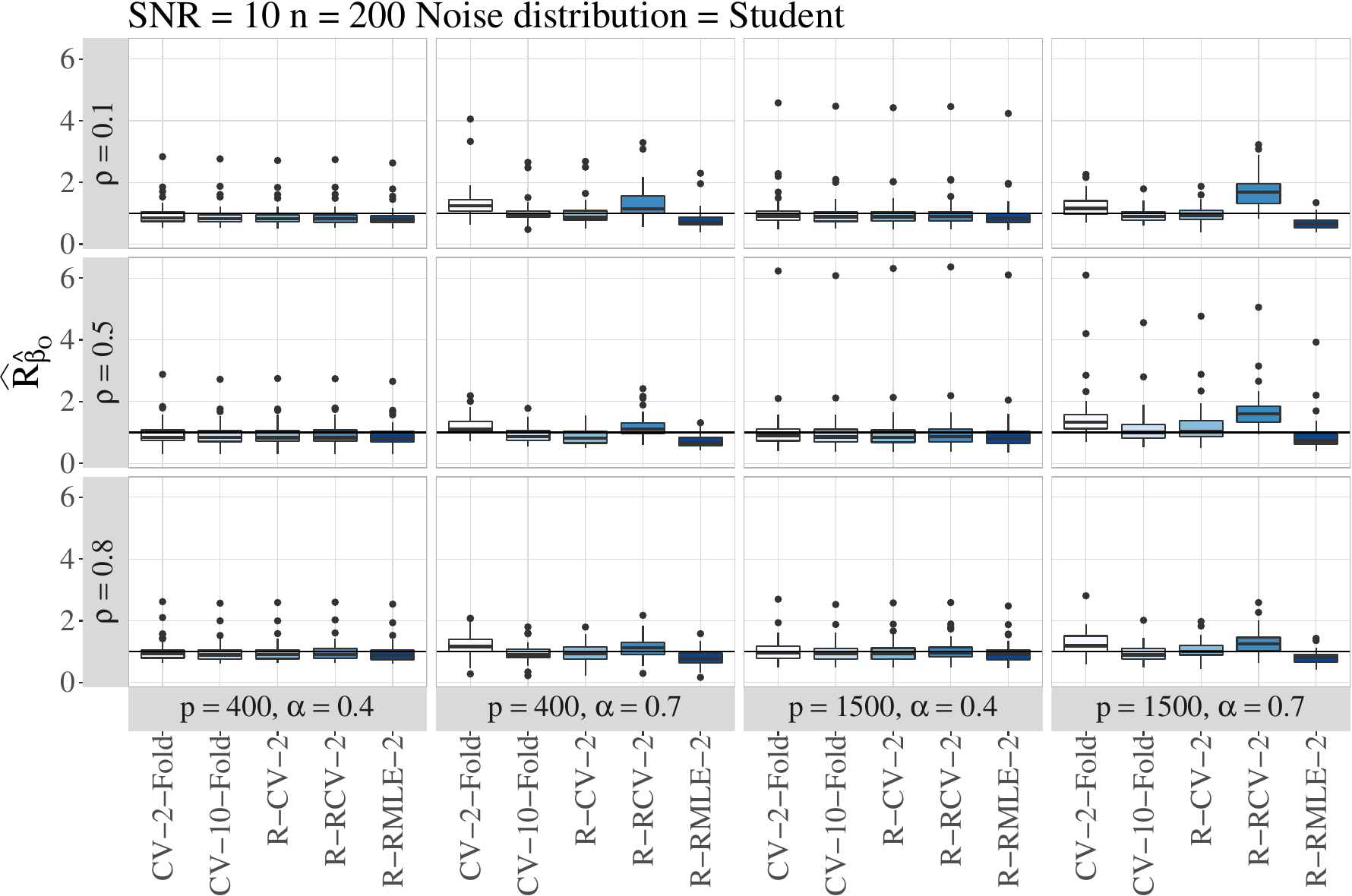}
\end{center}

\end{document}